%% file: nnpdf21-nnlo.tex
\documentclass[11pt,a4paper]{article}

\usepackage{graphicx}
\usepackage{afterpage}
\usepackage{epsfig,cite}
\usepackage{amssymb}
\usepackage{amsmath}
\usepackage{dsfont}
\usepackage{multirow}
\usepackage{url,hyperref}

\usepackage{url}
\usepackage{hyperref}

\def\smallfrac#1#2{\hbox{${{#1}\over {#2}}$}}
\newcommand{\be}{\begin{equation}}
\newcommand{\ee}{\end{equation}}
\newcommand{\bea}{\begin{eqnarray}}
\newcommand{\eea}{\end{eqnarray}}
\newcommand{\bi}{\begin{itemize}}
\newcommand{\ei}{\end{itemize}}
\newcommand{\ben}{\begin{enumerate}}
\newcommand{\een}{\end{enumerate}}
\newcommand{\la}{\left\langle}
\newcommand{\ra}{\right\rangle}
\newcommand{\lc}{\left[}
\newcommand{\rc}{\right]}
\newcommand{\lp}{\left(}
\newcommand{\rp}{\right)}

\def\frac#1#2{{{#1}\over {#2}}}
\def\gsim{\mathrel{\rlap{\lower4pt\hbox{\hskip1pt$\sim$}}
    \raise1pt\hbox{$>$}}}         
\def\lsim{\mathrel{\rlap{\lower4pt\hbox{\hskip1pt$\sim$}}
    \raise1pt\hbox{$<$}}}         

\newcommand{\dat}{\mathrm{dat}}

\newcommand{\net}{\mathrm{net}}

\newcommand{\tot}{\mathrm{tot}}

\newcommand{\draft}[1]{}

\def\beq{\begin{equation}}  
\def\eeq{\end{equation}}  

\def \n0{N_j^{(0)}}

\def\lapprox{\lower .7ex\hbox{$\;\stackrel{\textstyle <}{\sim}\;$}}
\def\gapprox{\lower .7ex\hbox{$\;\stackrel{\textstyle >}{\sim}\;$}}

\begin{document}
\begin{flushright}
Edinburgh 2011/14\\
IFUM-979-FT\\
FR-PHENO-2011-010\\
RWTH TTK-11-24\\
\end{flushright}
\begin{center}
{\Large \bf Unbiased global determination of parton distributions}
{\Large \bf and their uncertainties at NNLO and at LO}
\vspace{0.8cm}

{\bf  The NNPDF Collaboration:}\\
Richard~D.~Ball$^{1,5}$, Valerio~Bertone$^2$, Francesco~Cerutti$^3$,
 Luigi~Del~Debbio$^1$,\\ Stefano~Forte$^4$, Alberto~Guffanti$^{2,5}$, 
Jos\'e~I.~Latorre$^3$, Juan~Rojo$^4$ and Maria~Ubiali$^6$.

\vspace{1.cm}
{\it ~$^1$ Tait Institute, University of Edinburgh,\\
JCMB, KB, Mayfield Rd, Edinburgh EH9 3JZ, Scotland\\
~$^2$  Physikalisches Institut, Albert-Ludwigs-Universit\"at Freiburg,\\ 
Hermann-Herder-Stra\ss e 3, D-79104 Freiburg i. B., Germany  \\
~$^3$ Departament d'Estructura i Constituents de la Mat\`eria, 
Universitat de Barcelona,\\ Diagonal 647, E-08028 Barcelona, Spain\\
~$^4$ Dipartimento di Fisica, Universit\`a di Milano and
INFN, Sezione di Milano,\\ Via Celoria 16, I-20133 Milano, Italy\\
~$^5$ The Niels Bohr International Academy and Discovery Center,\\
The Niels Bohr Institute, Blegdamsvej 17, DK-2100 Copenhagen, Denmark\\
~$^6$ Institut f\"ur Theoretische Teilchenphysik und Kosmologie, RWTH Aachen University,\\ 
D-52056 Aachen, Germany\\}
\end{center}

\vspace{0.8cm}

\begin{center}
{\bf \large Abstract:}
\end{center}

We present a determination of the parton distributions of
the nucleon from a global set of 
hard scattering data using the NNPDF 
methodology at LO and NNLO in perturbative QCD, 
thereby generalizing to these orders the
NNPDF2.1 NLO parton set.  Heavy quark masses are included using the
so-called FONLL method, which is benchmarked here at NNLO. 
We demonstrate the stability of PDFs upon inclusion of NNLO
corrections, and we investigate the convergence of the perturbative
expansion by comparing LO, NLO and NNLO results. We
show that the momentum sum rule can be tested
with increasing accuracy at LO, NLO and NNLO.
We discuss the impact of NNLO corrections on collider phenomenology,
specifically  by comparing to recent LHC data. We present PDF
determinations using a range of values of $\alpha_s$, $m_c$ and $m_b$. 
We also present PDF determinations based on various subsets of the global dataset, show that they
generally lead to less accurate phenomenology, and discuss the
possibility of future PDF determinations based on collider data only.

\clearpage

\tableofcontents

\clearpage

\include{sec-intro}

\include{sec-expdata}

\include{sec-pdfevol-nc}

\include{sec-implementation}

\include{sec-lopdfs}

\include{sec-results}

\include{sec-msr}

\include{sec-pheno}

\include{sec-modfits}

\include{sec-conclusions}

\appendix

\include{sec-massive-nc-mellin}

\include{sec-massive-nc}

\include{sec-pdfevol-nnlo}

\input{nnpdf21-nnlo.bbl}
\end{document}

%% file: sec-intro.tex
\section{Introduction}

\label{sec-intro}

In a series of previous
papers~\cite{Forte:2002fg,DelDebbio:2004qj,DelDebbio:2007ee,Ball:2008by,Ball:2009mk,Ball:2010de},
we have presented a novel methodology for the determination of parton distributions
which strives to minimize parametrization bias and ad hoc statistical assumptions by
using a Monte Carlo approach with neural networks as unbiased underlying
interpolating functions. The statistical consistency of this approach was confirmed in
Ref.~\cite{Ball:2010gb} by showing explicitly that it yields a probability distribution of PDFs
which, upon the inclusion of new data, behaves in accordance with Bayes' theorem. In
Ref.~\cite{Ball:2011mu} this NNPDF methodology was used to determine a set of
parton distributions based on a global dataset, using NLO QCD, with inclusion of
heavy quark mass effects. This PDF set, called NNPDF2.1, is arguably the most
accurate NLO PDF set currently available, from every point of view: dataset, theoretical
treatment, and statistical methodology. It has been made available for a variety of values of the
strong coupling and of the heavy quark masses.

In this paper, we provide companion PDF sets based on the same methodology and data,
but now using LO or NNLO theory: NNPDF2.1 LO and NNPDF2.1 NNLO. Both are needed for
collider phenomenology: the LO parton distributions are principally for use
with LO Monte Carlos, while the NNLO sets are needed for evaluation of LHC standard candle processes, 
some of which (such as Higgs production) are characterized by large NNLO QCD corrections
and are either measured or measurable to an accuracy which may be comparable to the
size of NNLO effects.

The full theoretical framework that is necessary in order to construct NNLO (and of
course LO) PDFs is already available. There are however several implementation
issues which must be dealt with.  At LO, parton distributions can be
interpreted as probability distributions, and they are therefore non-negative: to
ensure this, it will be advantageous to introduce a modification of the
neural network parametrization of Ref.~\cite{Ball:2010de} such that positivity is
hard-wired. Also, it has been suggested~\cite{Sherstnev:2007nd,Lai:2009ne} that it may
be useful to 
relax the momentum sum rule at leading order, and use a next-to-leading order form of the 
strong coupling in the determination of LO PDFs. All these issues will be investigated.

As we move now to NNLO, we have to address the issue of implementing higher
order corrections in a numerically efficient way.  In Ref.~\cite{Ball:2010de} we
developed a method, dubbed FastKernel, for the inclusion of NLO
corrections to parton evolution and to the computation of deep-inelastic (DIS) and
Drell-Yan (DY) observables, without the use of $K$-factors. Here, we will use the same method for the computation of evolution and DIS to NNLO. Heavy quark mass effects will be included
using the so-called FONLL method, first developed for hadronic processes in
Ref.~\cite{Cacciari:1998it} and extended to DIS in Ref.~\cite{Forte:2010ta}: the
implementation of the FONLL method up to NNLO (called FONLL-C in
Ref.~\cite{Forte:2010ta}) requires the computation of some hitherto unknown Mellin
transforms, and its implementation in the FastKernel framework must be
benchmarked. For Drell-Yan we will rely on the NLO FastKernel implementation of
Ref.~\cite{Forte:2010ta}, with NNLO corrections to it included by means of
$K$-factors (note that in other global PDF fits such as
Refs.~\cite{Martin:2009iq,Lai:2010vv} both NLO and NNLO corrections to Drell-Yan are
included using $K$-factors). This computation of the Drell-Yan process to NNLO will
also be benchmarked. For the inclusive jet cross-section we will employ the threshold approximation 
to the NNLO corrections, since exact results are as yet unknown: these will be implemented using 
the FastNLO code\cite{Kluge:2006xs}.

While we will refer to our previous papers for a general introduction to the NNPDF
methodology and for a detailed description of the NNPDF2.1 NLO PDF set, here we will
document all the new issues that arise in the determination of LO and NNLO PDFs,
specifically those mentioned above. With LO, NLO and NNLO results at our disposal,
we will be able to investigate the perturbative stability of PDFs.  We will thus be able to show that for 
PDFs in the kinematic range currently accessible the
convergence of the perturbative expansion is very good: in particular NNLO PDFs are 
quite close to NLO ones.  In particular, we
will perform a study of the momentum sum rule at LO, NLO and NNLO, based on PDF
determinations in which the sum rule is not imposed as a constraint
and check that indeed the sum rule follows from the experimental data.
We will then perform some phenomenological NNLO
studies, in particular for LHC standard candles. Finally, we will
discuss, in the context of the NNLO determination --- which is
theoretically the most accurate --- the dependence of
results on the value of the strong coupling and the
size of the dataset, which are the main potential sources of
uncertainty. 

The outline of this paper is the following.  In Sect.~\ref{sec:expdata} we present
the experimental data used in the analysis: these only differ from those used in the
NNPDF2.1 NLO determination of Ref.~\cite{Ball:2011mu} in that the inclusion of NNLO
heavy quark corrections allows for looser kinematic cuts on charm structure function
data. In Sect.~\ref{sec:evolution} we summarize our computation of all NNLO physical
observables that enter the PDF fit, and specifically discuss the NNLO heavy quark
mass implementation, and the implementation of NNLO corrections to the Drell-Yan
process. Mellin transforms of the NNLO heavy quark coefficient functions are given in 
Appendix~\ref{sec:massive-nc-comp}. 
In Sect.~\ref{sec:impl} we discuss modifications to the PDF
parametrization and minimization which have been performed at LO and NNLO, in
particular to optimize the requirement of positivity at LO, and to
obtain accurate minimizations at LO and NNLO.  The NNPDF2.1 LO and NNPDF2.1 NNLO sets
are presented in Sect.~\ref{sec:lopdfs} and Sect.~\ref{sec:results} respectively,
where they are also compared to other available PDF sets.  In
Sect.~\ref{sec:pertstab} we examine the convergence of the perturbative expansions
for individual PDF flavours, and perform a precision determination of the momentum
carried by quarks and gluons in the nucleon.  The implications of NNPDF2.1 NNLO PDFs
for LHC phenomenology are reviewed in Sect.~\ref{sec:pheno}, where, after discussing
the relevant parton luminosities, we present predictions for LHC standard candles
and compare them to the LHC data which are available at present. We finally turn in
Sect.~\ref{sec:modfits} to the issues of the dependence of
results on the  value of the strong coupling, and the size of the dataset, which we
will study  by constructing PDFs based on
various subsets of data (HERA only, DIS only, collider only, DIS+Drell-Yan).
Technical details on the implementation and benchmarking of 
DIS structure functions and NNLO PDF evolution are collected in Appendices~\ref{sec:massive-nc} and
\ref{sec:pdfevol-nnlo} respectively.

%% file: sec-expdata.tex
\section{Experimental data}
\label{sec:expdata}

The experimental data on which the LO and NNLO PDF sets are based are
the same as those used for the NNPDF2.1 NLO set of
Ref.~\cite{Ball:2011mu} and discussed there, with some minor
differences in data and kinematic cuts which we discuss here.

\begin{table}
  \scriptsize
  \centering
  \begin{tabular}{|c|c|c|c|c|c|c|c|}
    \hline
    Experiment & Set & Ref. & $N_{\rm dat}$ & $x_{\rm min}$ & $x_{\rm max}$
    & $Q^2_{\rm min}$ & $Q^2_{\rm max}$  \\ \hline
    \hline
    ZEUSF2C         & &    &       69 (62) & & & & \\ \hline
    &  ZEUSF2C99       & \cite{Breitweg:1999ad} &  21 (18) &      $5\,10^{-5}$ 
    ($1.3\,10^{-4}$) &      0.02 &        1.8 (4) &      130 \\
    \hline
    &  ZEUSF2C03       &  \cite{Chekanov:2003rb}& 31 (27) &      $3\,10^{-5}$
    ($7\,10^{-5}$)&      0.03 &        2.0 (4.0) &      500\\
    \hline
    &  ZEUSF2C08       & \cite{Chekanov:2008yd}  & 9 &      $2.2\,10^{-4}$ &      0.032 &        7.0 &      112  \\
    \hline
    &  ZEUSF2C09       &  \cite{Chekanov:2009kj} & 8 &      $8\,10^{-4}$ &      0.03 &       30 &     1000 \\
    \hline
    \hline
    H1F2C           &  &     &      47 (45)  & & & & \\ 
    \hline
    &  H1F2C01         & \cite{Adloff:2001zj} & 12 (10) &     $5\,10^{-5}$  ($1.3\,10^{-4}$) &      $3.2\,10^{-3}$ &        1.5 (3.5) &       60   \\
    \hline
    &  H1F2C09         & \cite{Collaboration:2009jy}  & 6 &      $2.4\,10^{-4}$ &      0.025 &      120 &      400  \\
    \hline
    &  H1F2C10         & \cite{H1F2c10:2009ut}   &29 &      $2\,10^{-4}$ &      0.05 &        5.0 &     2000  \\
    \hline
    \hline\hline
    LO Total &\multicolumn{1}{c}{}    &     &3330     &\multicolumn{3}{c}{}& \\
    \cline{1-1}\cline{4-4}
    NLO Total &\multicolumn{1}{c}{} &   &3338 &\multicolumn{3}{c}{}&\\
    \cline{1-1}\cline{4-4}
    NNLO Total &  \multicolumn{1}{c}{} &  &3357 & \multicolumn{3}{c}{}&   \\
    \hline
  \end{tabular}
  \caption{\label{tab:exps-sets} \small 
    Charm structure function 
    datasets included in the NNPDF2.1 NNLO analysis. All other data are
    the same as in the NNPDF2.1 NLO analysis, given in Table~2 of Ref.~\cite{Ball:2011mu}. 
    The number of data points 
    after kinematic cuts are shown in parentheses. In the last three lines we  give 
    the total number of
    datapoints included in the NNPDF2.1 LO, NLO and NNLO fits.}
\end{table}

\subsection{Data sets}
\label{sec:dataset}

\begin{figure}[t]
\begin{center}
\epsfig{width=\textwidth,figure=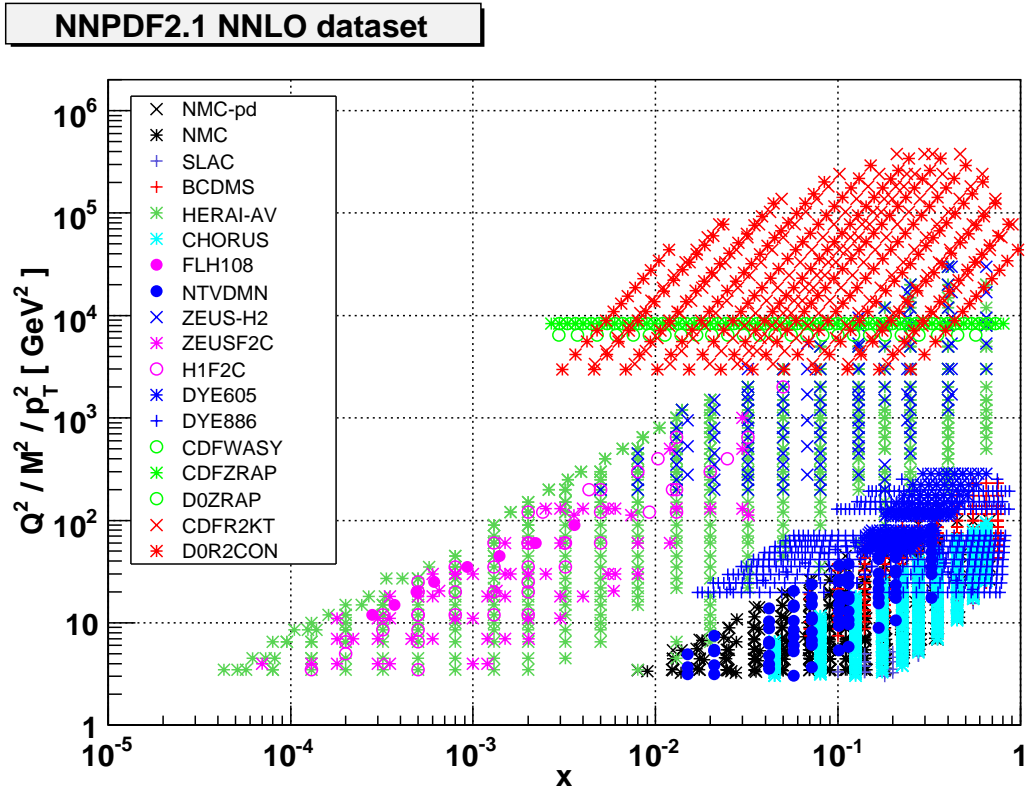}
\caption{ \small The experimental data
which enter the NNPDF2.1 PDF determination with NNLO
kinematic cuts. 
\label{fig:dataplottot}} 
\end{center}
\end{figure}

The NNPDF2.1 NLO dataset includes 
NMC~\cite{Arneodo:1996kd,Arneodo:1996qe}, BCDMS~\cite{bcdms1,bcdms2}
and SLAC~\cite{Whitlow:1991uw} deep--inelastic scattering (DIS) fixed
target data; the combined HERA-I DIS dataset~\cite{H1:2009wt}, HERA
$F_L$~\cite{h1fl} and
$F_2^c$ structure function
data~\cite{Breitweg:1999ad,Chekanov:2003rb,Chekanov:2008yd,Chekanov:2009kj,Adloff:2001zj,
  Collaboration:2009jy,H1F2c10:2009ut},
ZEUS HERA-II DIS cross-sections~\cite{Chekanov:2009gm,Chekanov:2008aa},
CHORUS~\cite{Onengut:2005kv} inclusive neutrino DIS, and
NuTeV~\cite{Goncharov:2001qe,MasonPhD} dimuon production data;
fixed-target E605~\cite{Moreno:1990sf} and E866~\cite{Webb:2003ps,Webb:2003bj,Towell:2001nh} Drell-Yan production data;
CDF~\cite{Aaltonen:2009ta} $W$ asymmetry and   
CDF~\cite{Aaltonen:2010zza} and D0~\cite{Abazov:2007jy} $Z$ rapidity
distributions; and CDF~\cite{Aaltonen:2008eq} and D0~\cite{D0:2008hua} Run-II one-jet inclusive
cross-sections. A scatter plot of these data  
in the $x,Q^2$ plane is displayed in Fig.~\ref{fig:dataplottot}, with
the values of $x$ determined using LO kinematics.

In the NNPDF2.1 LO fit, the dataset is modified in comparison to the
NLO dataset in that the $F_L$
structure function data are removed, since this observable vanishes at LO.

In the NNPDF2.1 NNLO fit, the dataset is modified in two
respects in comparison to the NLO dataset. 
First, the E866 data, published as $x_F$
distributions, have been converted into rapidity distributions, since 
the use of rapidity as kinematic variable makes the
inclusion of NNLO corrections simpler. This was done following
the procedure discussed in Ref.~\cite{Bonvini:2010tp}, by using 
the average $p_T$ of the lepton pair in each bin. We have verified
explicitly that if this procedure is also applied at NLO, the fit
results are unchanged. Second,
the NMC proton data are now included as data for reduced cross-sections,
rather than for structure functions. 
It was shown in Ref.~\cite{Ball:2011we} that  the impact of this different
treatment is almost negligible at NLO. However the use of
cross-section data is in principle preferable, as they are closer to
what is actually measured. In Ref.~\cite{Alekhin:2011ey} it was
claimed that the treatment of these data may have a significant impact
on NNLO PDFs, though this claim is not supported by preliminary
investigations with NNPDF2.1 NNLO~\cite{Cerutti:2011au}, 
or with MSTW08~\cite{Thorne:2011kq} PDFs.

\subsection{Kinematic cuts}
\label{sec:kincuts}

All data included in the NNPDF2.1 LO, NLO and NNLO fits are subject to
cuts on the invariant mass $W^2$ and the scale $Q^2$ of the DIS final state
$W^2_{\rm min}> 12.5$ GeV$^2$ and
$Q^2>3$ GeV$^2$. In the NLO fit, the $F_2^c$ data were subject to the
further cuts $Q^2>4$ GeV$^2$  and $Q^2>10$ GeV$^2$ if
$x< 10^{-3}$, due to the fact that in this region NNLO massive
corrections are so large that a NLO approximation is not acceptable. 
These cuts will be removed for the NNLO fit, 
in which the $F_2^c$ data will only be subject to
the cuts which are common to all other DIS data. The LO fit instead
will use the same cuts as the NLO one.
 The charm structure function data included in the NNLO fit are listed
in Table~\ref{tab:exps-sets}; all other data are the same as in the
NLO fit, Table~2 of Ref.~\cite{Ball:2011mu}. The total numbers of
datapoints used at LO, NLO and NNLO are also given in Table~\ref{tab:exps-sets}.

%% file: sec-pdfevol-nc.tex
\section{Physical observables}
\label{sec:evolution}

As mentioned in the introduction, there are two aspects of the
theoretical implementation of QCD corrections which require some
discussion away from NLO: the treatment of heavy quark masses, and the
fast implementation of NNLO corrections to hadronic observables and
the corresponding benchmarking. The former will be discussed in
Sect.~\ref{sec:hq}, both at LO and NNLO, while the analytic expression
of various hitherto unknown NNLO coefficient functions in Mellin space
are listed in Appendix~\ref{sec:massive-nc-comp}; the general-mass
deep-inelastic coefficient functions will then be benchmarked in
Appendix~\ref{sec:massive-nc}.  The latter will be discussed and
benchmarked in Sect.~\ref{sec:hadrnnlo}.  Perturbative evolution at
NNLO will be benchmarked in Appendix~\ref{sec:pdfevol-nnlo}. Given
that for jets we will rely on the FastNLO code~\cite{Kluge:2006xs},
this provides us with a full benchmarking of all NNLO expressions used.

\subsection{LO and NNLO structure functions with heavy quark mass effects}
\label{sec:hq}

We will include heavy quark masses using the FONLL method of
Ref.~\cite{Cacciari:1998it}, extended to DIS in
Ref.~\cite{Forte:2010ta}
(see Ref.~\cite{Ball:2011mu} for  charged-current DIS expression). 
In the NNPDF2.1 NLO fits \cite{Ball:2011mu} we used the FONLL-A scheme, which combines NLO 
massless perturbative evolution with the $\mathcal{O}\lp\alpha_s\rp$ massive coefficient functions.

At LO, the massive neutral-current 
coefficient function vanishes, and thus, for neutral current DIS, the FONLL 
scheme of Ref.~\cite{Forte:2010ta} only
differs from a naive zero-mass scheme by a
damping factor which suppresses dynamically generated  charm
contributions below threshold. The same is true for charged-current
DIS when the heavy quark is the struck quark, while for contributions
in which a heavy quark is produced from a struck light quark the FONLL
expression simply reduces to the parton-model (${\mathcal
  O}(\alpha_s^0)$) massive coefficient function.

At NNLO, it is possible to combine NNLO massless perturbative evolution with
the ${\mathcal O}(\alpha_s^2)$ massive coefficient functions: this was called the
FONLL-C scheme in Ref.~\cite{Forte:2010ta}. It is also possible to instead combine 
the ${\mathcal O}(\alpha_s^2)$ massive coefficient functions with NLO massless perturbative 
evolution, which was called FONLL-B scheme in Ref.~\cite{Forte:2010ta}; a comparison of FONLL-B with 
FONLL-A for NLO fits will be performed elsewhere.

Different schemes for the inclusion of the heavy quark mass in DIS
structure functions were
benchmarked in Ref.~\cite{LHhq}, with common input toy PDFs and 
common choices of all other settings, such as the values of the heavy
quark masses. Preliminary comparisons of FONLL-C with S-ACOT-$\chi$ at 
NNLO~\cite{guzzitalk} suggest that
the two GM-VFN schemes are numerically similar.

\begin{figure}[t!]
\begin{center}
\includegraphics[width=0.49\textwidth,angle=90]{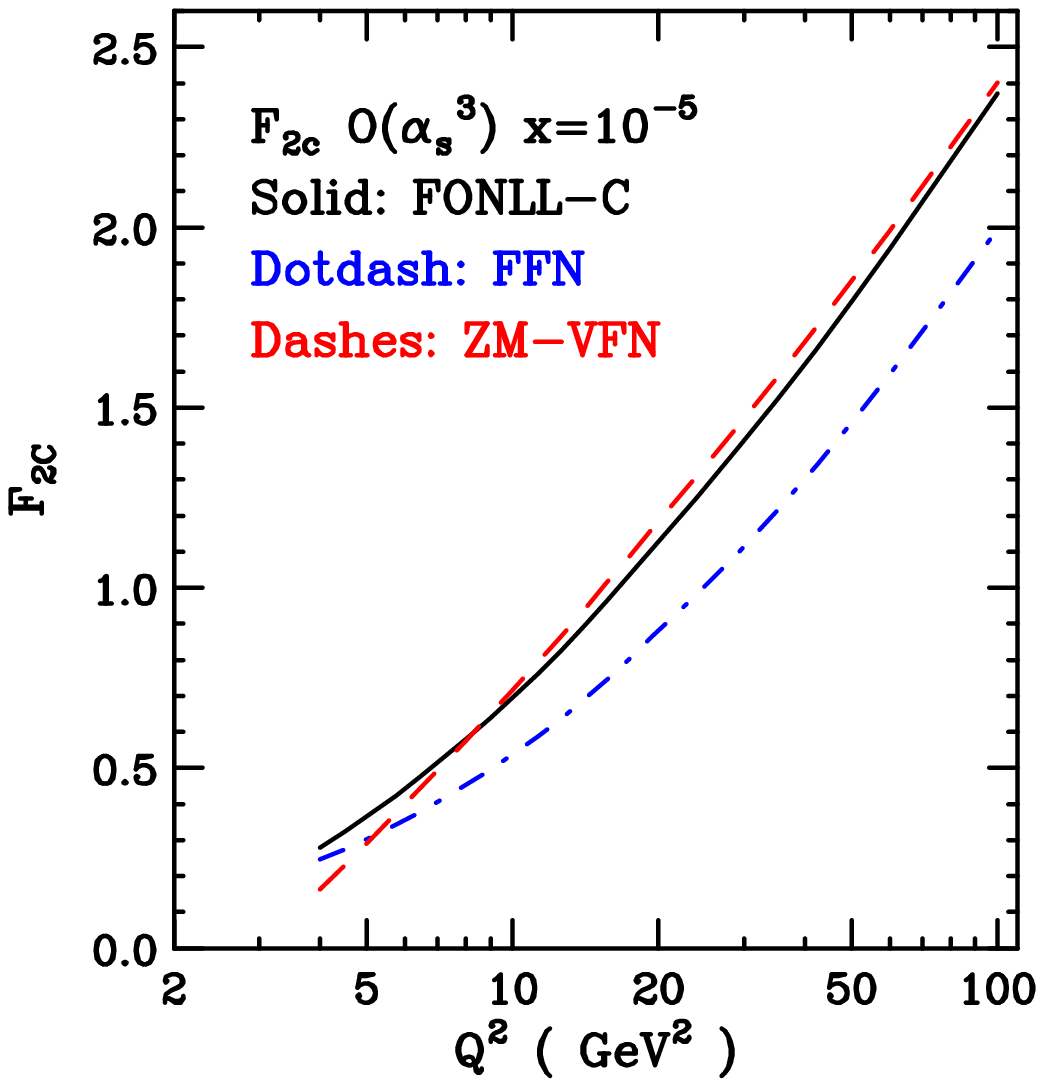}
\includegraphics[width=0.49\textwidth,angle=90]{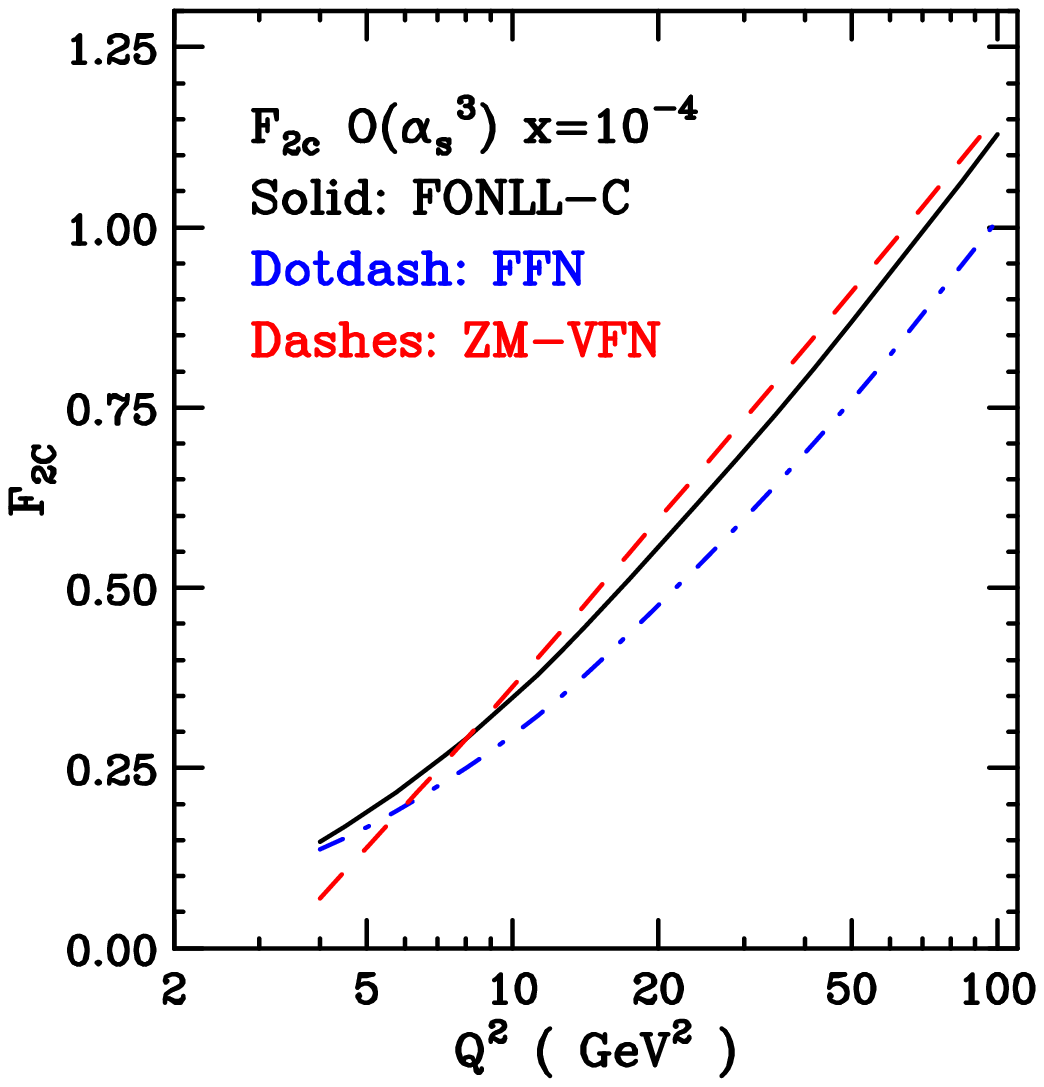}\\
\includegraphics[width=0.49\textwidth,angle=90]{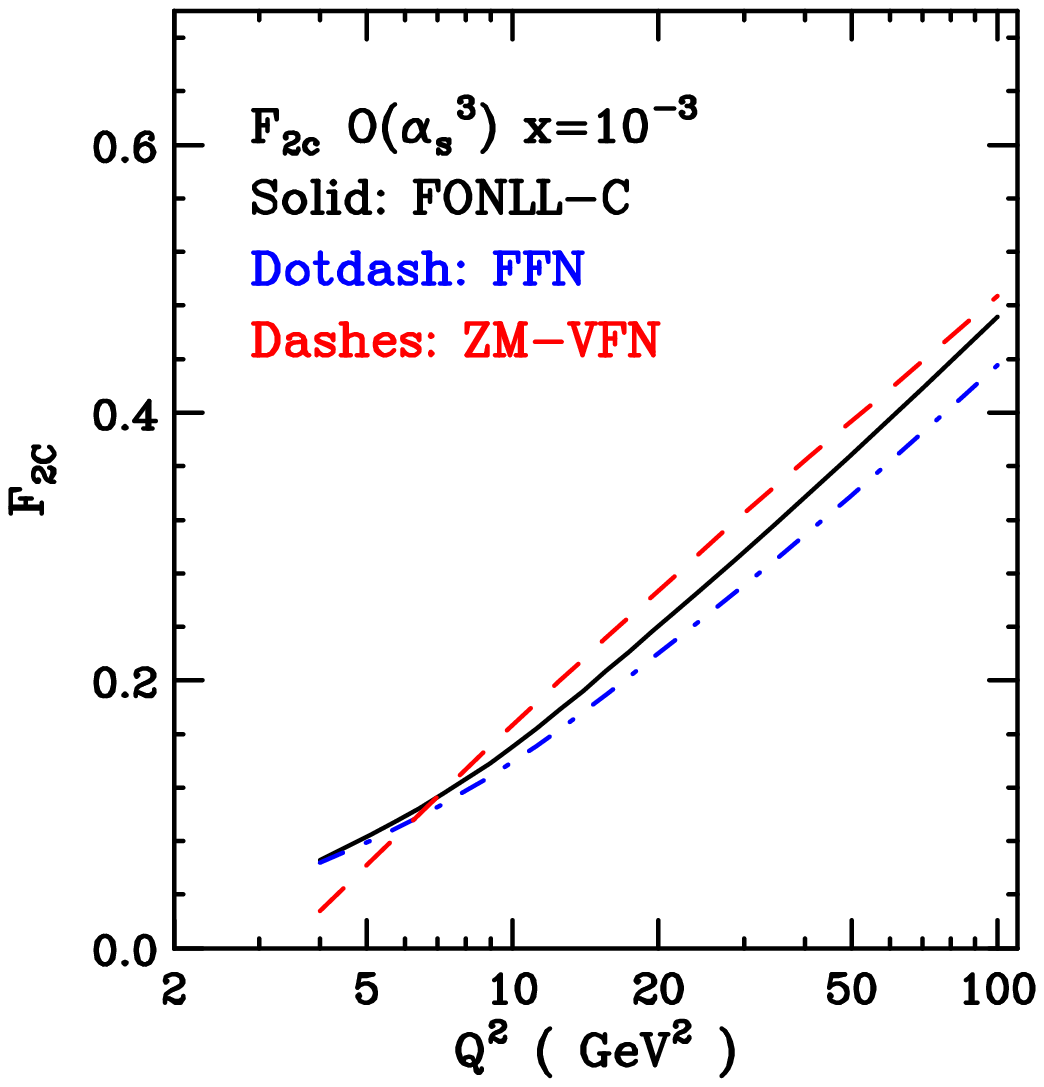}
\includegraphics[width=0.49\textwidth,angle=90]{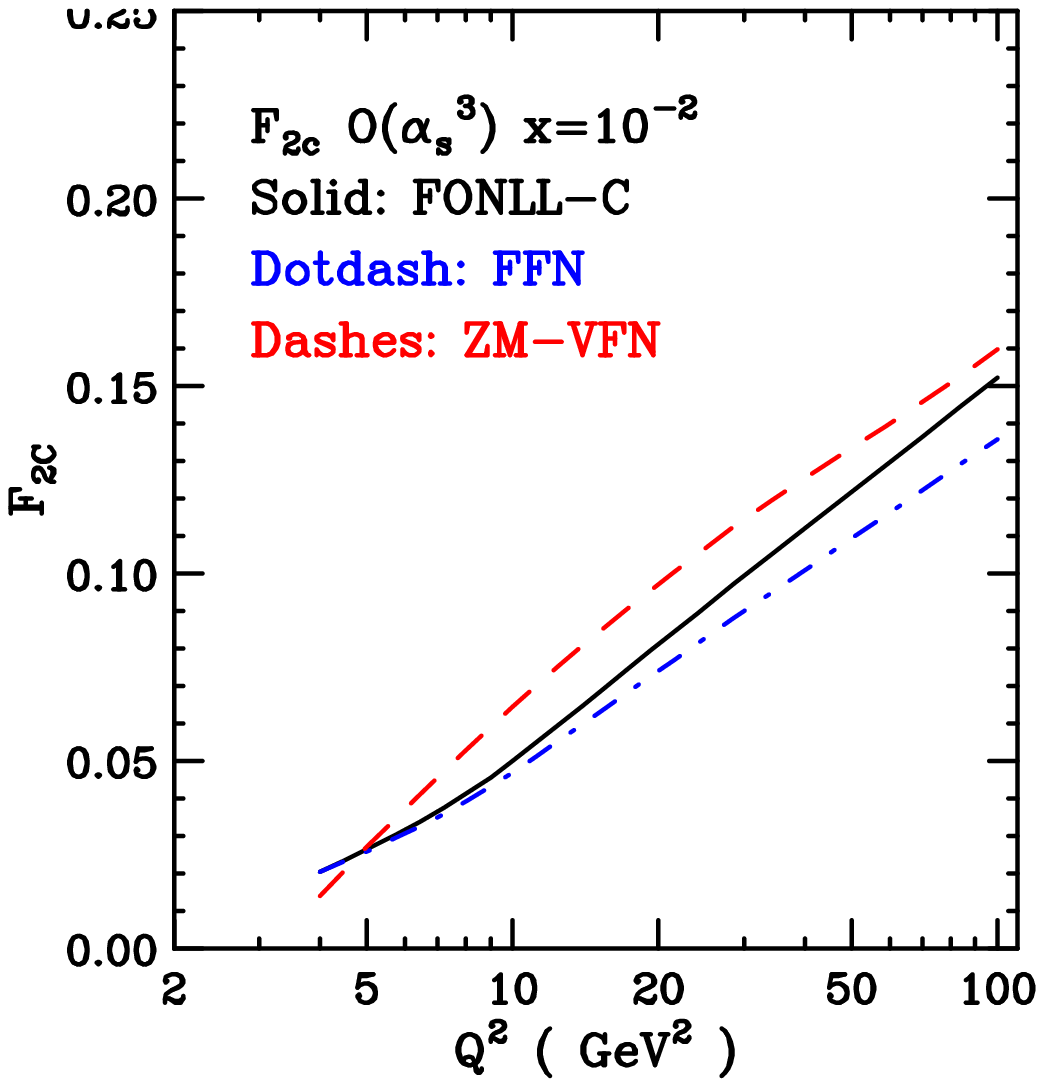}
\end{center}
\caption{\small The charm structure function $F_{2,c}(x,Q^2)$ 
as a function of $Q^2$ for different values of $x$
  from $x=10^{-5}$ to $x=10^{-2}$ computed in the FONLL-C scheme, and
  compared to the zero mass (ZM-VFN) and fixed-flavour number (FFN)
  results. The reference input PDFs  and settings of  Les
  Houches benchmarks~\cite{LHhq} are used throughout.}
\label{fig:f2c_vs_q2}
\end{figure}

\begin{figure}[t!]
\begin{center}
\includegraphics[width=0.49\textwidth,angle=90]{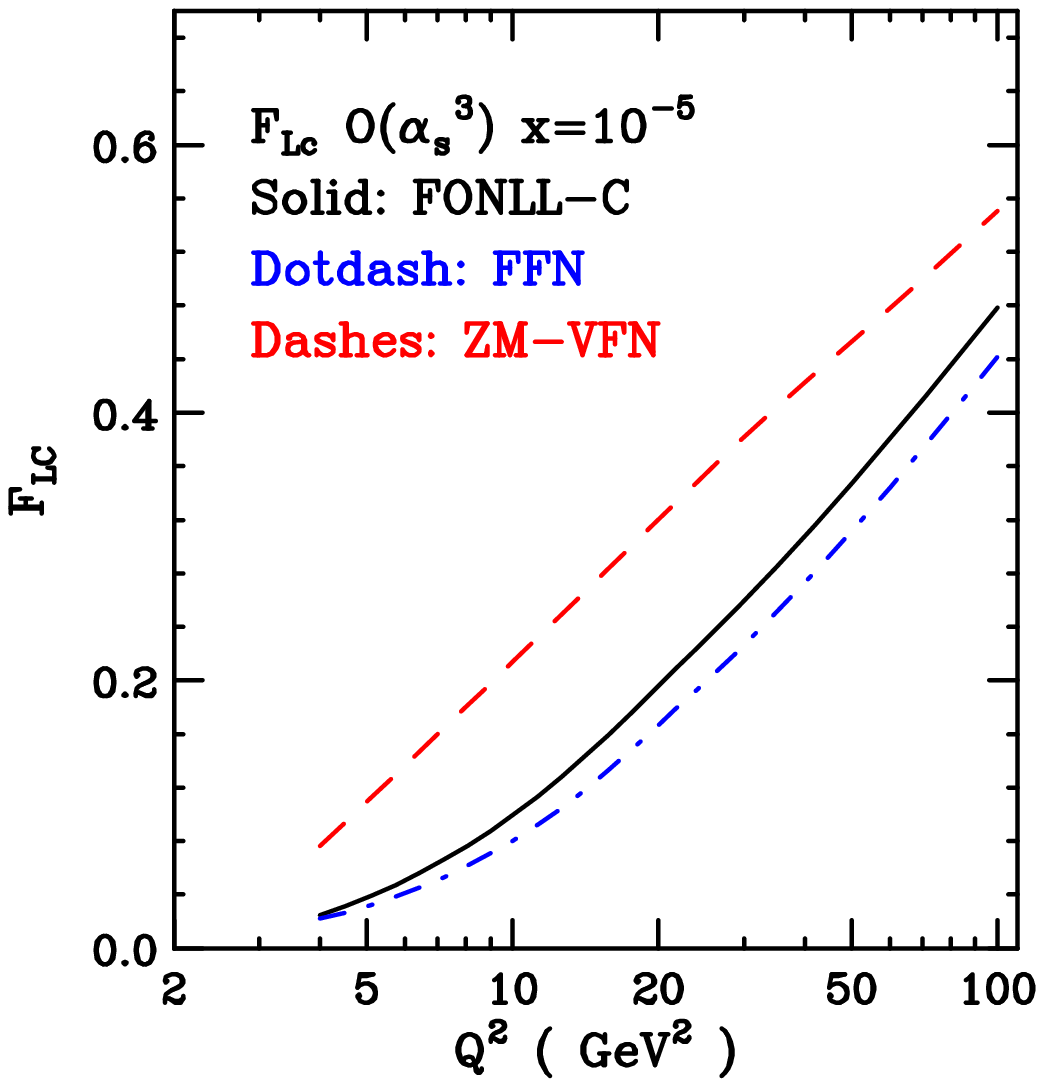}
\includegraphics[width=0.49\textwidth,angle=90]{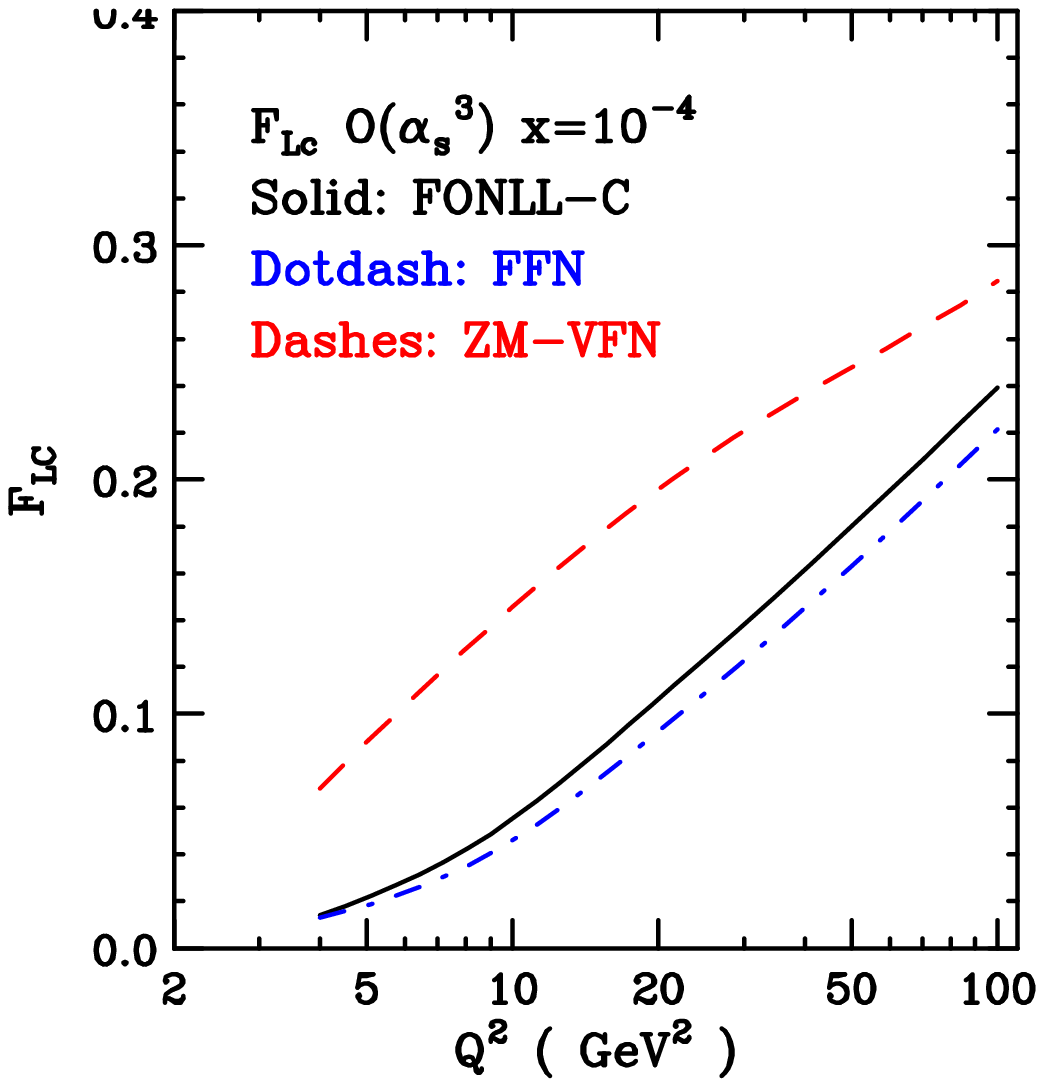}\\
\includegraphics[width=0.49\textwidth,angle=90]{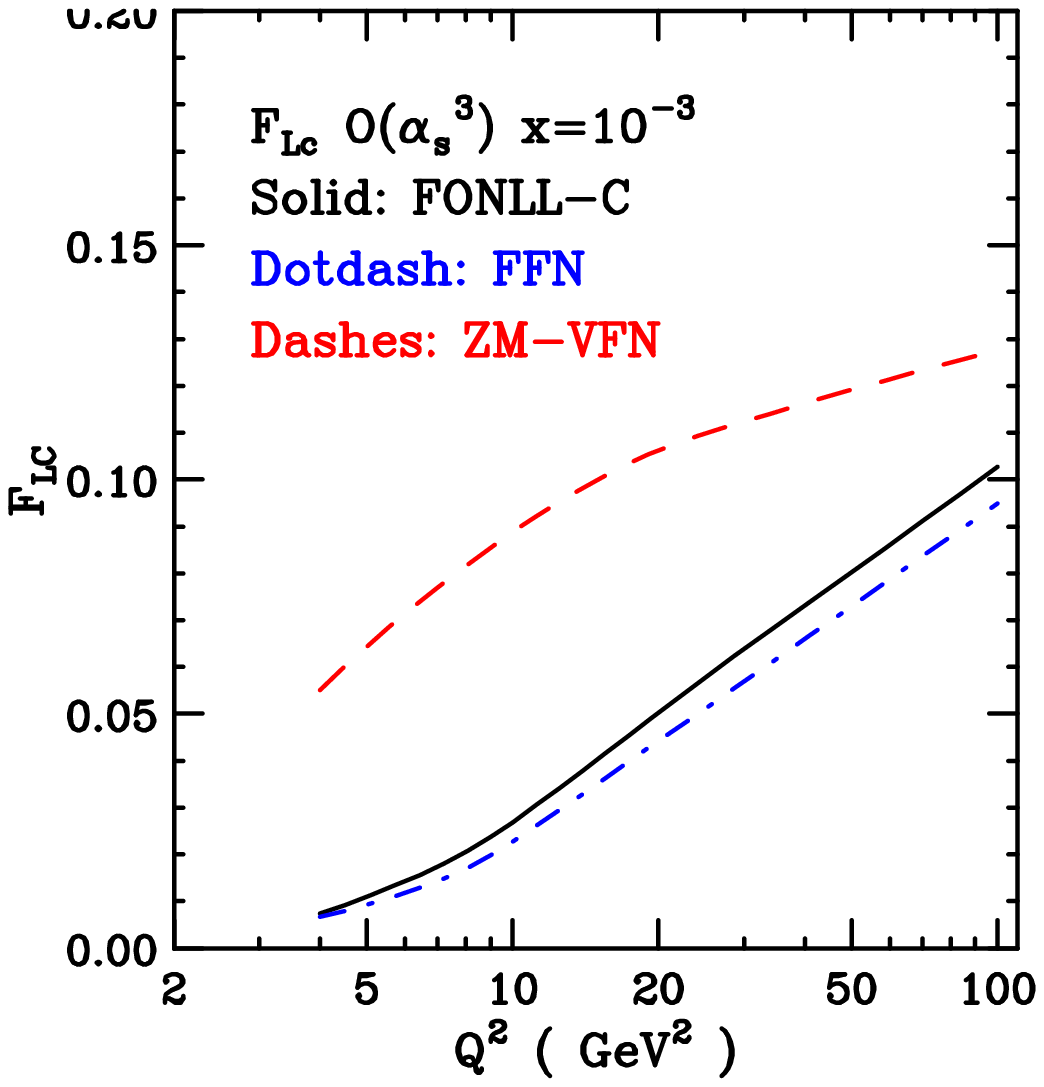}
\includegraphics[width=0.49\textwidth,angle=90]{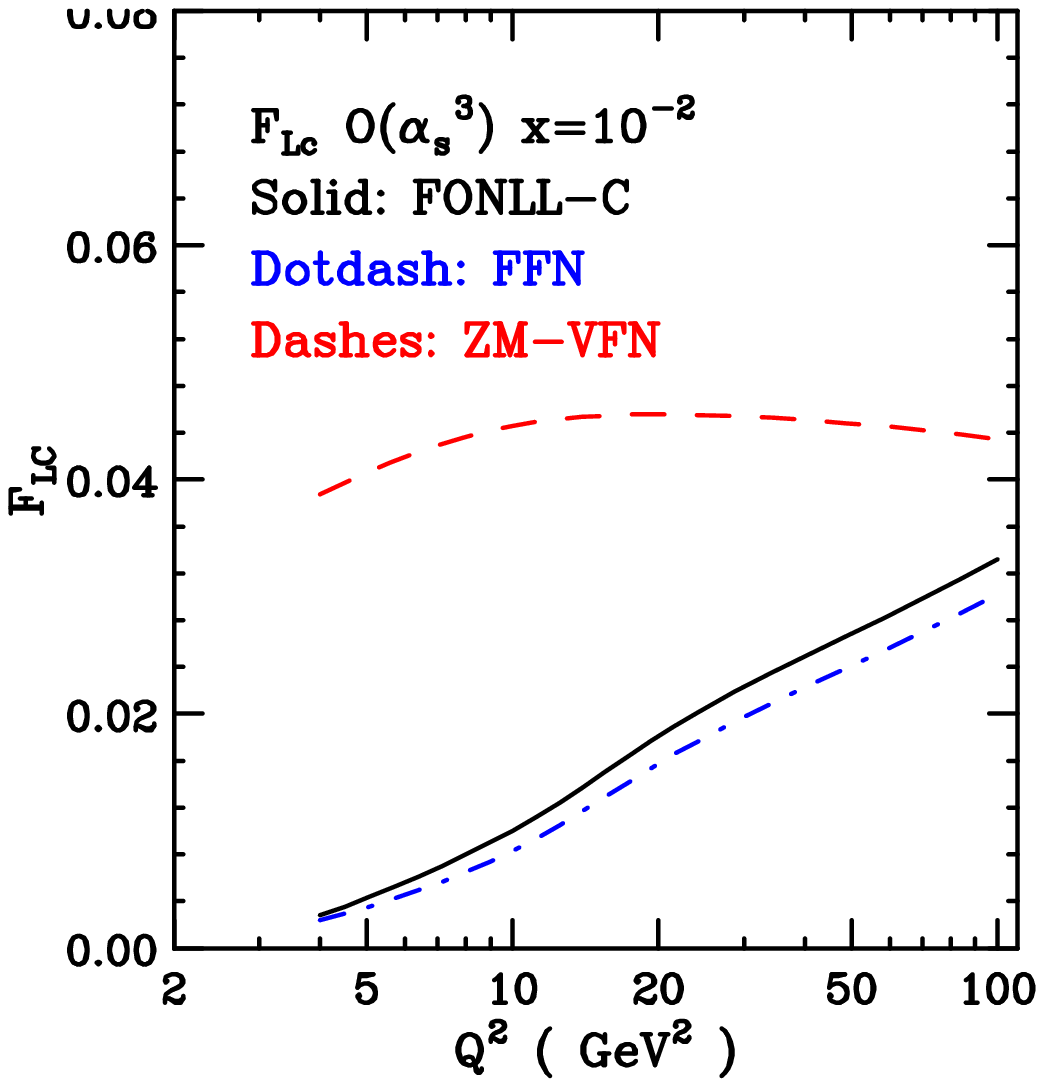}
\end{center}
\caption{\small Same as Fig.~\ref{fig:f2c_vs_q2} for the
longitudinal charm structure function $F_{L,c}(x,Q^2)$.}
\label{fig:flc_vs_q2}
\end{figure}

The charm structure functions $F_{2,c}$ and $F_{L,c}$ computed in the
FONLL-C scheme are shown in Fig.~\ref{fig:f2c_vs_q2} and
Fig.~\ref{fig:flc_vs_q2} respectively. They are compared to the NNLO
determination of the structure function in which the heavy quark mass
is neglected (ZM-VFN, or zero-mass variable-flavour number scheme),
and to the $O(\alpha_s^2)$ computation in a fixed $n_f=3$ scheme with
charm mass (FFN).  It is clear from the plots that the FONLL-C scheme
interpolates smoothly between the $O\lp\alpha_s^2\rp$ massive scheme
(FFN) near the heavy quark threshold, and the $O\lp\alpha_s^2\rp$
massless scheme (ZM-VFN) at large $Q^2$.  Mass effects are
much larger for the longitudinal structure function $F_{L,c}$ than for $F_{2,c}$, so there 
the ZM-VFN computation is completely unreliable.

In the sequel we will adopt FONLL-C with 
threshold damping factor~\cite{Forte:2010ta} as our default choice for the NNPDF2.1 NNLO fit. 
The heavy quark masses will also take the same default values as in the
NNPDF2.1 NLO fit, namely $m_c=1.414$~GeV and $m_b=4.75$~GeV. These
should be taken as pole masses, because this choice is adopted in
the construction of the FONLL-B and FONLL-C expressions given in
Ref.~\cite{Forte:2010ta}. The use of running $\overline{\rm MS}$ 
masses has been advocated
recently~\cite{Alekhin:2010sv} because of the greater perturbative
stability of the running mass: this possibility will be studied in
future NNPDF releases. 

All the above discussion applies to neutral-current structure
functions. In the case of charged-current DIS, a full implementation
of the FONLL-C scheme is not possible, because the massive $O(\alpha_s^2)$ 
heavy quark coefficient functions are not available
(only the asymptotic $Q^2\to\infty$ limit is
known~\cite{Buza:1997mg}). Consequently, in the FONLL-C charged
current structure functions the $ O(\alpha_s^2)$ massive contribution
is set to zero, while PDFs, the ZM structure functions and $\alpha_s$
are evaluated at NNLO. We have checked that the impact of NNLO
corrections in the charged current sector is very moderate, typically
well below 10\%. This choice achieves the best accuracy that can be
obtained from the available perturbative information without
introducing any modelling.

All NNLO structure functions have been implemented in the FastKernel
framework of Ref.~\cite{Ball:2010de}. The  benchmarking of the numerical accuracy
of the implementation is discussed in Appendix~\ref{sec:massive-nc}.

\subsection{The treatment of hadronic data}
\label{sec:hadrnnlo}

We now turn to the NNLO implementation of the  hadronic data, namely,
Drell-Yan, $W$ and $Z$ production, and inclusive jets.

For NNPDF2.1 NLO, Drell-Yan and vector boson production were treated
consistently at NLO in perturbative QCD in all the stages of the
PDF fit using the FastKernel framework~\cite{Ball:2010de}.  The
extension of the  FastKernel  method to NNLO is in principle
straightforward, but in practice challenging, in
particular because of the distribution structure and intricate 
choice of kinematic
dependence of the NNLO coefficient functions of
Ref.~\cite{Anastasiou:2003ds}. Therefore, here
we instead adopt an approximate NNLO computation, which leads to an
accuracy which is fully adequate for our purposes as we shall now
show.

\begin{figure}[t!]
\begin{center}
\includegraphics[width=0.80\textwidth]{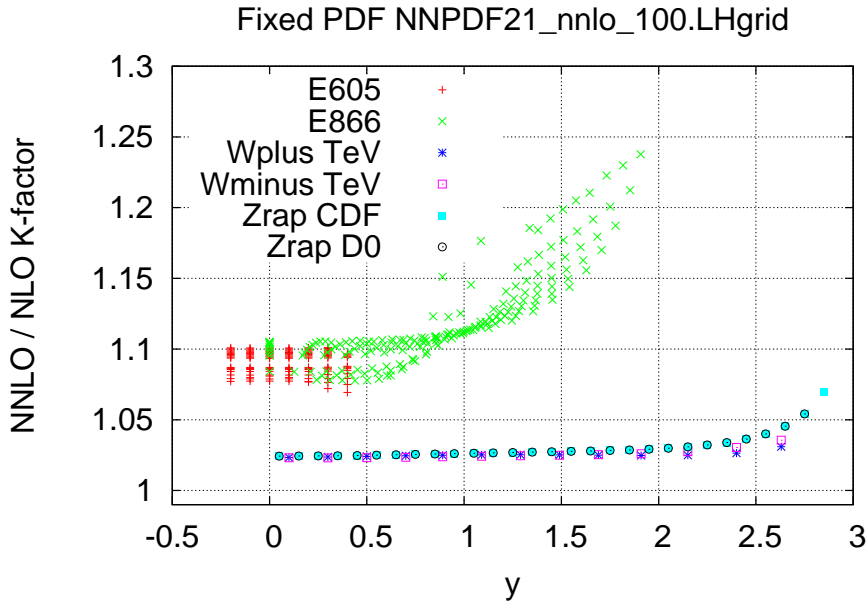}
\end{center}
\caption{\small The NNLO/NLO $K$-factors for the Drell-Yan, $W$ and $Z$
production data  included in the NNPDF2.1 fit.}
\label{fig:dy-kfactor}
\end{figure}
In this approximation, Drell-Yan observables are computed with NNLO PDF
evolution and NLO partonic cross-section supplemented by a $K$-factor
that accounts for the missing $O(\alpha_s^2)$ partonic coefficient
functions.  The $K$-factors are defined as the ratio of double
differential cross-sections $d^2\sigma/dydM^2$ in Drell-Yan production
where in the numerator we use the full NNLO expression and in the
denominator the same expression but with the $\mathcal{O}\lp\alpha_s^2\rp$ correction to the
partonic cross-section set to zero.  In this definition the same
NNLO PDFs and $\alpha_s$ are used both in the numerator and the
denominator. This minimizes the impact of the NNLO $K$-factor
corrections, which are then reduced to the missing
$\mathcal{O}\lp\alpha_s^2\rp$ partonic cross-sections. These are rather
small for most processes of interest, especially for collider
kinematics.

\begin{figure}[t!]
\begin{center}
\includegraphics[width=0.80\textwidth]{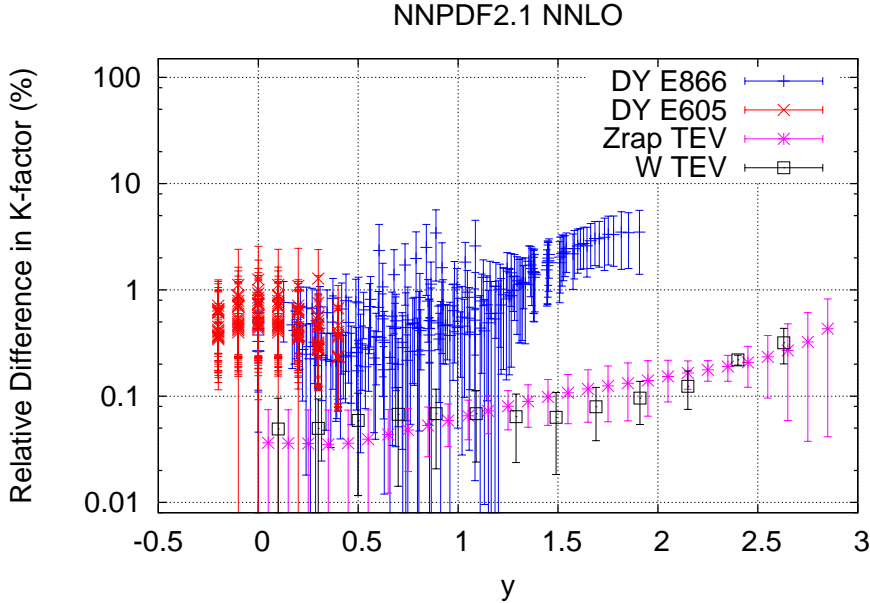}
\end{center}
\caption{\small Percentage difference between the values of the $K$
  factors used in the NNPDF2.1 PDF fit (shown in Fig.~\ref{fig:dy-kfactor})
  and their redetermination with the final  NNPDF2.1 PDFs. The error
  bar is obtained as  one sigma uncertainty over replicas.}
\label{fig:dy-kfactorbench}
\end{figure}
We have computed these $K$-factors using the VRAP code~\cite{Anastasiou:2003ds}, 
and cross-checked the results with the DYRAP~\cite{Anastasiou:2003yy}
program. The $K$-factors are computed iteratively using central PDFs from a previous 
NNPDF2.1 NNLO fit. Results are shown in Fig.~\ref{fig:dy-kfactor} for the 
different datasets as a function of the rapidity
$y$ of the produced electroweak boson. For collider observables like
$W/Z$ production at the Tevatron the $K$-factors are at the few percent level.
The NNLO $K$-factors are more important for fixed-target Drell-Yan data,
in particular for the E866 kinematics, where they are typically of
order 10\%, but sometimes as
large as 25\%. However, the average total experimental uncertainty on
these data is larger than 20\%.

The error incurred replica by replica through the use of the 
$K$-factor approximation is then subleading in the perturbative
expansion. To see this, note that
approximation in the computation of the
$K$ factor comes from its dependence on the PDF. 
However, the $K$-factor only enters at NNLO, while the NLO is exact, so the
error in the PDF used to compute the $K$-factor is $\mathcal{O}\lp\alpha_s^2\rp$. 
However the $K$-factor itself is $\mathcal{O}\lp\alpha_s^2\rp$, so the error in the cross
section is in fact $\mathcal{O}\lp\alpha_s^4\rp$. Note that this 
is not the case if $K$-factors are used for both NLO and NNLO 
corrections (as in Refs.~\cite{Martin:2009iq,Lai:2010vv}), since then
the error in the PDF is  $\mathcal{O}\lp\alpha_s\rp$, and thus
$\mathcal{O}\lp\alpha_s^2\rp$ in the cross section, hence
at least in principle of the same order as the NNLO correction to the cross section itself. 

In Fig.~\ref{fig:dy-kfactorbench} we show the percentage shift of the
$K$-factors if they are recomputed using the final NNPDF2.1 NNLO PDF
set. In order to
study the possible dependence of the accuracy of the $K$-factor
approximation on the choice of (central) PDF used to compute the
$K$--factors,  we have repeated this comparison for ten different
randomly chosen replicas. The standard deviation of the results is
also shown as an error bar in Fig.~\ref{fig:dy-kfactorbench}. We
conclude that the accuracy is always better than 3\%, which translates into an 
uncertainty of no more than $0.7\%$ in the cross-section. 

For the inclusive jet production data exact 
NNLO corrections are not known. 
However, an approximation to the full NNLO result based on threshold
resummation is available~\cite{Kidonakis:2000gi}. 
We will compute  inclusive jet observables using  
an approximation on which PDFs are evolved at NNLO, but coefficient
functions are computed using this threshold approximation of the full
NNLO result, as implemented in the
FastNLO code~\cite{Kluge:2006xs}.  
This provides us with  an approximate NNLO calculation
which combines the most accurate perturbative information available.
Fits in which the jet data are evaluated with NLO coefficient functions, or 
simply removed altogether, will be discussed
in Sect.~\ref{sect:reddataset}.

%% file: sec-implementation.tex
\section{Implementation issues at LO and NNLO}
\label{sec:impl}
The parton parametrization,  minimization algorithm, and determination
of the optimal fit  in the LO and NNLO fits presented here are the same
as in the NLO PDF determination 
of Ref.~\cite{Ball:2011mu}, including almost all settings
for the parameters which control the parametrization and
minimization. The small number of changes are discussed here: first,
we describe how at LO the parton parametrization is optimized in view
of LO positivity constraints, and then we examine some adjustments in the
choices of parameters for the genetic algorithm and the stopping of the
minimization.

\subsection{Parton parametrization: positivity constraints}
\label{sec:pos}

In all previous NNPDF fits, positivity of physical observables has been
imposed: beyond leading order,  PDFs depend  on the
factorization scheme and can be either positive or negative, however
cross-sections must remain non-negative~\cite{Altarelli:1998gn}. 
Positivity of physical
observables at NLO was enforced by means of Lagrange multipliers (see 
Ref.~\cite{Ball:2011mu}), and the same procedure will be used here for
the determination of NNLO PDFs. 

However, at leading order
parton distributions admit a probabilistic interpretation and are thus
non-negative. The positivity constraint can then be imposed directly on
all PDFs at the initial scale. Leading-order evolution preserves the
probabilistic interpretation of
PDFs~\cite{Durand:1986te,Collins:1988wj}, hence this is sufficient to
guarantee positivity at all scales. This can again be done by Lagrange
multipliers, i.e. adding to the $\chi^2$ a large penalty term whenever
any of the individual PDFs turns negative. 
However, in order to speed up the LO PDF fits, it is
advantageous to impose positivity directly at
the level of the PDF parametrization.  
Within the neural network PDF
parametrization which we adopt, this can be done as follows. Recall
that in the architecure that we adopt for neural networks, the
response function is a sigmoid
\be
\xi^{(l)}_i\,=\,g\Big( \sum_j\,\omega_{ij}^{(l)}\xi_j^{(l-1)}\,-\,\theta^{(l)}_i\Big),
\qquad g(x)=\frac{1}{1+e^{-x}} \, ,
\label{eq:posnna}
\ee
for hidden layer, but it is linear in the last layer. For the LO fits,
we adopt instead for the last layer a quadratic reponse function
\be
\xi^{(n_l)}_i\,=\,\Big(
\sum_j\,\omega_{ij}^{(n_l)}\xi_j^{(n_l-1)}\,-\,\theta^{(n_l)}_i\Big)^2\, .
\label{eq:posnnb}
\ee
The output of the neural network, and thus the PDF, is then guaranteed
to be non-negative.

The basis of PDFs that are parametrized by neural networks in NNPDF
fits~\cite{Ball:2008by,Ball:2009mk} includes the gluon, quark singlet,
and various other linear combinations of quark PDFs. Of these, only
the gluon and singlet must be positive, since all other combinations
contain differences of PDFs. However, in practice also the total
valence and isospin triplet combination are positive definite. Hence
the parametrization Eq.~(\ref{eq:posnna}-\ref{eq:posnnb}) for
simplicity is adopted for all PDFs: PDFs other than singlet,
gluon, valence and triplet are allowed to change sign by simply adding
to the above form a constant shift. It turns out that with the constraints 
from the data, this is sufficient in practice to ensure positivity of all PDFs:
we have checked a posteriori that for every replica the gluon and all 
individual quark and anti-quark flavours are positive for all values of $x$ and
$Q^2$ for which the NNPDF2.1 LO PDFs are provided.

\subsection{Minimization and stopping}
\label{sec:minstop}

\begin{table}
\begin{center}
  \begin{tabular}{|c||c|c|c|c|c|c|}
    \hline 
 &   $N_{\rm gen}^{\rm wt}$ & $N_{\rm gen}^{\rm mut}$
&   $N_{\rm gen}^{\rm max}$ & $E^{\mathrm{sw}}$ & $N_{\rm mut}^a$ 
&  $N_{\rm mut}^b $\\
    \hline
LO \& NLO &    $10000$ & 2500 & 30000 & 2.6 & 80 & 10\\
    \hline
NNLO &    $10000$ & 2500 & 30000 & 2.3 & 80 & 30\\
    \hline
  \end{tabular}\\

\bigskip
\bigskip

  \begin{tabular}{|c||c|c||c|c|}
\hline
&  LO \& NLO &&  NNLO &   \\
    \hline 
PDF &   $N_{\rm mut}$ &  $\eta^{\rm k}$ &  $N_{\rm mut}$ &  $\eta^{k}$  \\
    \hline
\hline 
$\Sigma(x)$   &2 & 10,1 & 2 & 10,1 \\
$g(x)$  & 2& 10,1& 3 & 10,3,0.4 \\
$T_3(x)$  &2 & 1,0.1 & 2 &  1,0.1 \\
$V(x)$  &2 & 1,0.1 & 3 &  8,1,0.1\\
$\Delta_S(x)$  &2 & 1,0.1 &3 & 5,1,0.1 \\
$s^+(x)$  & 2&  5,0.5 & 2 & 5,0.5 \\
$s^-(x)$  & 2&  1,0.1& 2 & 1,0.1\\
\hline 
  \end{tabular}
  \end{center}
  \caption{\small Parameter values for the genetic 
algorithm for the
NNLO fits compared to those of the LO and NLO fits (top). The number of mutations and the values of the mutation rates for the individual PDFs
in the NNLO fit as compared to the values of the LO and NLO fits are
also given (bottom). }
  \label{tab:gapars}
\end{table}

The poorer quality of the LO fit on the one hand,
and the greater complexity of NNLO coefficient functions on 
the other hand, require some
retuning of the parameters of the minimization algorithm. 

At leading order, the best-fit value 
of the figure of merit $E^{(k)}$
which is being minimized for each replica (which is essentially the
$\chi^2$ of the fit of each PDF replica to the given data replica) is
on average rather larger than in an NLO fit, because of the
poorer accuracy of the LO theory. This is particularly
true for the Drell-Yan  
observables, which have large NLO corrections with a $K$-factor of
order two. As a consequence, the minimum value that $E^{(k)}$ must
reach for each experiment 
in order for the fit to stop has been increased  from $E_{\rm th}=6$ 
to $E_{\rm th}^{\rm
  DY}=12$ for all Drell-Yan
experiments. 
Furthermore,  the   cross-validation method that we use to determine
the optimal fit stops the minimization when the moving average (over
iterations of the genetic algorithm)
of $E^{(k)}$ increases more than a fixed percentage threshold
value $r_v$, larger than a typical random fluctuation. Because the
size of fluctuations of $E^{(k)}$ remains fixed, while its  value at
best fit has increased, the typical values of $r_v$ are smaller at LO,
and thus 
it turns out to be necessary to reduce the value of $r_v$
required for stopping to $r_v-1=2\cdot 10^{-4}$, from  $r_v-1=3\cdot 10^{-4}$ used at NLO.

Even with these adjustments, for a sizable fraction 
of replicas the cross-validation algorithm fails
to stop dynamically the minimization even after a large number
of generations of the genetic algorithm. This reflects the poor accuracy of
LO theory, and it could only be obviated by letting the genetic
minimization run much longer. In view of
the large theoretical uncertainties inherent to
any LO PDF determination, as a practical compromise, we
discard replicas that do not stop dynamically after 50000 iterations
of the genetic algorithm, retaining only those
replicas for which the stopping criterion was fulfilled. We have checked 
that this leads to no significant statistical bias. 

At next-to-next-to-leading order the partonic cross-sections have
rather more structure than at lower orders, both because of the
opening of new partonic channels and because of the appearance of new
transcendental functions in the perturbative results
(such as higher order harmonic sums). This results in
somewhat more complex PDF shapes. As a consequence, it turns out to be 
necessary to increase
the number of mutants and mutations per PDF in the genetic
minimization in order to  fully explore
this more complex space of minima. The NNLO settings for the genetic algorithm
used for minimization are summarized  in Table~\ref{tab:gapars} and
compared to those used at LO and NLO.  The table is to be compared to
Table~6 of Ref.~\cite{Ball:2010de} and Table~5 of
Ref.~\cite{Ball:2008by}, 
to which we refer for a more
detailed discussion. We give the number of mutants in the two stages in which the minimization
is divided, the number of 
mutations per PDF, and the values of the mutation rates for
each PDF. We also show in a separate table the number 
of mutations that are applied
to each PDF (which at NLO was two for all PDFs), 
and the values of the mutation rates 
 $\eta^{\rm k}$  of each PDF (which at NLO were given in
Ref.~\cite{Ball:2008by} and kept unchanged in Ref.~\cite{Ball:2010de}).


%% file: sec-lopdfs.tex
\section{Leading order parton distributions}

\label{sec:lopdfs}

Parton distributions based on a leading-order QCD treatment of the
data are mostly used with
leading order   Monte Carlo event generators, and are also of interest for
comparison of QCD calculations at different perturbative
orders. Of course, nothing prevents the inclusion of some NLO terms in
a calculation which has LO accuracy, so in principle one could always
use NLO PDFs in these and related contexts. However, in practice
using NLO PDFs with LO matrix elements may lead to a poorly
behaved perturbative expansion and to bad phenomenology. Indeed, the
difference between the optimal PDFs determined from a LO analysis at
their standard NLO counterparts is typically rather larger than NLO
PDF uncertainties. Hence, the dominant uncertainty on LO PDFs is
theoretical, and there is 
a certain latitude in their definition. Therefore, we will at first
discuss various options for the construction of LO PDFs, then turn to
results and comparisons, with the statistical aspects of the PDF
determination now playing a relatively less important role.

\begin{table}[t]
\centering
\small
\begin{tabular}{|c||c|c|c|c|c|}
\hline
PDF set  & .LHgrid file  & $\alpha_s\lp M_Z\rp$ & Momentum SR \\
\hline
\hline
NNPDF2.1 LO    & NNPDF21\_lo\_as\_0119\_100.LHgrid   & 0.119  & Yes  \\
NNPDF2.1 LO    & NNPDF21\_lo\_as\_0130\_100.LHgrid  & 0.130  & Yes  \\
NNPDF2.1 LO*   & NNPDF21\_lostar\_as\_0119\_100.LHgrid  & 0.119  & No   \\
NNPDF2.1 LO*   & NNPDF21\_lostar\_as\_0130\_100.LHgrid  & 0.130  & No   \\
\hline
\end{tabular}
\caption{\small Summary of  NNPDF2.1 Leading Order PDF sets.\label{tab:fits}}
\end{table}


\subsection{Definition of leading order PDFs}
\label{sec:lochoices}

The issue of choosing  the optimal parton sets to be used
in combination with LO event generators has been discussed
extensively. On the one hand, the possibility of using standard LO QCD
theory (including the running of $\alpha_s$) seems theoretically
simplest and most consistent. On the other hand, it could be that this
leads to unacceptably poor fit quality for some datasets included in the
global fit and thus modifications of the standard LO framework should
be considered.

The simplest of these, 
advocated in~\cite{Campbell:2006wx}, is to just use NLO PDFs within
the  LO Monte Carlo.  This choice at hadron colliders 
can be  justified by arguing that
the difference between LO and NLO PDFs is driven by the difference in
DIS matrix elements used in the PDF determination, but the LO and NLO matrix
elements for hadron collider processes are much closer to each other,
so NLO PDFs with LO collider matrix element may provide a reasonable
approximation to the exact NLO result. However, it turns out
that this choice  requires a substantial retuning of the 
parameters in event generators.

An intermediate possibility consists of including some dominant NLO
corrections to  the LO  matrix elements: for instance, in
Ref.~\cite{Martin:2009iq} it was pointed out that a sizable fraction of
the large NLO and NNLO $K$ factor for Drell-Yan comes from
contributions which have the same kinematics as the LO, and thus can
be simply absorbed in a rescaling of the LO cross-section. The
MSTW08LO PDFs of Ref.~\cite{Martin:2009iq} where determined by
rescaling the Drell-Yan cross-section in this way. 

More general  modifications of the standard LO were
suggested in Ref.~\cite{Sherstnev:2007nd} and adopted in the
construction of the  MRST2007lomod PDFs.
These PDFs are based on the observation that the LO fit quality mostly
deteriorates because of the faster 
gluon evolution at small $x$ and the slower quark
density evolution at large $x$. A possible way to improve this is to 
use the NLO value for the strong coupling constant
together with its two-loop running. This leads to
smaller values of $\alpha_s$ in the low $Q^2$ region where the
small $x$ data are concentrated,
and thus to slower PDF evolution. Of course, use of
 NLO $\alpha_s$ within an
otherwise LO framework is a subleading and thus legitimate change.
Another possibility is to relax the momentum sum rule. This alleviates
another possible problem of the LO fit, namely the fact that 
faster small $x$ gluon evolution leads through
the momentum sum rule to depletion of the gluon content at
medium/large $x$, which may cause a  poor description of large $x$ 
fixed-target data. Of course, a  violation of the  momentum sum rule 
 is in principle forbidden by first principles, however it  
can be justified as an {\it ad
  hoc} phenomenological patch.

Finally, one may take the point of view~\cite{Lai:2009ne} 
that the  goal of LO PDFs
is to be used with Monte Carlo generators, and thus they should be
determined by optimizing the agreement with the data
of the predictions obtained by
using them in combination with a generator. This involves 
considering all the various modifications of the minimal LO framework
discussed above, and also introducing suitable pseudodata  to optimize
the agreement with Monte Carlo generators. The
CT09MC1/MC2~\cite{Lai:2009ne} PDFs were constructed in this way.

\begin{table}[t]
\footnotesize
\centering
\begin{tabular}{|c||c|c|c|c|c|}
\hline
              &    NLO   &  LO $\alpha_s=$0.119    & LO* $\alpha_s=$0.119
&  LO 
$\alpha_s=$0.130  & LO* 
$\alpha_s=$0.130 \\
\hline
\hline
Total $\chi^2$     &     1.16   & 1.74   &   1.76    &   1.68  & 1.74 \\
$\la \chi^{2\,(k)}\ra$   &  $1.25 \pm 0.07$  & $1.95 \pm 0.21$       &     $1.89\pm 0.22$     & 
$1.95\pm 0.19$    &  $1.94\pm 0.18$\\
\hline
NMC-pd  &    0.97   & 1.43  &    1.13   &    1.18  &  1.12\\
NMC     &    1.72   & 2.05  &    1.68   &    1.74  &  1.72\\
SLAC    &    1.29   & 3.77  &    3.00   &    2.91  &  2.70 \\
BCDMS    &   1.24   & 1.87  &    1.82   &    1.76  &  1.75 \\
HERAI-AV  &  1.07   & 1.70  &    1.55   &    1.58 &  1.59\\
CHORUS    &  1.15   & 1.51  &    1.67   &    1.53  & 1.67 \\
NTVDMN    &  0.45   & 0.69  &    0.71   &    0.71  &  0.78\\
ZEUS-H2   &  1.29   & 1.51  &    1.42   &    1.43  &  1.44\\
ZEUSF2C   &  0.78   & 1.75  &    1.26   &    1.56 &  1.34\\
H1F2C     &  1.51   & 1.77  &    2.00   &    1.81  & 2.02 \\
\hline
DYE605    &  0.85   & 1.86   &   2.02    &   1.70  &  1.83\\
DYE886    &  1.26   & 1.99   &   2.52    &   2.59  &  3.11\\
CDFWASY   &  1.83   & 1.80   &   2.50    &   2.16  & 2.29 \\
CDFZRAP   &  1.64   & 2.88   &   3.89    &   2.08  & 2.58 \\
D0ZRAP    &  0.59   & 1.07   &   1.29    &   0.87  & 1.02 \\
\hline
CDFR2KT   &  0.96   & 2.60   &   3.22    &   2.45  &  2.76\\
D0R2CON   &  0.83   & 1.18   &   1.56    &   1.17  & 1.35 \\
\hline
\hline
$\lc M \rc $ & 1  &  1 & $1.16 \pm 0.03$ & 1 & $1.09\pm 0.03$ \\ 
\hline
\end{tabular}
\caption{\small Fit quality for the global fit and for all  experiments included
in it for each of the  NNPDF2.1 LO PDF sets. The corresponding values
for the NNPDF2.1 NLO set of Ref.~\cite{Ball:2009qv} are given for
comparison. The value of the momentum integral $\lc M \rc $
Eq.~(\ref{eq:momint}) 
 is also shown. All the fits 
have $N_{\rm rep}=100$ replicas. \label{tab:chi2-lopdfs}}
\end{table}

\subsection{Quality of the fit}
\label{sec:loqual}

We have produced four NNPDF2.1 LO PDF determinations: with two different
values of $\alpha_s\lp M_Z\rp$, 0.119 and 0.130 (but always with LO
running of $\alpha_s$), and 
with and without imposing the momentum sum rule. The various 
PDF sets, including the name of
the corresponding LHAPDF grid files,
are summarized in Table~\ref{tab:fits}.

The $\chi^2$ of the four LO NNPDF2.1 sets, both for
the global fit and for
individual experiments, are collected in 
Table~\ref{tab:chi2-lopdfs},
and compared to the corresponding results of the NNPDF2.1 NLO
set. The value of $\chi^2$ corresponds to the central PDF set (replica zero),
obtained as the average over replicas, while   
$\langle\chi^{2\,(k)}\rangle$ is the average over the replica sample 
of the  $\chi^2$ of each PDF
replica. We refer to Sect.~4 of Ref.~\cite{Ball:2011mu} for a more detailed
discussion of the various statistical indicators: here it will suffice
to say that all $\chi^2$ values are computed including the full
covariance matrix of each experiment, with normalization uncertainties
included using the method of Ref.~\cite{Ball:2009qv}.

 The fit quality is the same within uncertainties in all four
cases: the values of $\langle\chi^{2\,(k)}\rangle$ differ from each other by less than 
a standard deviation.
The fit with a larger value $\alpha_s\lp M_Z\rp=0.130$ seems
to be slightly favored, but the difference in $\chi^2$ as the value of
$\alpha_s$ is varied is so small
that we have not pursued further the option of also using NLO running
of the strong coupling.  
The behaviour of the fit when the 
momentum sum rule (MSR)  is not imposed is interesting: while the global fit
quality is the same as in  the fit with the MSR,  the
$\chi^2$ of individual experiments changes significantly: the fit
quality improves for some  sets (like for example HERA), but relaxing the MSR leads to a 
worse description of the hadronic data. 

Note that we can fit simultaneously the
Drell-Yan and deep-inelastic data without having 
to rescale the Drell-Yan data  
as discussed in Sect.~\ref{sec:lochoices} (unlike
Ref.~\cite{Martin:2009iq}). Of course, such a rescaling would likely
lead to an improvement of the agreement of quality of the fit to
Drell-Yan  data. However, the ensuing PDFs would then be optimized for
use in conjunction with codes (Monte Carlo or otherwise) in which
similar corrections are also included. Optimizing LO PDFs in view of
their use with some specific code such as a Monte Carlo event
generator,  as was done in Ref.~\cite{Lai:2009ne}, is an 
interesting task; however, we will not pursue it here, where we are
rather mostly interested in constructing PDFs based on pure LO theory, with
the clear limitations that this implies.

In Table~\ref{tab:chi2-lopdfs} we also give the value of the momentum integral
\be
\label{eq:momint}
\lc M \rc \equiv  \int_0^1 dx \, x\Sigma\lp x,Q^2\rp+ 
\int_0^1 dx \, x g\lp x,Q^2\rp  \ ,
\ee
for each of LO PDF sets. These are determined at the starting scale
$Q^2_0=2$~GeV$^2$, but note that
the momentum integral $\lc M\rc$ does not depend on scale. A
discussion of the behaviour of the momentum integral at LO, NLO and
NNLO will be given in Sect.~\ref{sec:msr} below.

\begin{figure}[t]
\begin{center}
\epsfig{width=0.99\textwidth,figure=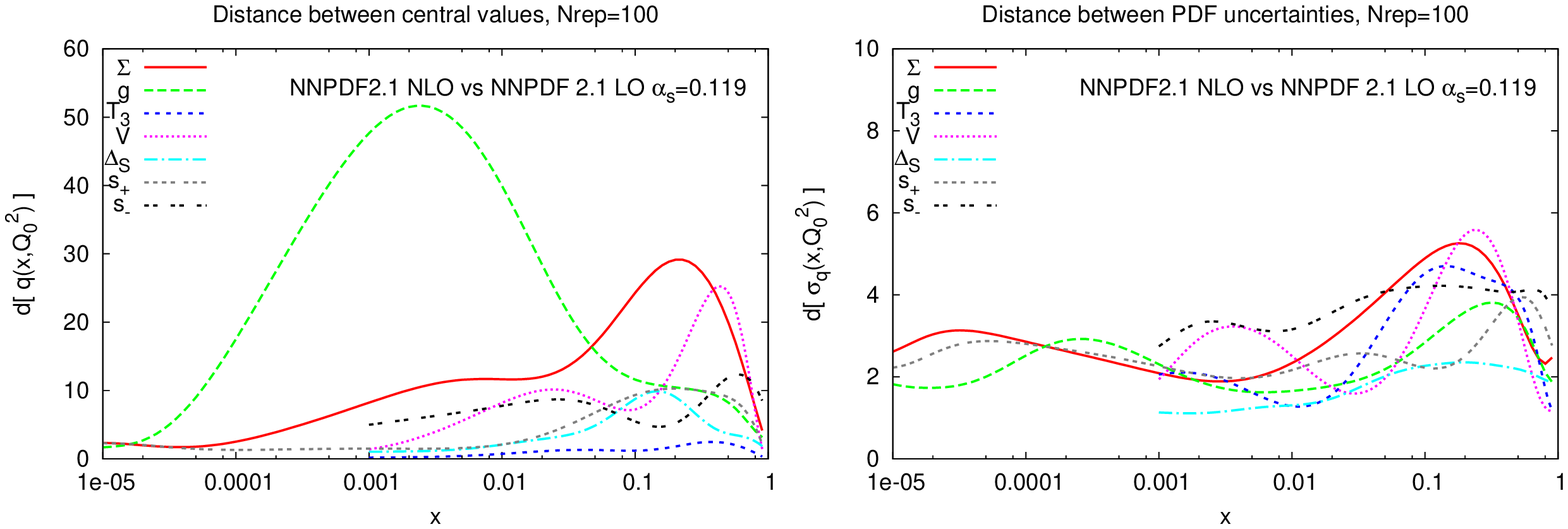}
\caption{\small Distances between the reference LO and NLO NNPDF2.1
  sets. Here and in subsequent figures in this section, the left plot
  shows the distance between central values, while the right plot
  shows the distance between the uncertainties. Both have
  $\alpha_s(M_z)=0.119$.  \label{fig:nlo-vs-lo-distances}}
\end{center}
\end{figure}

In summary, while we do find a non-negligible deterioration in fit
quality in comparison to the NLO fit, we do not find that this can be
improved by either relaxing the momentum sum rule or changing the
value of $\alpha_s$. Preliminary investigations using the NLO running
of $\alpha_s$ also did not show significant improvements in fit
quality. We did, however, find a significant improvement in fit
quality if the positivity constraint on PDFs is removed: the $\chi^2$
of the LO fit then becomes only about $10$\% higher than in the NLO
case. The price to pay for this is that the gluon becomes rather
negative at large $x$.  However, negative LO PDFs are not acceptable,
as they might lead to negative cross-sections; therefore we have not
pursued this possibility further.

\begin{figure}[t]
\begin{center}
\epsfig{width=0.48\textwidth,figure=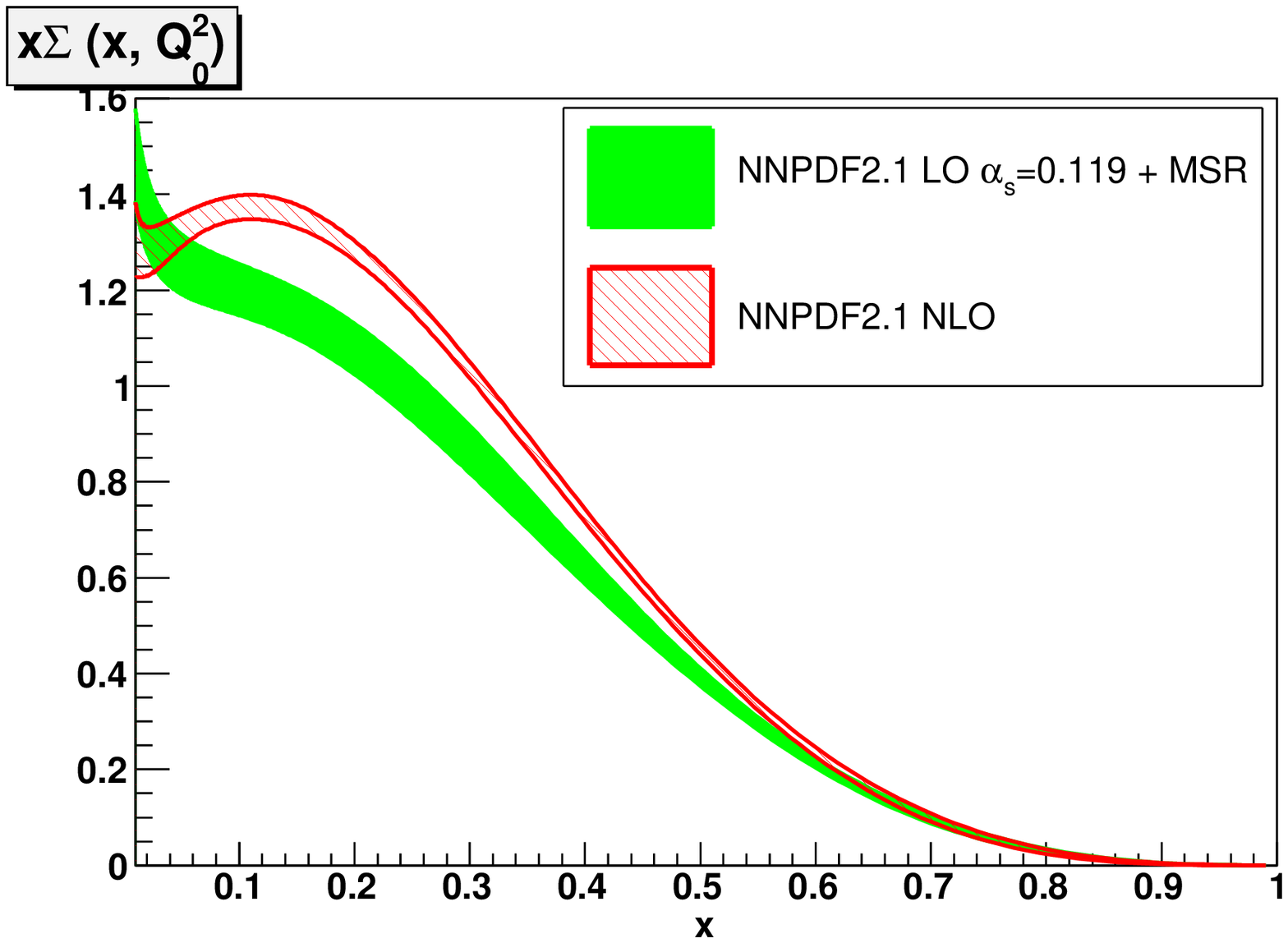}
\epsfig{width=0.48\textwidth,figure=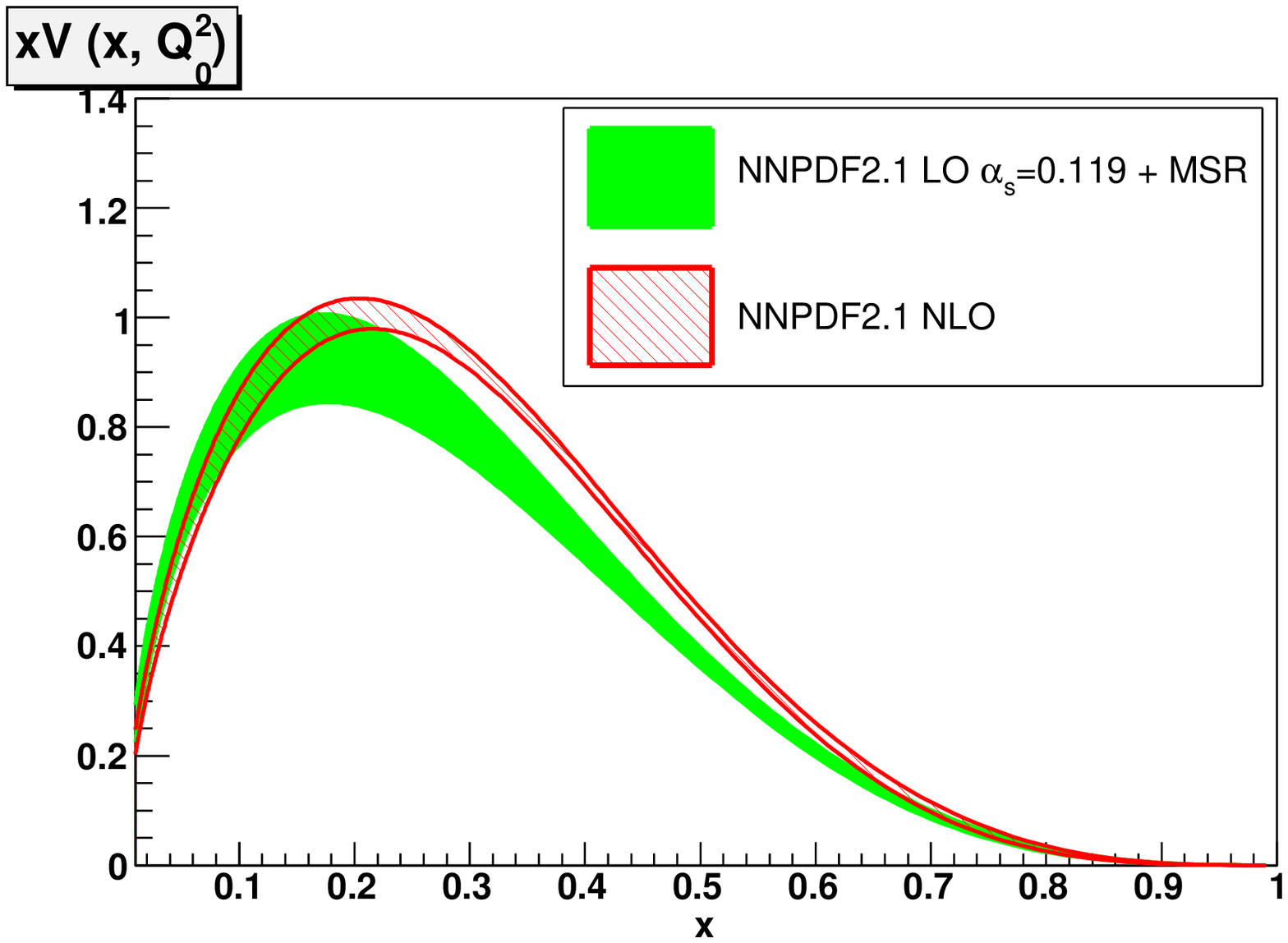}
\epsfig{width=0.48\textwidth,figure=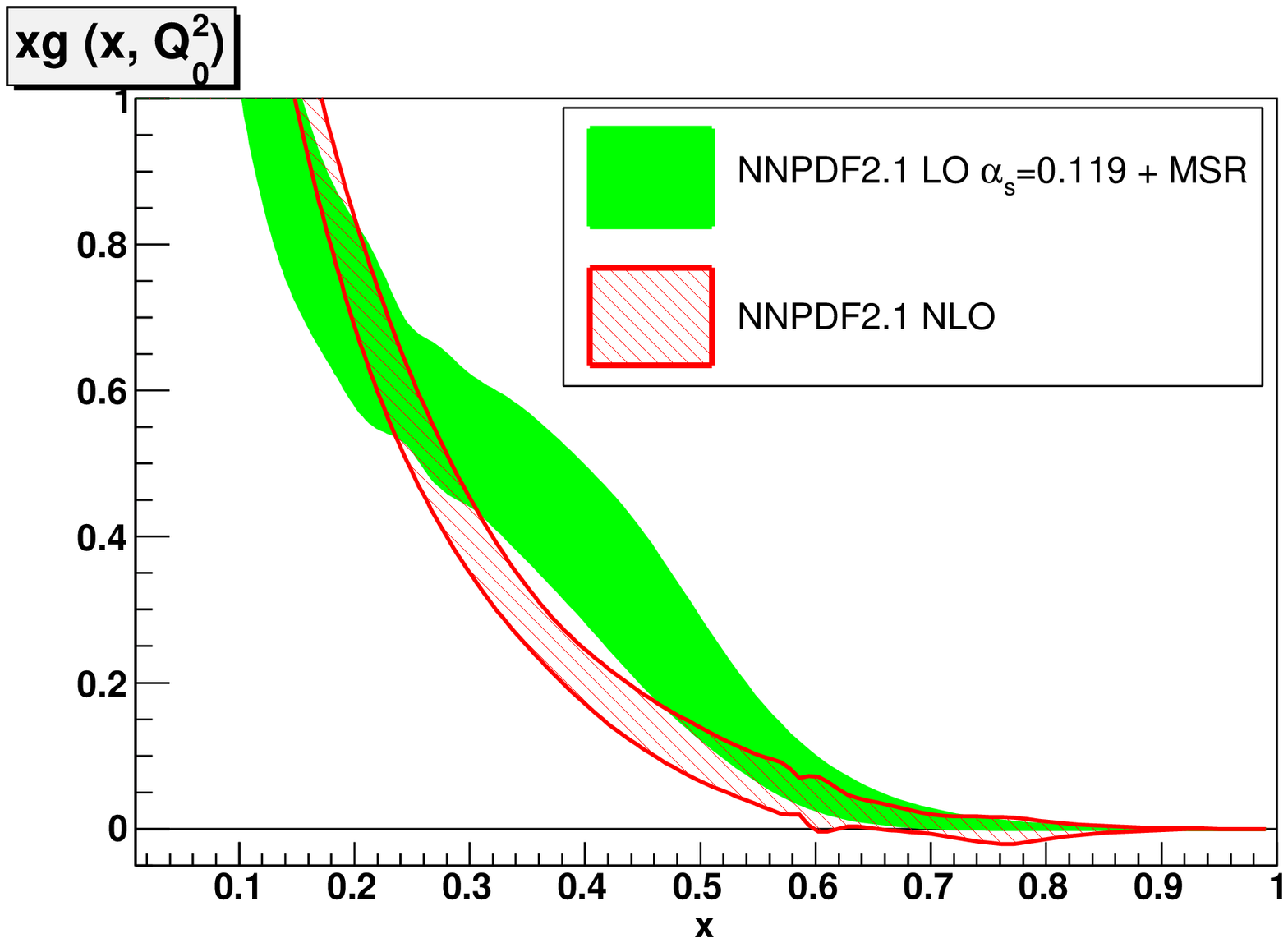}
\epsfig{width=0.48\textwidth,figure=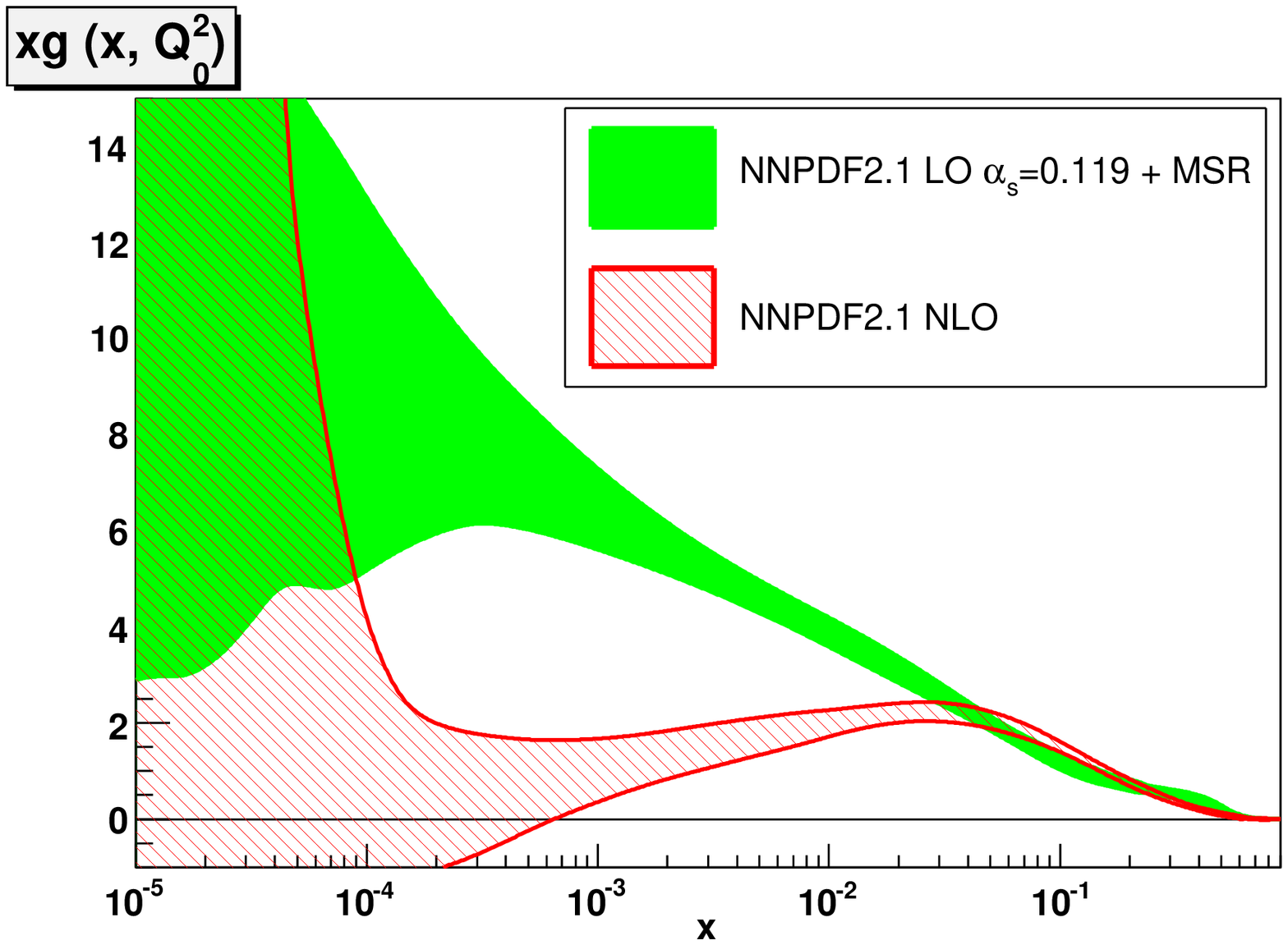}
\caption{\small Comparison of the quark singlet, valence and gluon
  distributions for the pair of PDF fits whose distances are plotted in
  Fig.~\ref{fig:nlo-vs-lo-distances}. 
\label{fig:lo-vs-nlo}} 
\end{center}
\end{figure}

\begin{figure}[t]
\begin{center}
\epsfig{width=0.99\textwidth,figure=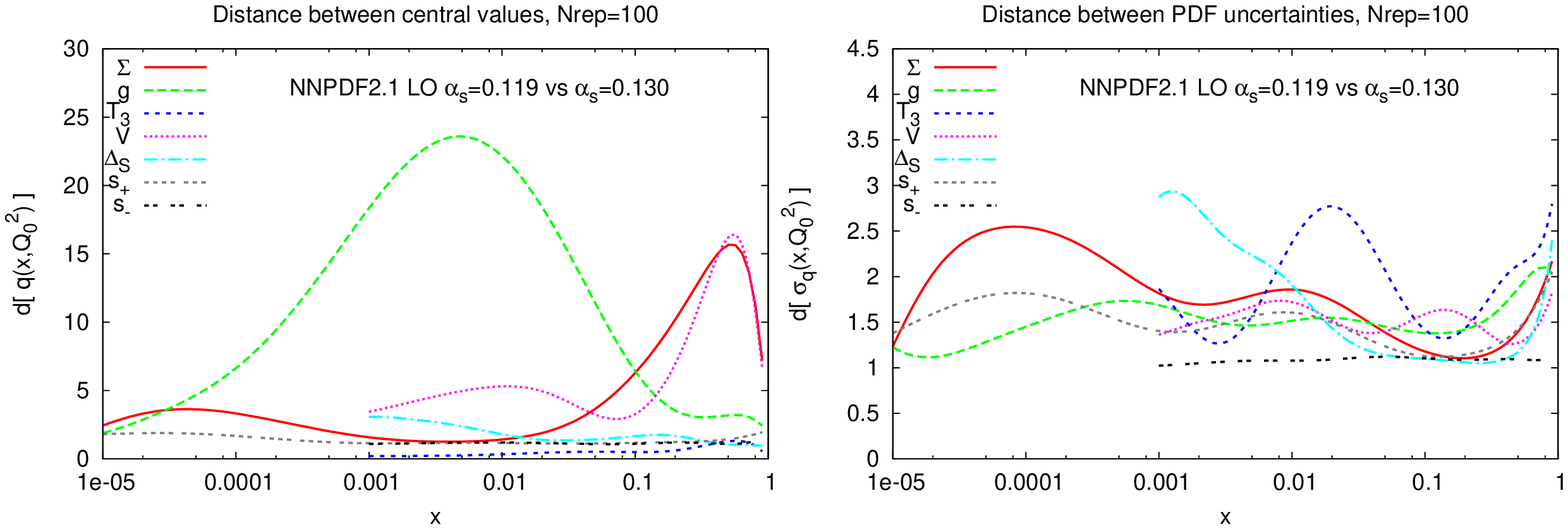}
\caption{\small Distances between the NNPDF2.1 LO sets
with $\alpha_s$=0.119 and $\alpha_s$=0.130.
\label{fig:lo-vs-loas0130-distances}} 
\end{center}
\end{figure}

\subsection{Parton distributions}
\label{sec:lopdfres}

We now compare the four LO PDF sets with each other, with 
NLO PDFs, and with other available LO PDF sets.
We will compute the distance 
 between central values and uncertainties of the various pairs of PDFs
 which are being compared, defined as in Appendix~A of 
Ref.~\cite{Ball:2010de}. Recall that with $N_{\rm rep}=100$ replicas a
distance $d=1$ corresponds to central values which differ by
$\frac{1}{\sqrt{50}}\sigma$, with $\sigma$ the sum in quadrature of
the uncertainties of the two sets.  If the sets which are being
compared are statistically equivalent, then all distances are of order
one, while if they are statistically inequivalent but consistent at
the $n$ sigma level, then distances are of order of $d\sim 7 n$.

\begin{figure}[t]
\begin{center}
\epsfig{width=0.48\textwidth,figure=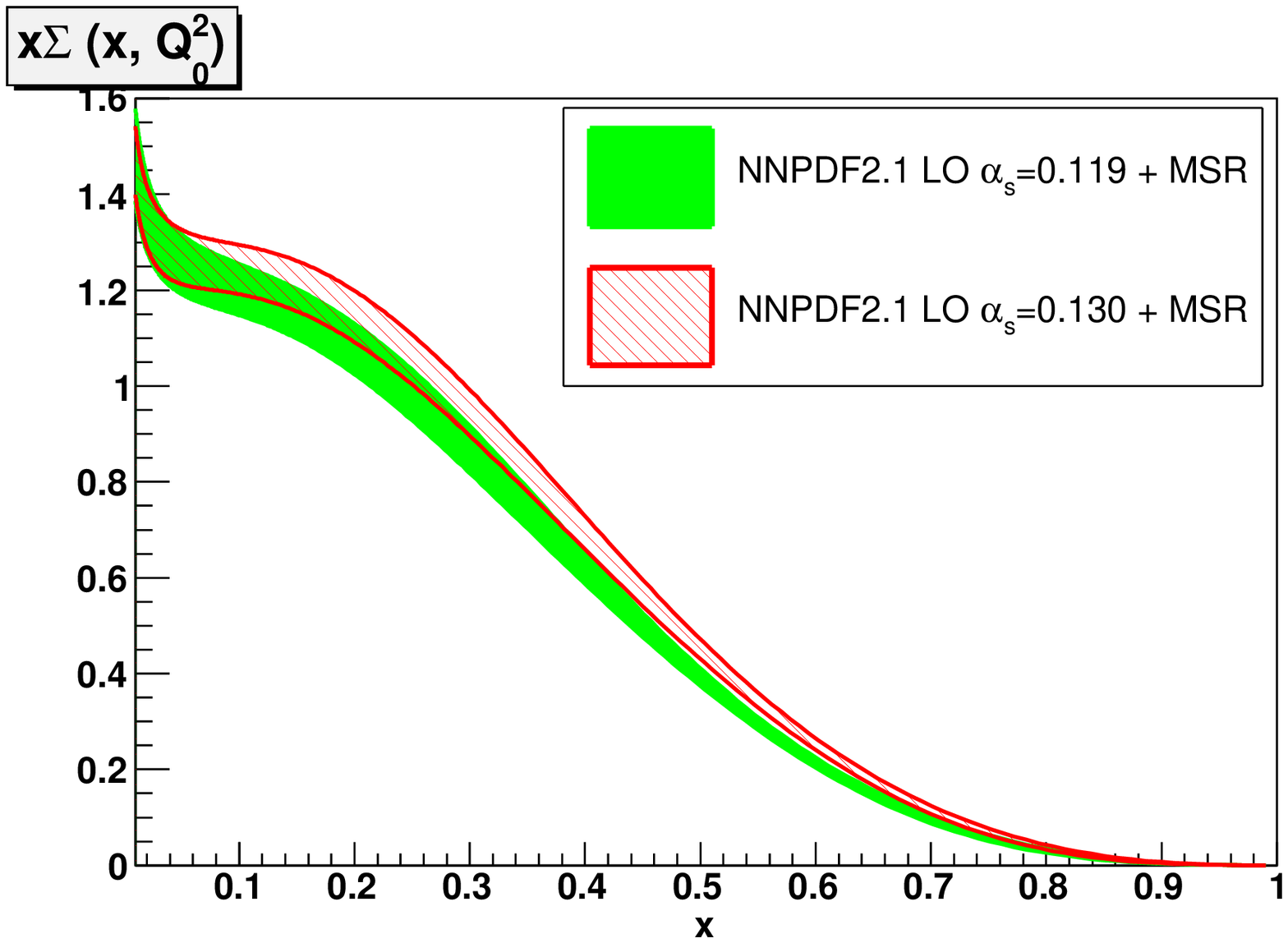}
\epsfig{width=0.48\textwidth,figure=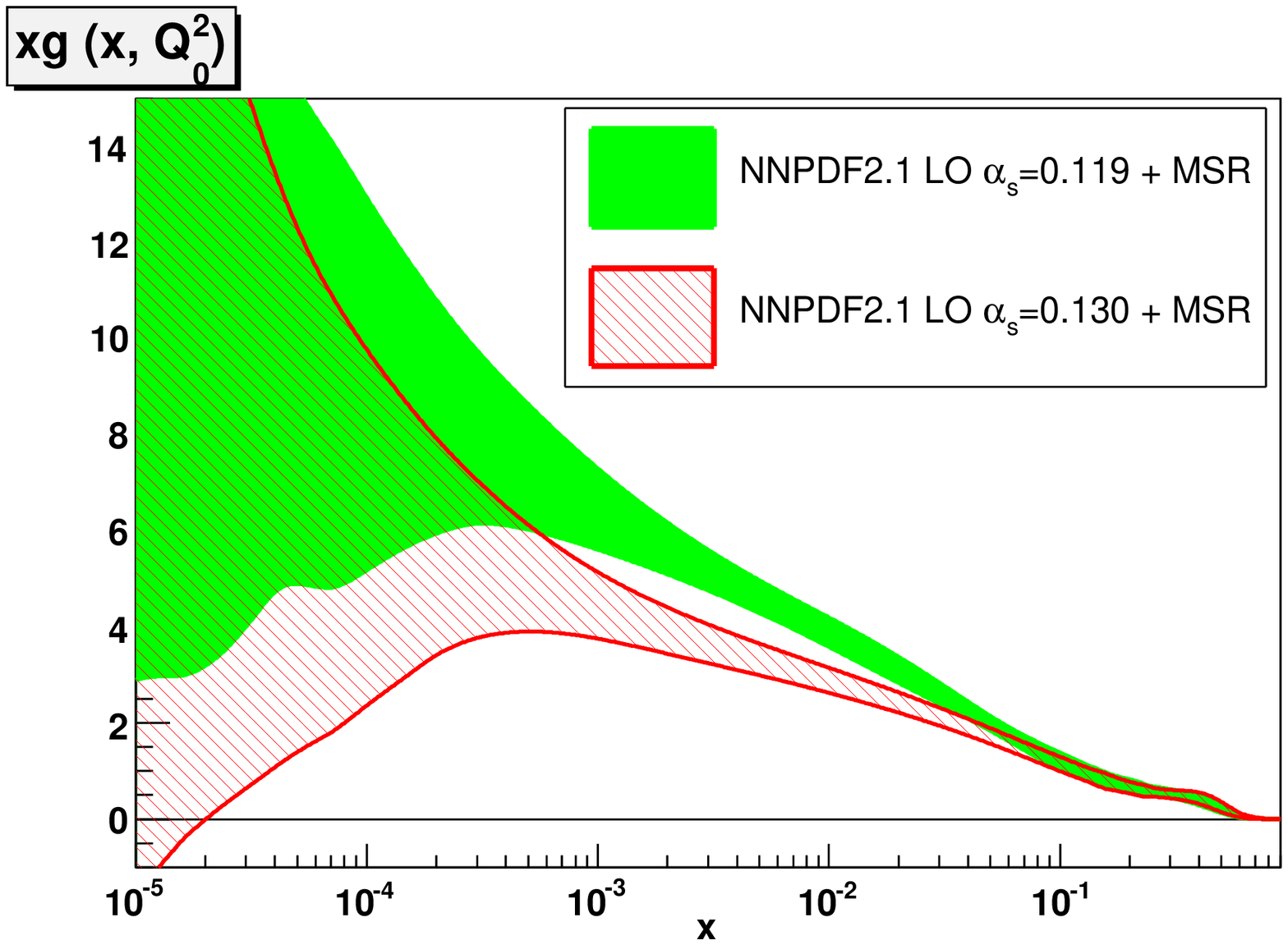}
\caption{\small Comparison of the quark singlet, and gluon
  distributions for the pair of PDF fits whose distances are plotted in
  Fig.~\ref{fig:lo-vs-loas0130-distances}.
\label{fig:lo-vs-loas0130}} 
\end{center}
\end{figure}

\begin{figure}[t]
\begin{center}
\epsfig{width=0.99\textwidth,figure=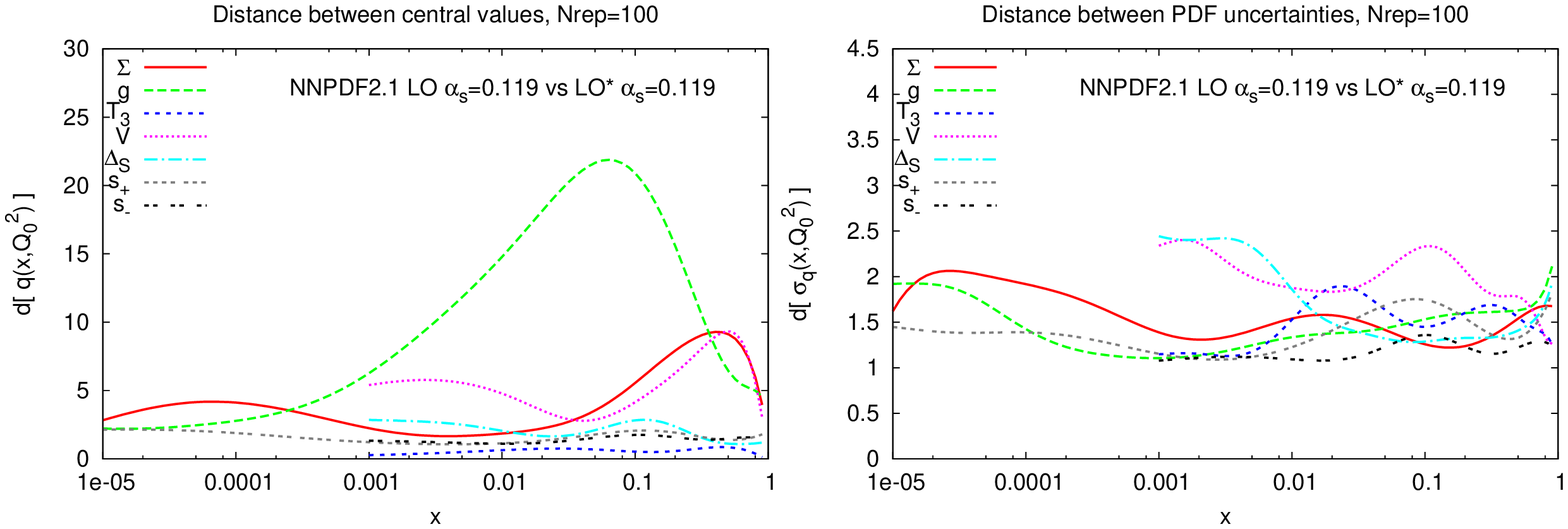}
\caption{\small Distances between the NNPDF2.1 LO and LO$^*$ sets
with $\alpha_s$=0.119.
\label{fig:lo-vs-lostar-distances}} 
\end{center}
\end{figure}

We begin by comparing the NNPDF2.1 LO set with $\alpha_s$=0.119, which
we take as the LO reference, to the reference NNPDF2.1 NLO set.  The
corresponding distances are plotted in
Fig.~\ref{fig:nlo-vs-lo-distances}, while the singlet,
 valence  and gluon PDFs are compared in Fig.~\ref{fig:lo-vs-nlo}.  A
full comparison of all LO, NLO and NNLO PDFs will be presented in
Sect.~\ref{sec:pdfcomp}. It is clear from
Fig.~\ref{fig:nlo-vs-lo-distances} that LO and NLO PDF uncertainties,
though clearly not statistically equivalent, are consistent at the
one sigma level: this shows that these uncertainties essentially
reflect the uncertainty of the underlying data, which are the same in
the two PDF determinations. On the other hand, 
central values differ by many sigma: this means that, as already
mentioned, the difference between LO and NLO PDFs is much larger than
the uncertainty on either, and thus the dominant uncertainty on LO PDF
is the theoretical uncertainty due to the lack of inclusion of higher
order corrections. 

The largest shift from LO to NLO, more than five
times larger than the PDF uncertainty, is observed for the gluon at
medium-small $x$ ($10^{-4} \le x \le 0.05$), consistent with the fact
that the gluon decouples from LO observables, but also the large $x$
quark singlet and valence distributions change by more than three
sigma.
Generally, the LO gluon
is larger than the NLO one.  
However, for $x\le 10^{-4}$, where
there are no data to constrain the fits, the LO and NLO gluons become
consistent within the large PDF uncertainties. At larger $x$, the LO
and NLO gluons are quite similar and compatible within the respective
uncertainties.  The LO quark is rather smaller (by more than one
sigma) than the NLO one for large $x>0.1$, but it becomes compatible
with it at the one sigma level for smaller $x$.
Finally, the light sea
and strangeness asymmetries are minimally affected and quite close at
LO and NLO.

It is interesting to observe that the missing large NLO
$K$-factors in Drell-Yan data should enhance the LO quark
distributions in comparison to the NLO ones: the fact that they end up
being instead either smaller or comparable suggests that the Drell-Yan
data actually have relatively little effect on the LO fit, other than through 
the determination of the $\bar u-\bar d$ light flavor asymmetry. This is less
sensitive to the $K$ factors (being mostly determined by a cross-section
ratio), and indeed turns out to be almost the same at LO and NLO.

\begin{figure}[t]
\begin{center}
\epsfig{width=0.48\textwidth,figure=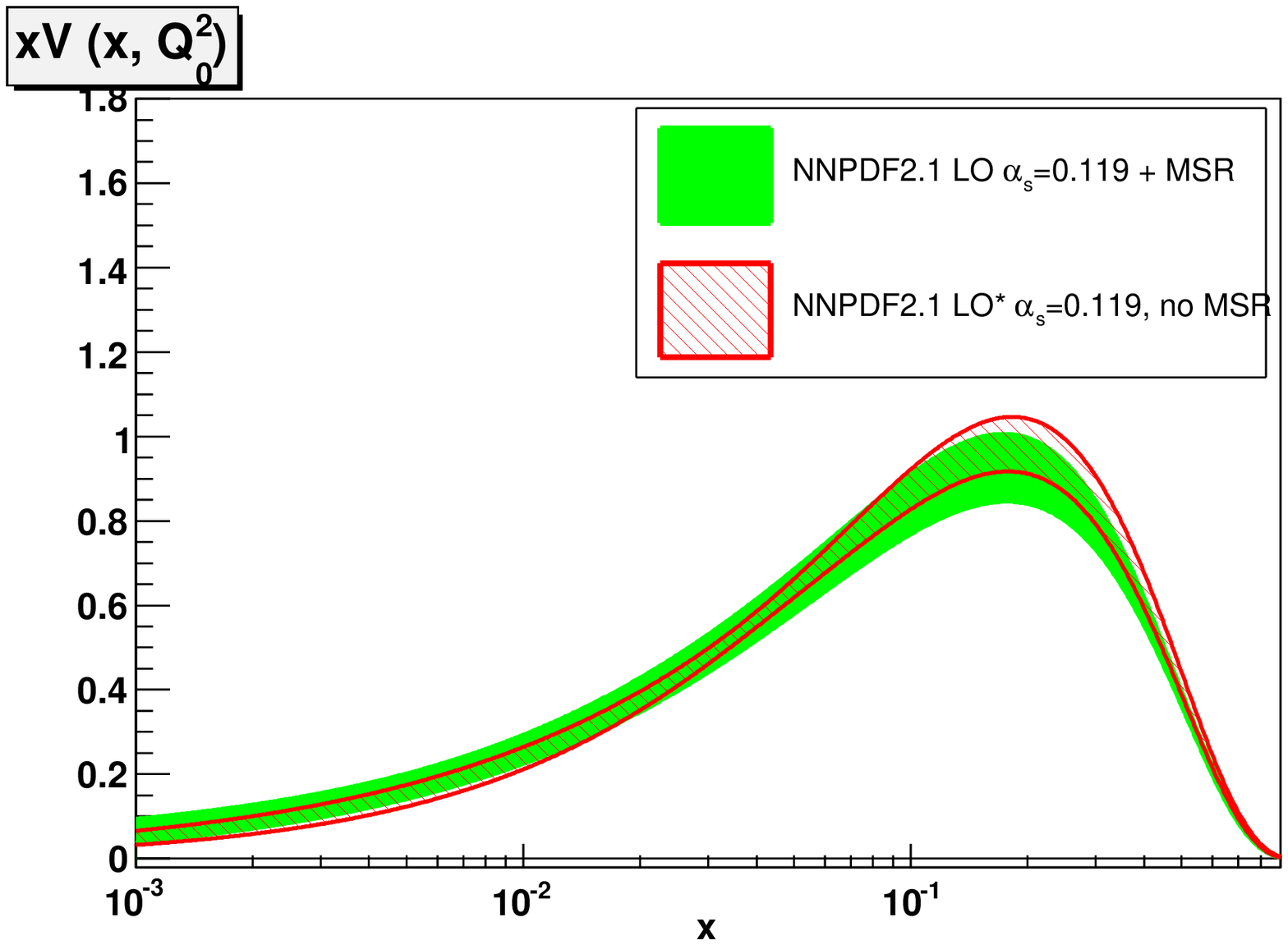}
\epsfig{width=0.48\textwidth,figure=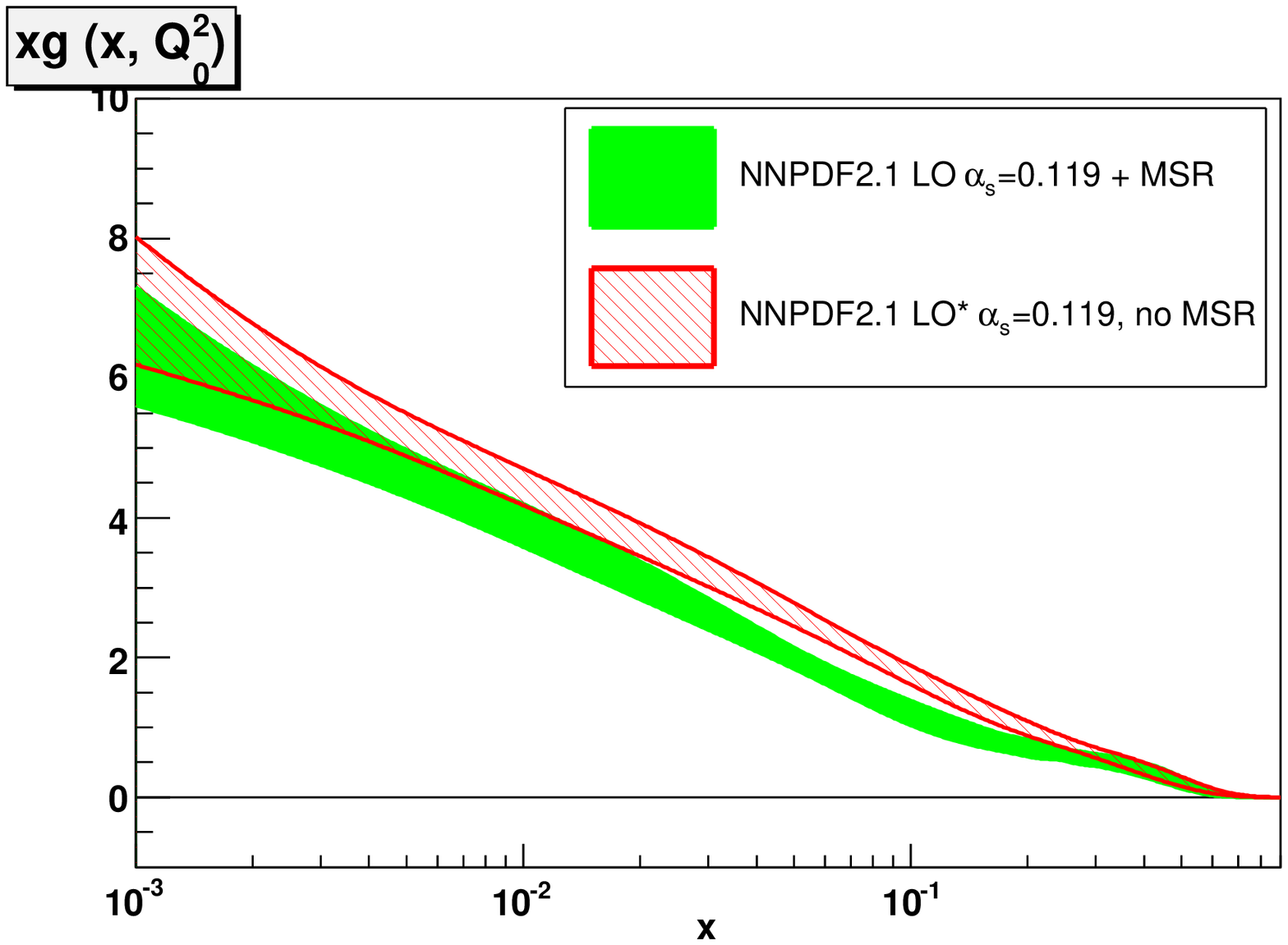}
\caption{\small Comparison of the valence and  gluon
  distributions for the pair of PDF fits whose distances are plotted in
  Fig.~\ref{fig:lo-vs-lostar-distances}.
\label{fig:lo-vs-lostar}} 
\end{center}
\end{figure}

Next we compare the various LO PDF sets to each other.
First we compare the two LO sets which differ in value of the strong
coupling,  $\alpha_s(M_Z)=0.119$ vs.
$\alpha_s(M_Z)=0.130$. The larger value of the strong coupling, when
evolved  down to a scale $Q^2\sim 10$~GeV$^2$ using LO evolution, 
leads to a value of
$\alpha_s$ close to that preferred by data in this region. Hence, the
larger value leads to a better description of  scaling
violations at low scale, and conversely, as it is apparent 
from Table~\ref{tab:chi2-lopdfs}.

\begin{figure}[t]
\begin{center}
\epsfig{width=0.48\textwidth,figure=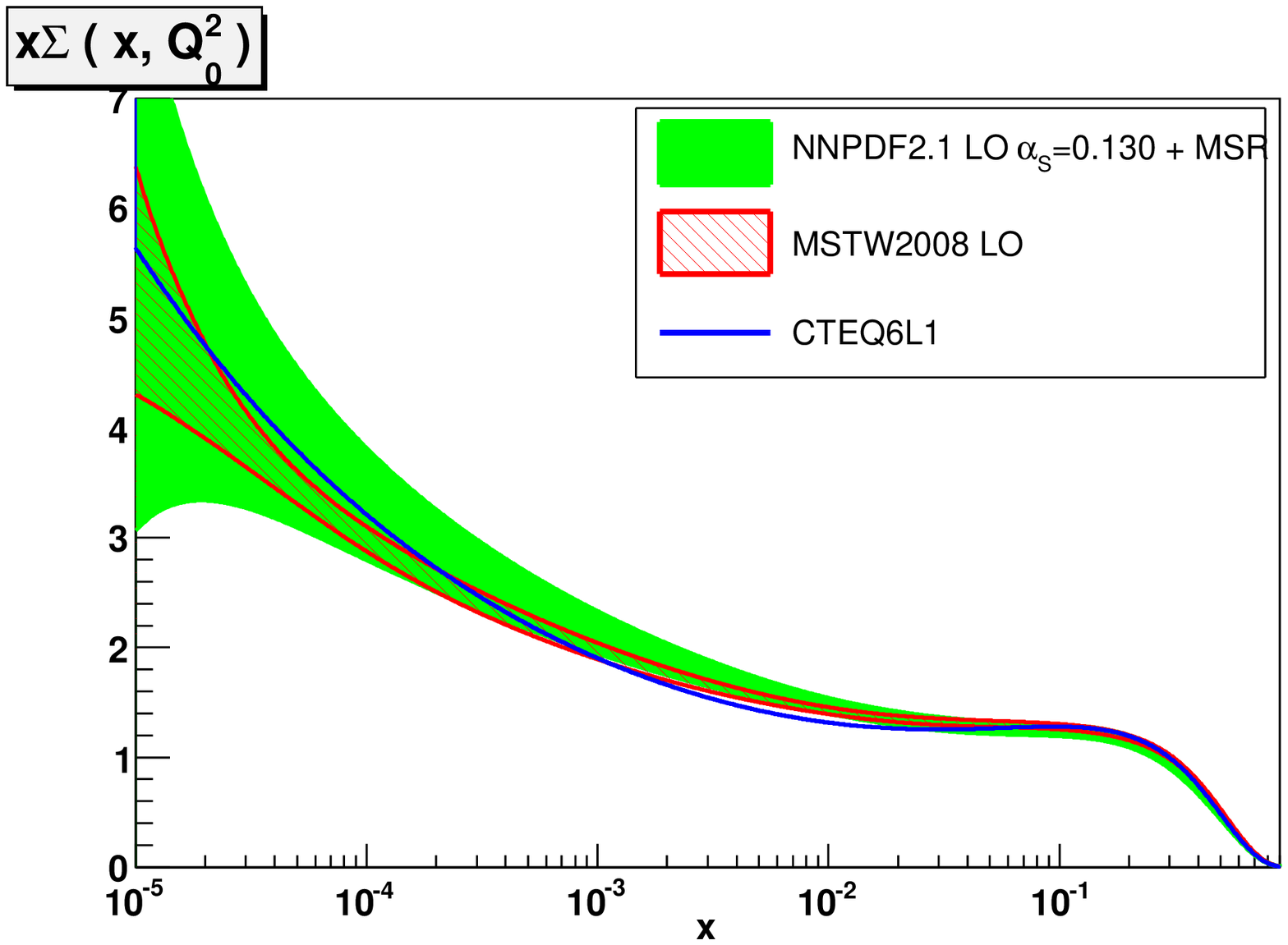}
\epsfig{width=0.48\textwidth,figure=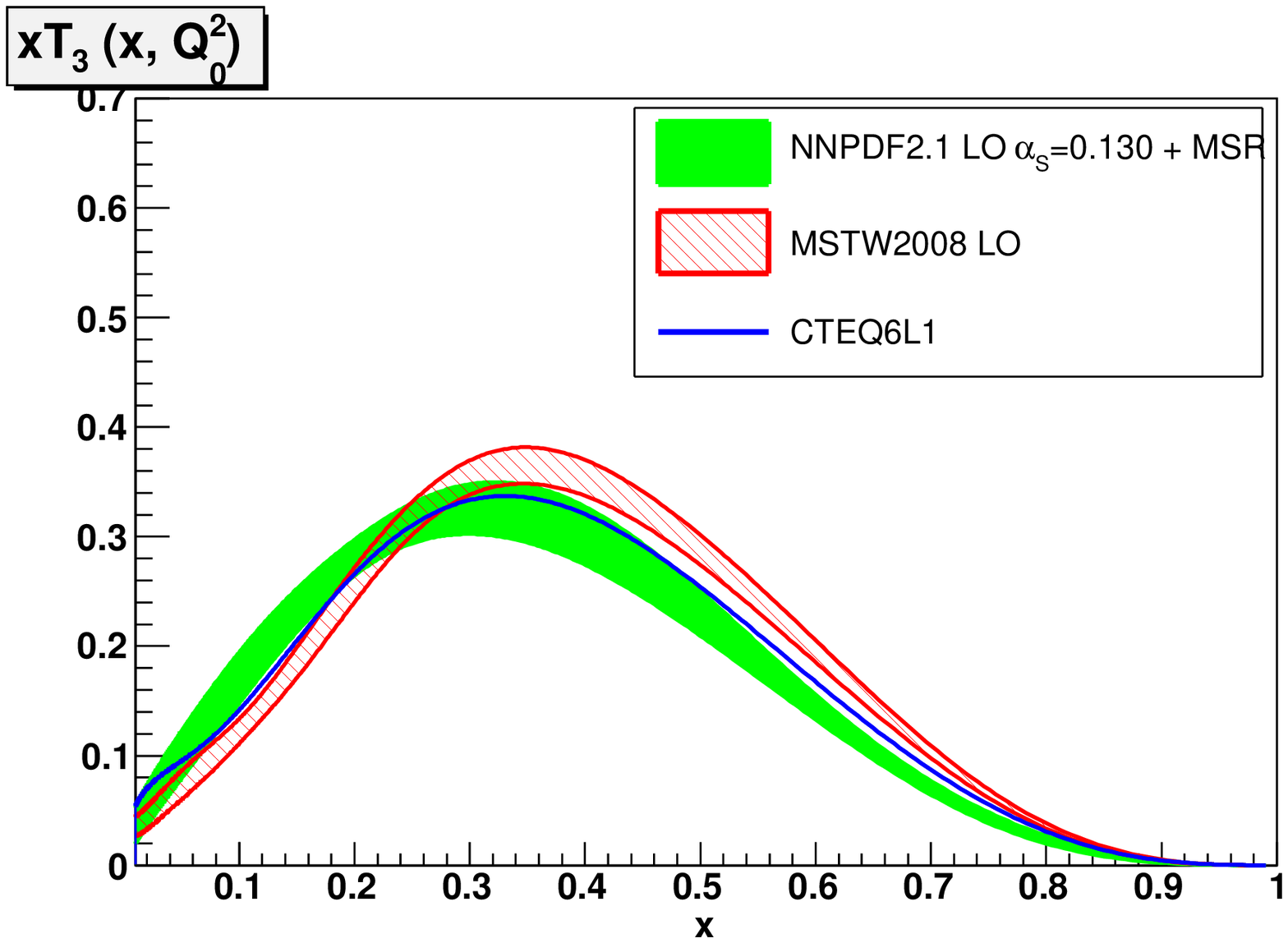}
\epsfig{width=0.48\textwidth,figure=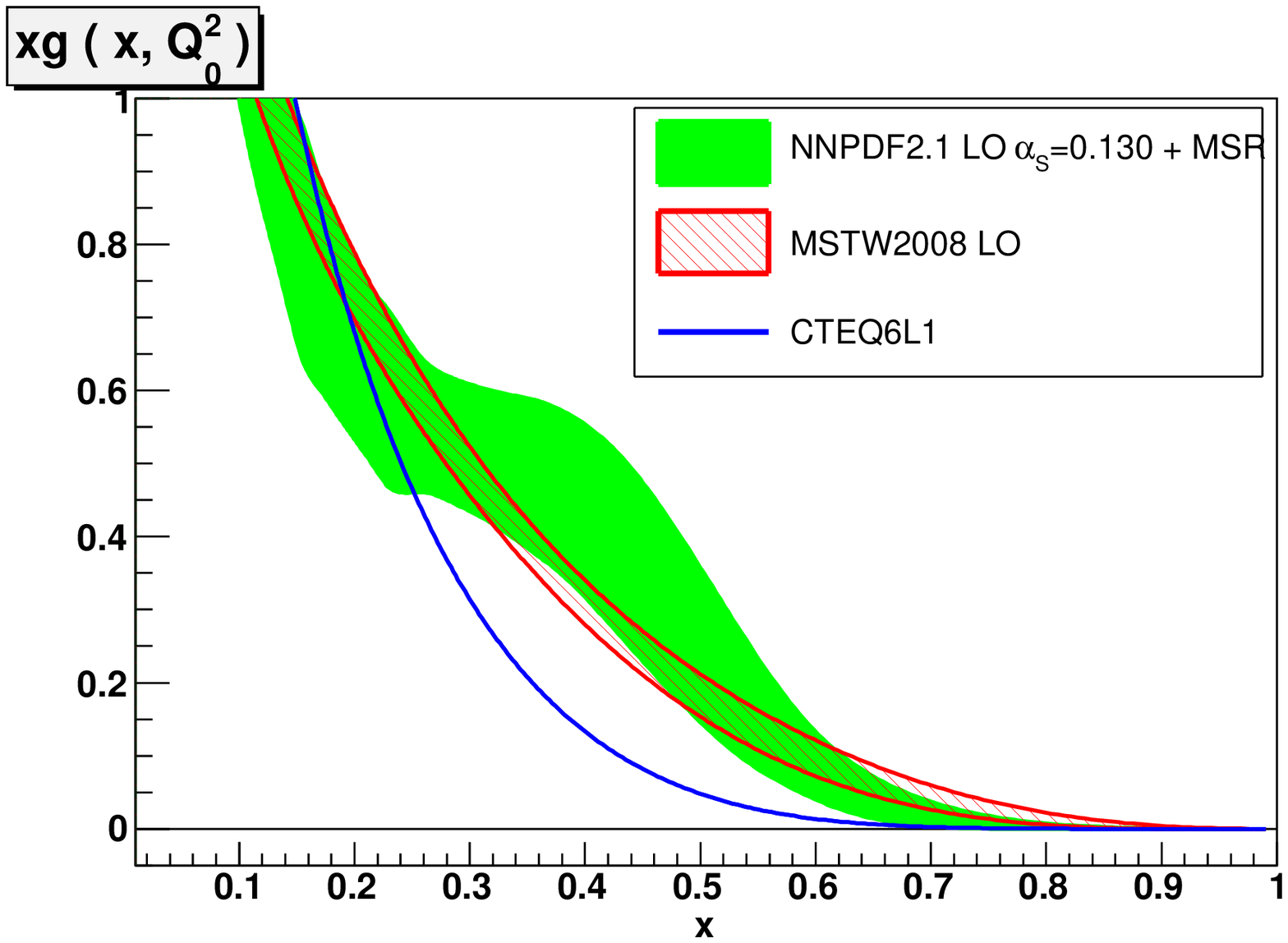}
\epsfig{width=0.48\textwidth,figure=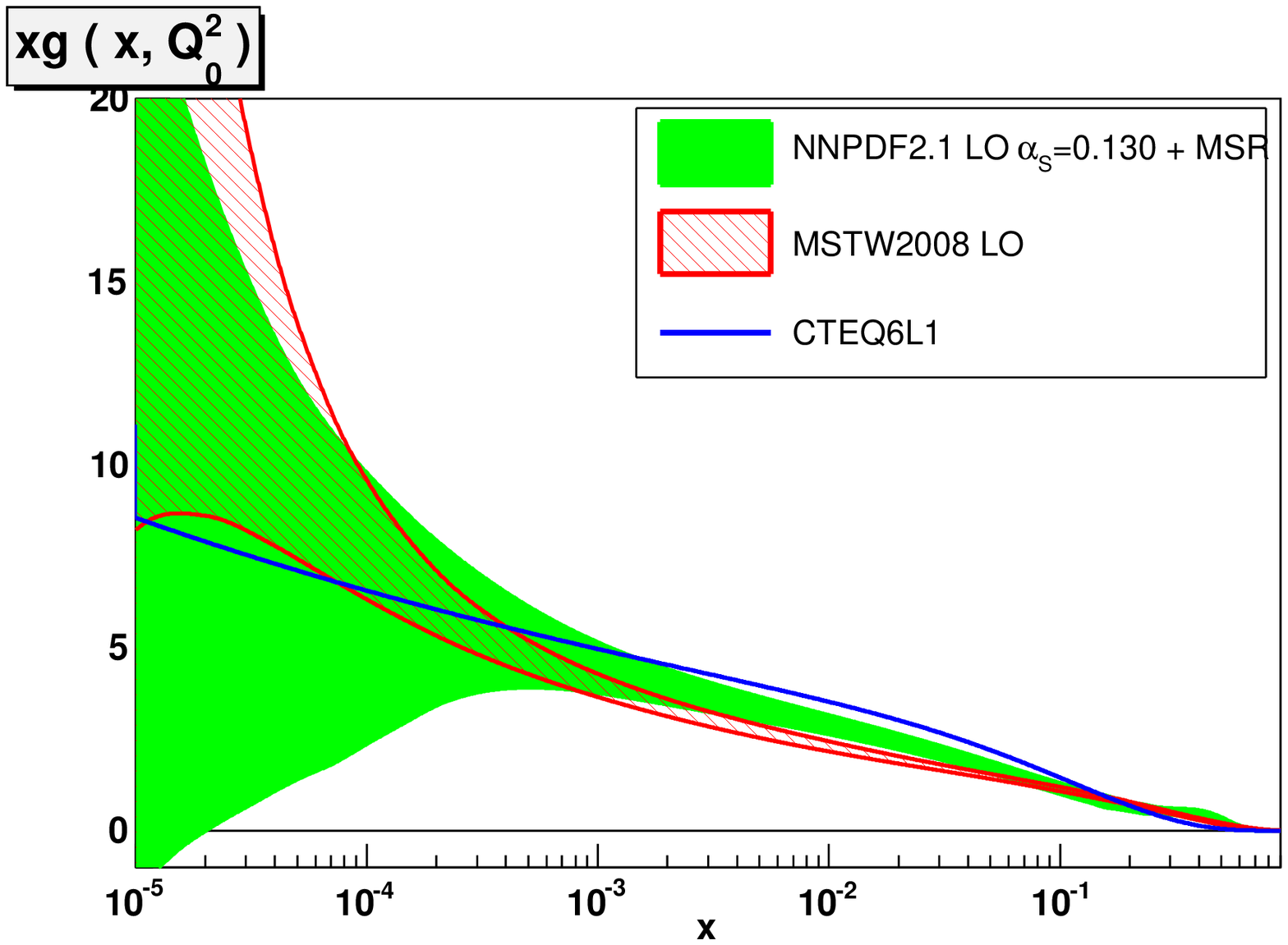}
\caption{\small Comparison of  LO PDFs: the 
quark
singlet, triplet and gluon PDFs are shown for the NNPDF2.1, MSTW08 and
CTEQ6L1 sets.
\label{fig:globalcomp1}} 
\end{center}
\end{figure}
The distances for central values and uncertainties between the LO fits
with different $\alpha_s$ are plotted in 
Fig.~\ref{fig:lo-vs-loas0130-distances}. The only PDF which is
significantly affected by the value of $\alpha_s$ is the gluon, which,
as shown in Fig.~\ref{fig:lo-vs-loas0130}, becomes smaller at
medium-small $x$ and thus, by the momentum sum rule, somewhat larger at
large $x$ when $\alpha_s$ is increased. This makes the LO gluon
with larger $\alpha_s$ closer to the NLO gluon. However, the shift as
$\alpha_s$ is varied in this range is comparable to the PDF
uncertainty. Also, the large $x$ singlet and valence
quark PDFs increase somewhat when
$\alpha_s$ is raised, especially at  large $x\sim0.3$,
where a shift of about two sigma is observed.

The effect of relaxing the momentum sum rule is studied by
comparing the LO and LO* sets. Those with  $\alpha_s(M_Z)=0.119$ are
compared in Fig.~\ref{fig:lo-vs-lostar-distances}, where the distance
between them is displayed. The main difference is seen in the
medium $x$ gluon, as shown in
Fig.~\ref{fig:lo-vs-lostar}:  the LO* gluon is rather larger  than the
LO one. However,  the central values for all quark PDFs are very close
to the  standard LO ones.

In conclusion, we compare the NNPDF2.1 LO PDFs  to other available
LO sets.
First, we compare the  NNPDF2.1 LO set with $\alpha_s=0.130$
to 
MSTW08~LO~\cite{Martin:2009iq} ($\alpha_s=0.139$)  and 
CTEQ6L1~\cite{Pumplin:2002vw} ($\alpha_s=0.130$) in Fig.~\ref{fig:globalcomp1}.
Differences are especially large for the gluon distribution, both at
small 
and large $x$, and for the isospin triplet distribution and large $x$, 
though differences between the NNPDF and MSTW sets are mostly compatible
with the large uncertainties, while the difference between CTEQ and
other sets is more
difficult to quantify more precisely because CTEQ LO PDFs
come without an uncertainty
estimate.  
\begin{figure}[t]
\begin{center}
\epsfig{width=0.48\textwidth,figure=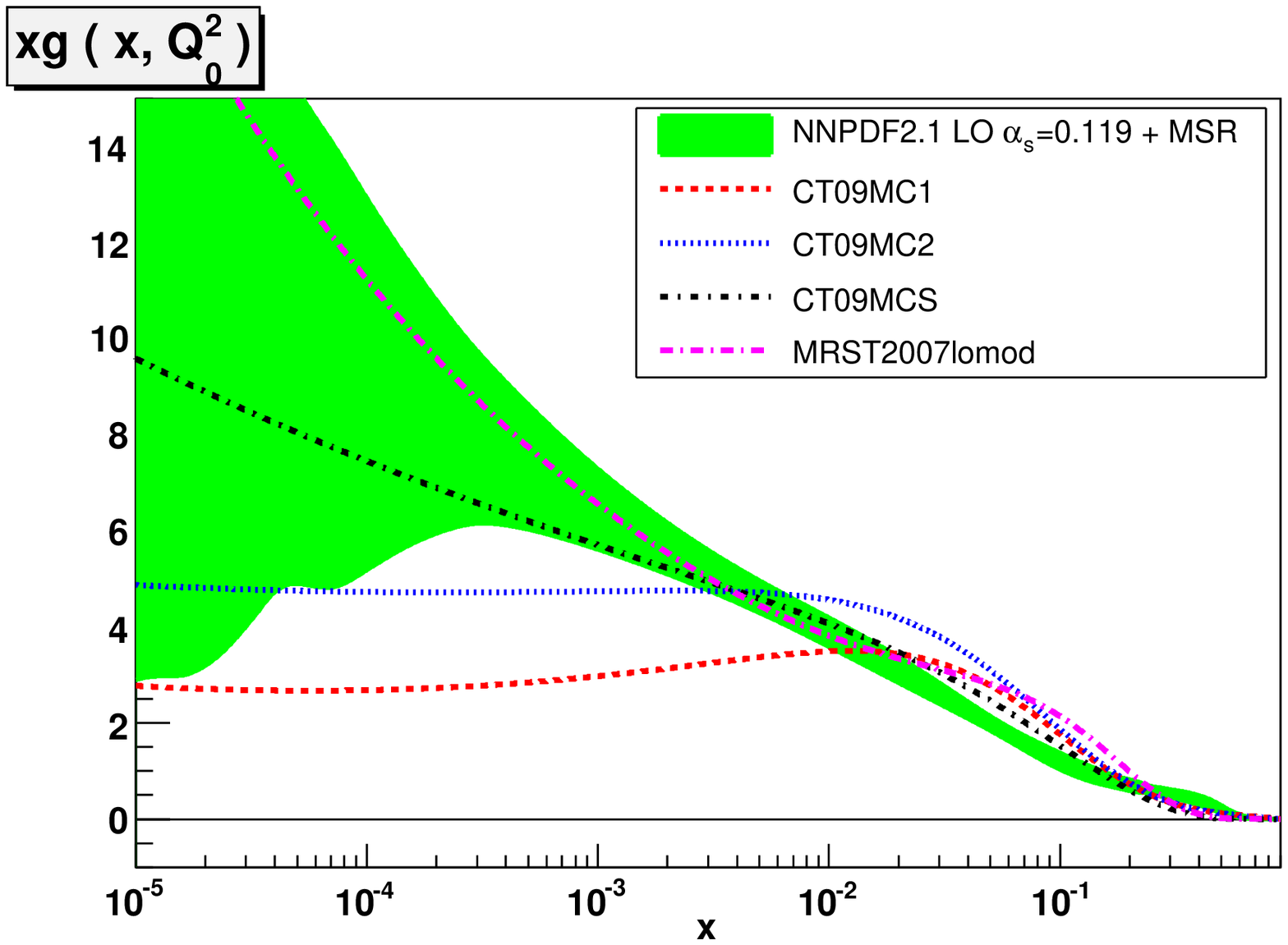}
\epsfig{width=0.48\textwidth,figure=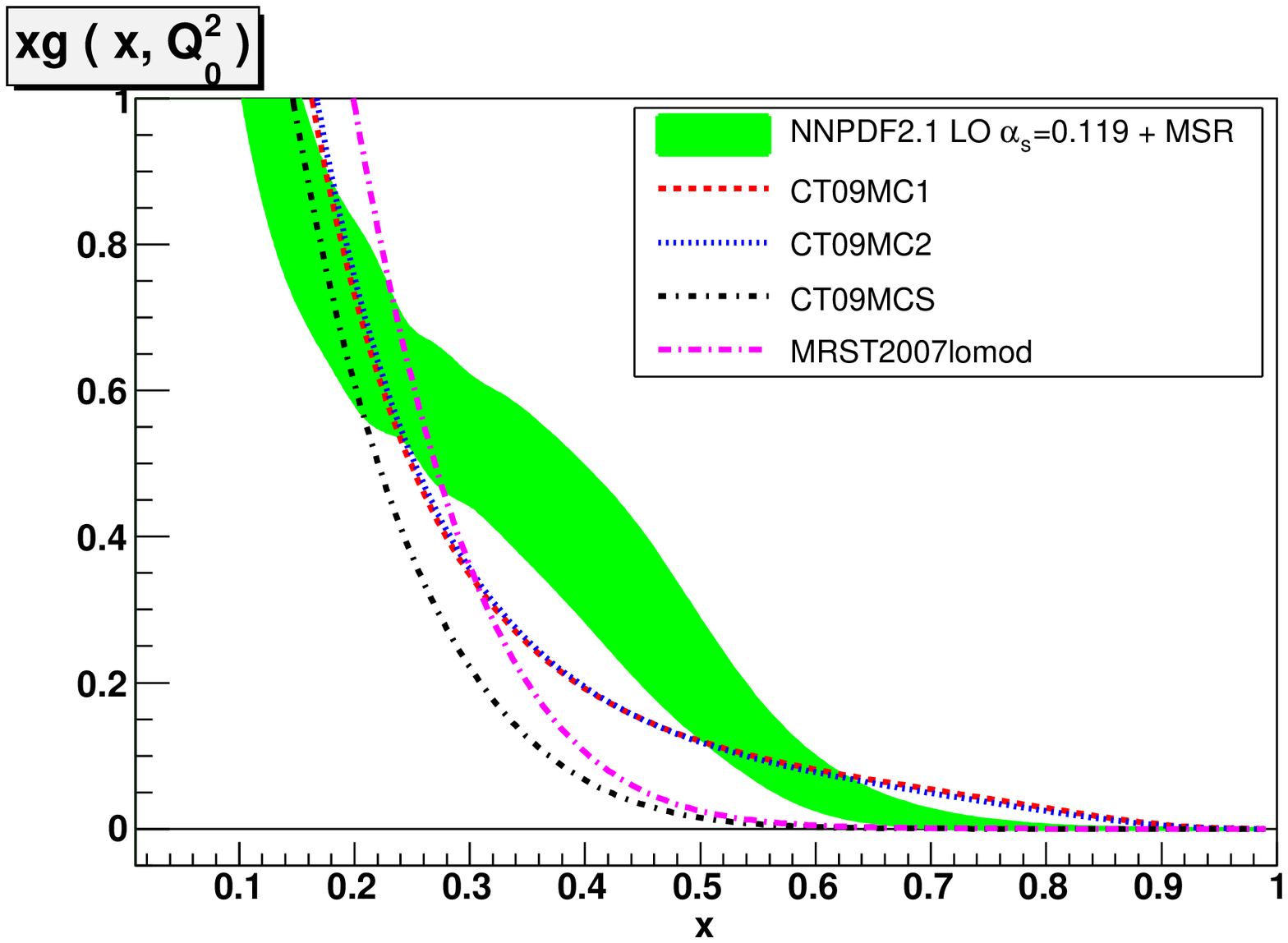}
\epsfig{width=0.48\textwidth,figure=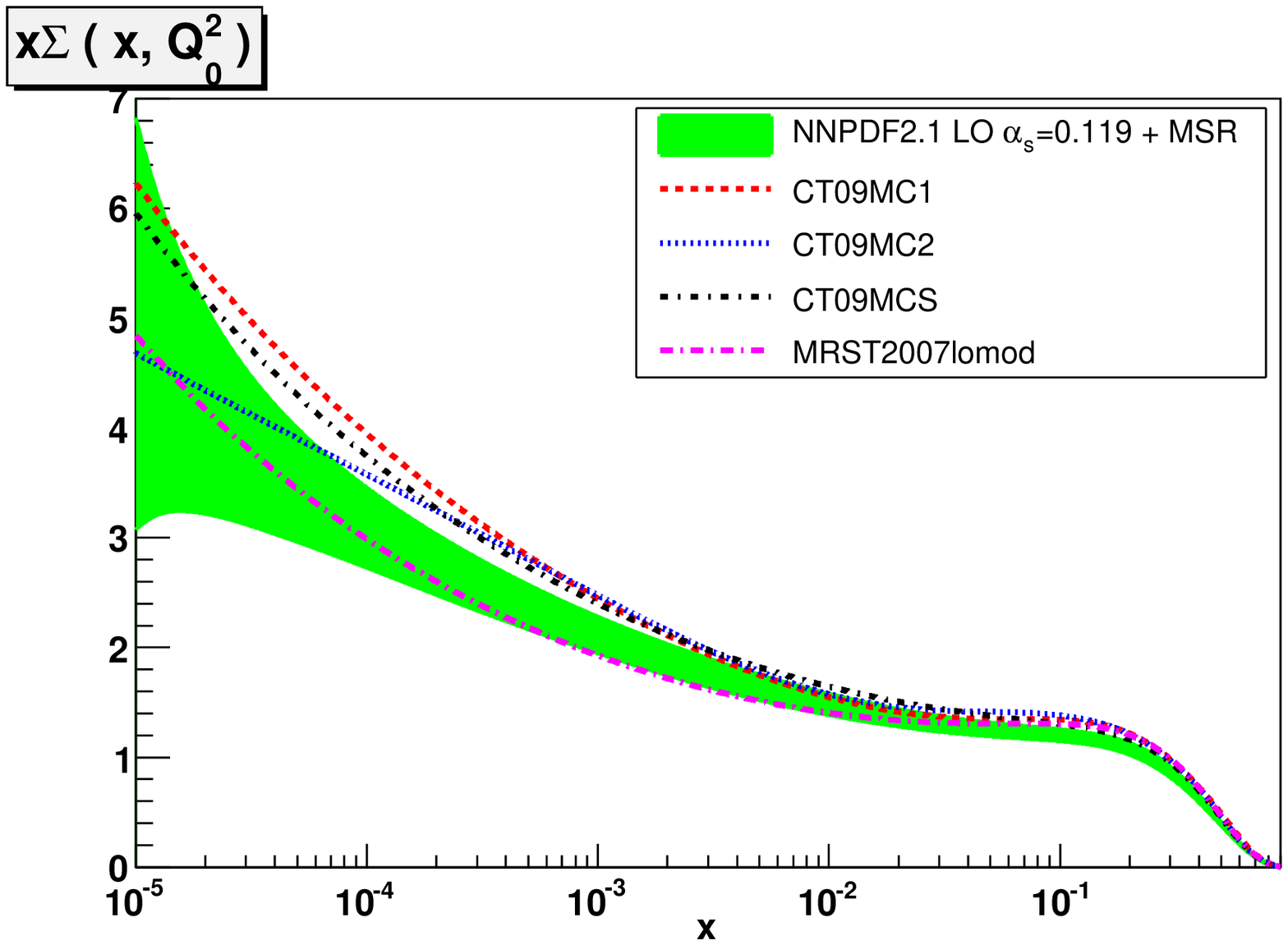}
\epsfig{width=0.48\textwidth,figure=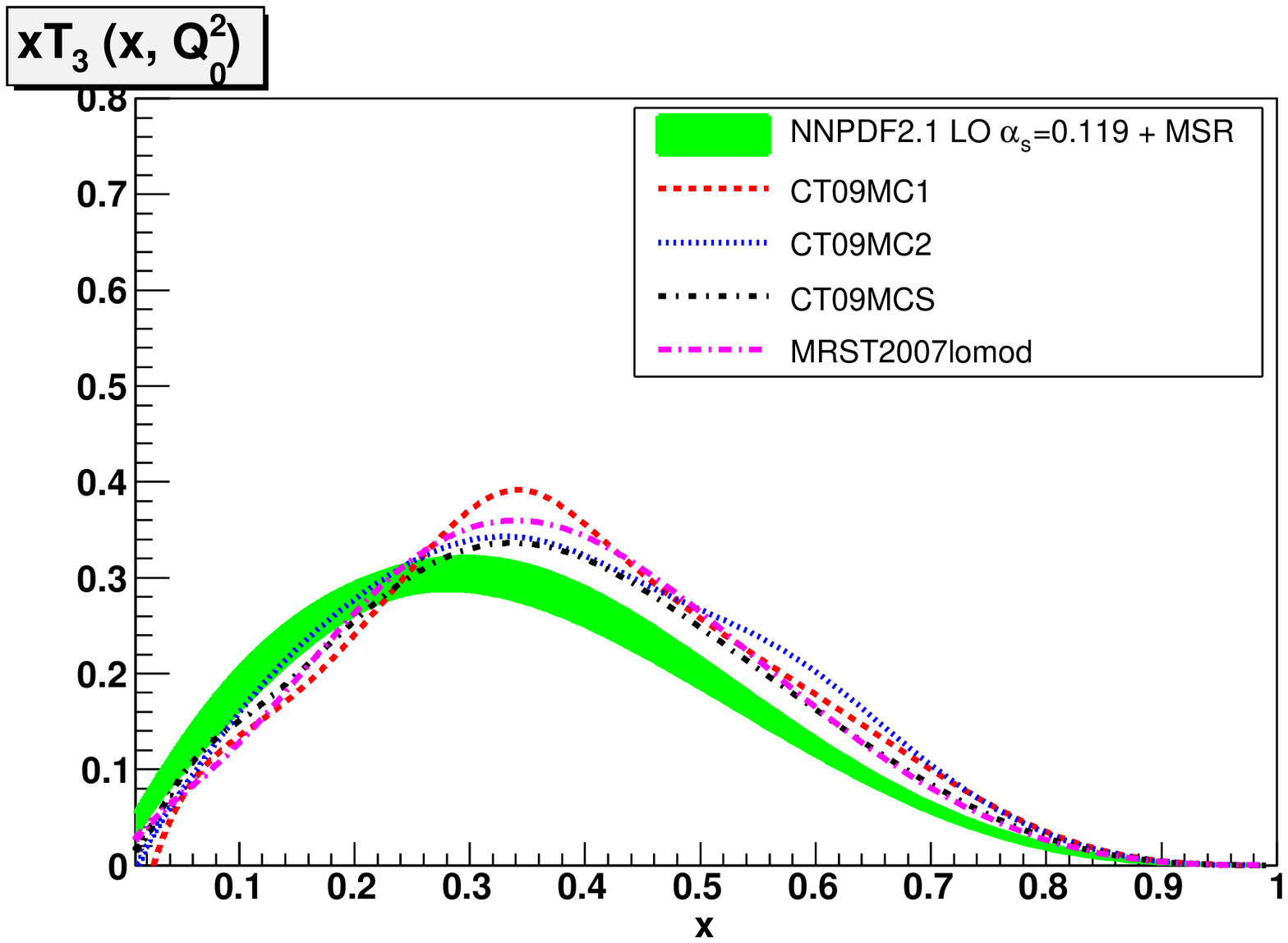}
\caption{\small Comparison of the NNPDF2.1 LO PDF to modified LO PDF sets:
MRST2007lomod, CT09MC1,
CT09MC2 and CT09MCS.
\label{fig:globalcomp2}} 
\end{center}
\end{figure}

Finally we compare with the modified LO
PDF sets MRST2007lomod~\cite{Sherstnev:2007nd},
and with the dedicated  Monte Carlo
sets of the CTEQ/TEA collaboration~\cite{Lai:2009ne}, CT09MC1, 
CT09MC2 and CT09MCS. The MRST2007lomod set is obtained relaxing the
momentum sum rule and using two-loop running of $\alpha_s$, with
$\alpha_s(M_z)=0.121$. The CT sets
are based on an LO QCD analysis framework of data which,  on top of the
standard global dataset used for the NLO PDF determination, also
includes a set of LHC pseudo-data generated using NLO PDFs. 
The normalization of the LO calculation for each pseudo-data
set is allowed to float to reach the best agreement with the NLO cross-section. 
The CT09MCS is extracted from an analysis in which the two-loop strong coupling is used 
and the momentum sum rule is imposed during the fit. The CT09MC1 and CT09MC2 are fits 
in which the momentum sum rule is relaxed and one- and two-loop expressions for $\alpha_s$ 
are used respectively. All these sets are compared to the default
NNPDF2.1 LO set in Fig.~\ref{fig:globalcomp2}.
Differences between these modified PDF sets are significant, and typically
larger than the difference between the NNPDF2.1 LO and LO$^*$ sets,
consistent with the fact that they are based on different
methodologies and assumptions.

%% file: sec-results.tex
\section{Next-to-next-to-leading order parton distributions}
\label{sec:results}

Next-to-next-to-leading order PDFs are mostly of interest for their
use in the computation of standard candle processes such as $W$, $Z$,
top and Higgs production at hadron colliders. In this section we will 
discuss the statistical features
of the NNLO fit, then present the NNPDF2.1 NNLO PDFs and compare them
to other available NNLO sets.
The implications of the NNPDF2.1 NNLO set for LHC observables
are discussed in Sect.~\ref{sec:pheno}.

\subsection{Statistical features}

\begin{table}
\centering
\begin{tabular}{|c|c|}
\hline 
$\chi^{2}_{\tot}$ &      1.16 \\
$\la E \ra \pm \sigma_{E} $   &   $2.22 \pm 0.07$     \\
$\la E_{\rm tr} \ra \pm \sigma_{E_{\rm tr}}$&    $2.19 \pm 0.09$     \\
$\la E_{\rm val} \ra \pm \sigma_{E_{\rm val}}$&     $2.27 \pm 0.10$      \\
$\la{\rm TL} \ra \pm \sigma_{\rm TL}$   &  $\lp 17 \pm 7\rp\,10^{3}$    \\
\hline
$\la \chi^{2(k)} \ra \pm \sigma_{\chi^{2}} $  &  $1.23\pm 0.05$    \\
\hline
 $\la \sigma^{(\exp)}
\ra_{\dat}$(\%) & 11.9 \\
 $\la \sigma^{(\net)}
\ra_{\dat} $(\%)&  3.2 \\
\hline
 $\la \rho^{(\exp)}
\ra_{\dat}$ & 0.18 \\
 $\la \rho^{(\net)}
\ra_{\dat}$& 0.53 \\
\hline
\end{tabular}
\caption{\small \label{tab:estfit1} Table of statistical estimators
  for the NNPDF2.1 NNLO fit with $N_{\rm rep}=
1000$ replicas.  }
\end{table}

{
\begin{table}
\centering
\small
\begin{tabular}{|c||c|c||c|c|c|c|c|}
\hline 
Experiment    & $\chi^2 $& $\chi^2_{\rm nlo}$& $\la E\ra $   & $\la \sigma^{(\exp)}\ra_{\dat}$(\%) & $\la \sigma^{(\net)}\ra_{\dat}$(\%) & $\la \rho^{(\exp)}\ra_{\dat}$ & $\la \rho^{(\net)}\ra_{\dat}$ \\
\hline 
NMC-pd    & 0.93   & 0.97   &  1.98 & 1.8 & 0.5   &  0.03    & 0.34  \\
\hline
NMC       & 1.63  & 1.73  & 2.67  & 5.0  & 1.8   &  0.16  & 0.75     \\
\hline
SLAC      & 1.01 & 1.27  & 2.05 & 4.4 & 1.8  &  0.31 & 0.78   \\
\hline
BCDMS &  1.32  & 1.24  & 2.38& 5.7  & 2.6  &  0.47 & 0.58     \\
\hline
HERAI-AV & 1.10 & 1.07  & 2.16 & 7.6  & 1.3   & 0.06  & 0.44     \\
\hline
CHORUS   & 1.12 & 1.15 & 2.18 & 15.0   & 3.5  & 0.08  & 0.37    \\
\hline
FLH108   & 1.26 & 1.37  & 2.25 & 72.1 &  4.8  & 0.65  & 0.68     \\
\hline
NTVDMN   & 0.49 & 0.47  & 1.74 & 21.0  & 14.0  & 0.04  & 0.64     \\
\hline
ZEUS-H2  & 1.31 & 1.29   & 2.33 &  14.0 & 1.3  &  0.28 & 0.55     \\
\hline
ZEUSF2C & 0.88 & 0.78  & 1.89 & 23.0  & 3.7  &  0.07 & 0.40    \\
\hline
H1F2C  &  1.46& 1.50  & 2.48 & 18.0  & 3.5  & 0.27  & 0.36     \\
\hline
DYE605  & 0.81  & 0.84   & 1.88 & 25.0  & 7.2   & 0.55  & 0.76     \\
\hline
DYE866  & 1.32  & 1.27  & 2.40 & 21.0 & 8.7  & 0.23  & 0.48    \\
\hline
CDFWASY & 1.65  &  1.86  & 2.80 & 6.0  & 4.3  & 0.52   & 0.61    \\
\hline
CDFZRAP &  2.12 & 1.65  & 3.21 & 12.0   & 3.6  & 0.82  & 0.67     \\
\hline
D0ZRAP   & 0.67 & 0.60  & 1.69 &10.0 & 3.0  & 0.54  & 0.70     \\
\hline
CDFR2KT  & 0.74 & 0.97   & 1.84 & 23.0  & 4.8  &  0.77 &  0.61    \\
\hline
D0R2CON  & 0.82 & 0.84  & 1.89 & 17.0 & 5.5  & 0.78  & 0.62    \\
\hline
\end{tabular}
\caption{\small \label{tab:estfit2} Same as Table \ref{tab:estfit1}
  for individual experiments. All estimators
have been obtained with
$N_{\rm rep}= 1000$ replicas. Note that
experimental uncertainties are always given in percentage. For
reference we also provide the NNPDF2.1 NLO 
 $\chi^2$ for the various experiments.  }
\end{table}
}

Statistical estimators for the NNPDF2.1 NNLO
fit are shown in Table~\ref{tab:estfit1} for the global fit and in
Table~\ref{tab:estfit2} for individual experiments, with, in the latter
case, the NLO $\chi^2$ values also
shown for comparison. While referring to
Refs.~\cite{Ball:2010de,Ball:2011mu,phystat,Ball:2009qv} 
for a detailed discussion of
statistical indicators and their meaning, we recall that
$\chi^{2}_{\tot}$ is computed comparing the
central (average) NNPDF2.1 fit  to the original experimental data, 
$\la \chi^{2(k)} \ra$ is computed comparing to the data 
each NNPDF2.1 replica and averaging over replicas, while $\la E \ra$
is the quantity which is minimized, i.e. it coincides with the
$\chi^{2}$ computed comparing each NNPDF2.1 replica to the data
replica it is fitted to, with the three values given corresponding to
the total, training, and validation datasets. 
\begin{figure}[t]
\begin{center}
\epsfig{width=0.49\textwidth,figure=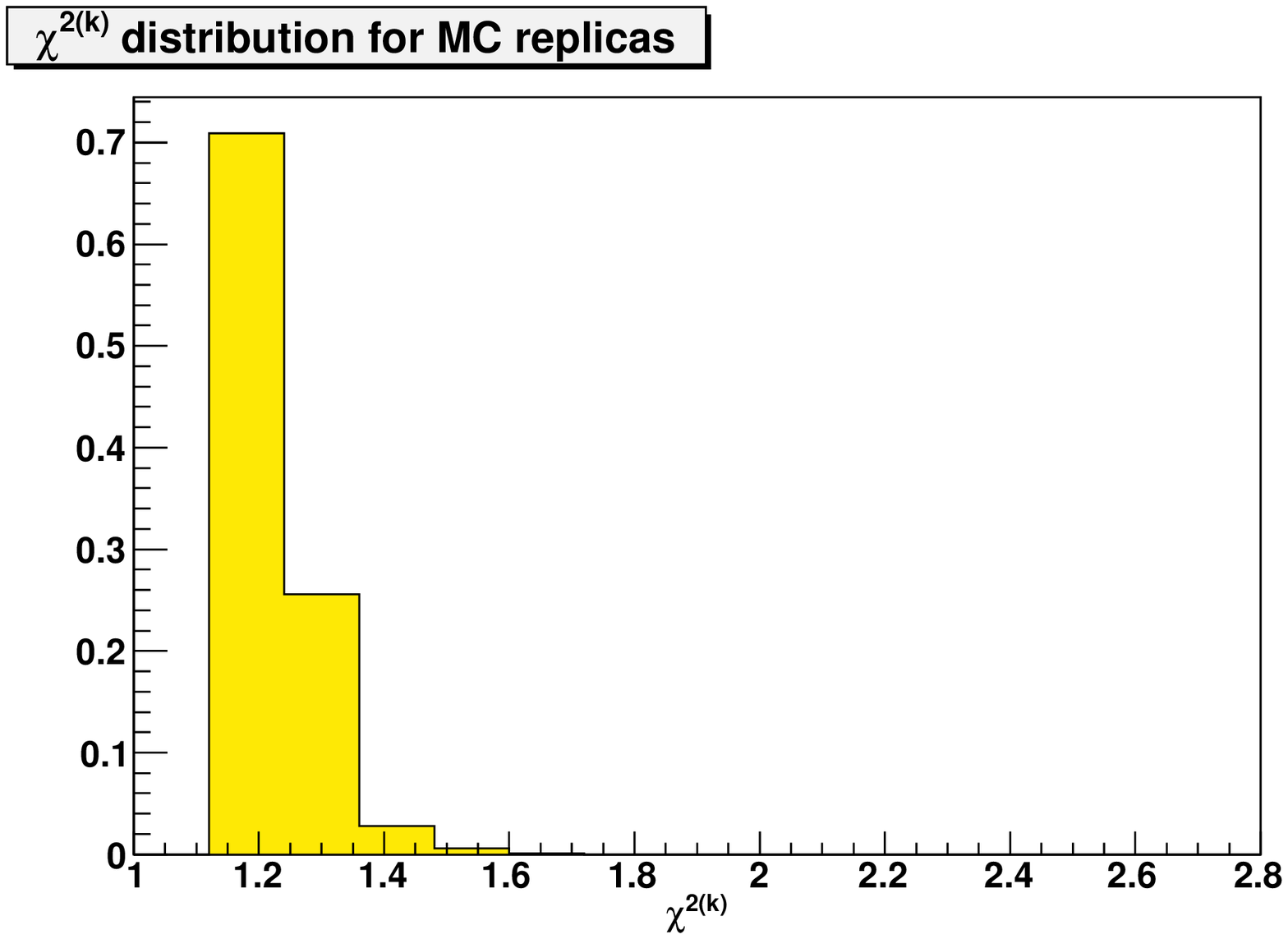}
\epsfig{width=0.49\textwidth,figure=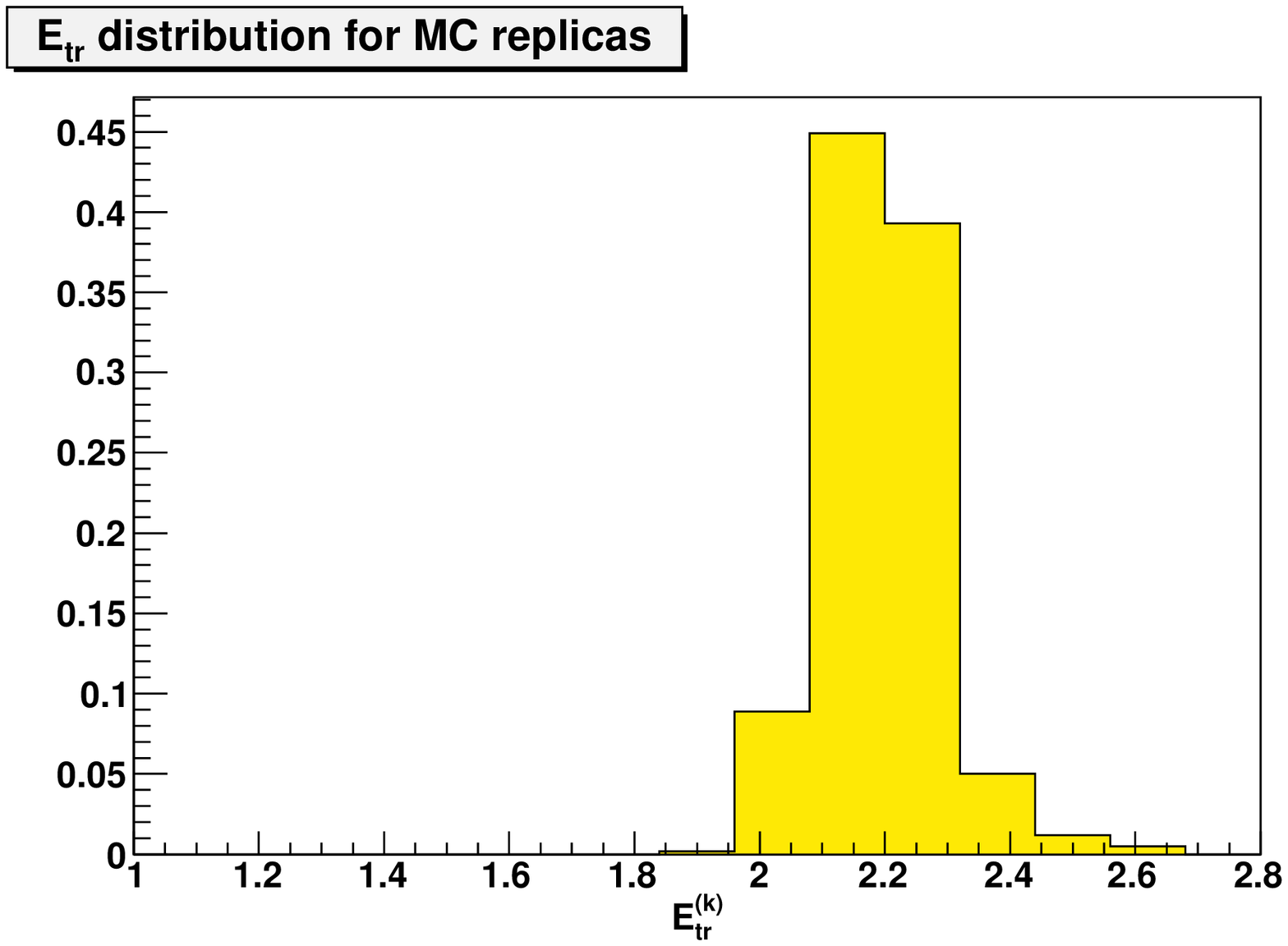}
\caption{\small Distribution of $\chi^{2(k)}$ (left) and  $E^{(k)}_{\rm tr}$ (right),
over the sample of $N_{\mathrm rep}=1000$ replicas. \label{chi2histoplots}} 
\end{center}
\end{figure}

\begin{figure}[ht!]
\begin{center}
\epsfig{width=0.60\textwidth,figure=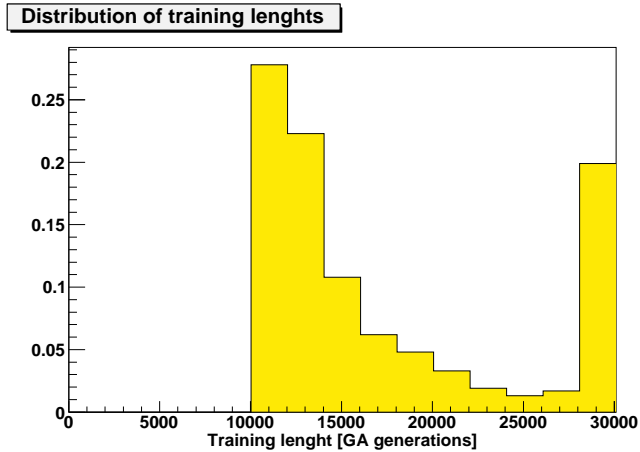}
\caption{\small Distribution of training lengths over the sample of
  $N_{\mathrm rep}=1000$ replicas.  
\label{fig:tl}} 
\end{center}
\end{figure}

All statistical indicators (including the training length), and 
in particular the quality of the global fit as measured by the value
  $\chi^{2}_{\tot}=1.16$ are quite similar to those of the NLO fit.
Specifically, the NLO and NNLO $\chi^2$ differ
  by less than 10\% for all experiments, except SLAC, the $W$ 
  asymmetry and CDF jet data (for which NNLO is better) 
and the $Z$ rapidity distribution (for which it is worse). It is
interesting to observe that  an 
excellent description of the HERA $F_2^c$ data is obtained
without the need of any \emph{ad hoc} cut or tuning of the treatment of heavy
quarks (the NNLO $\chi^2$ is somewhat worse than the NLO one, but at
NNLO the dataset is considerably wider, as discussed in Sect.~\ref{sec:kincuts}).

The distribution of 
$\chi^{2(k)}$, $E^{(k)}_{\rm tr}$, and
  training lengths among the $N_{\rm
  rep}=1000$ NNPDF2.1 NNLO replicas are shown in Fig.~\ref{chi2histoplots} and
Fig.~\ref{fig:tl} respectively.  While
most of the replicas fulfill the stopping criterion, a  fraction
($\sim 20\%$)
of them stops at the maximum training length $N_{\rm gen}^{\rm max}$
which has been introduced in
order to avoid unacceptably long fits. This fraction is comparable but
somewhat larger than the corresponding NLO one. In order to check that
this causes no significant loss of accuracy, we have verified that if
all replicas that do not stop dynamically are discarded, the PDF
change by an amount which is smaller than a statistical fluctuation.
We have also verified that this fraction is reduced if the maximum
training length is raised, thereby showing that the issue is merely
one of
computational efficiency, rather than principle.

\begin{figure}[t]
\begin{center}
\epsfig{width=0.49\textwidth,figure=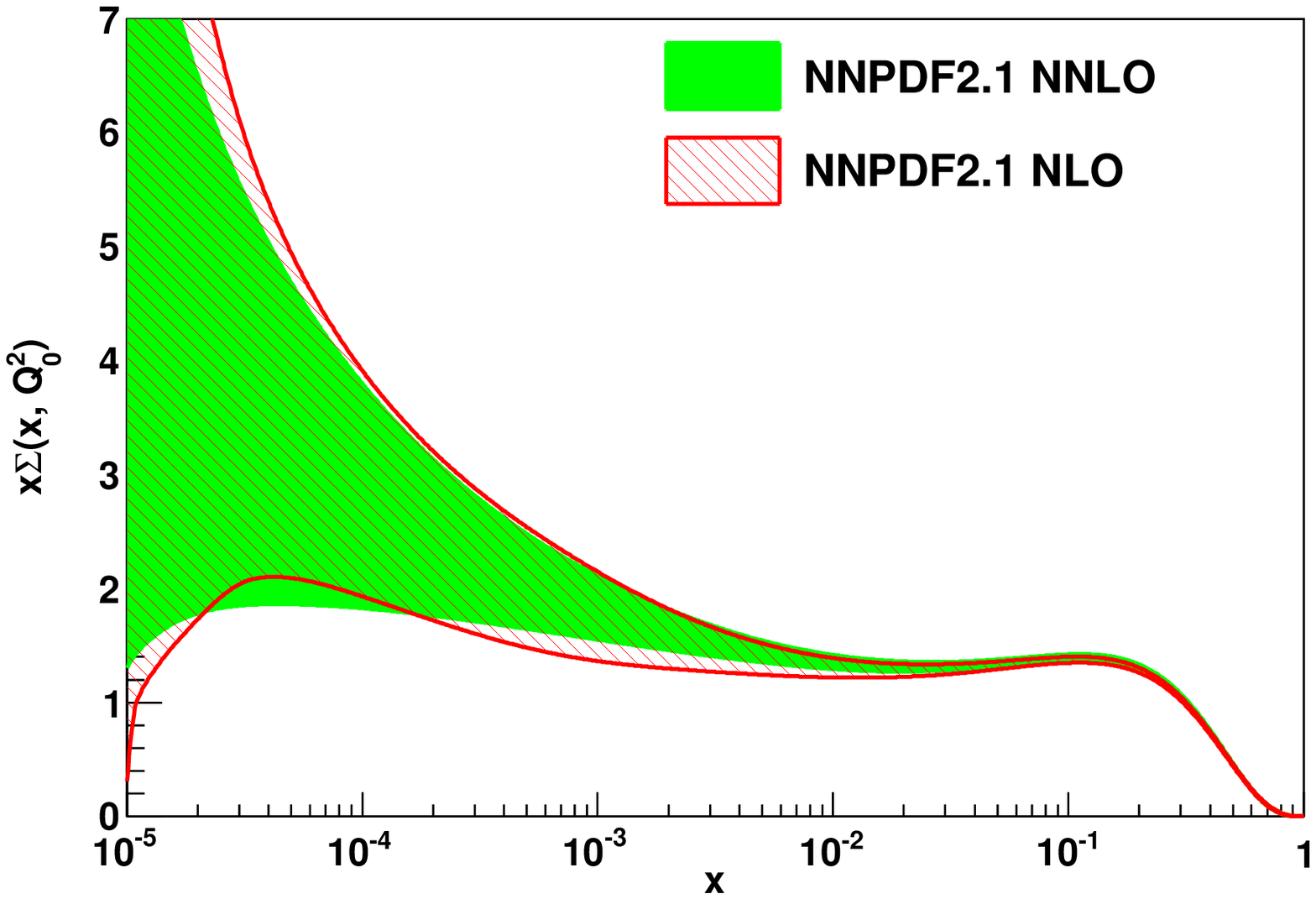}
\epsfig{width=0.49\textwidth,figure=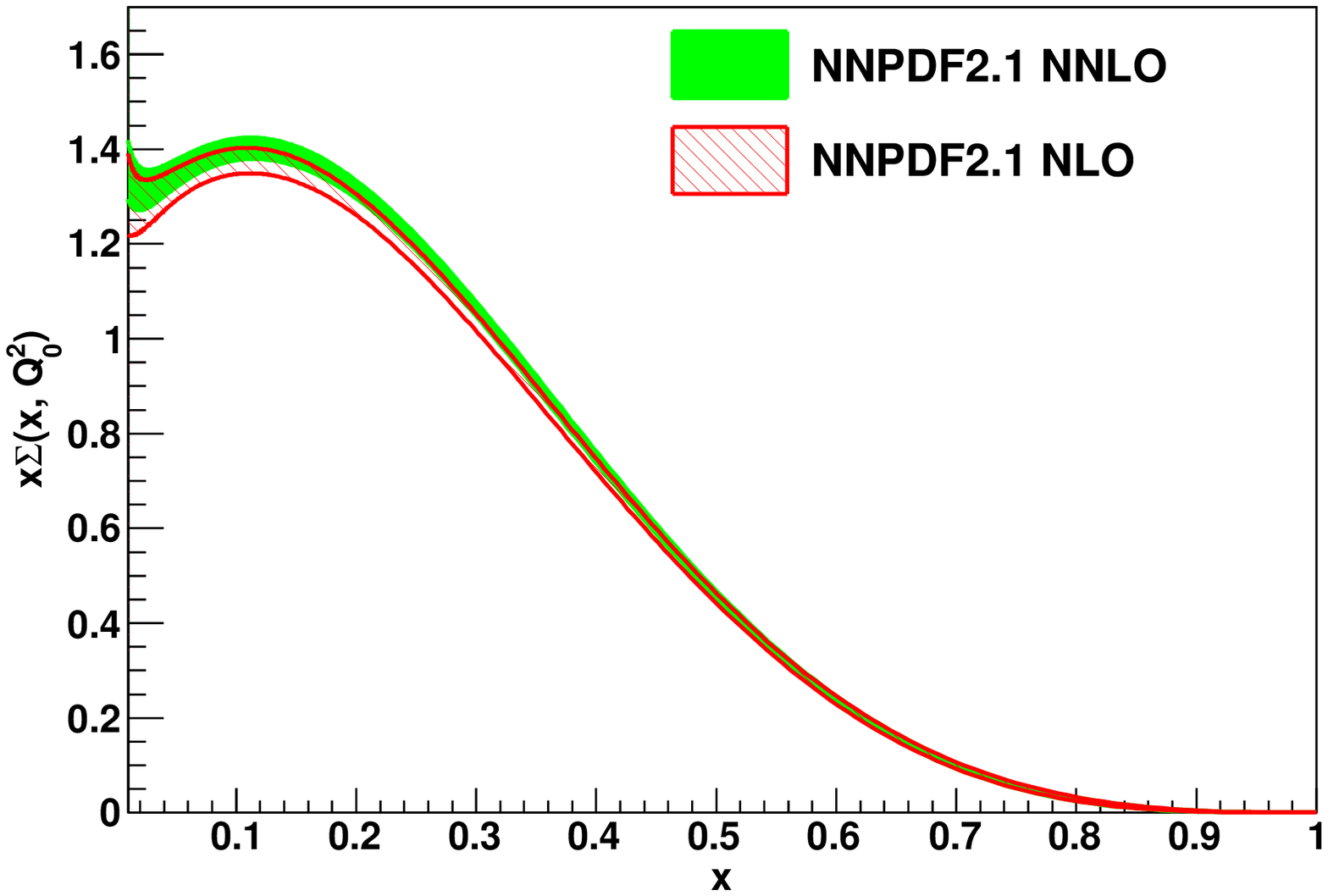}
\epsfig{width=0.49\textwidth,figure=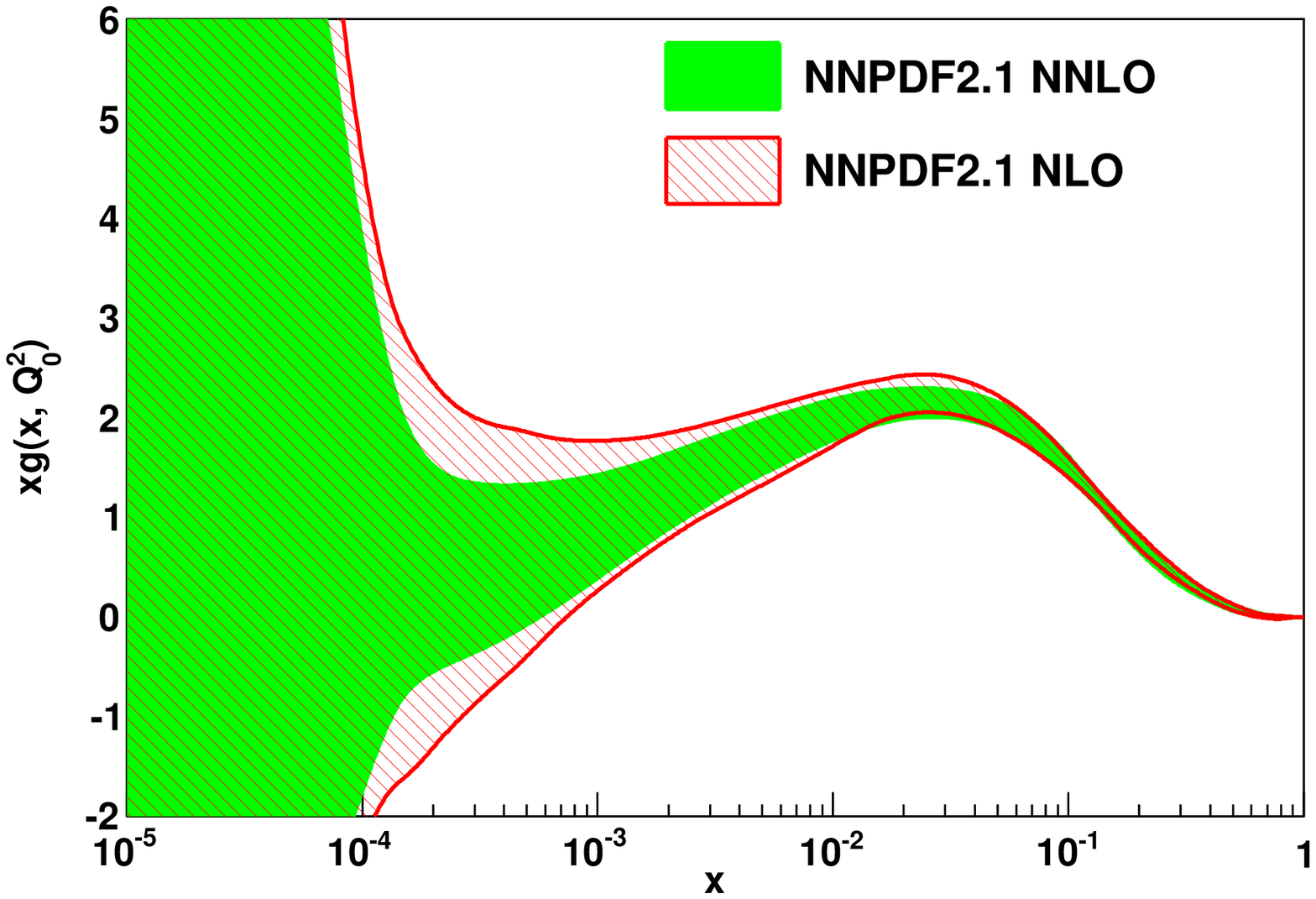}
\epsfig{width=0.49\textwidth,figure=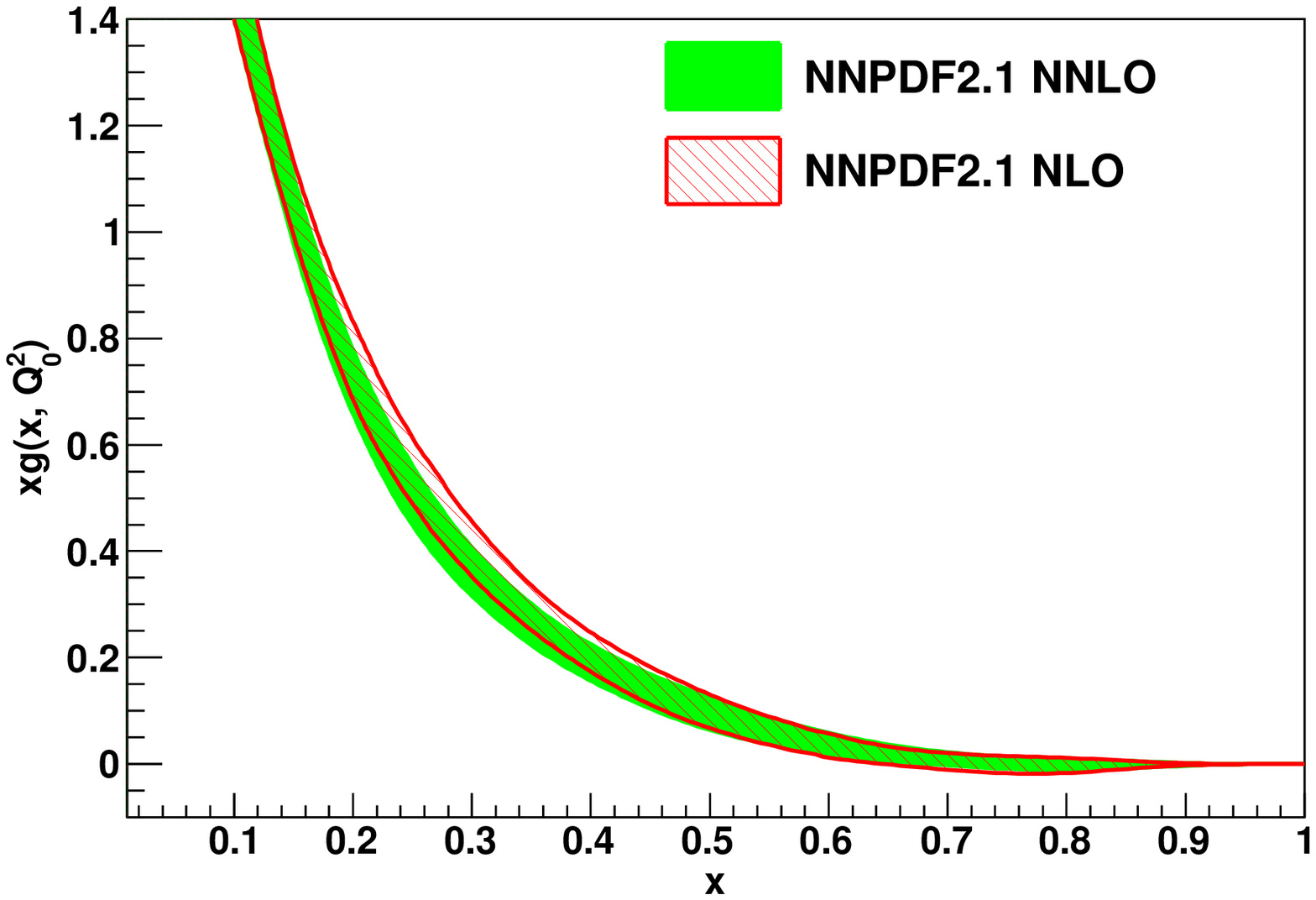}
\caption{\small Comparison of  NNPDF2.1 NLO and NNLO  singlet sector PDFs, computed using $N_{\rm rep}=1000$ replicas from both sets. All
  error bands shown correspond to one sigma.
 \label{fig:singletPDFs}} 
\end{center}
\vskip-0.5cm
\end{figure}

\begin{figure}[t!]
\begin{center}
\epsfig{width=0.49\textwidth,figure=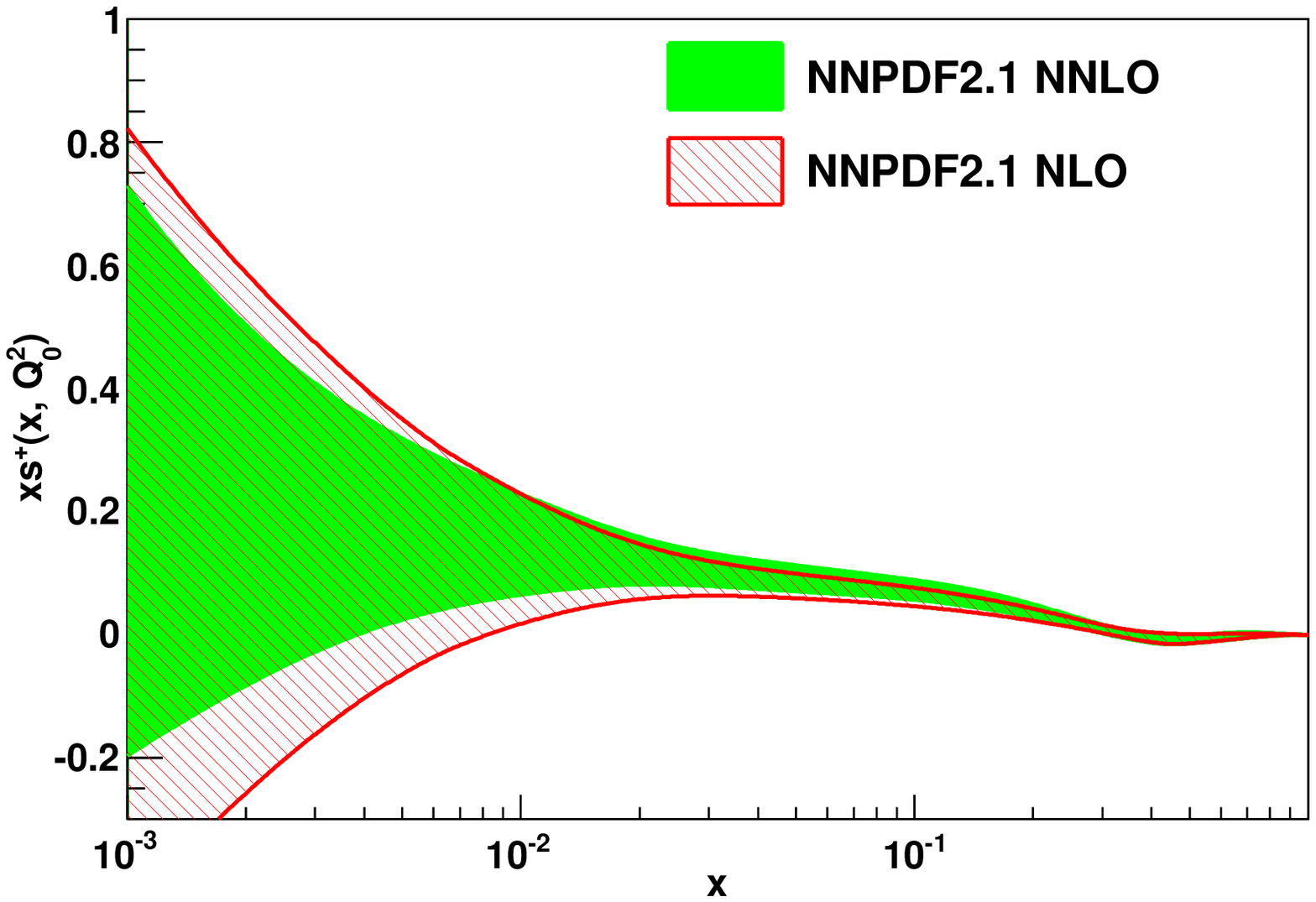}
\epsfig{width=0.49\textwidth,figure=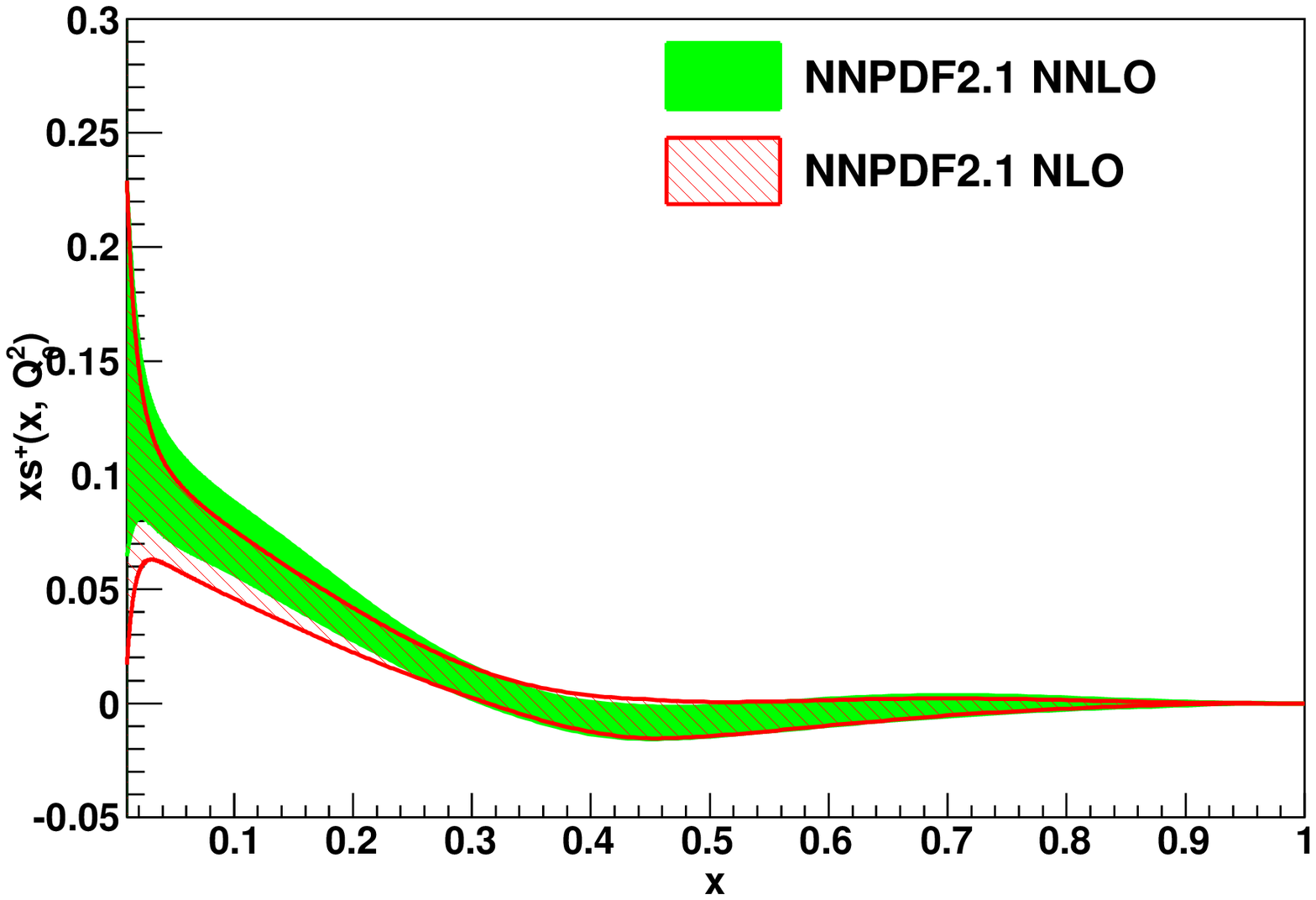}
\epsfig{width=0.49\textwidth,figure=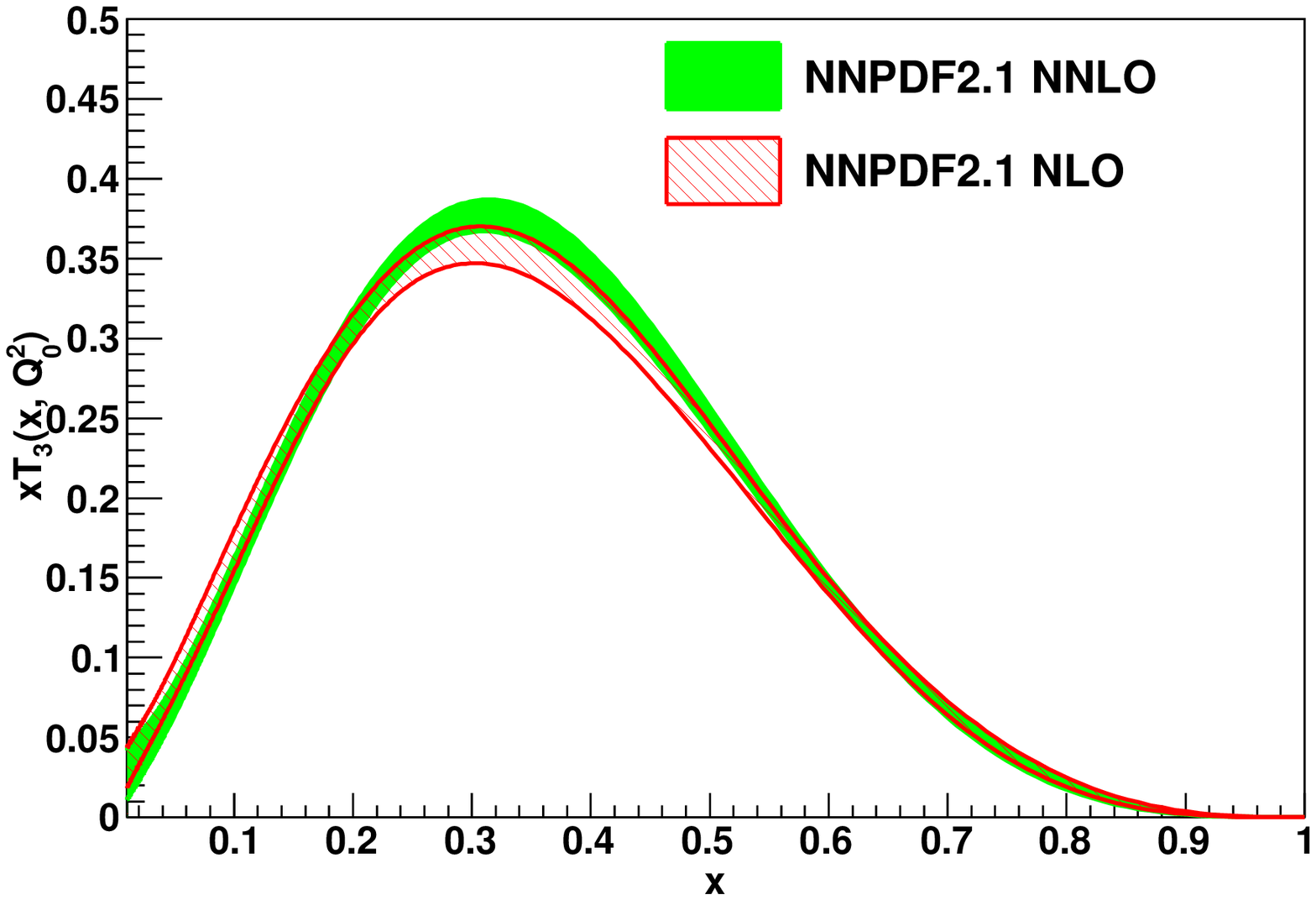}
\epsfig{width=0.49\textwidth,figure=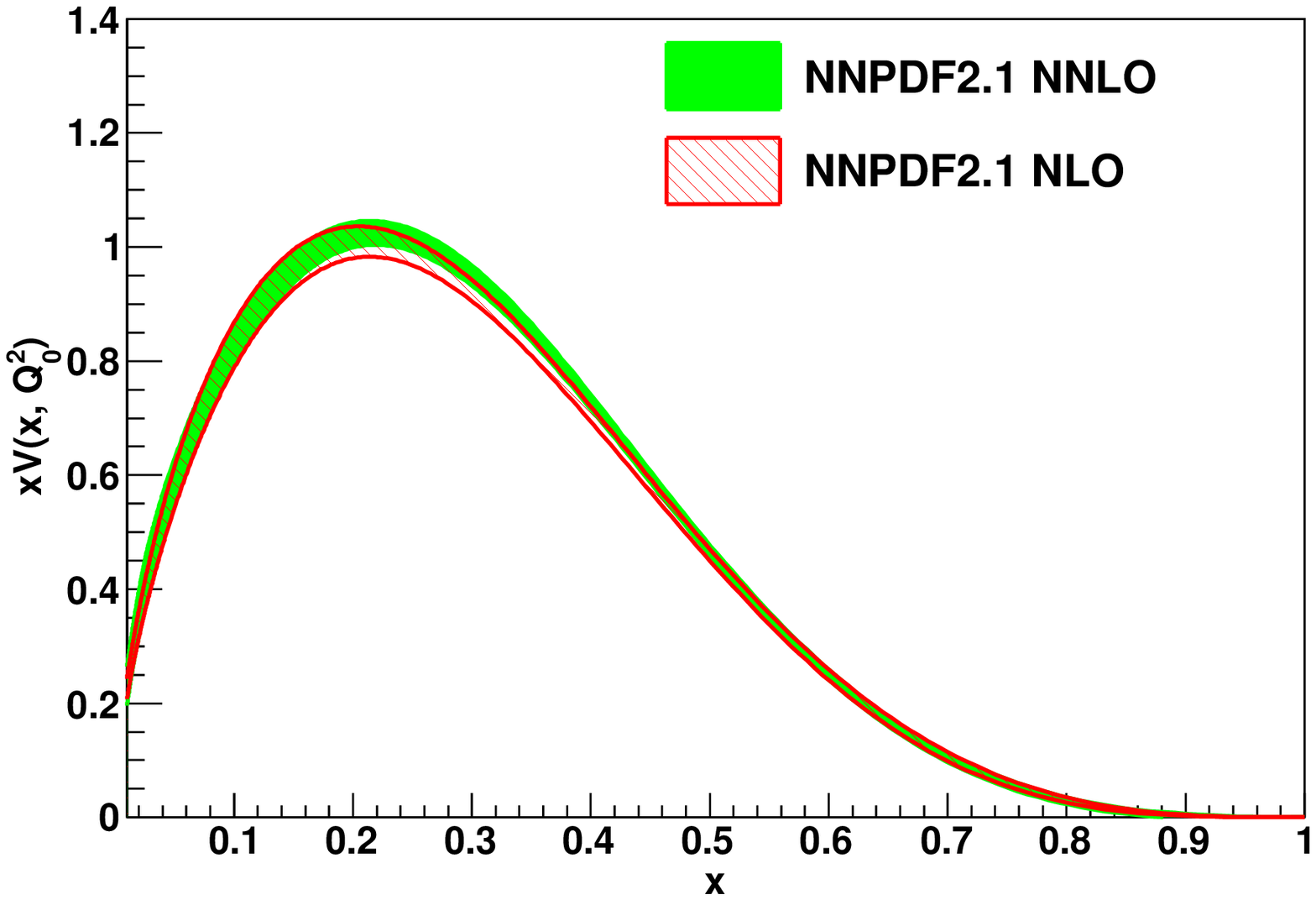}
\epsfig{width=0.49\textwidth,figure=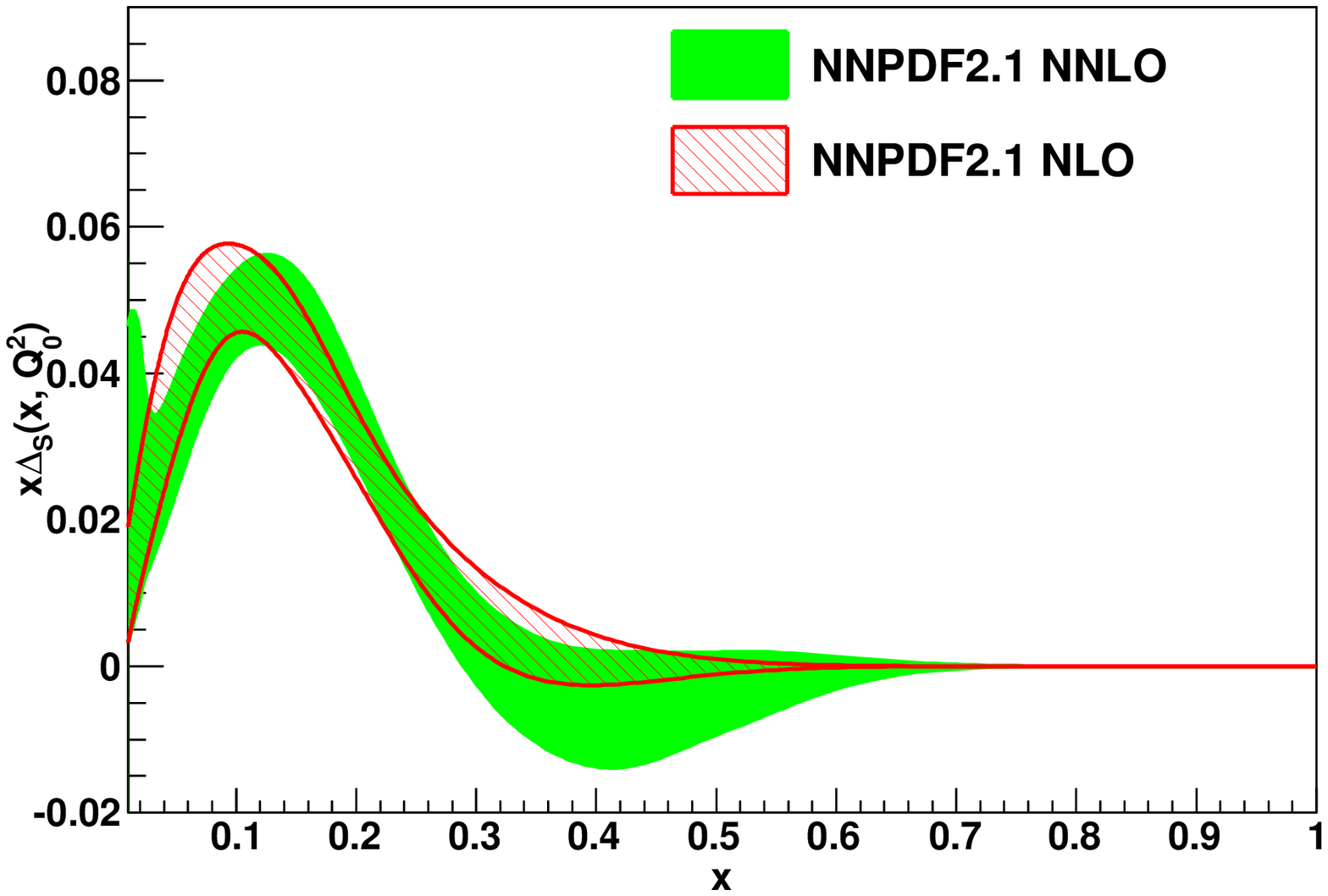}
\epsfig{width=0.49\textwidth,figure=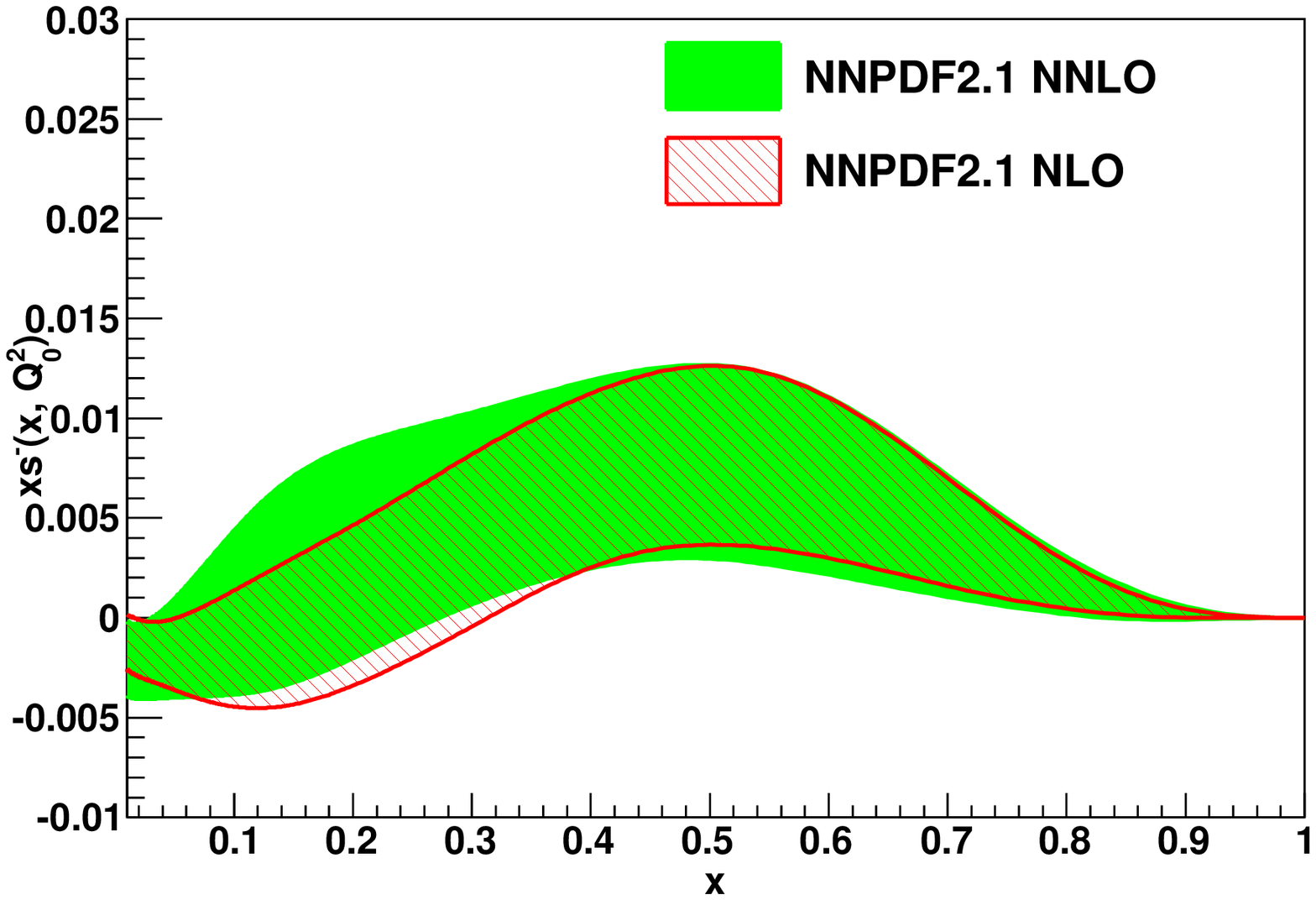}
\caption{\small Same as Fig.~\ref{fig:singletPDFs}
for the non--singlet sector PDFs.
 \label{fig:valencePDFs}} 
\end{center}
\vskip-0.5cm
\end{figure}

\begin{figure}[t!]
\begin{center}
\epsfig{width=0.99\textwidth,figure=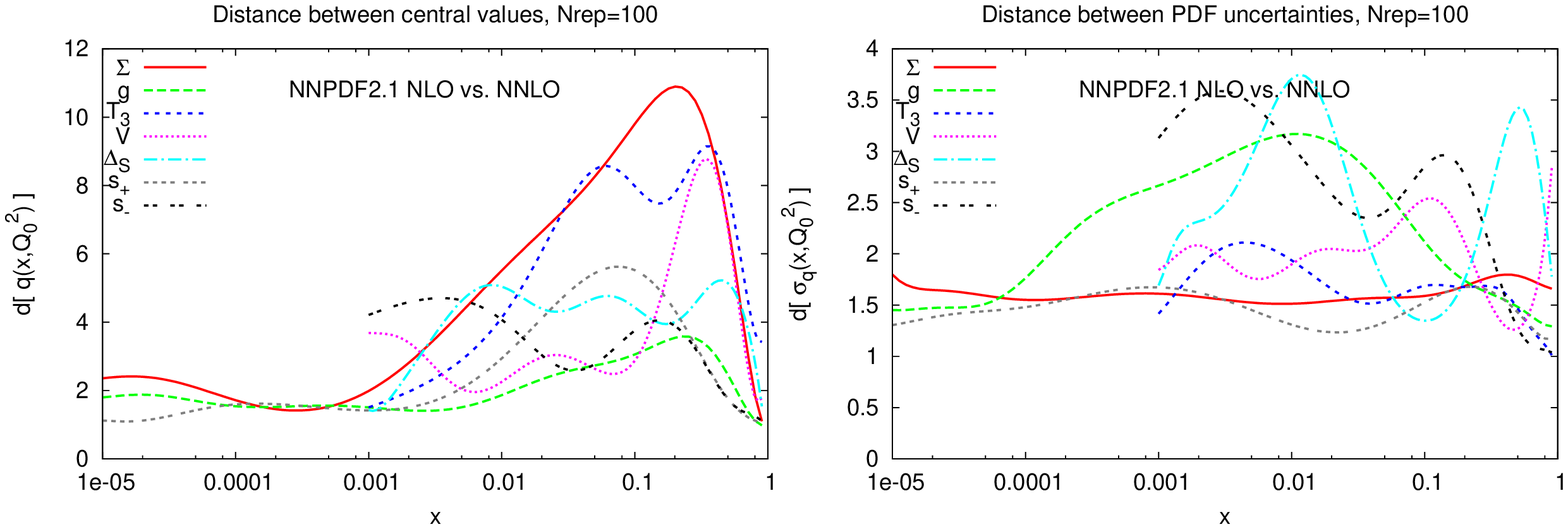}
\caption{\small Distances between the NNPDF2.1 NLO and
NNLO parton sets shown in Figs.~\ref{fig:singletPDFs}
and~\ref{fig:valencePDFs}.  All distances are computed from
sets of $N_{\rm rep}=100$ replicas.  \label{fig:dist_20_21}}
\end{center}
\end{figure}

\subsection{Parton distributions}

The NNPDF2.1 NNLO parton distributions are shown along with their NLO
counterparts in  
Figs.~\ref{fig:singletPDFs} and~\ref{fig:valencePDFs} at 
the 
input scale $Q_0^2=2$ GeV$^2$, in the basis in which they are parametrized.
The distances (defined as in Appendix~A of Ref.~\cite{Ball:2010de})
between the NLO and NNLO sets are shown in
Fig.~\ref{fig:dist_20_21}.

Recalling that a distance $d\sim 1$ corresponds to statistical
equivalence, while (with 100 replicas) $d\sim 7$ is a one sigma shift,
it is apparent that the NLO and NNLO sets are statistically
inequivalent, but differ by typically less than one sigma.
This in particular means that PDFs in the NNPDF2.1 set 
are quite  stable when going from NLO to NNLO. The largest
variations are observed for quarks at $x\sim 0.1$, while
the small $x$ PDFs (gluon and light quark sea) are very similar
to their NLO counterparts. It is worth noting that the distances for the PDF
uncertainties in Fig.~\ref{fig:dist_20_21} are particularly small.
 This is as it should be, consistently with the fact that the quality
 of the NLO and NNLO fits are similar, given that theory uncertainties are not
 included in PDF uncertainties.

\begin{figure}[ht!]
\begin{center}
\epsfig{width=0.49\textwidth,figure=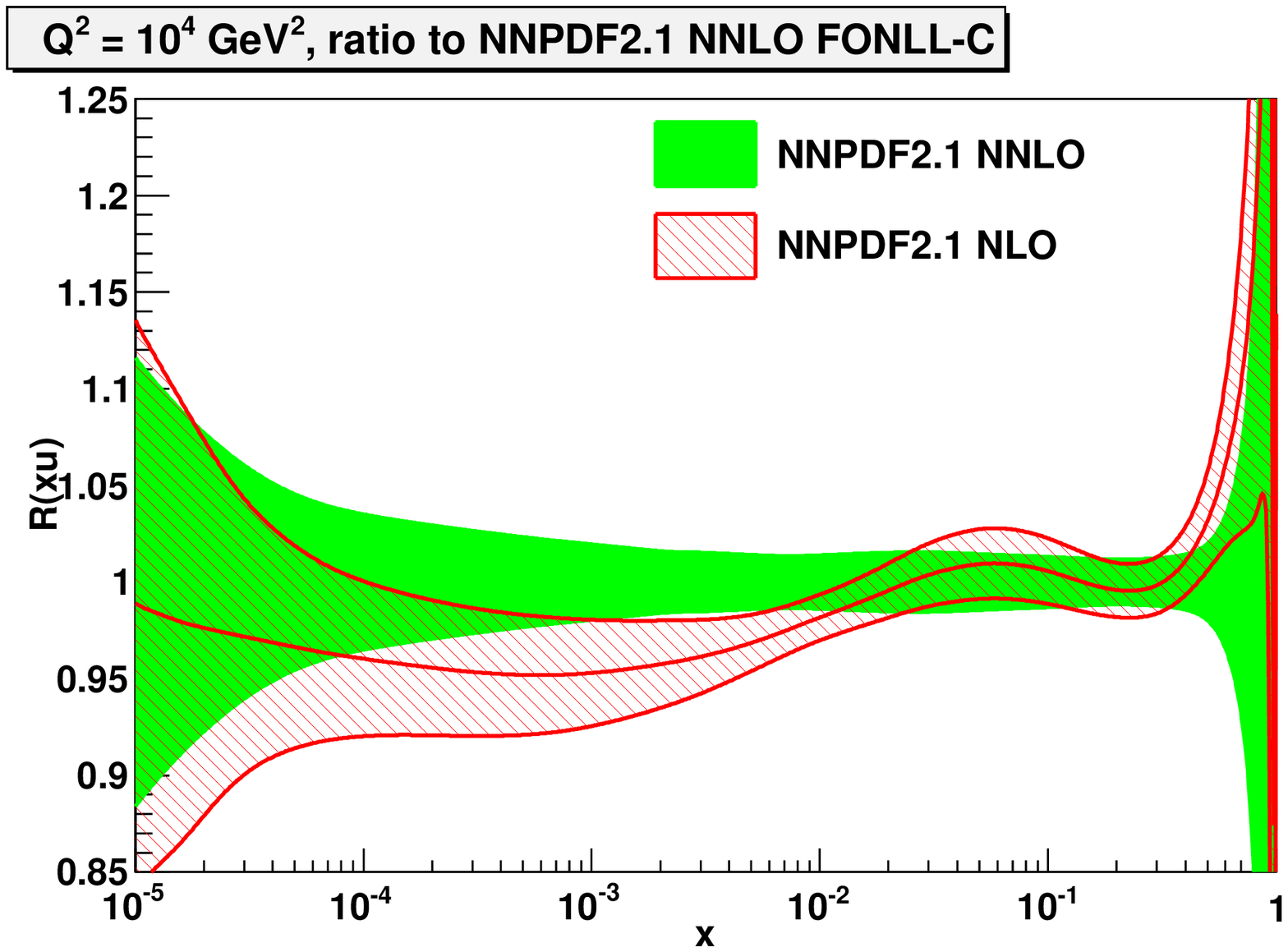}
\epsfig{width=0.49\textwidth,figure=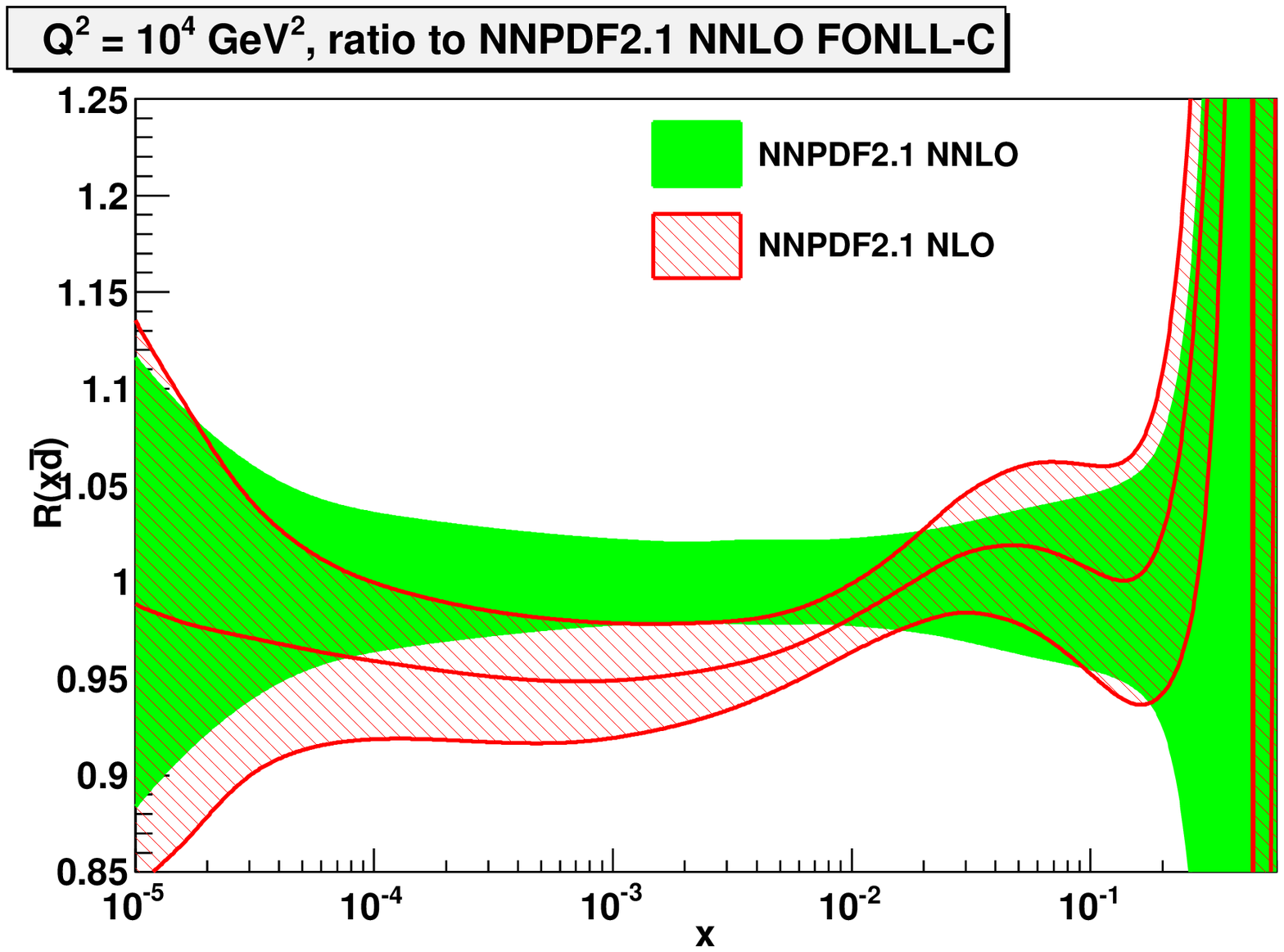}
\epsfig{width=0.49\textwidth,figure=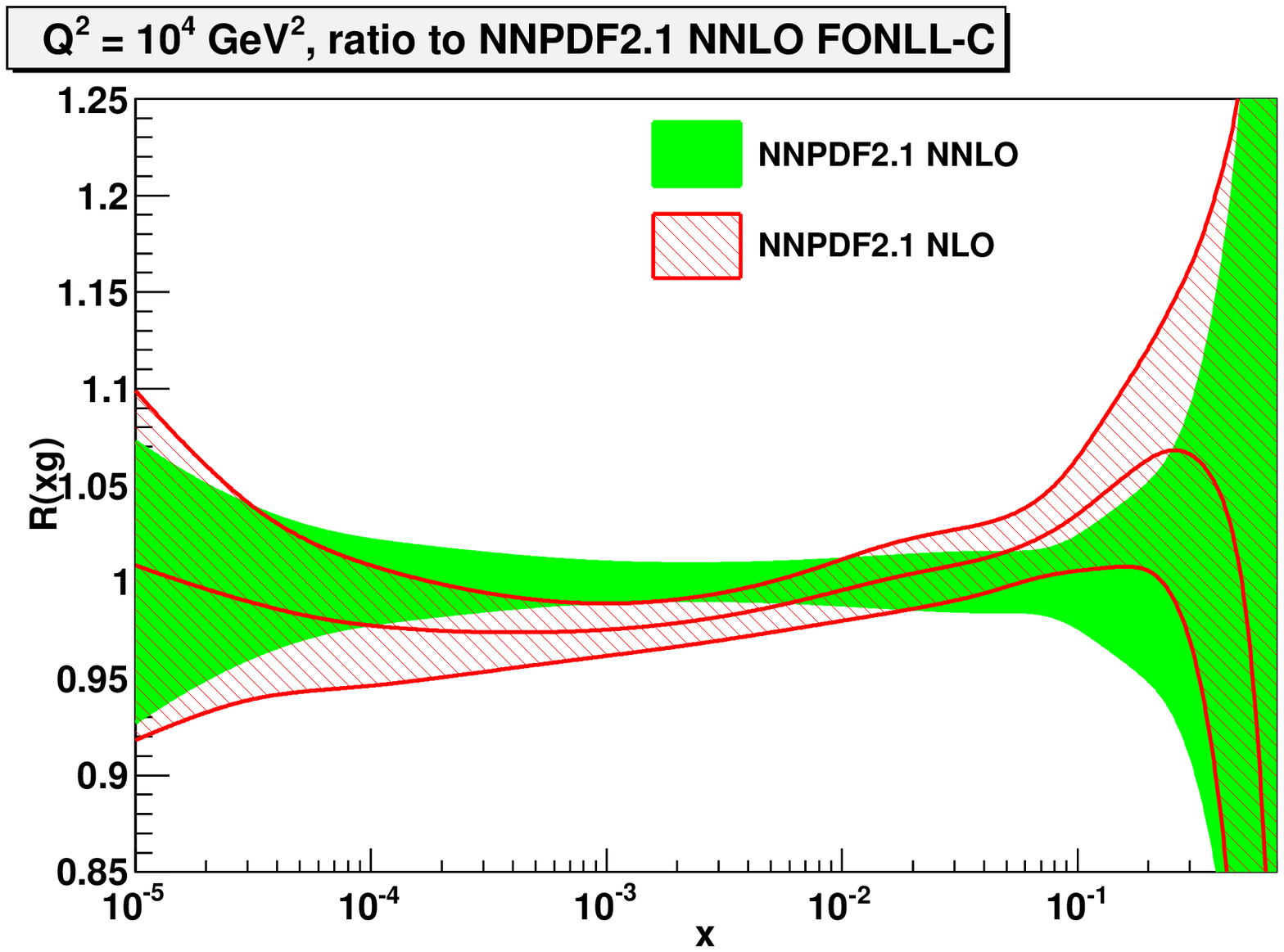}
\epsfig{width=0.49\textwidth,figure=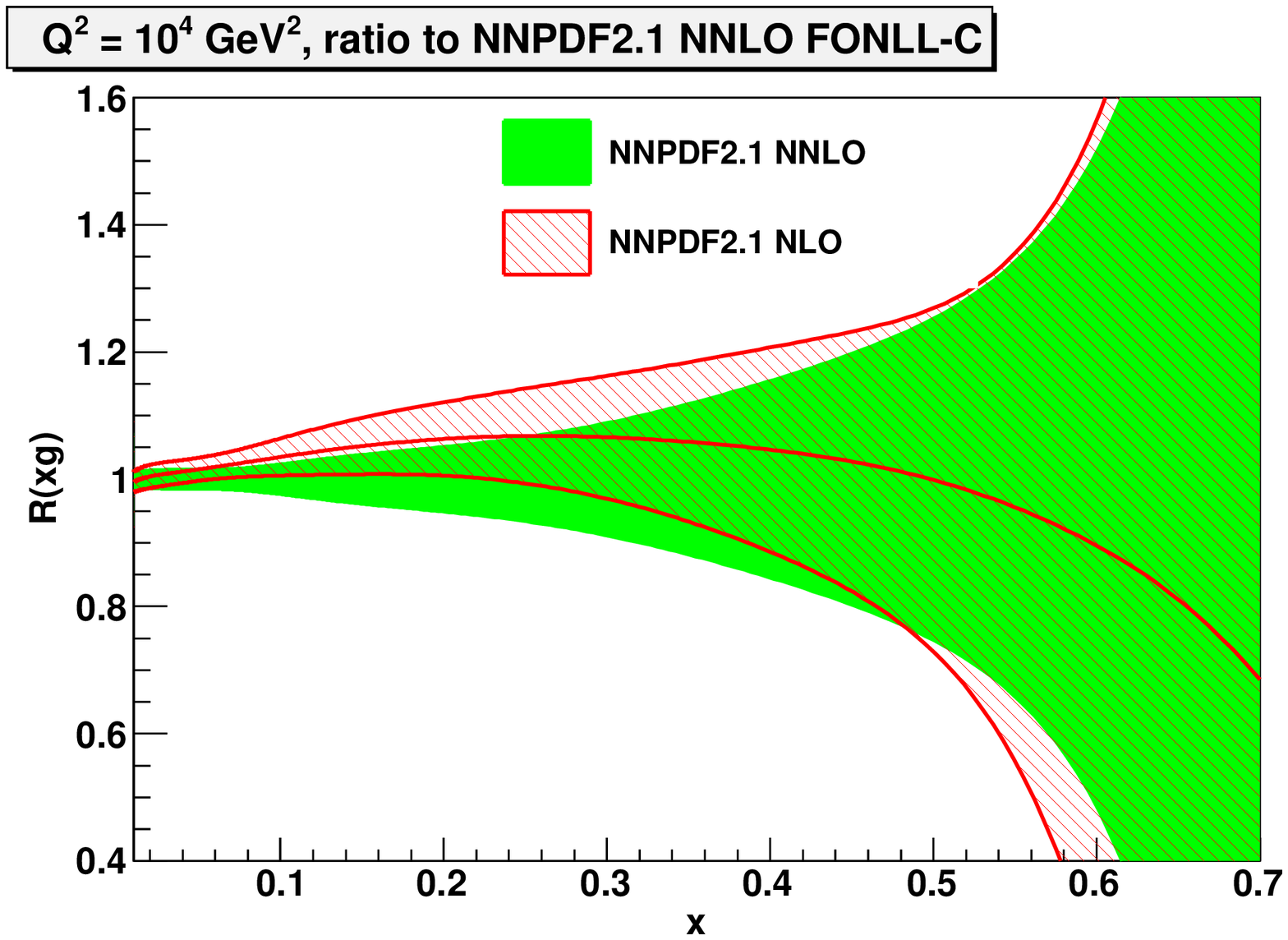}
\caption{\small Comparison between NNPDF2.1 NLO and NNLO 
light quark and gluon PDFs at $Q^2=10^4$~GeV$^2$. The results have been obtained with
$N_{\rm rep}=1000$ replicas.
All curves are shown as ratios to the central NNPDF2.1 NNLO result.
\label{fig:nnpdf21ratcomp}} 
\end{center}
\end{figure}
In order to assess the impact of NNLO corrections on physical observables it is
useful to compare NNLO and NLO PDFs for individual flavours at a
typical hard scale. This is done in Fig.~\ref{fig:nnpdf21ratcomp},
where the NNLO/NLO ratio is shown as a function of $x$ at  $Q^2=10^4$ GeV$^2$.
The most noticeable changes are
larger  small $x$ quarks (and correspondingly, due
to evolution, larger small $x$ gluons) and smaller  large $x$ quarks.
The biggest differences are observed for the light quark sea
at $x\sim 10^{-3}$, where the NNLO and NLO bands just about miss each other. 

\begin{figure}[t]
\begin{center}
\epsfig{width=0.49\textwidth,figure=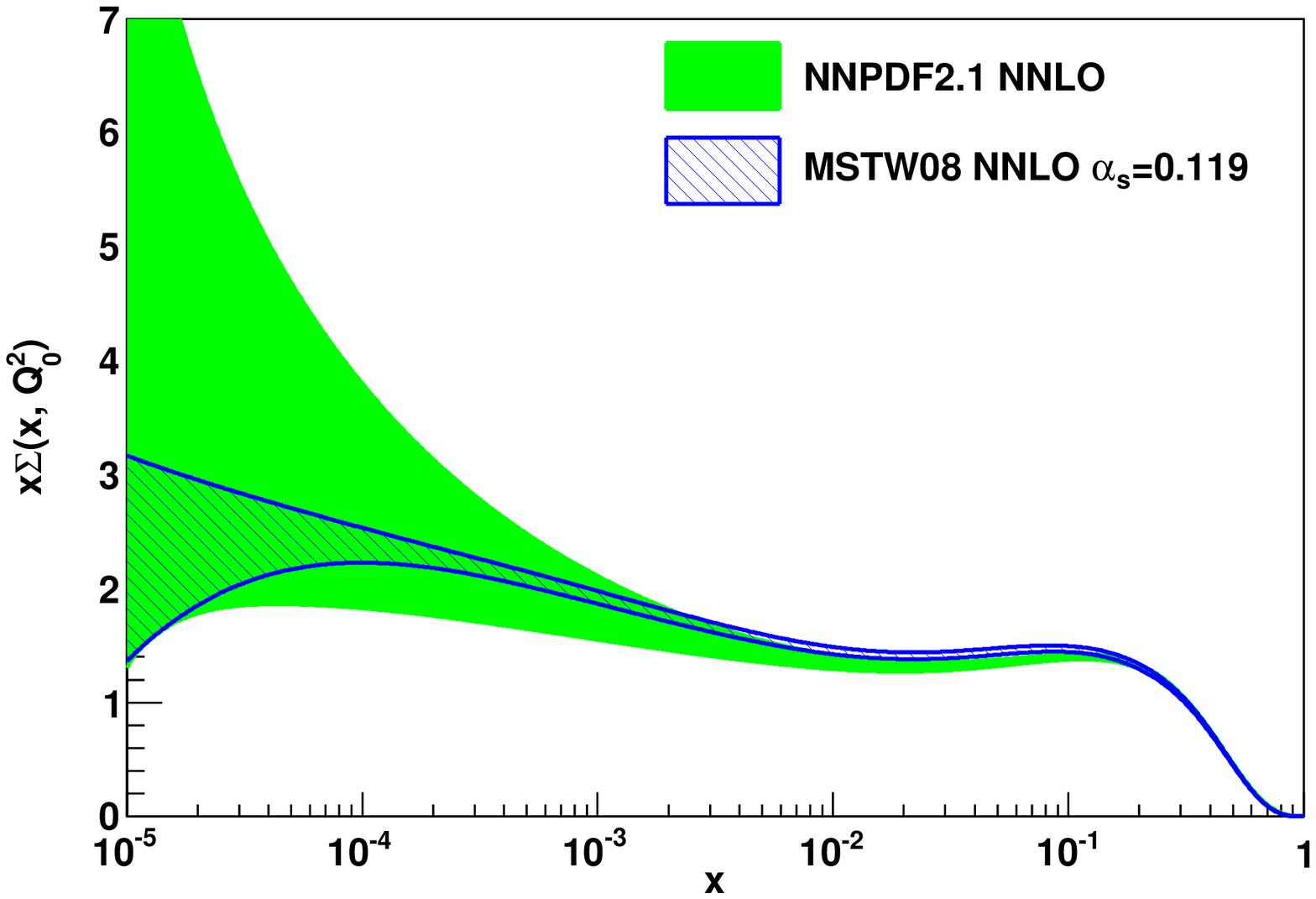}
\epsfig{width=0.49\textwidth,figure=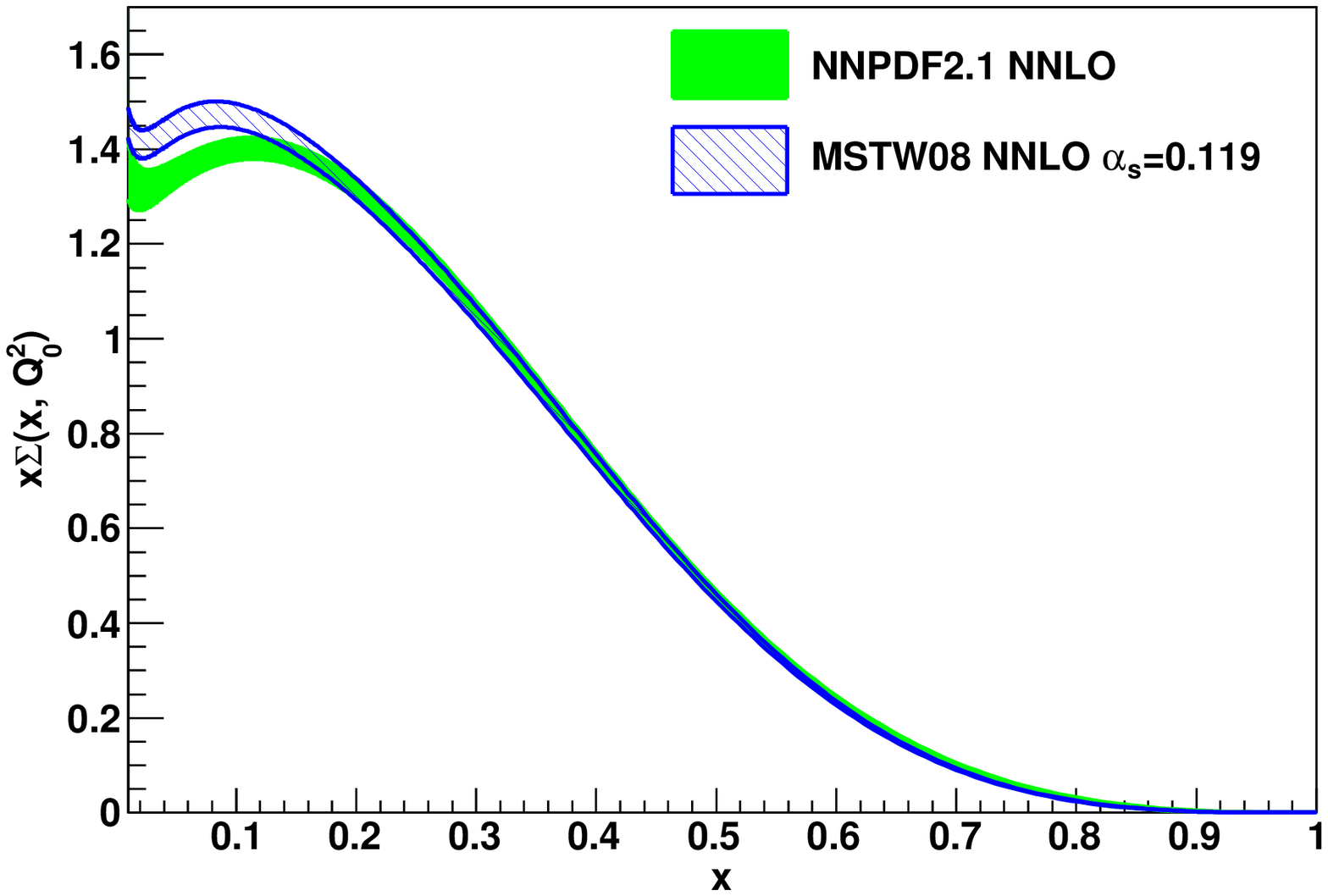}
\epsfig{width=0.49\textwidth,figure=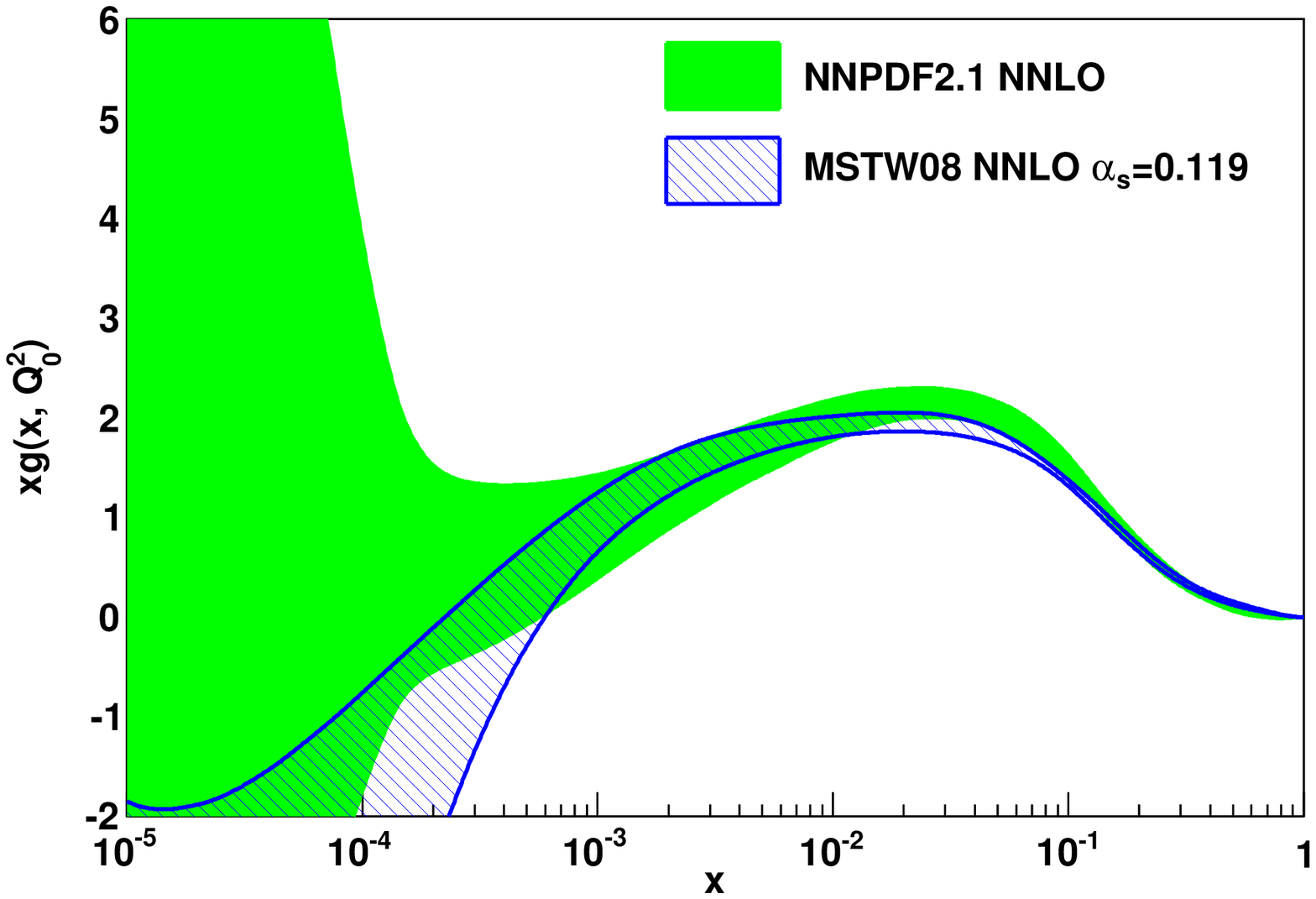}
\epsfig{width=0.49\textwidth,figure=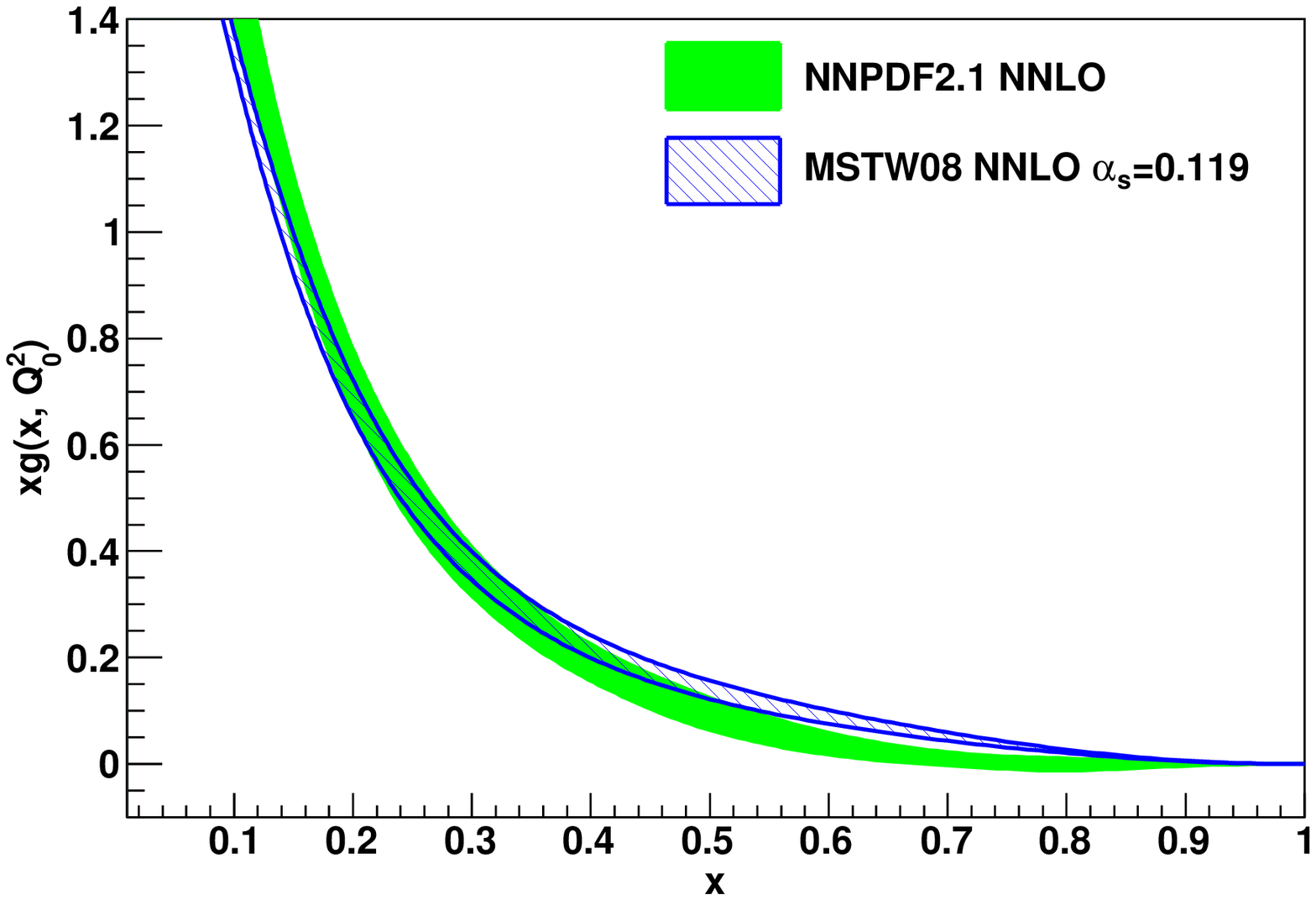}
\caption{\small The NNPDF2.1 NNLO singlet sector PDFs, compared
to MSTW08 PDFs. The results for NNPDF2.1 NNLO have been obtained with
$N_{\rm rep}=1000$ replicas. All PDF errors are given 
as one sigma uncertainties. In the comparison 
a common value of $\alpha_s\lp M_Z\rp$=0.119 has been used.
 \label{fig:singletPDFs-lhapdf}} 
\end{center}
\end{figure}

\begin{figure}[t]
\begin{center}
\epsfig{width=0.49\textwidth,figure=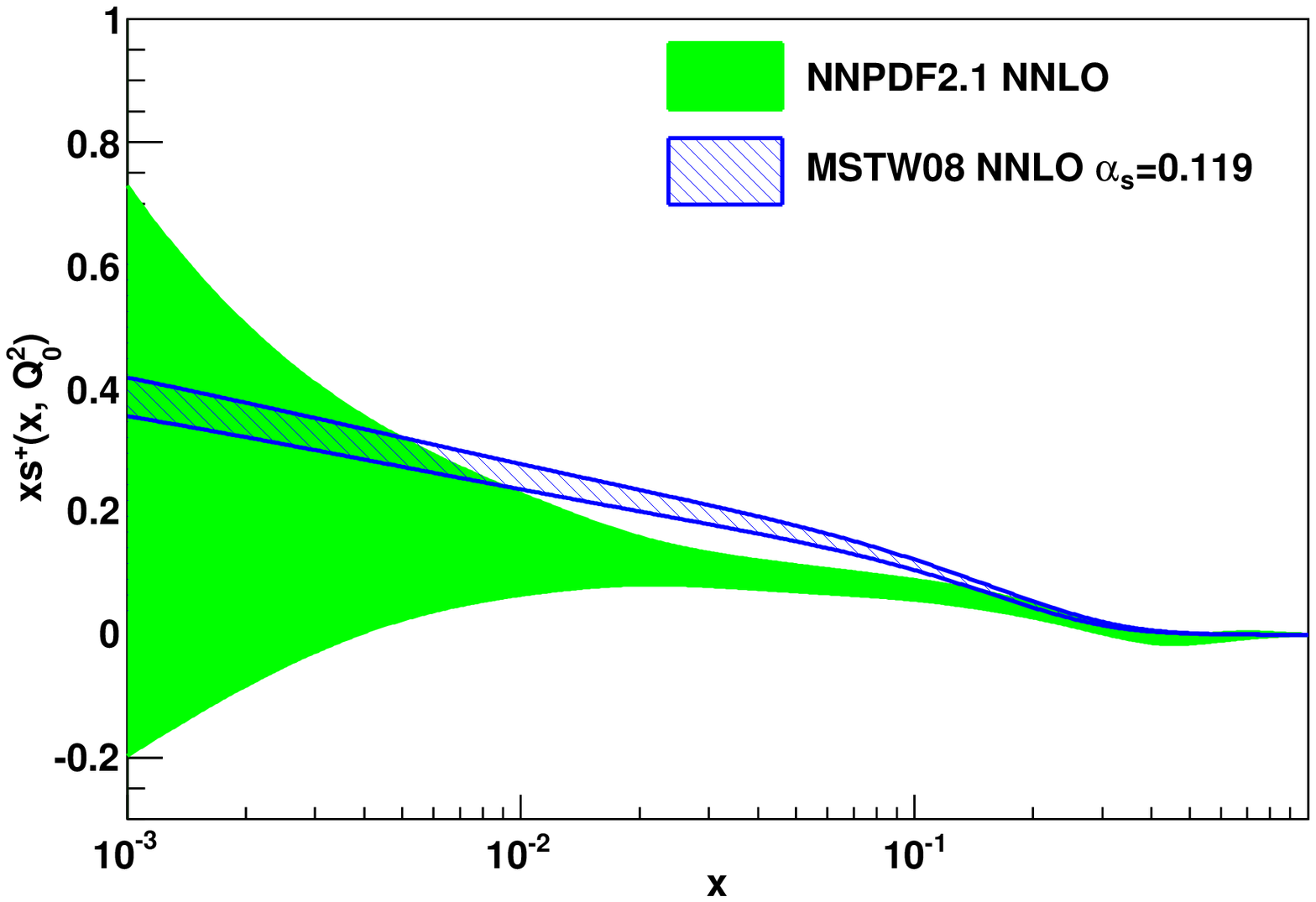}
\epsfig{width=0.49\textwidth,figure=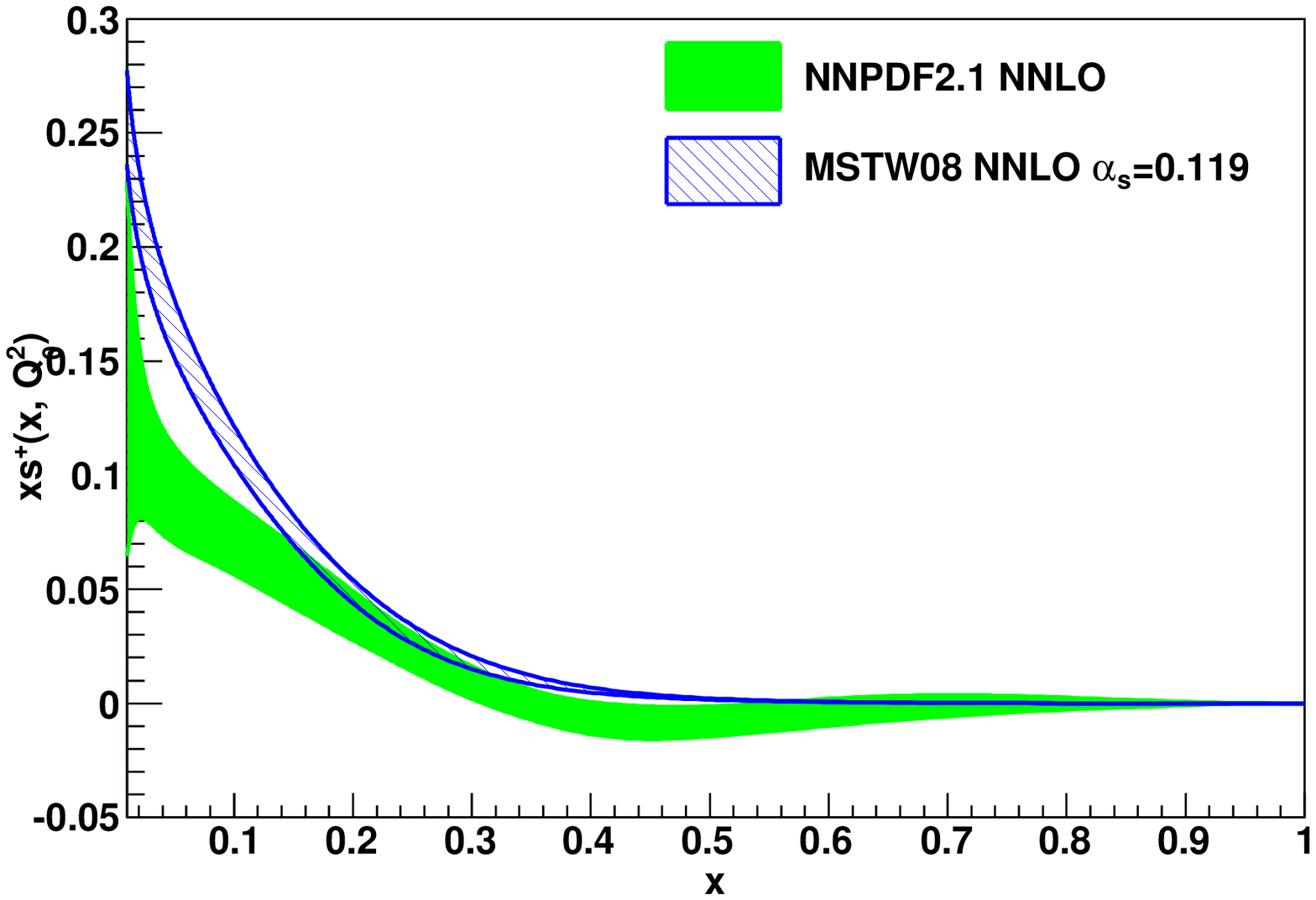}
\epsfig{width=0.49\textwidth,figure=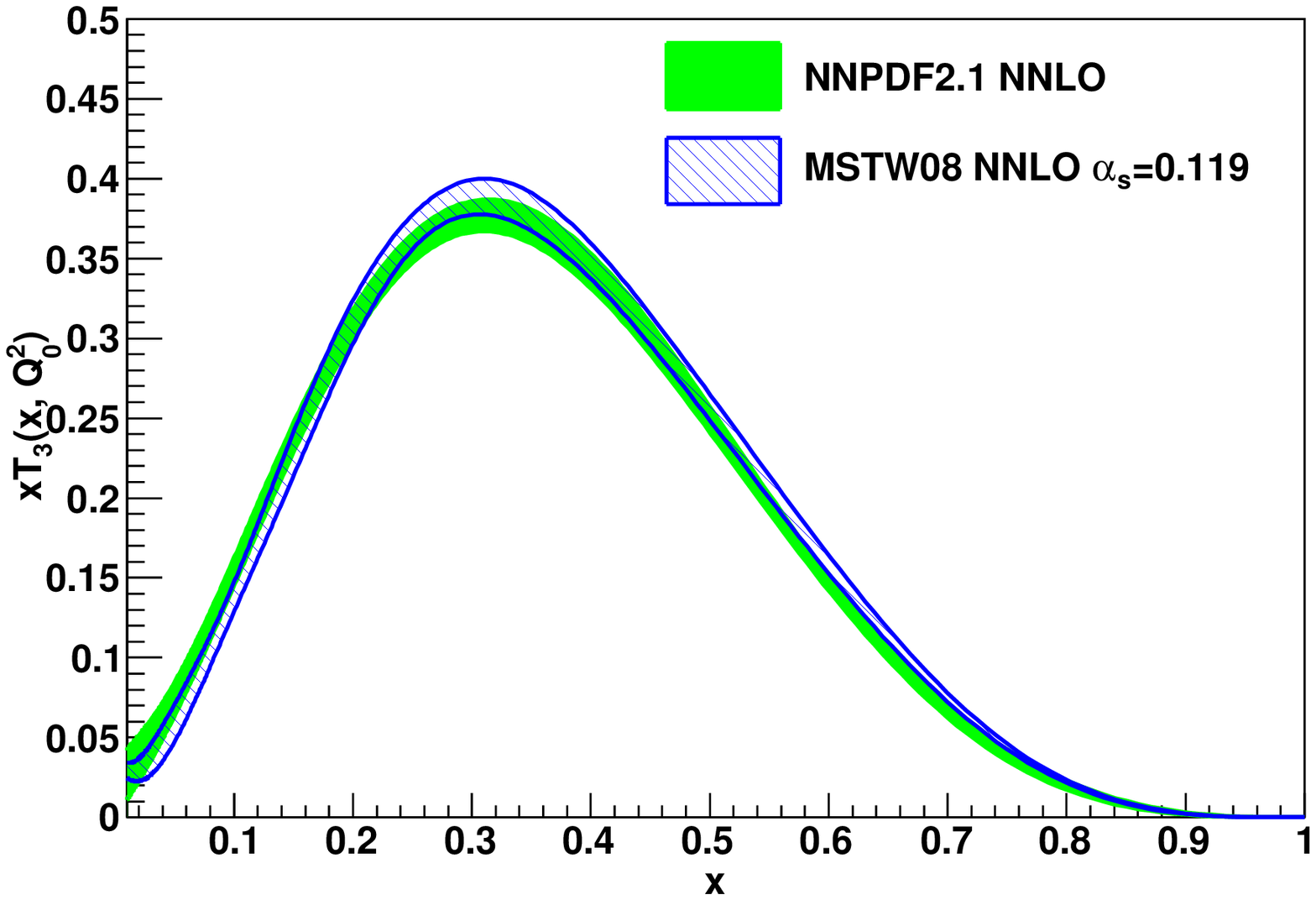}
\epsfig{width=0.49\textwidth,figure=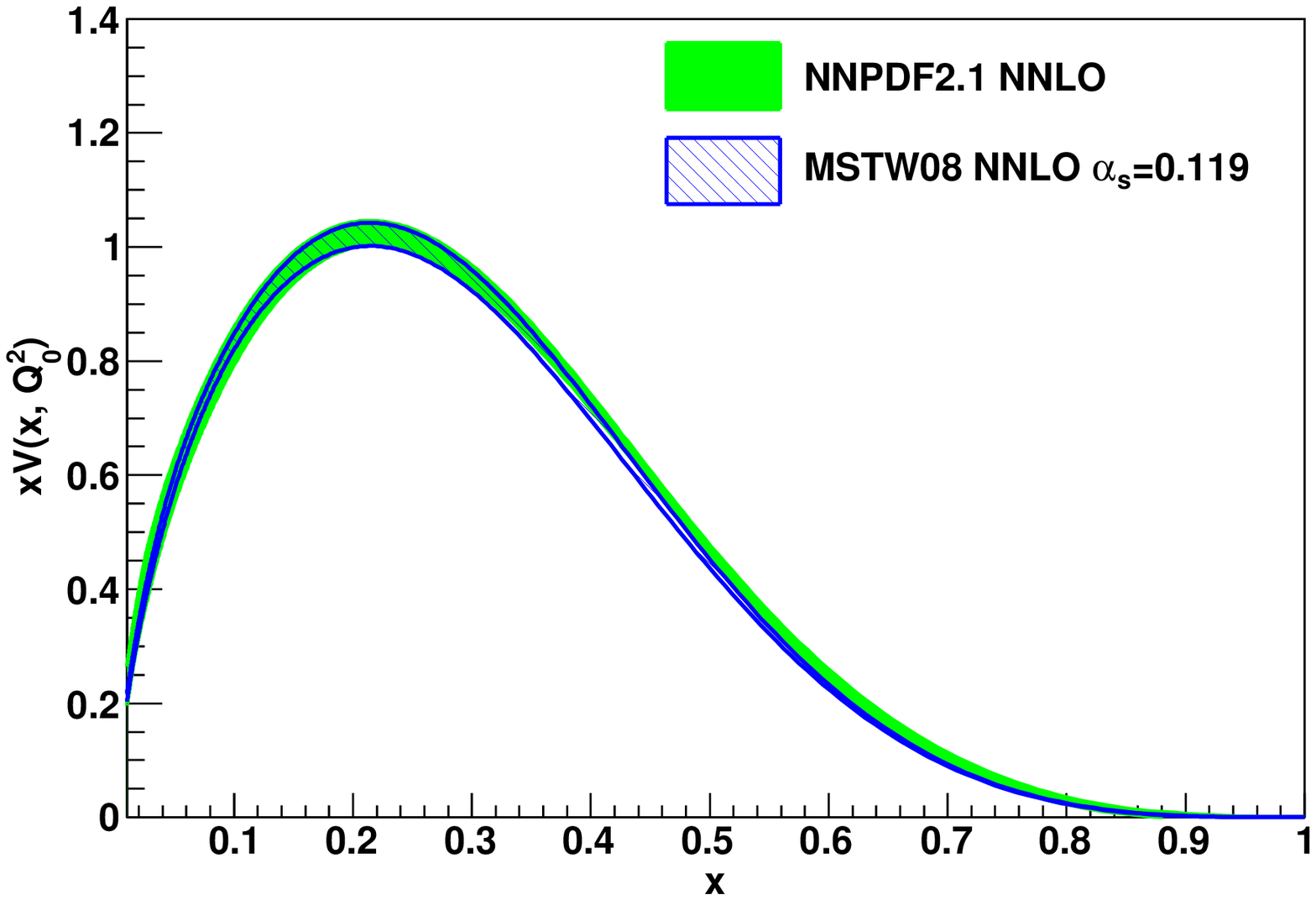}
\epsfig{width=0.49\textwidth,figure=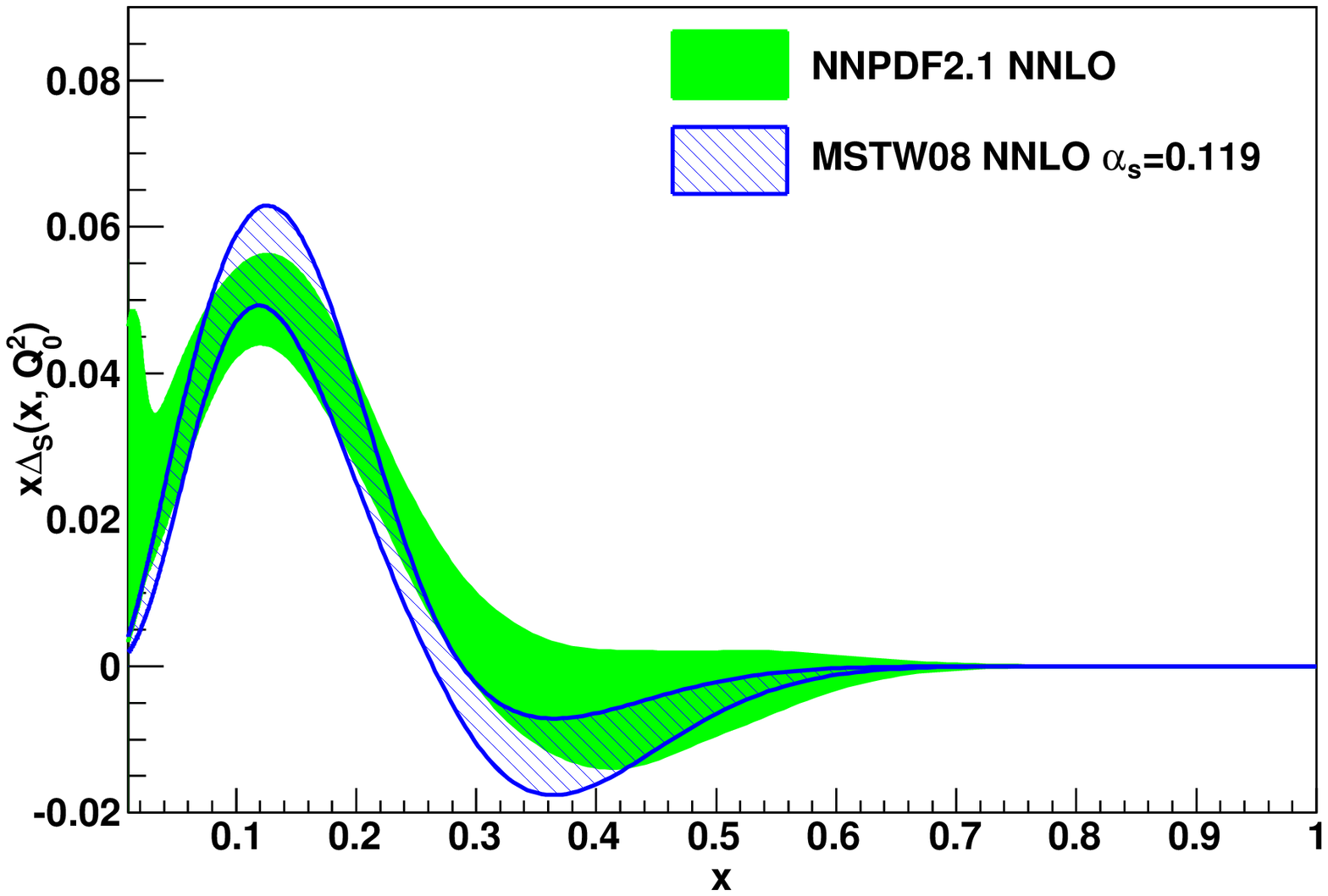}
\epsfig{width=0.49\textwidth,figure=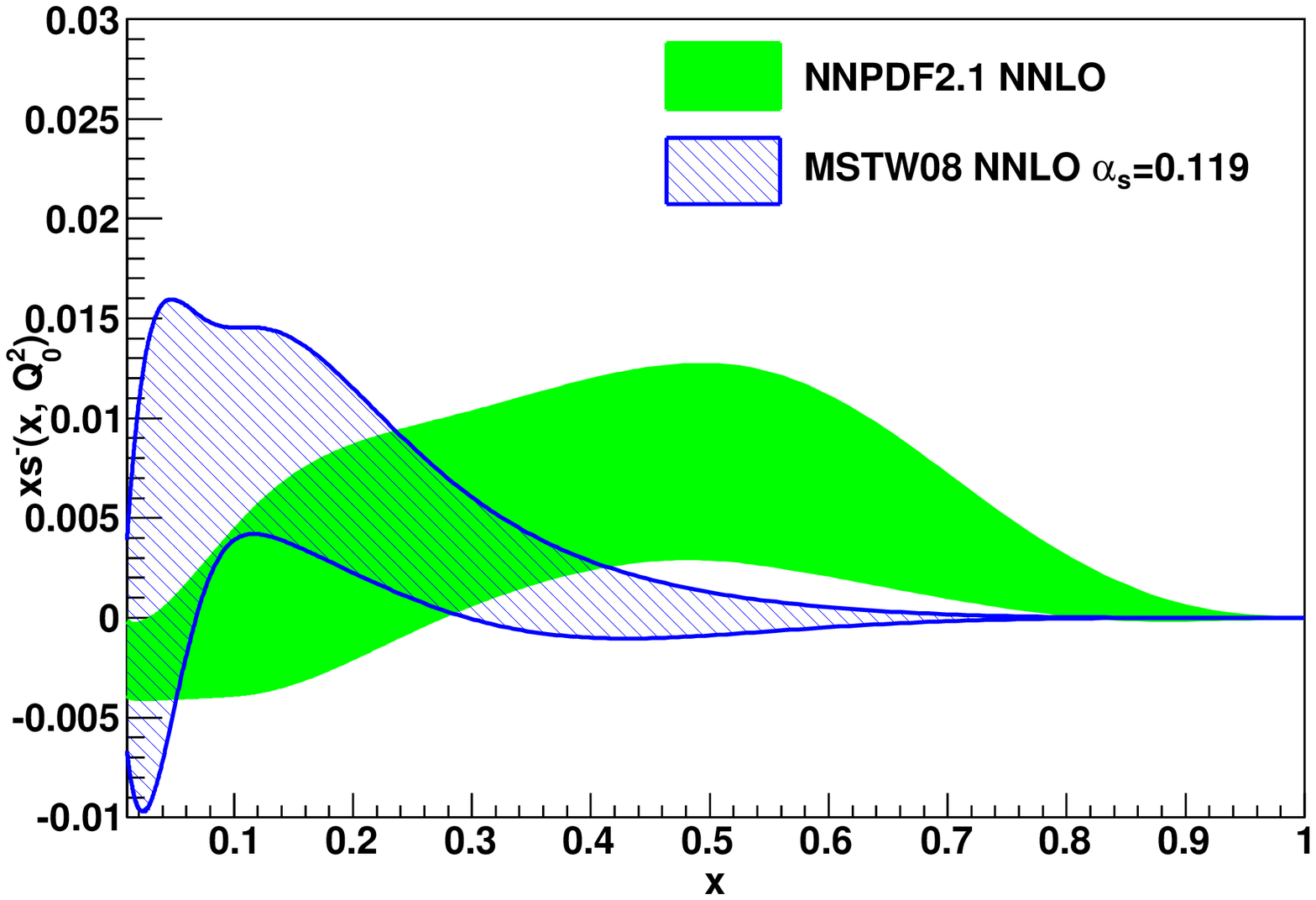}
\caption{\small Same as Fig.~\ref{fig:singletPDFs-lhapdf}
for the non--singlet sector PDFs.
 \label{fig:valencePDFs-lhapdf}} 
\end{center}
\end{figure}

Next, in Figs.~\ref{fig:singletPDFs-lhapdf}
and~\ref{fig:valencePDFs-lhapdf} we compare the NNPDF2.1 NNLO PDFs to 
those from the MSTW08 NNLO set.
For consistency we use in the comparison a common value
of $\alpha_s\lp M_Z\rp$=0.119. The MSTW08 
NNLO gluon, unlike its NNPDF2.1 counterpart, is unstable at small $x$,
where it becomes very negative.
For other PDFs there is reasonable agreement for central values, 
although the uncertainty bands from MSTW often seem unusually small. Sizable differences are
observed in the strange distribution, but it should be recalled
that in MSTW08 the parametrization of the $s+\bar s$ and
especially $s-\bar s$ PDF is extremely restrictive, while in 
NNPDF2.1 they are treated on the same footing as the other PDFs.

\begin{figure}[t]
\begin{center}
\epsfig{width=0.49\textwidth,figure=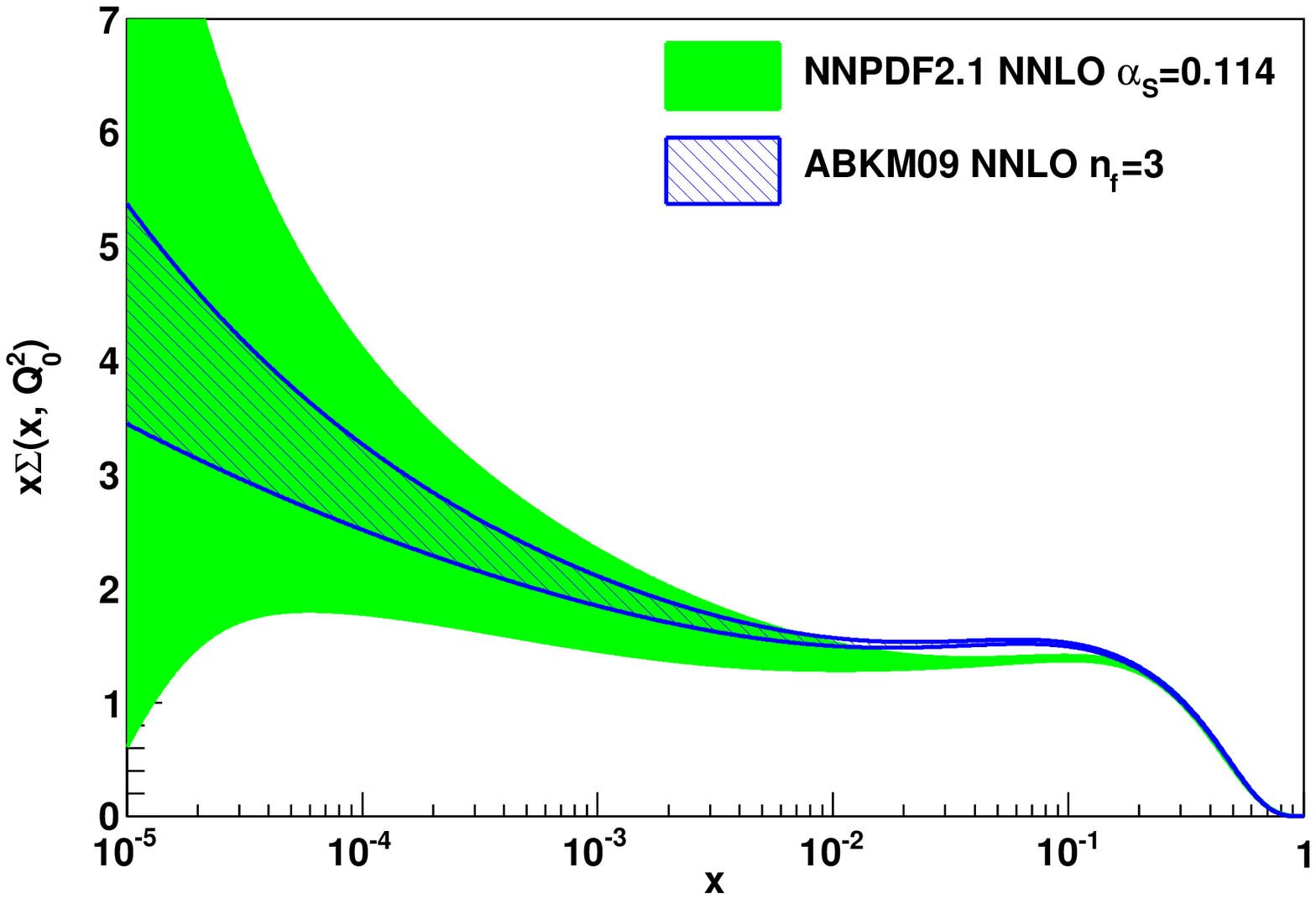}
\epsfig{width=0.49\textwidth,figure=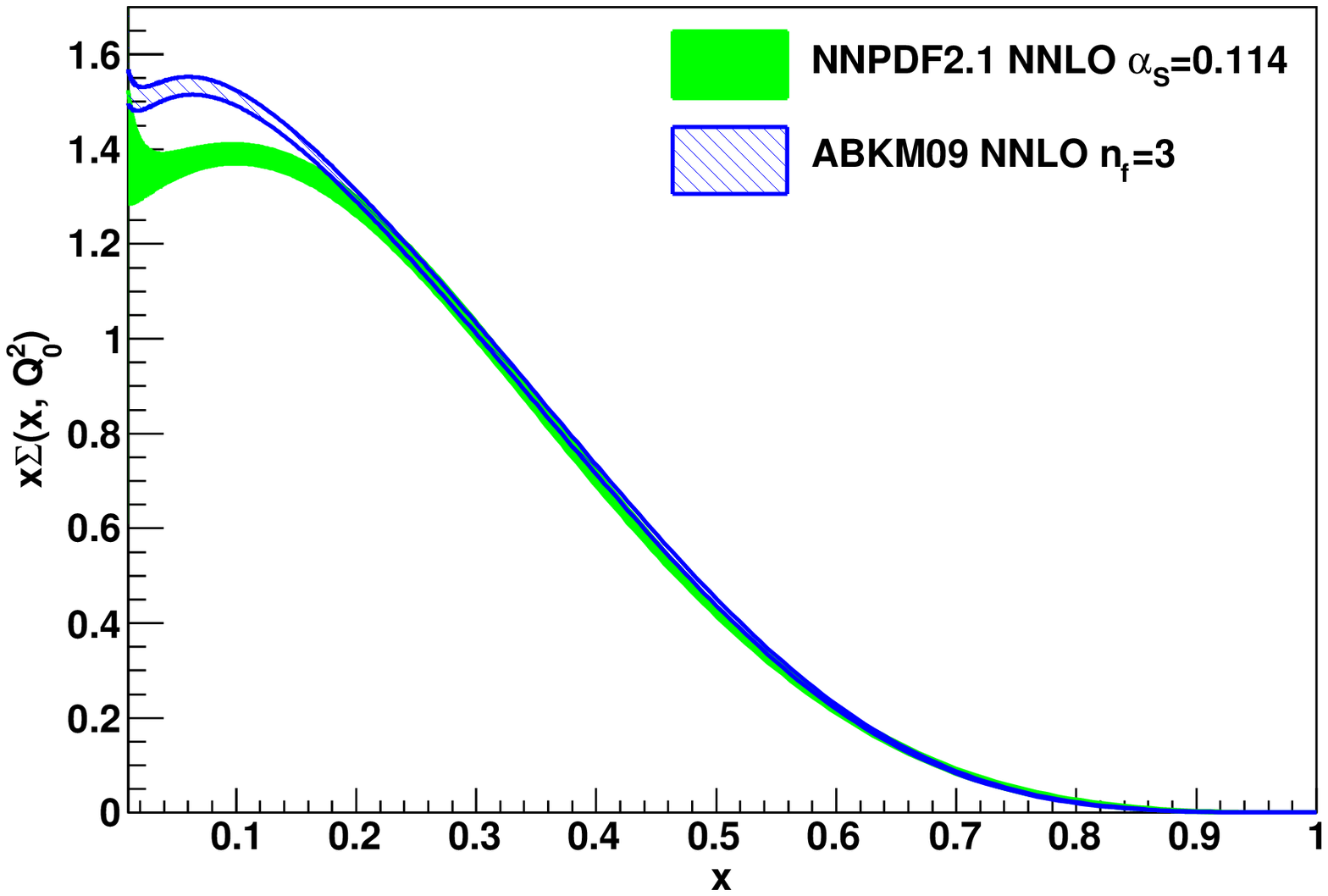}
\epsfig{width=0.49\textwidth,figure=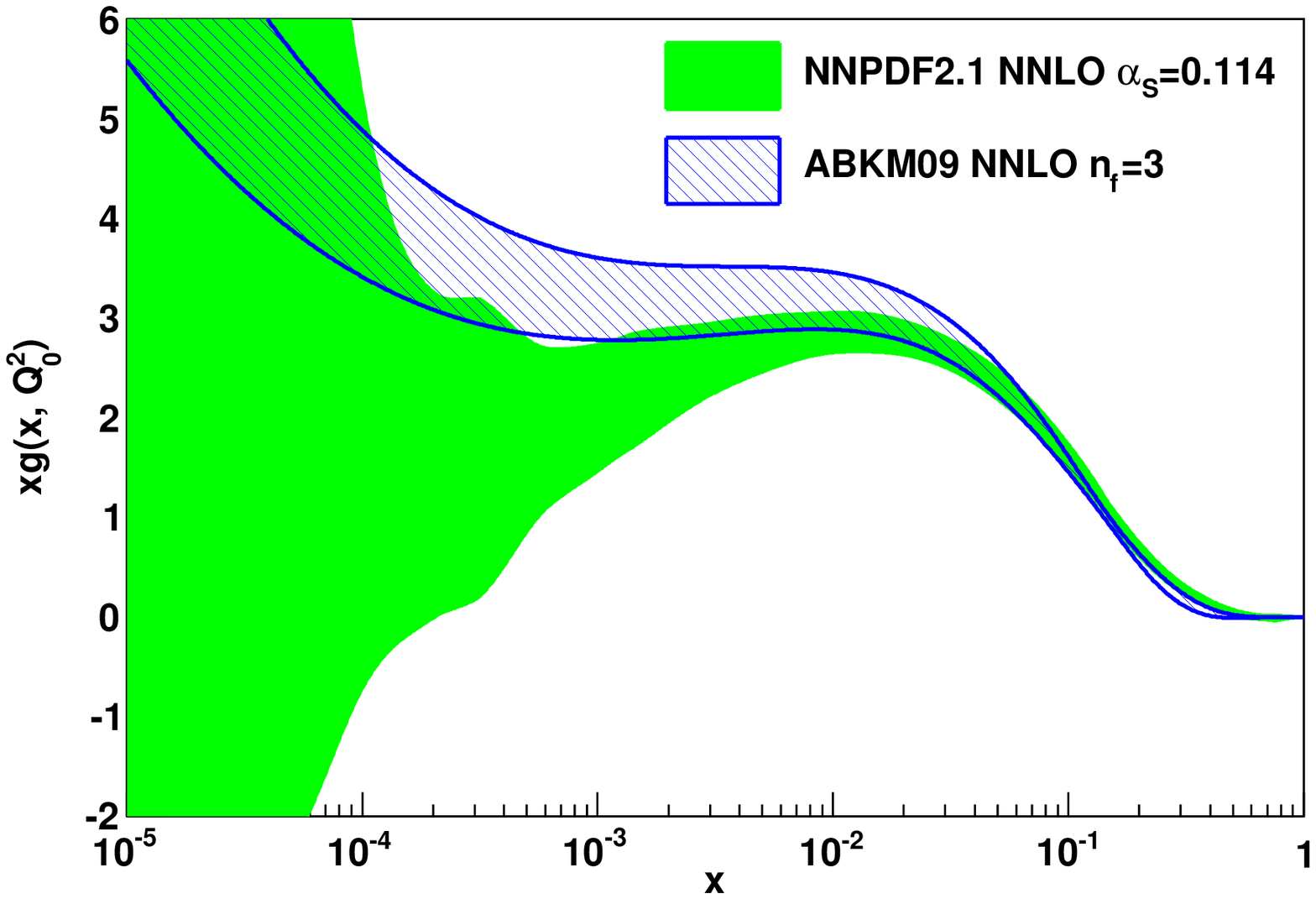}
\epsfig{width=0.49\textwidth,figure=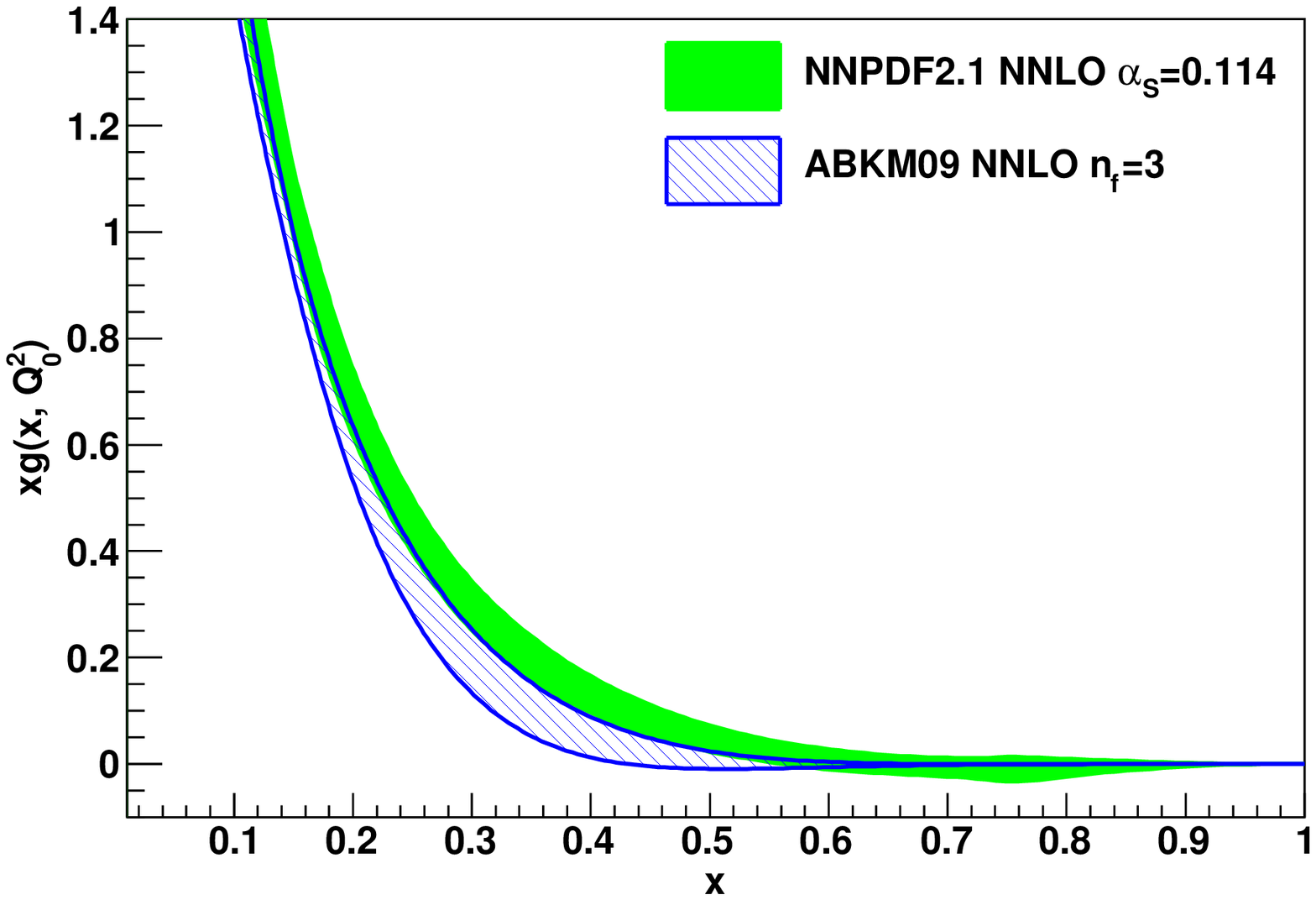}
\caption{\small The NNPDF2.1 NNLO singlet sector PDFs, compared
to the ABKM09 three-flavor set. 
The results for  NNPDF2.1 NNLO  have been obtained with
$N_{\rm rep}=100$ replicas. The NNPDF2.1 set with $\alpha_s=0.114$ is
shown because ABKM PDFs are only available for this value of
$\alpha_s$. Note that for ABKM uncertainties also
include the uncertainty on $\alpha_s$ while for NNPDF they are pure
PDF uncertainties.
 \label{fig:singletPDFs-abkm}} 
\end{center}
\end{figure}

\begin{figure}[t]
\begin{center}
\epsfig{width=0.49\textwidth,figure=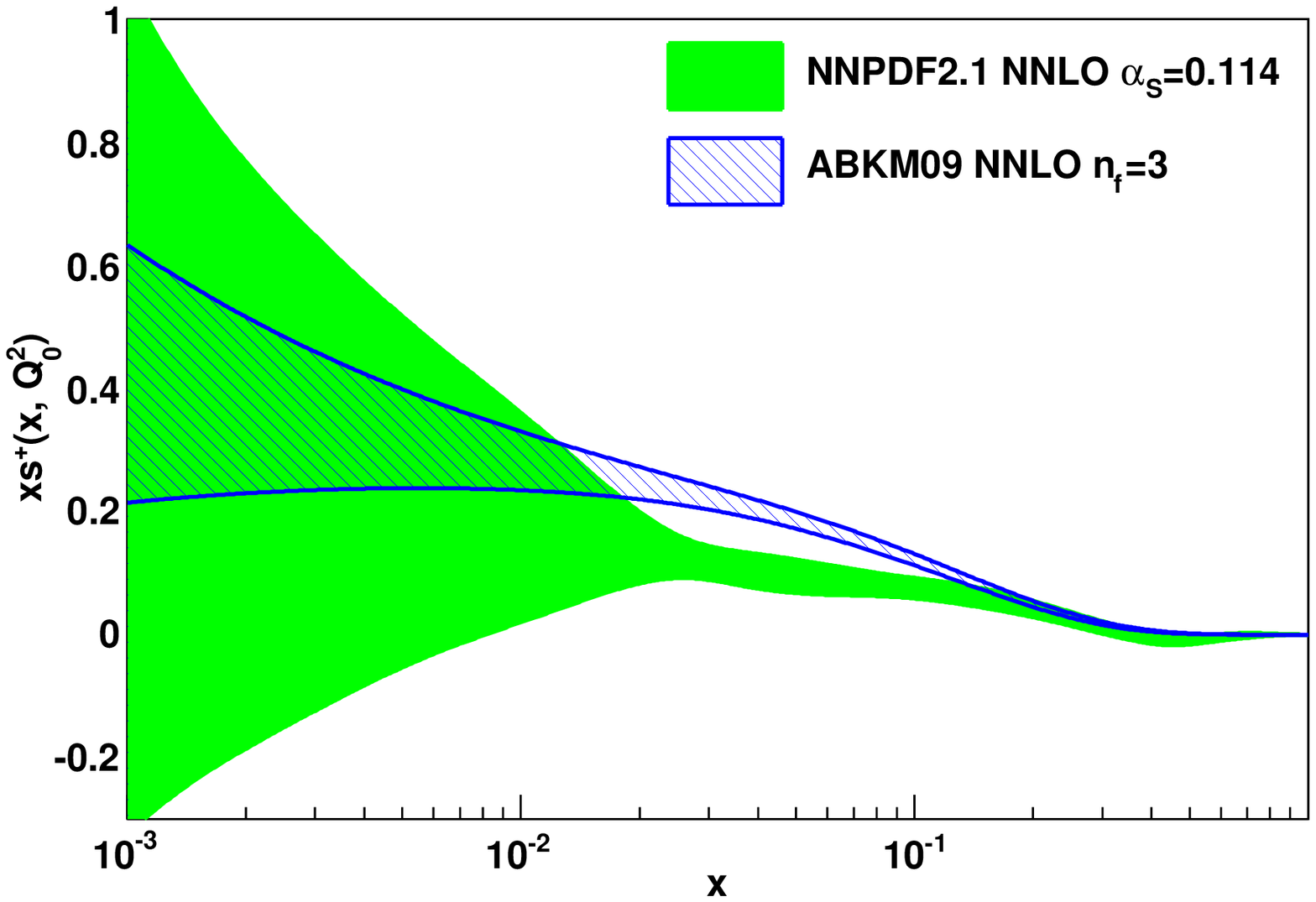}
\epsfig{width=0.49\textwidth,figure=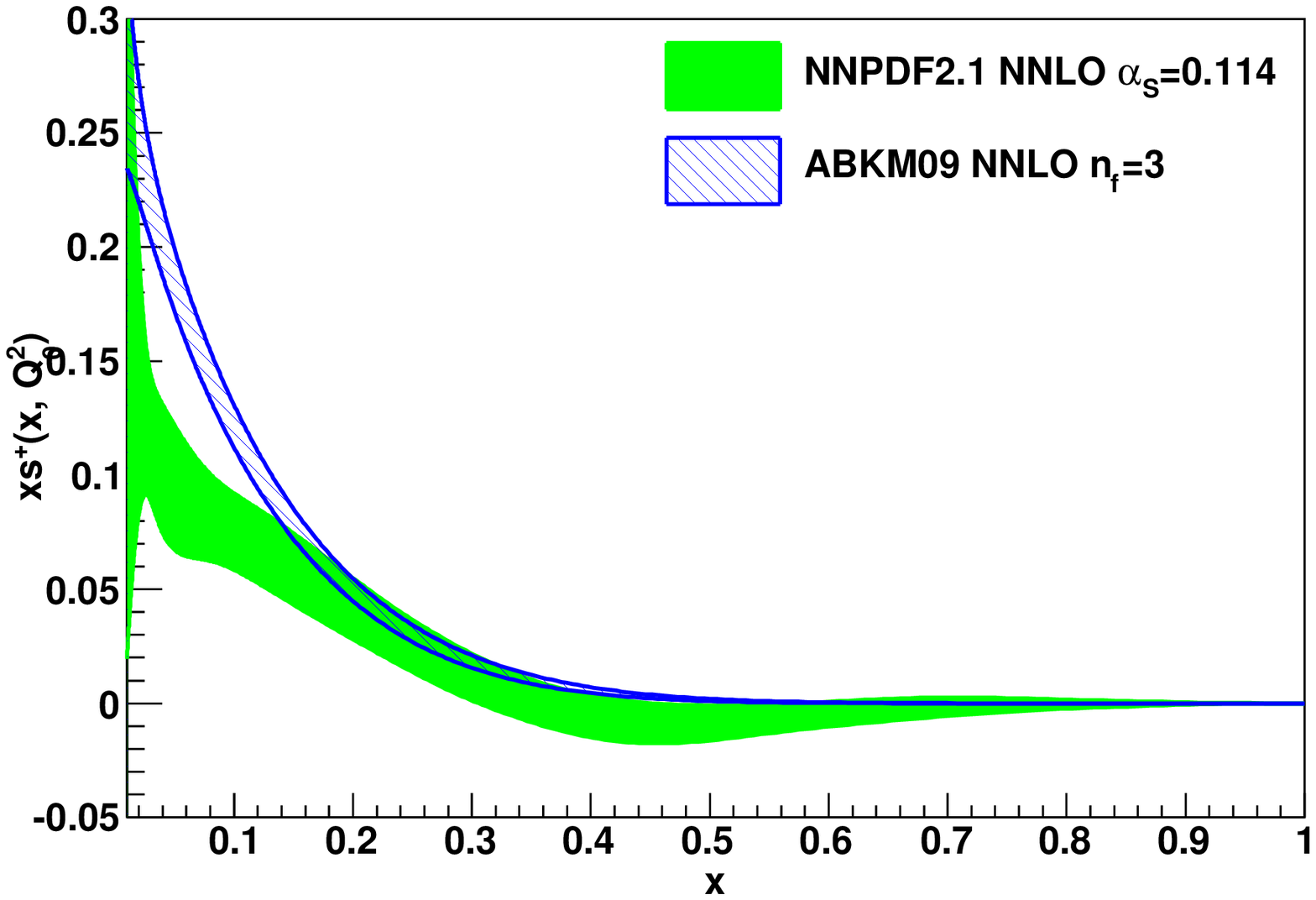}
\epsfig{width=0.49\textwidth,figure=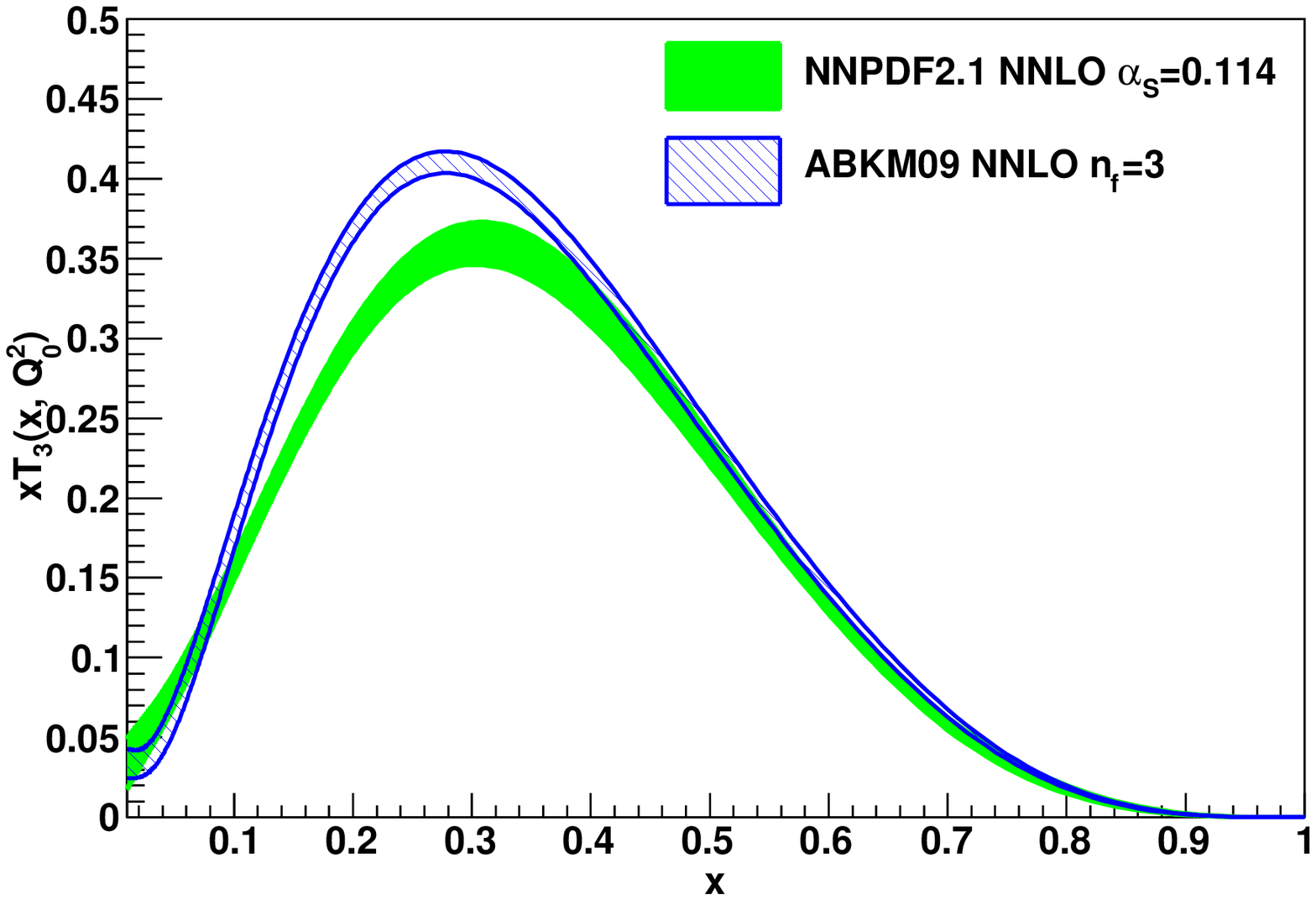}
\epsfig{width=0.49\textwidth,figure=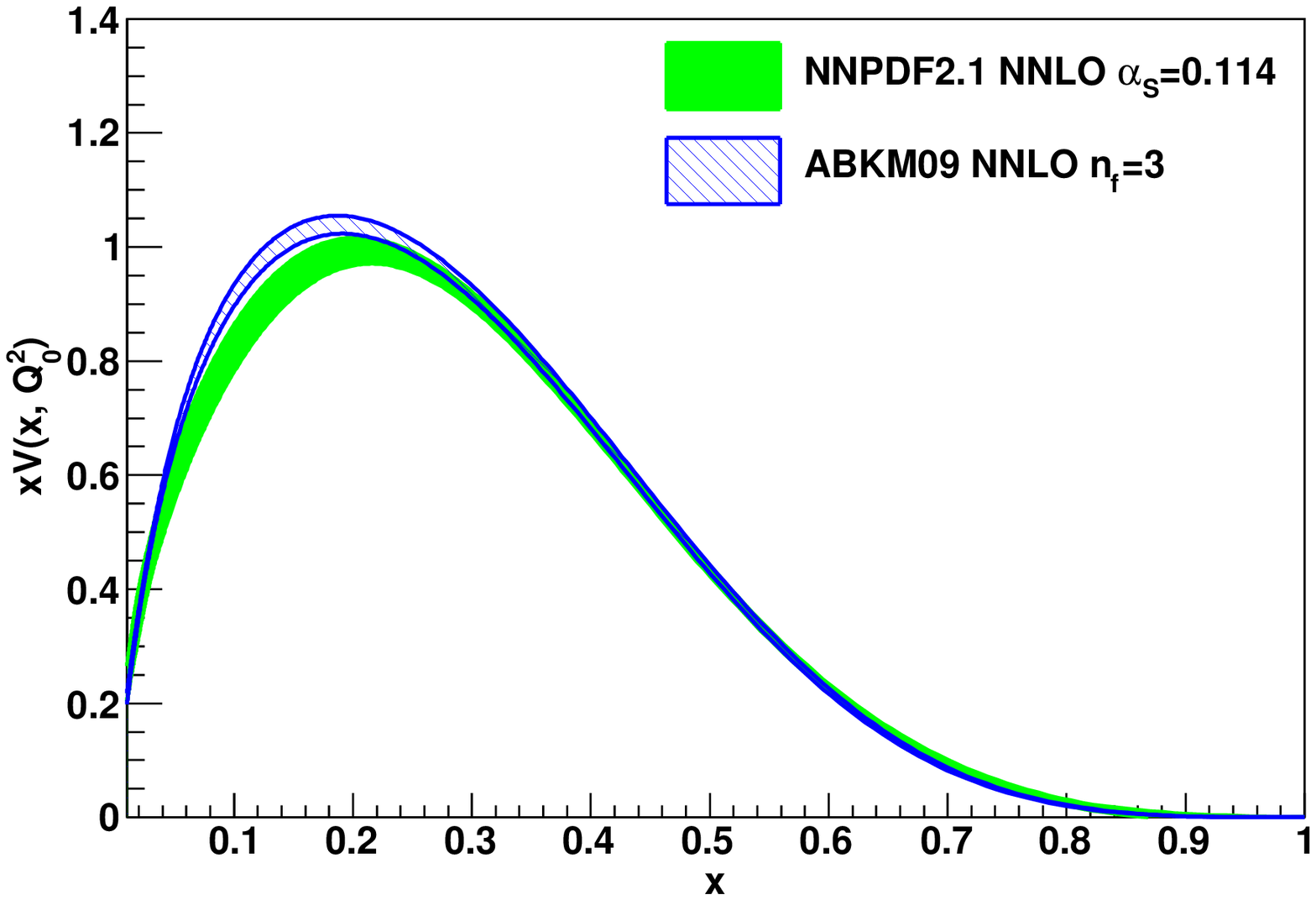}
\epsfig{width=0.49\textwidth,figure=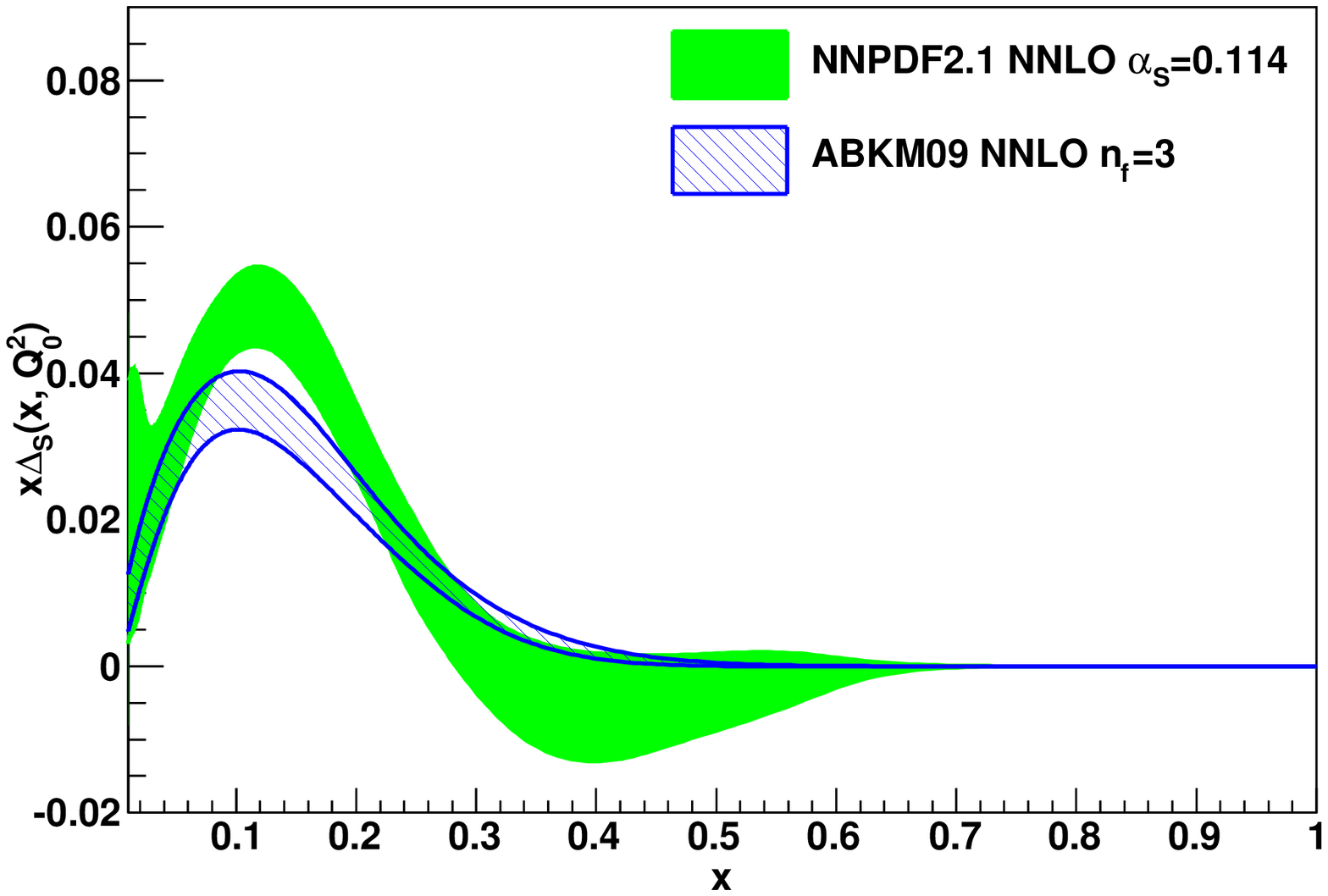}
\epsfig{width=0.49\textwidth,figure=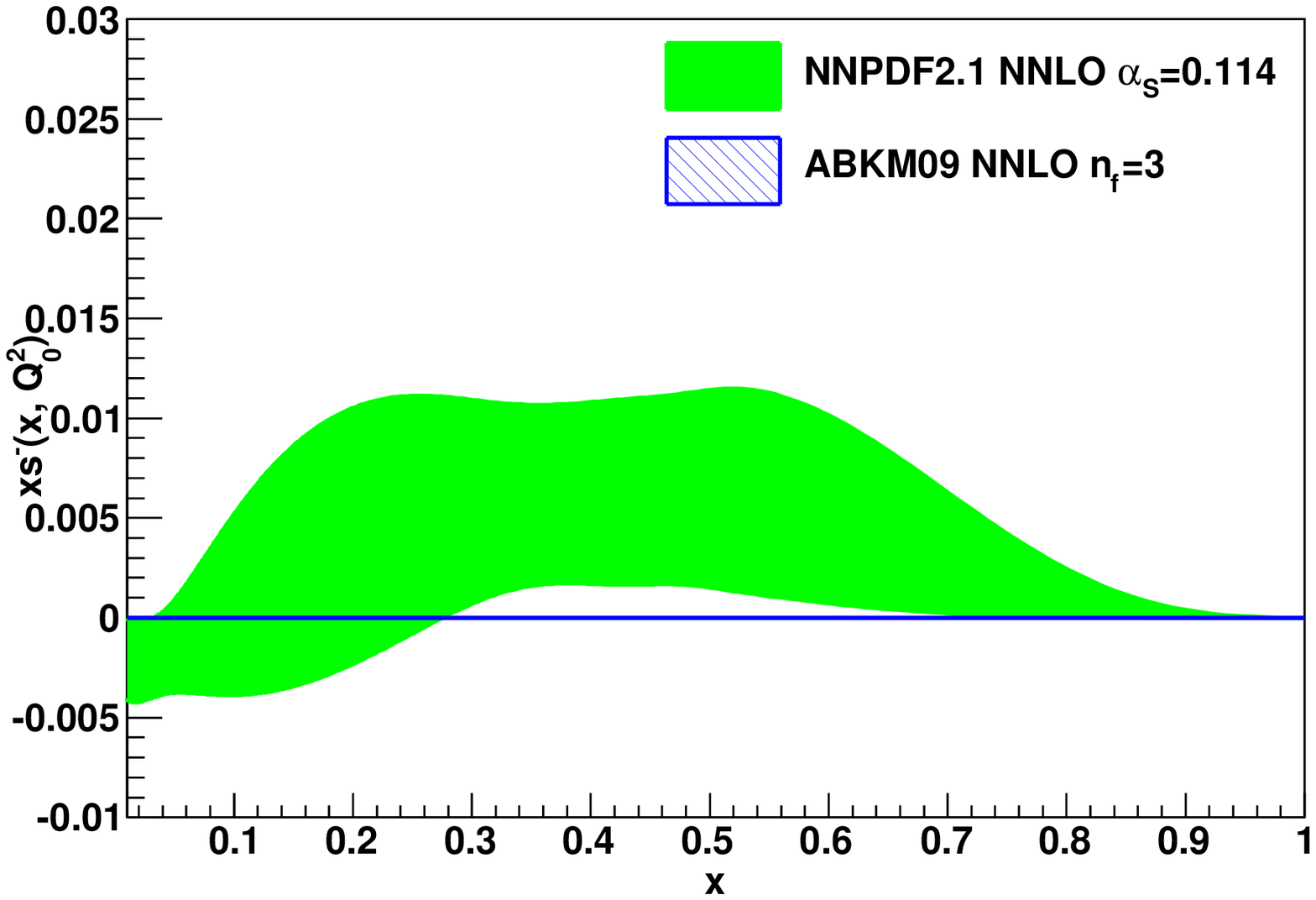}
\caption{\small Same as Fig.~\ref{fig:singletPDFs-abkm}
for the non--singlet sector PDFs.
 \label{fig:valencePDFs-abkm}} 
\end{center}
\end{figure}

At present, MSTW08 is the only NNLO PDF set which is publicly
available through the LHAPDF~\cite{Bourilkov:2006cj,LHAPDFurl}
interface for a variety of values of $\alpha_s$. However, it may also
be interesting to compare the NNPDF2.1 NNLO PDFs to the ABKM09 NNLO 
set (with fixed flavour number $n_f=3$)~\cite{Alekhin:2009ni}. This
set
is only provided for $\alpha_s\lp
M_Z\rp=0.1135\pm0.0014$, furthermore  for this set (and its NLO
counterpart) only combined PDF+$\alpha_s$ uncertainties 
can be determined, unlike
other sets for which PDF uncertainties with fixed $\alpha_s$ may also
be computed.  The comparison is shown in
Figs.~\ref{fig:singletPDFs-abkm} and ~\ref{fig:valencePDFs-abkm} at
$Q_0^2=$ 2 GeV$^2$, where we have chosen the NNPDF2.1 set with
$\alpha_s=0.114$ in order to make the comparison more
significant. Even so,   the agreement is generally not very good.

%% file: sec-msr.tex
\begin{figure}[t]
\begin{center}
\epsfig{width=0.49\textwidth,figure=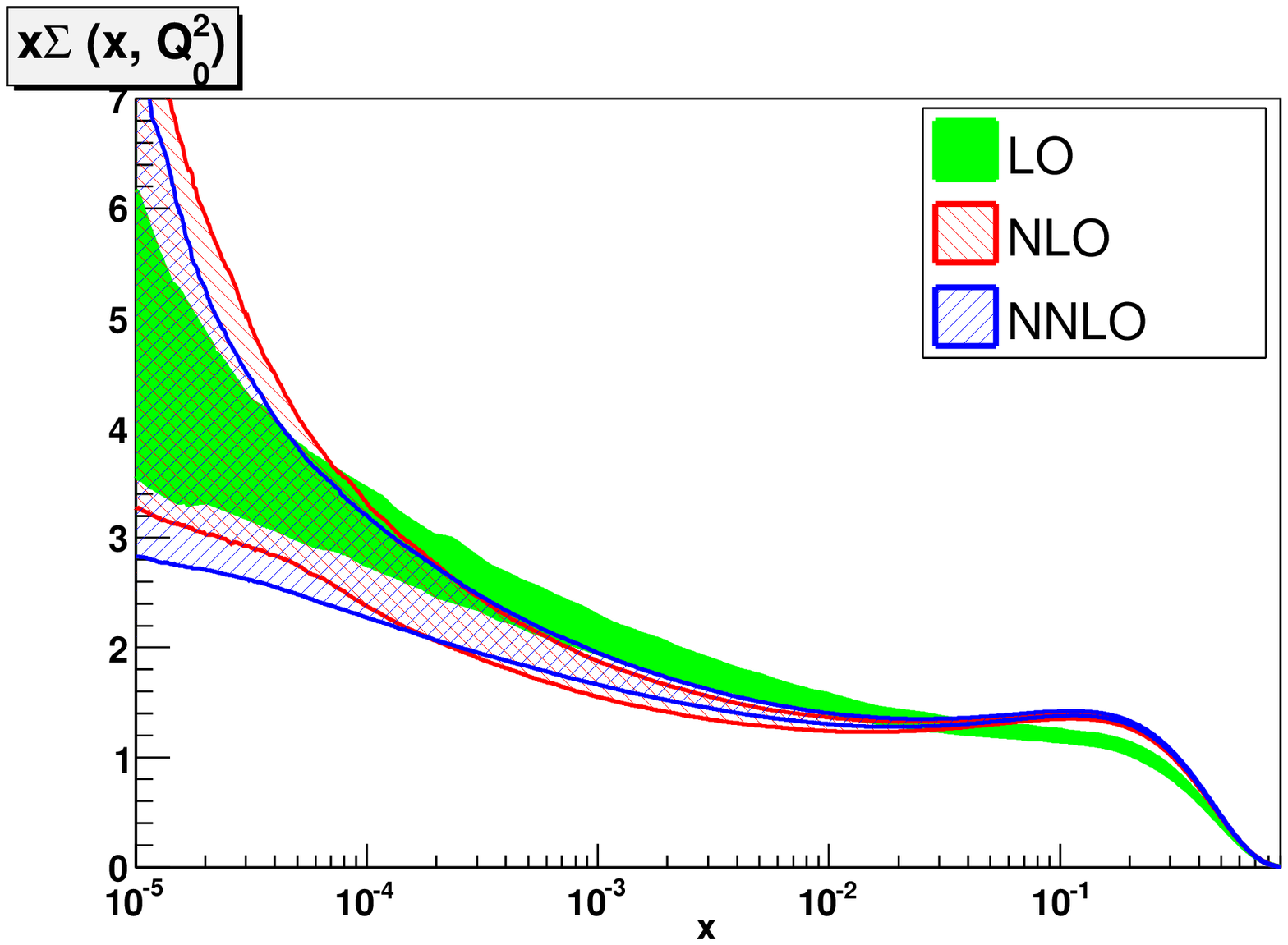}
\epsfig{width=0.49\textwidth,figure=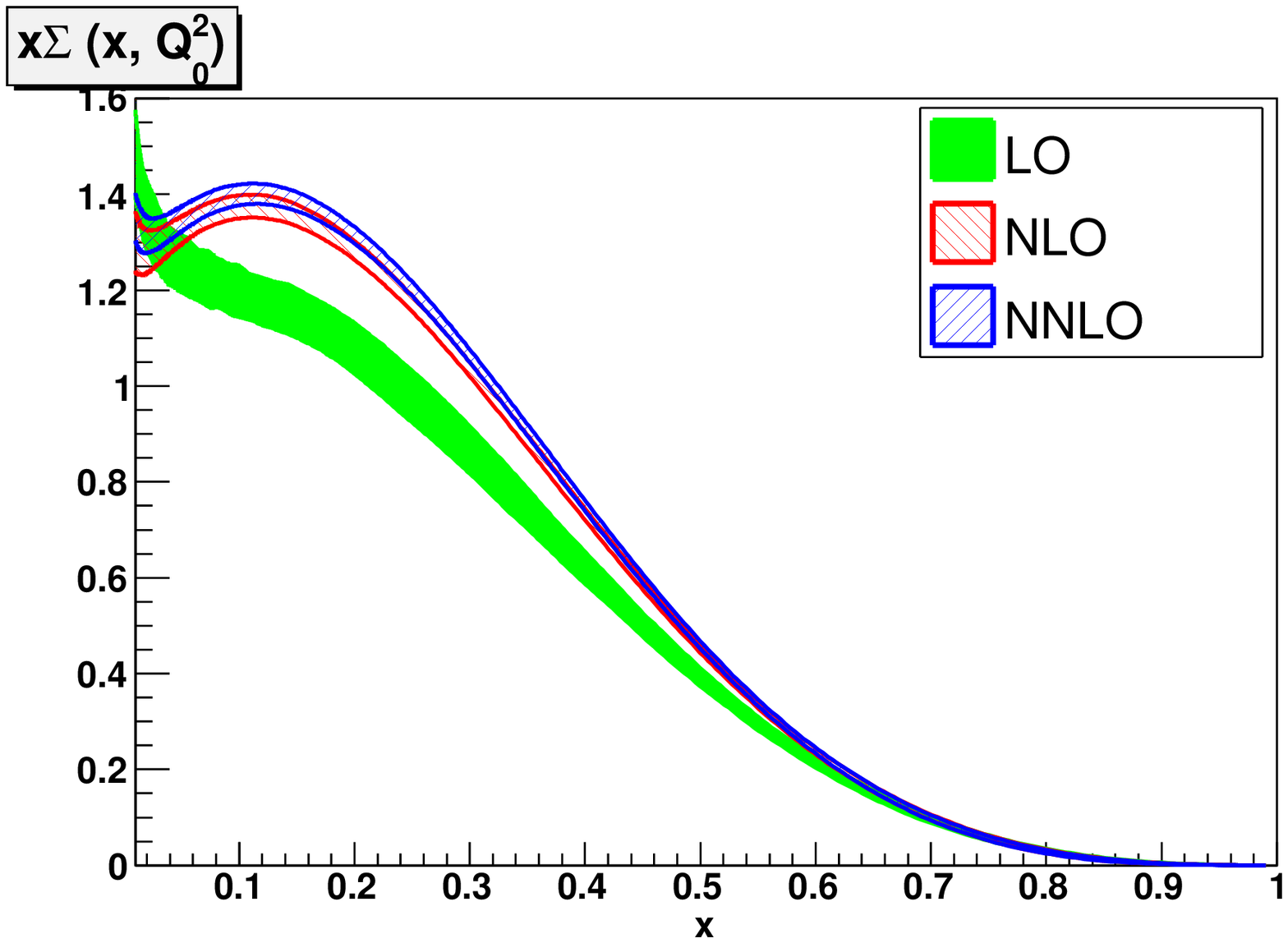}
\epsfig{width=0.49\textwidth,figure=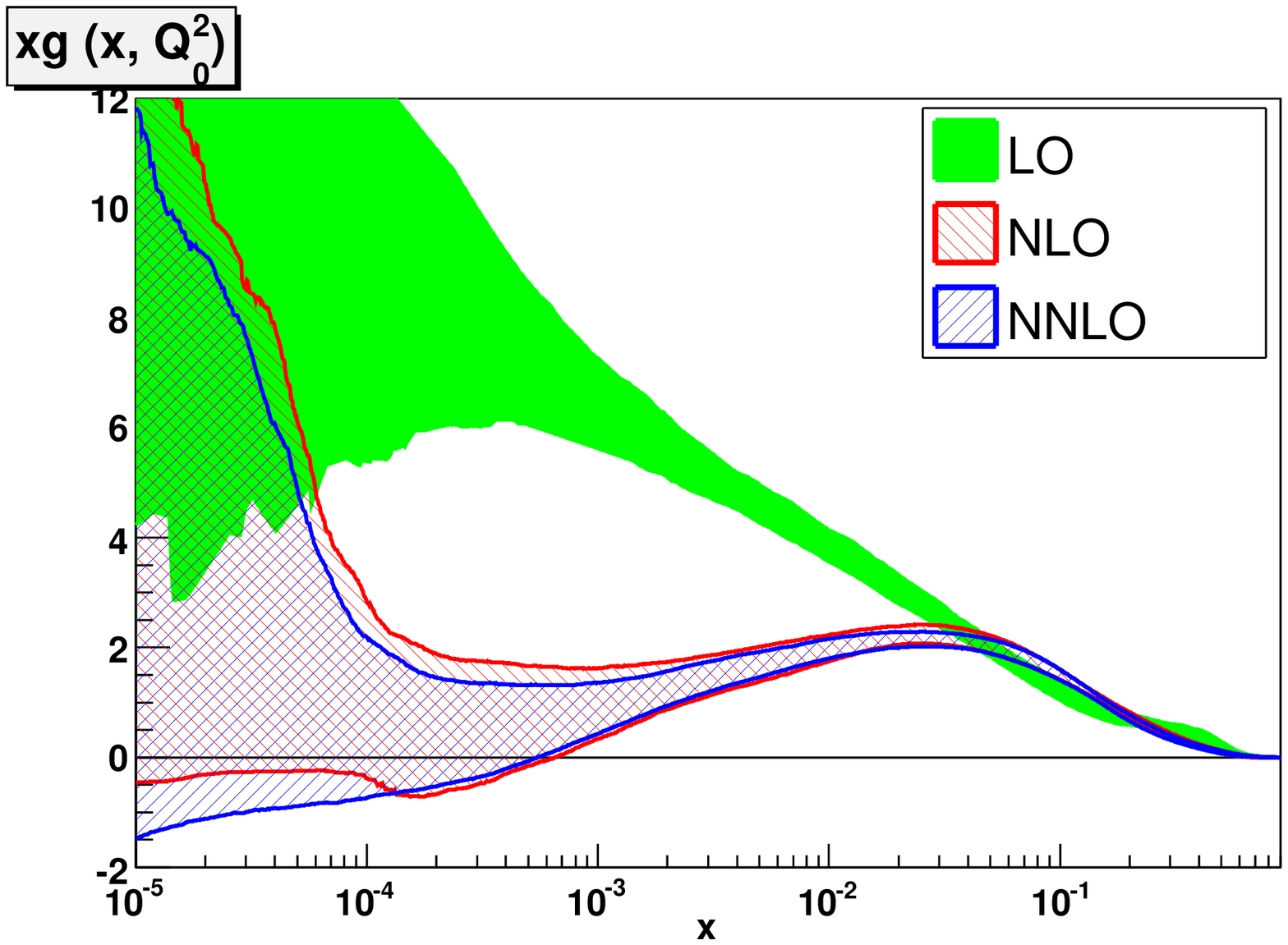}
\epsfig{width=0.49\textwidth,figure=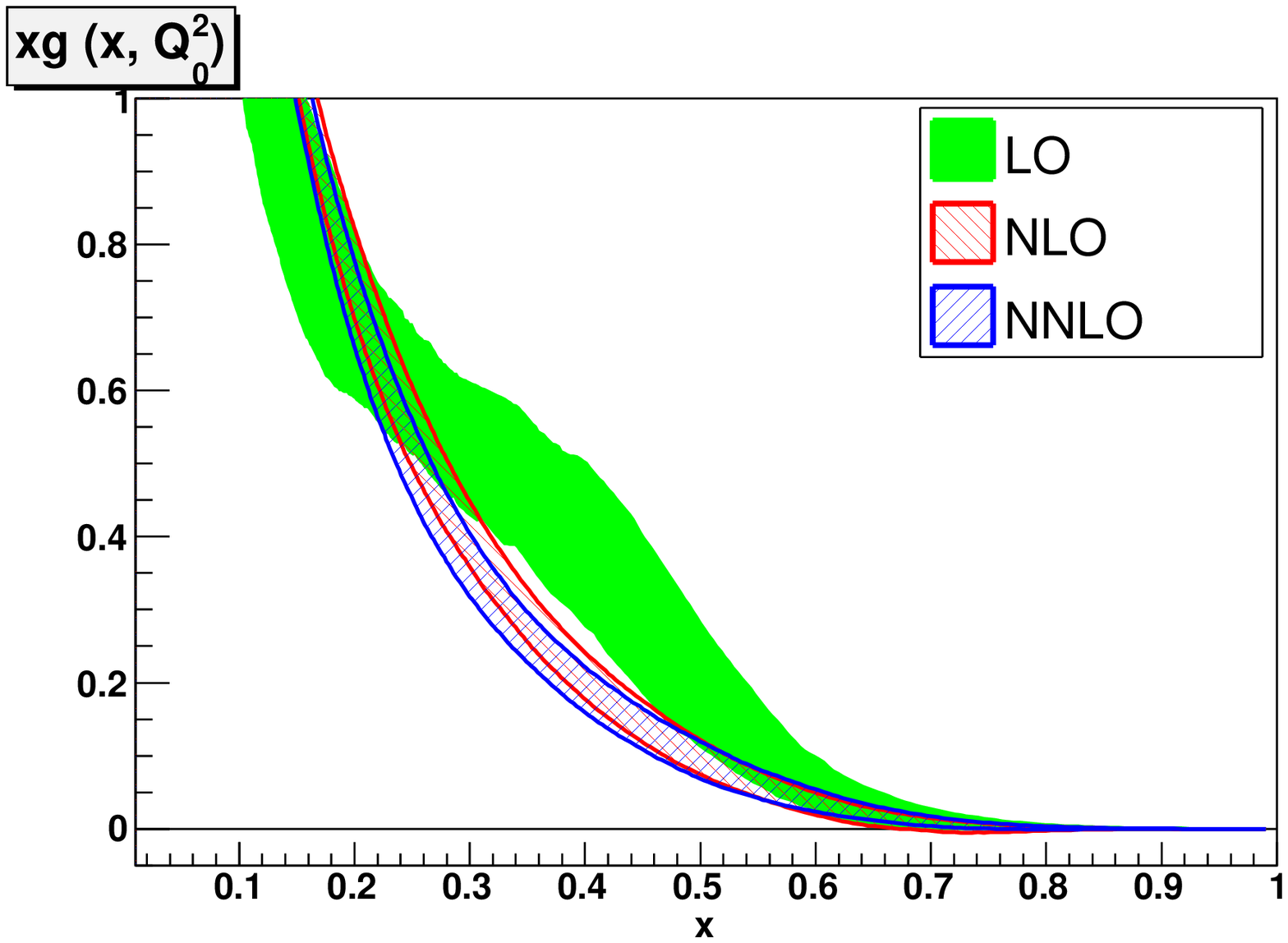}
\caption{\small Comparison of  NNPDF2.1 LO, NLO and NNLO  singlet
  sector PDFs at $Q_0^2$=2~GeV$^2$.
All  uncertainty bands are defined as 68\% confidence levels.
 \label{fig:singletPDFs-summary}} 
\end{center}
\end{figure}

\section{Perturbative stability}
\label{sec:pertstab}

With PDF sets at LO, NLO and NNLO determined from the same data and
using a uniform methodology we can address issues of perturbative
stability. We will do this first by comparing individual PDFs, and
then by looking at the behaviour of the total momentum fraction
carried by partons.

\subsection{Parton distributions}

\label{sec:pdfcomp}

\label{sec:summary-plots}

\begin{figure}[t]
\begin{center}
\epsfig{width=0.49\textwidth,figure=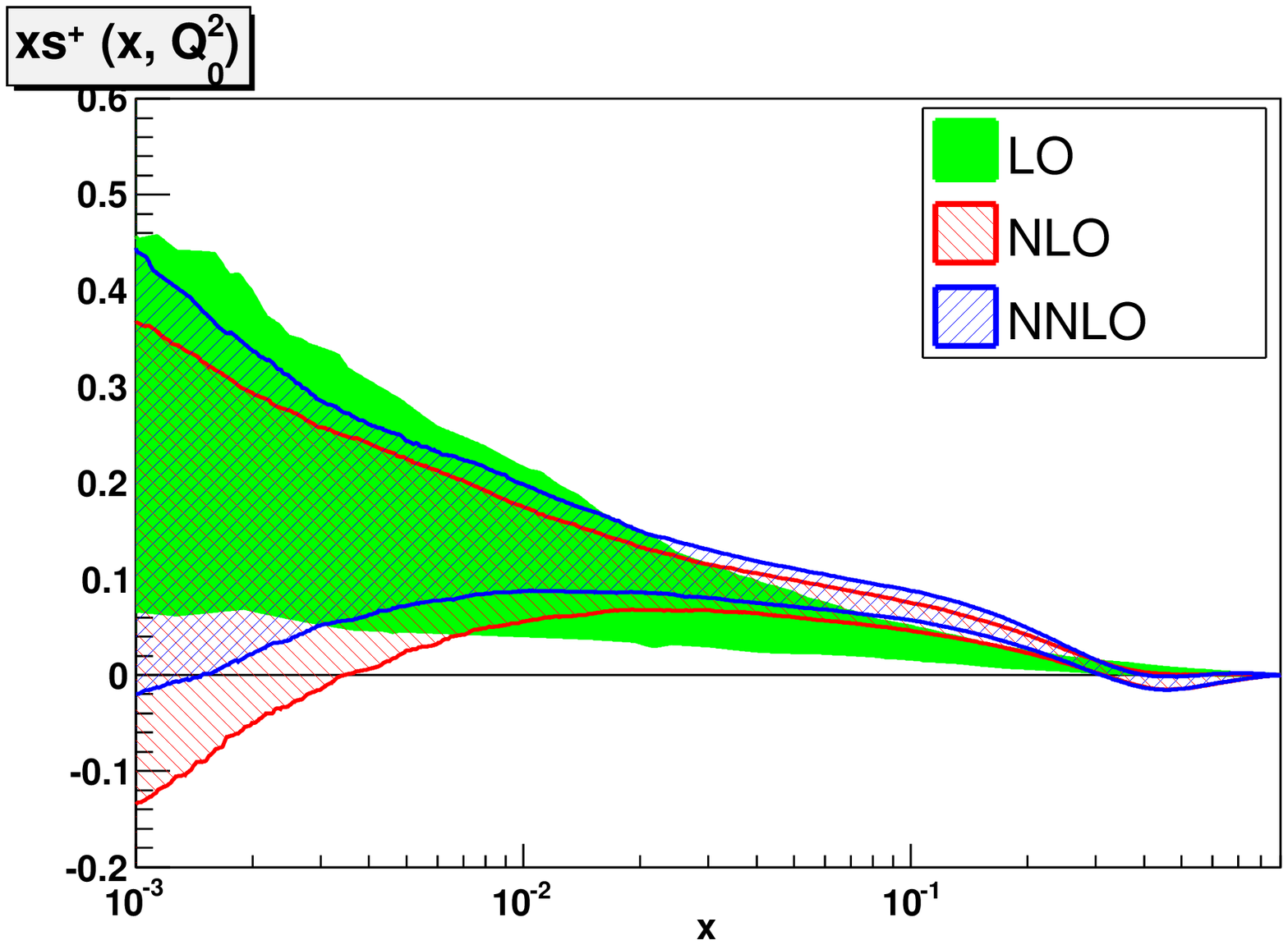}
\epsfig{width=0.49\textwidth,figure=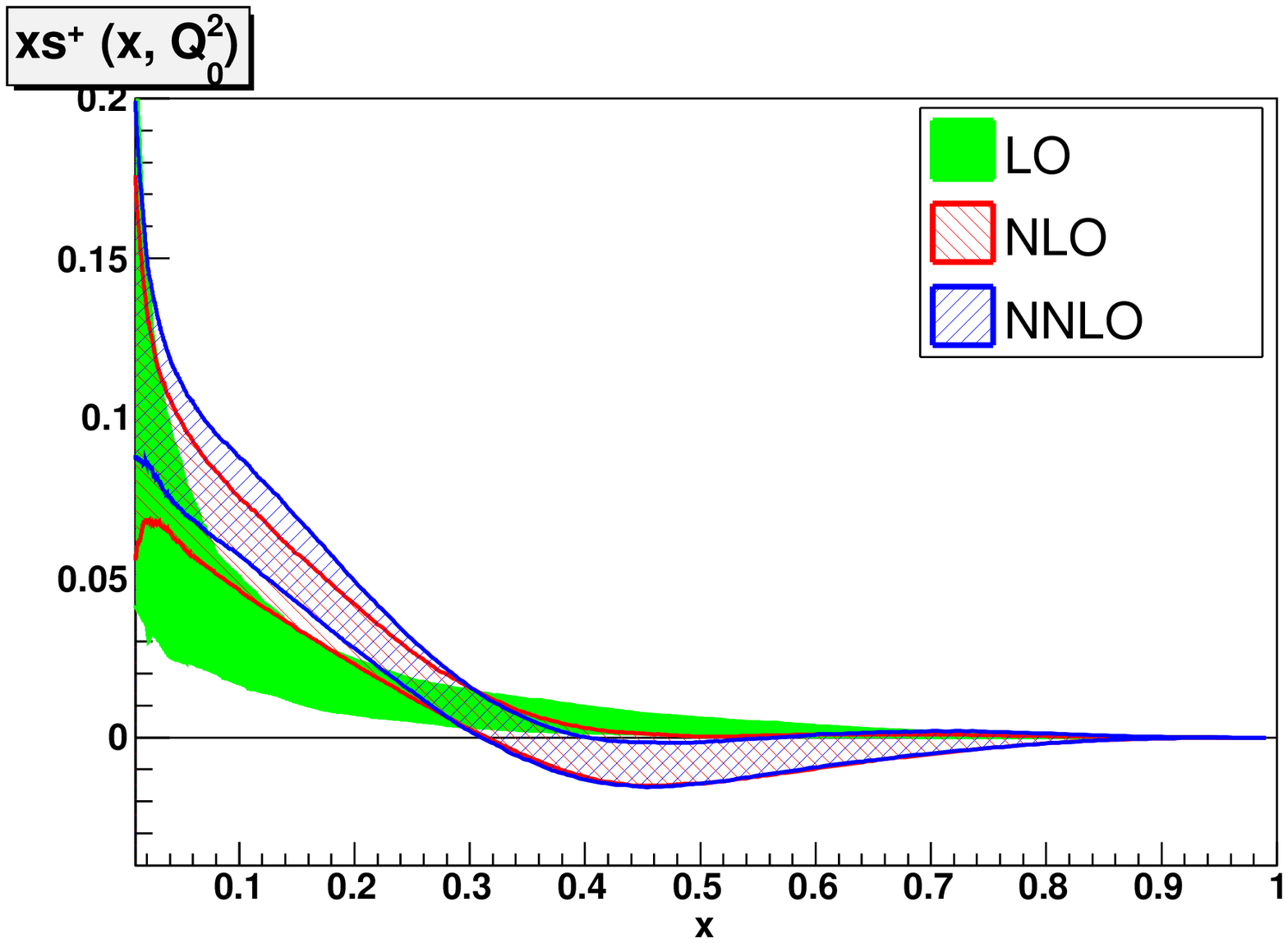}
\epsfig{width=0.49\textwidth,figure=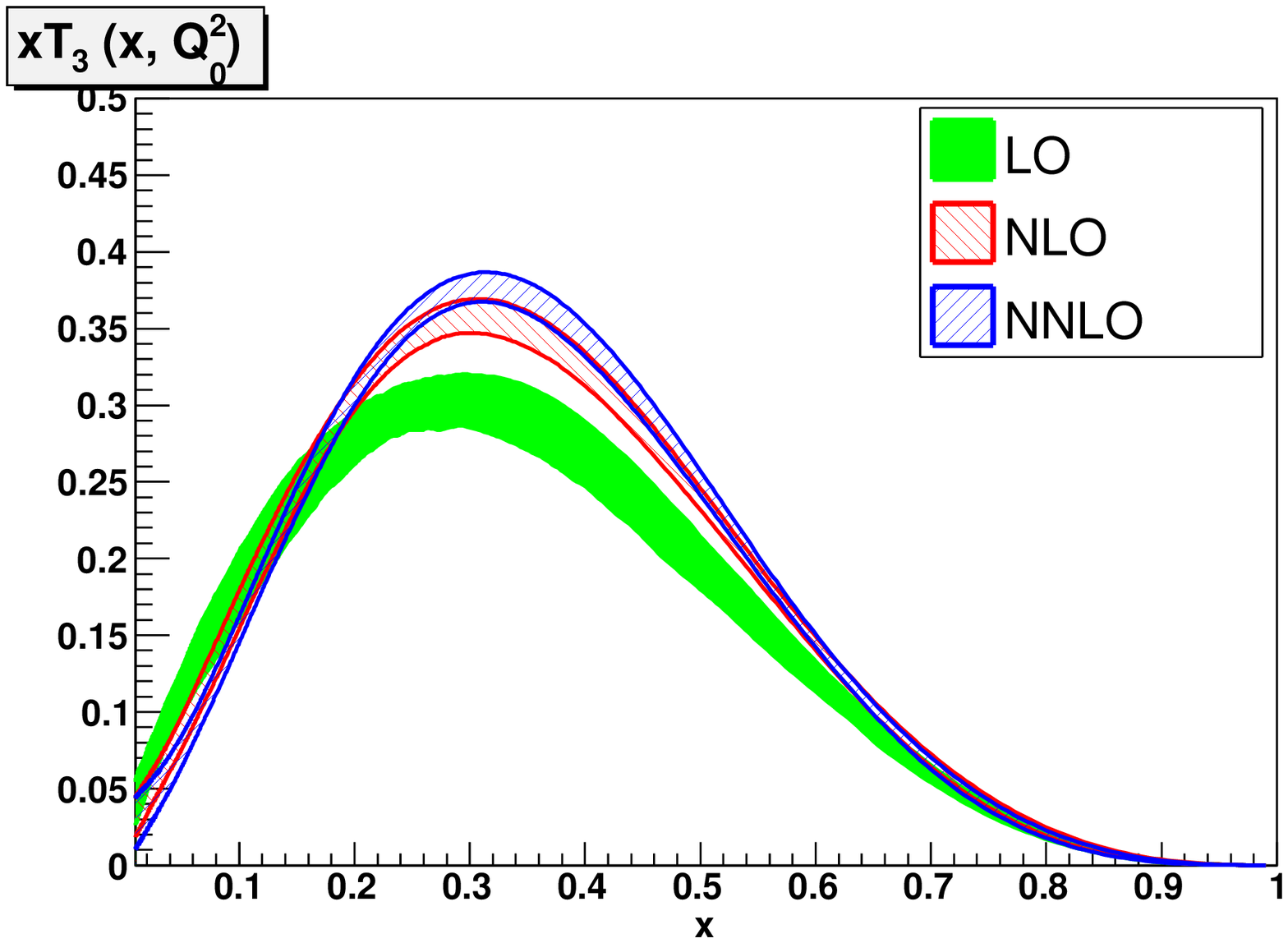}
\epsfig{width=0.49\textwidth,figure=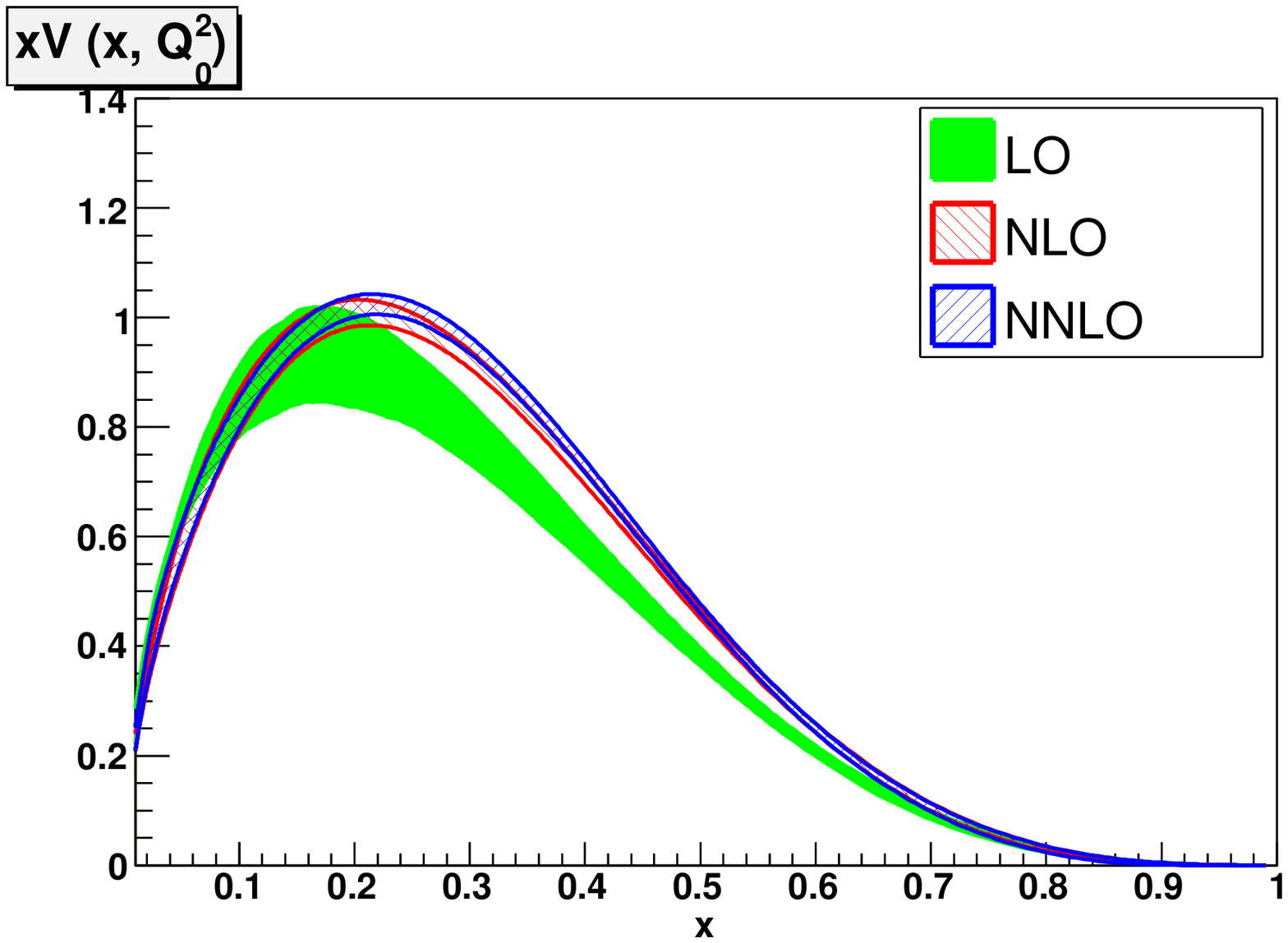}
\epsfig{width=0.49\textwidth,figure=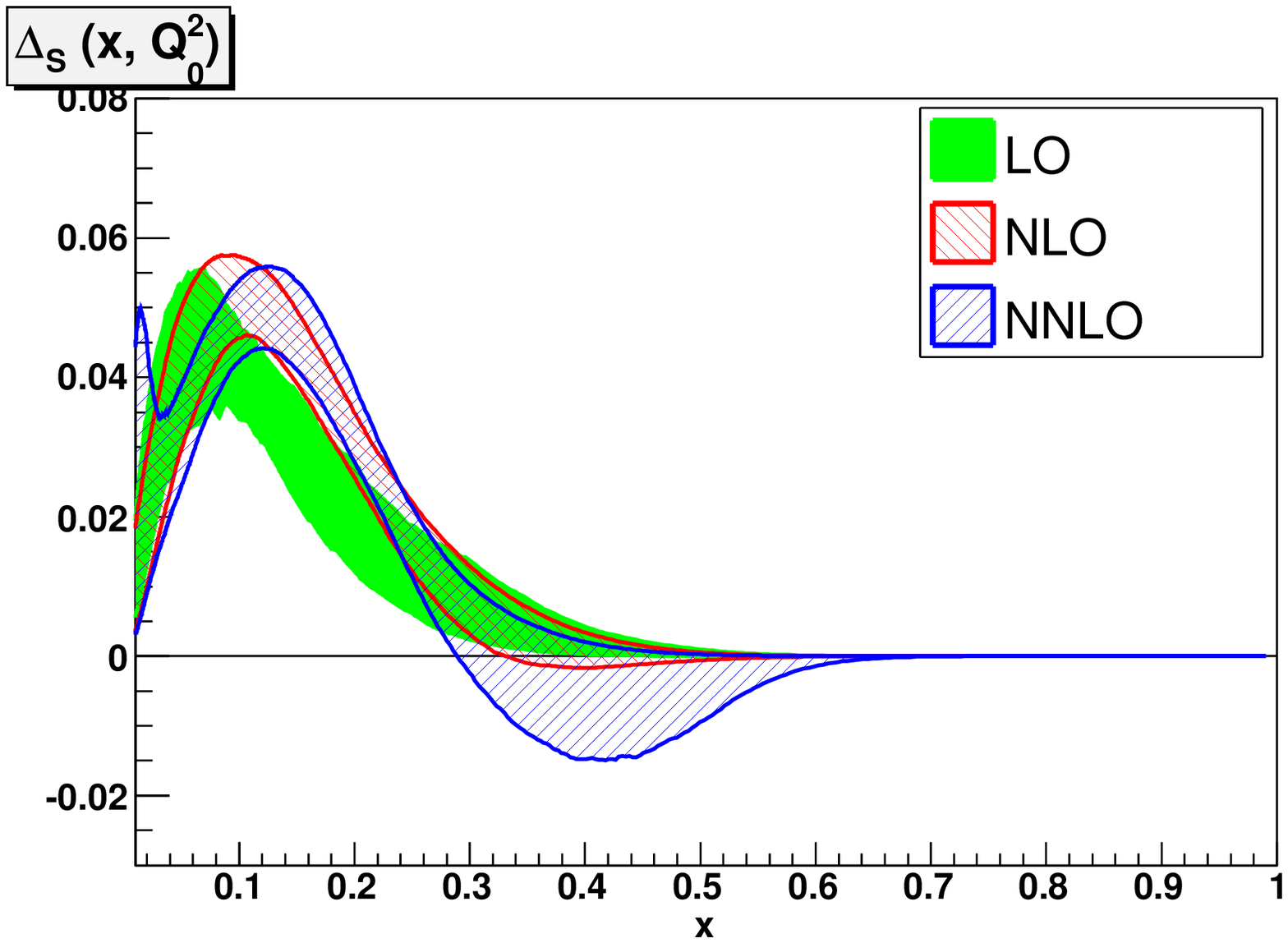}
\epsfig{width=0.49\textwidth,figure=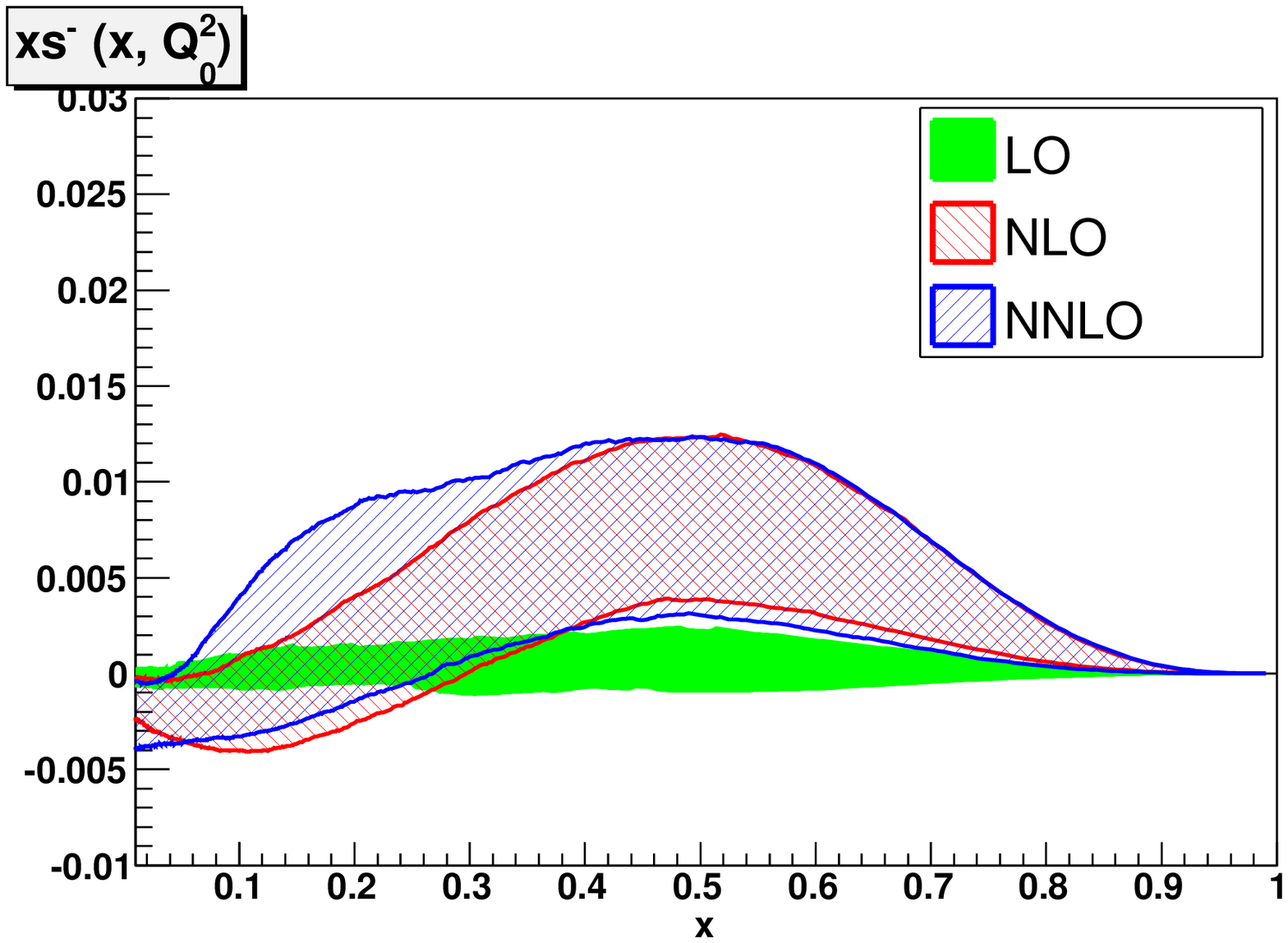}
\caption{\small Same as Fig.~\ref{fig:singletPDFs-summary}
for the non-singlet sector PDFs.
 \label{fig:valencePDFs-summary}} 
\end{center}
\end{figure}

\begin{figure}[t]
\begin{center}
\epsfig{width=0.49\textwidth,figure=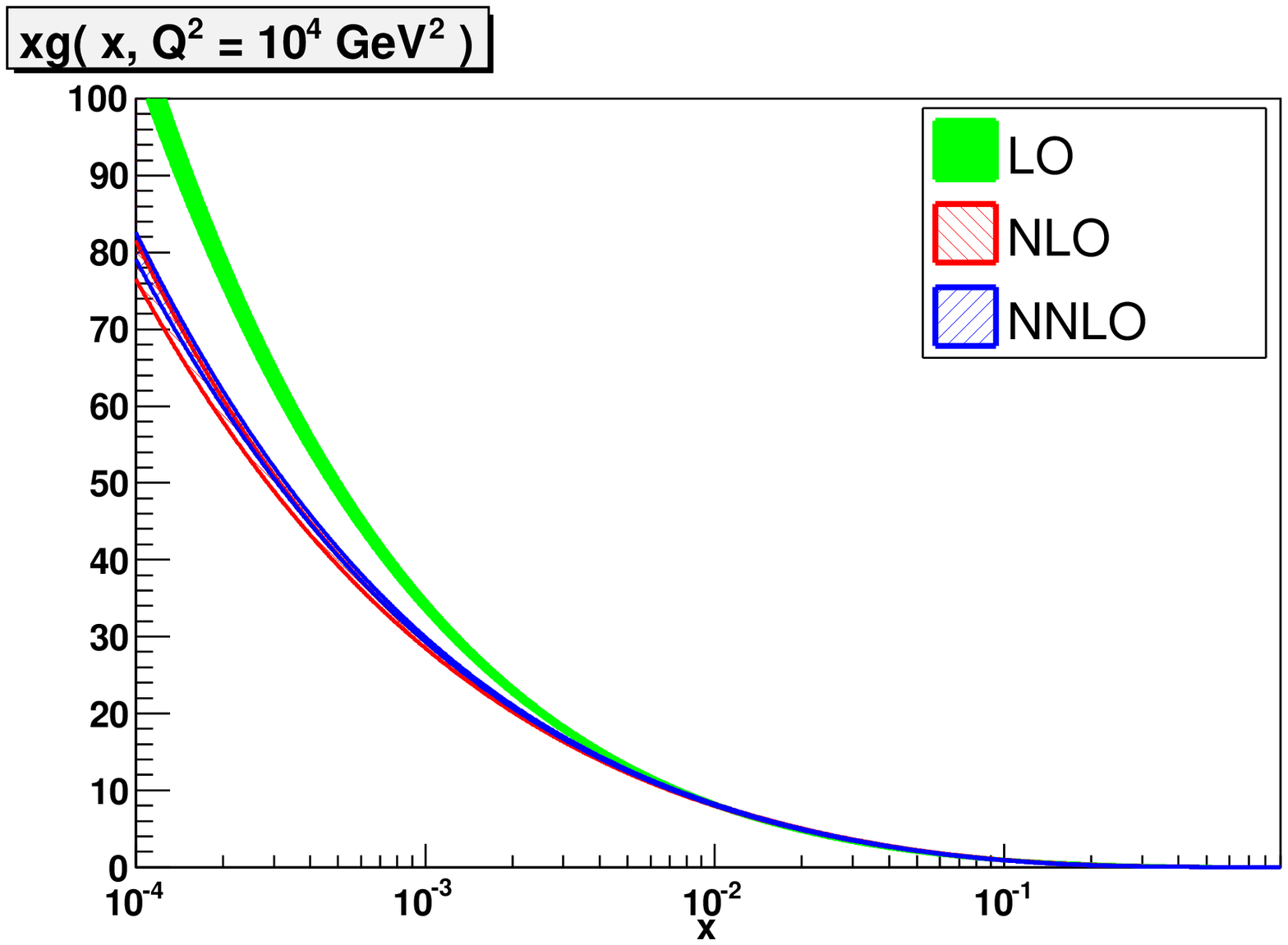}
\epsfig{width=0.49\textwidth,figure=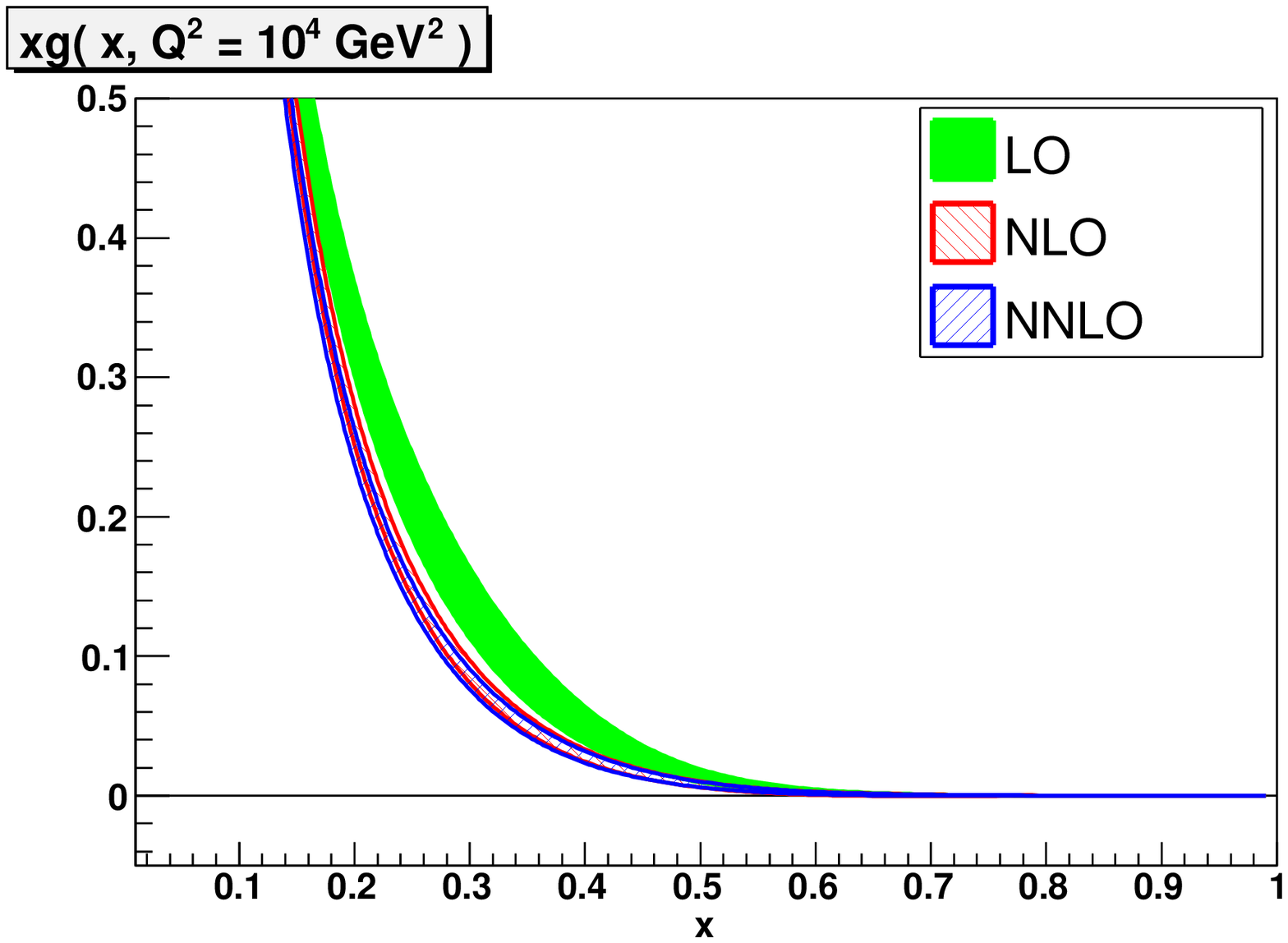}
\epsfig{width=0.49\textwidth,figure=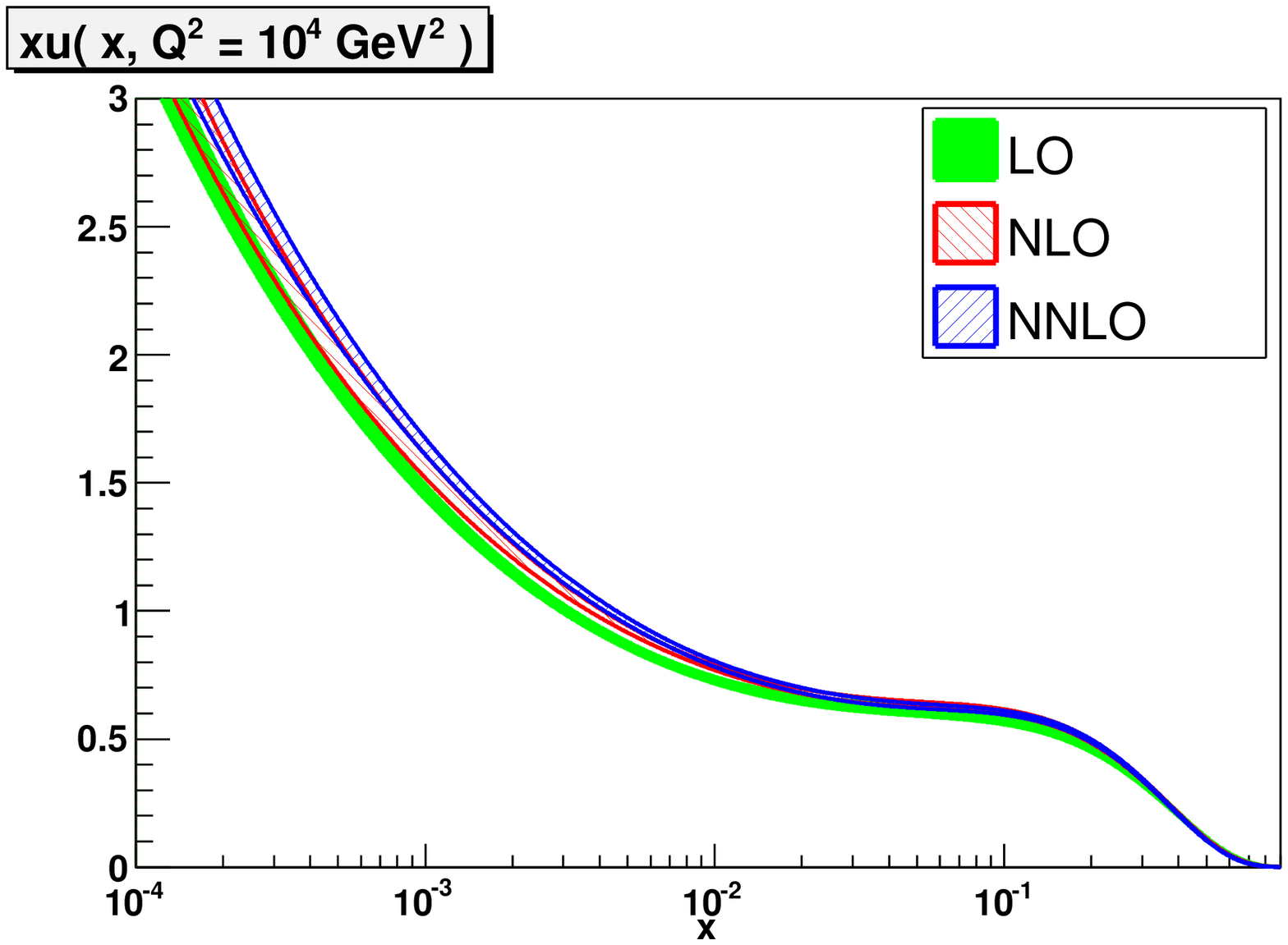}
\epsfig{width=0.49\textwidth,figure=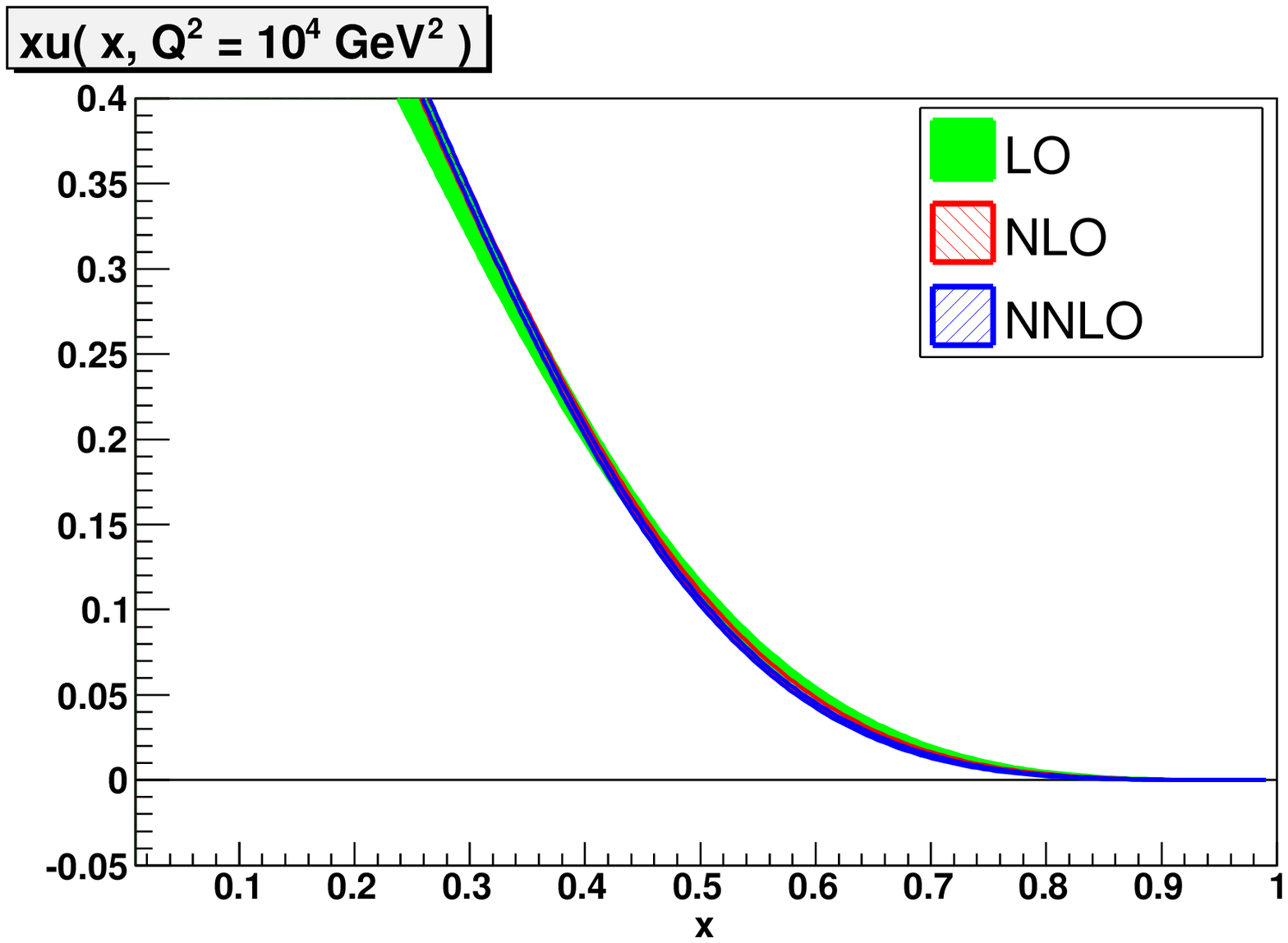}
\epsfig{width=0.49\textwidth,figure=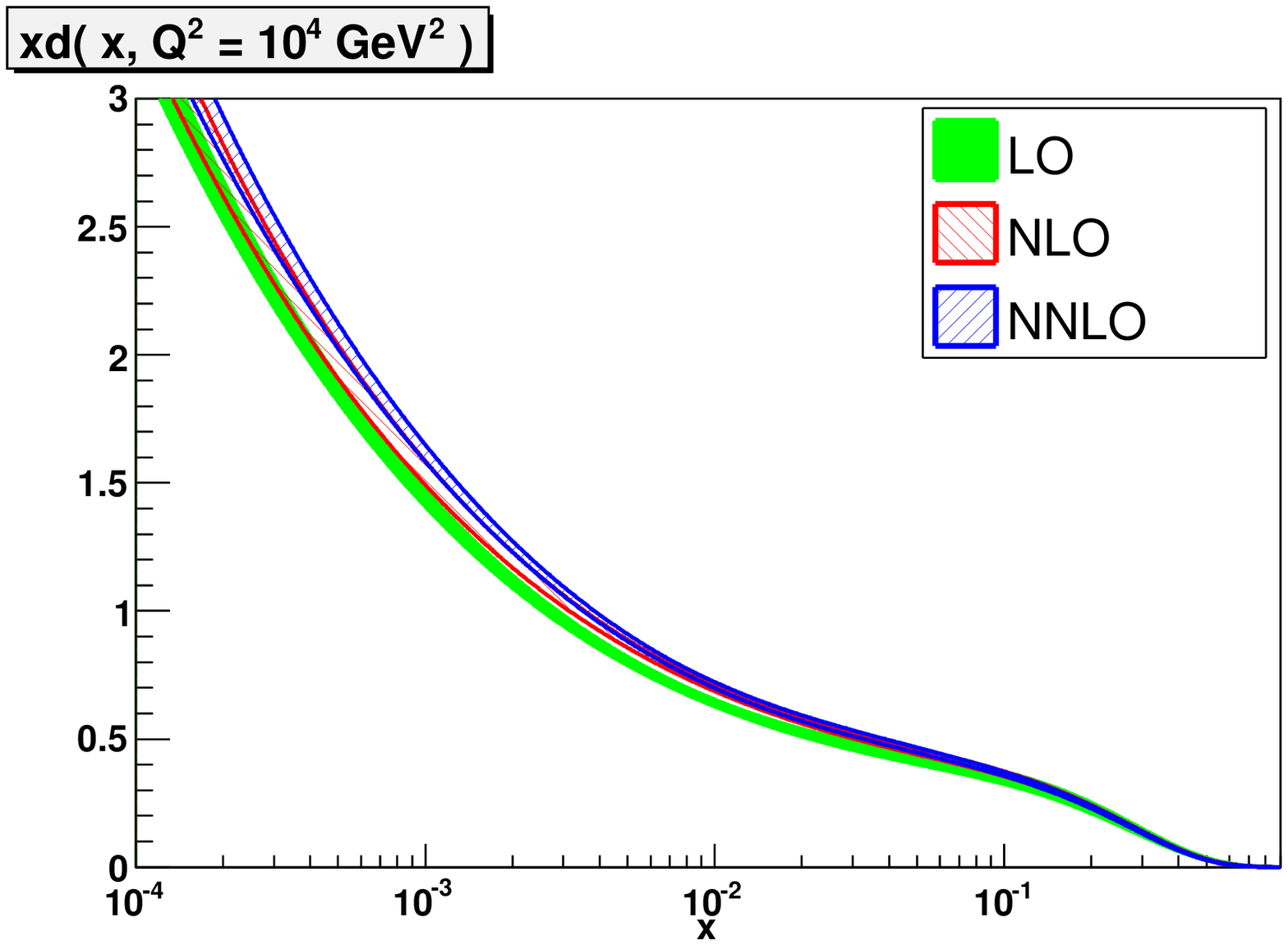}
\epsfig{width=0.49\textwidth,figure=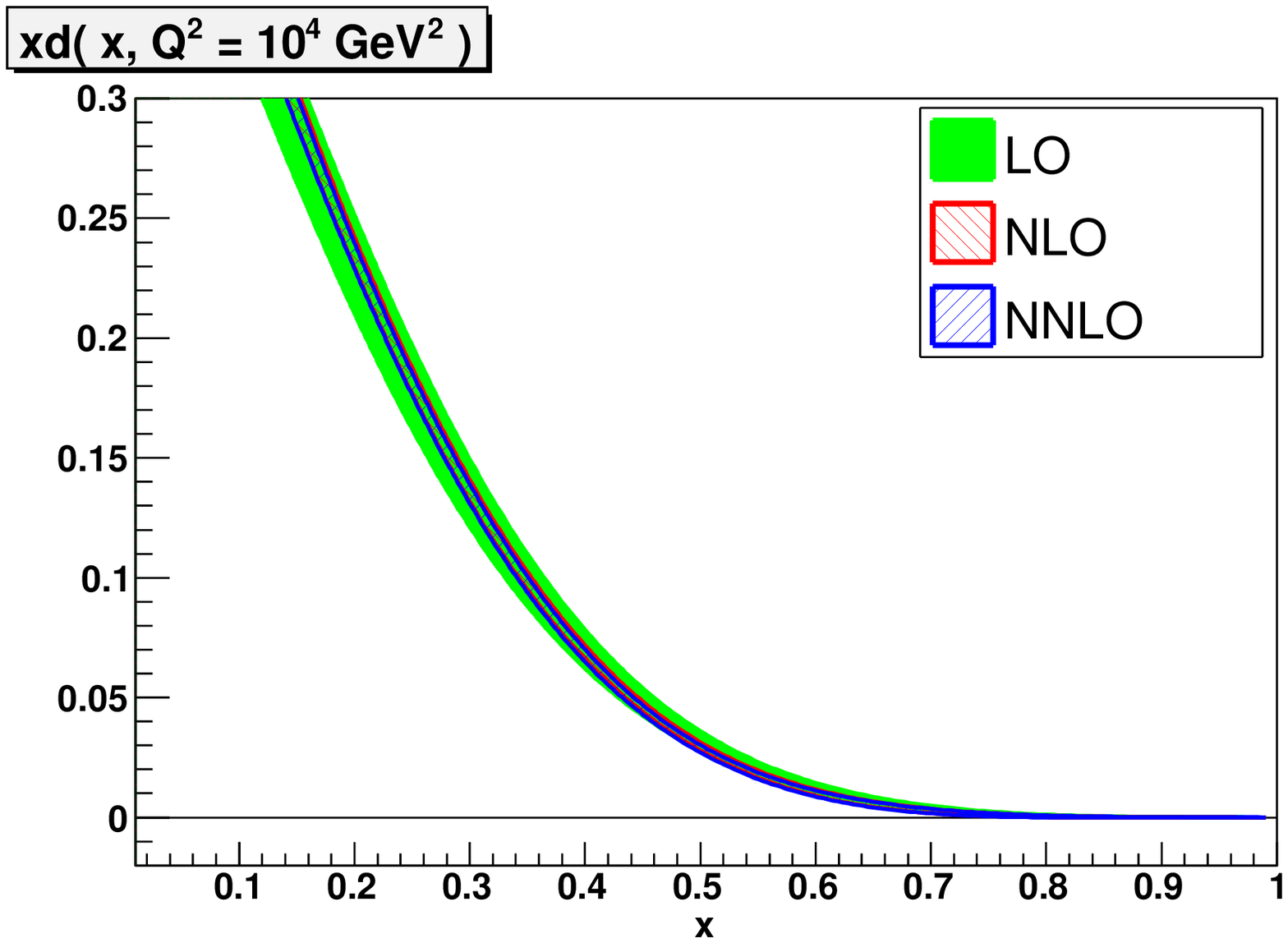}
\epsfig{width=0.49\textwidth,figure=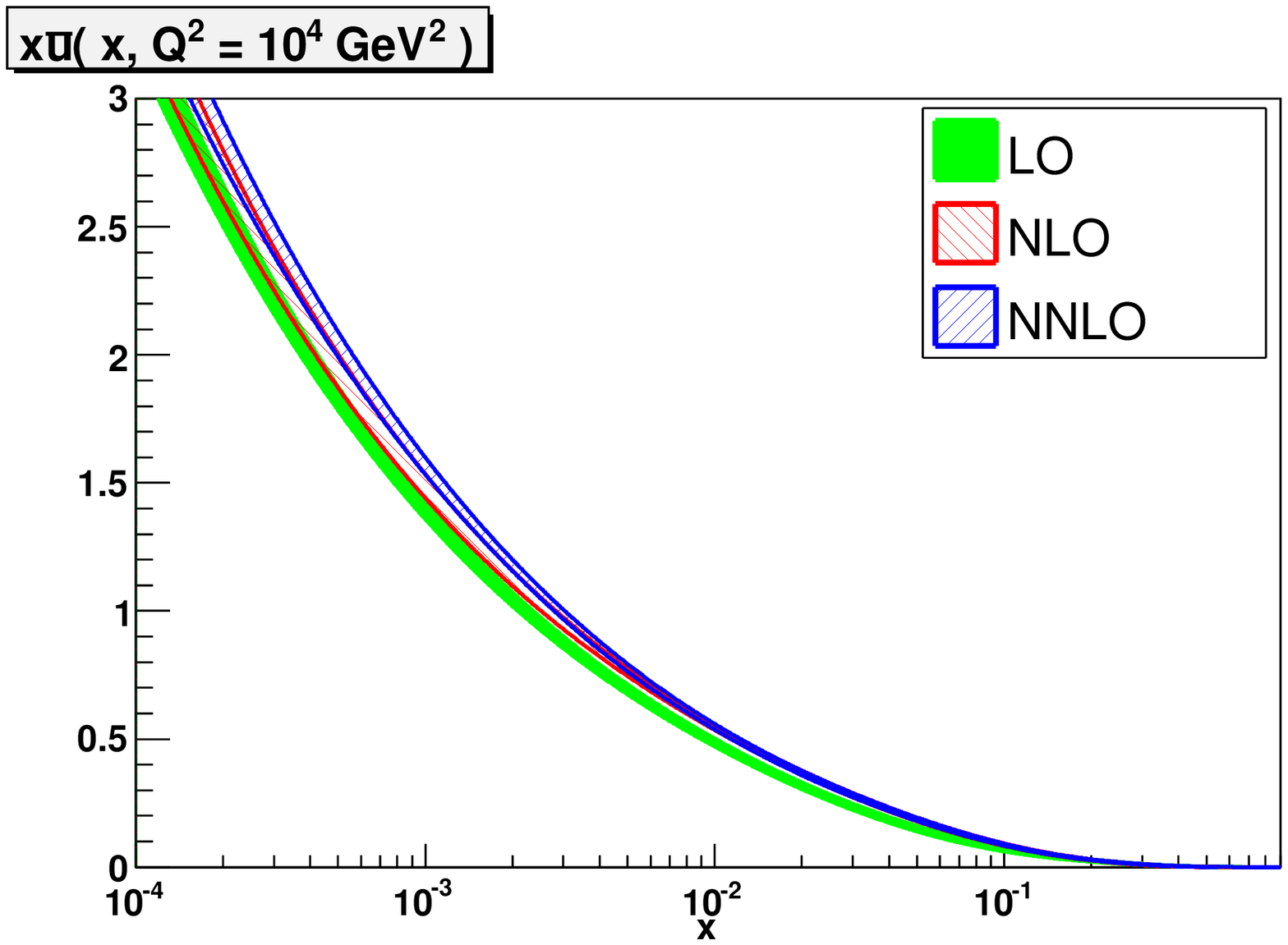}
\epsfig{width=0.49\textwidth,figure=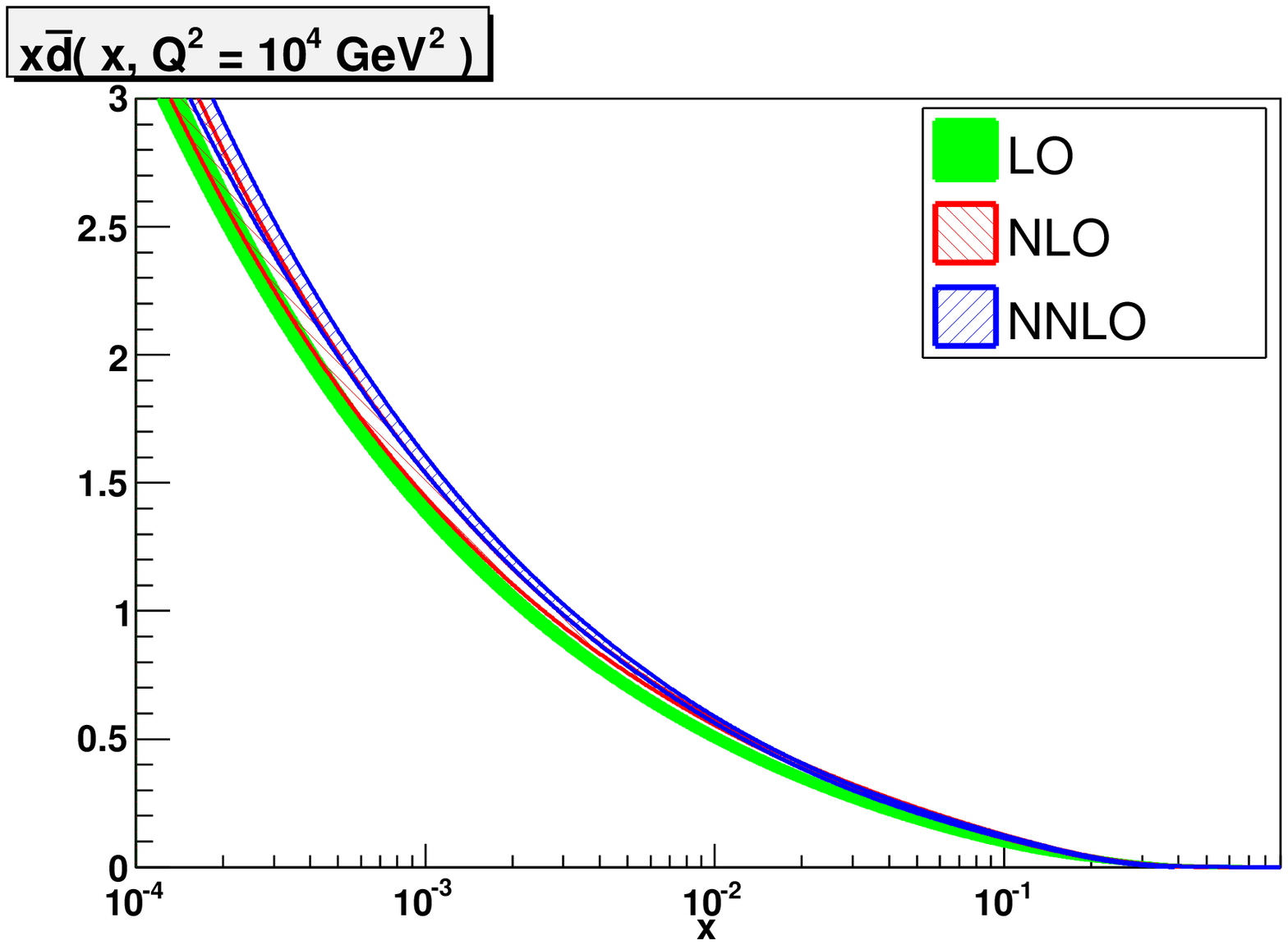}
\caption{\small Comparison of  NNPDF2.1 LO, NLO and NNLO   PDFs
in the flavour basis at $Q^2=10^4$~GeV$^2$: light quarks and gluon.
 \label{fig:flavbasisPDFs-summary}} 
\end{center}
\end{figure}

\begin{figure}[t]
\begin{center}
\epsfig{width=0.49\textwidth,figure=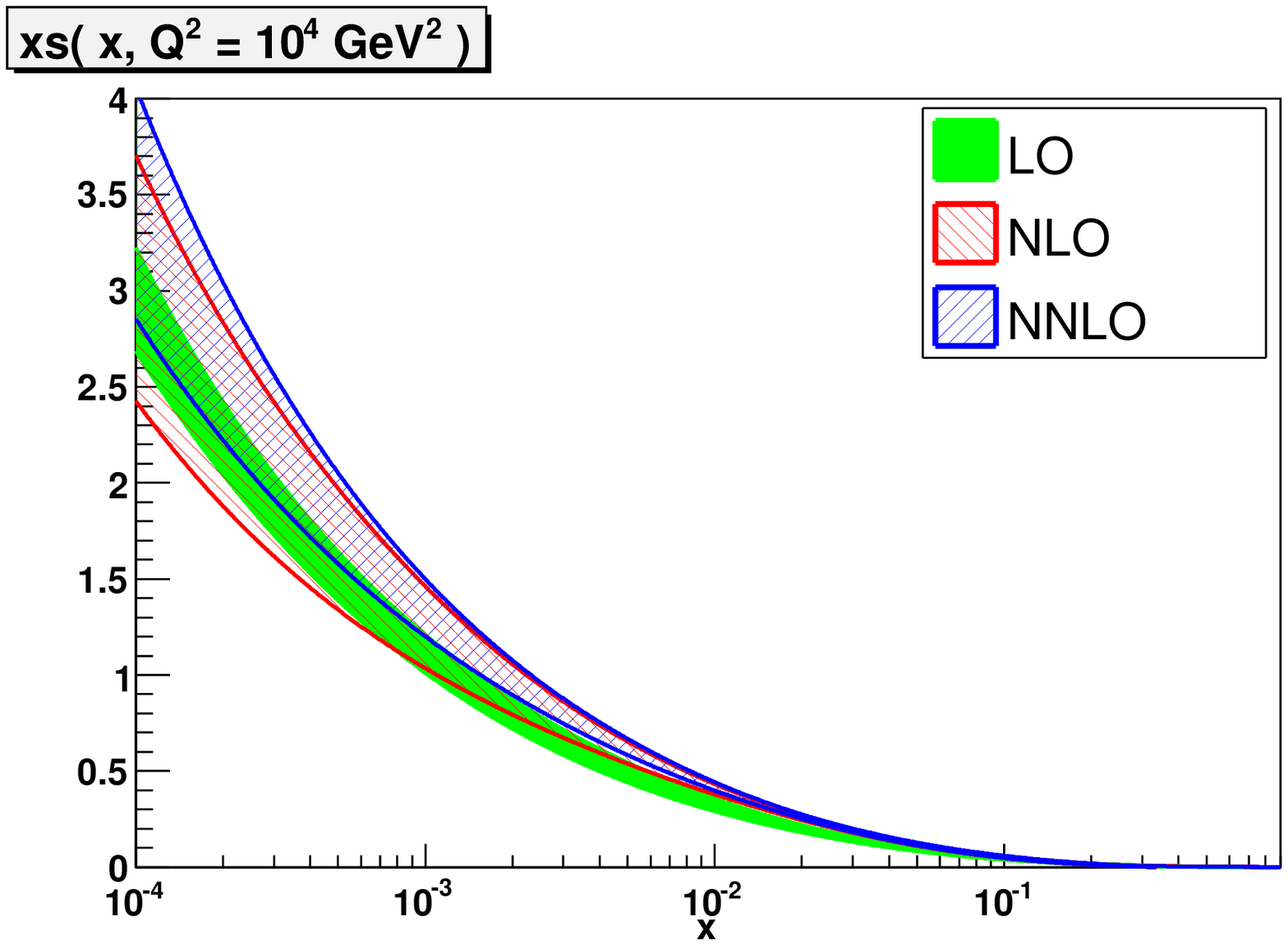}
\epsfig{width=0.49\textwidth,figure=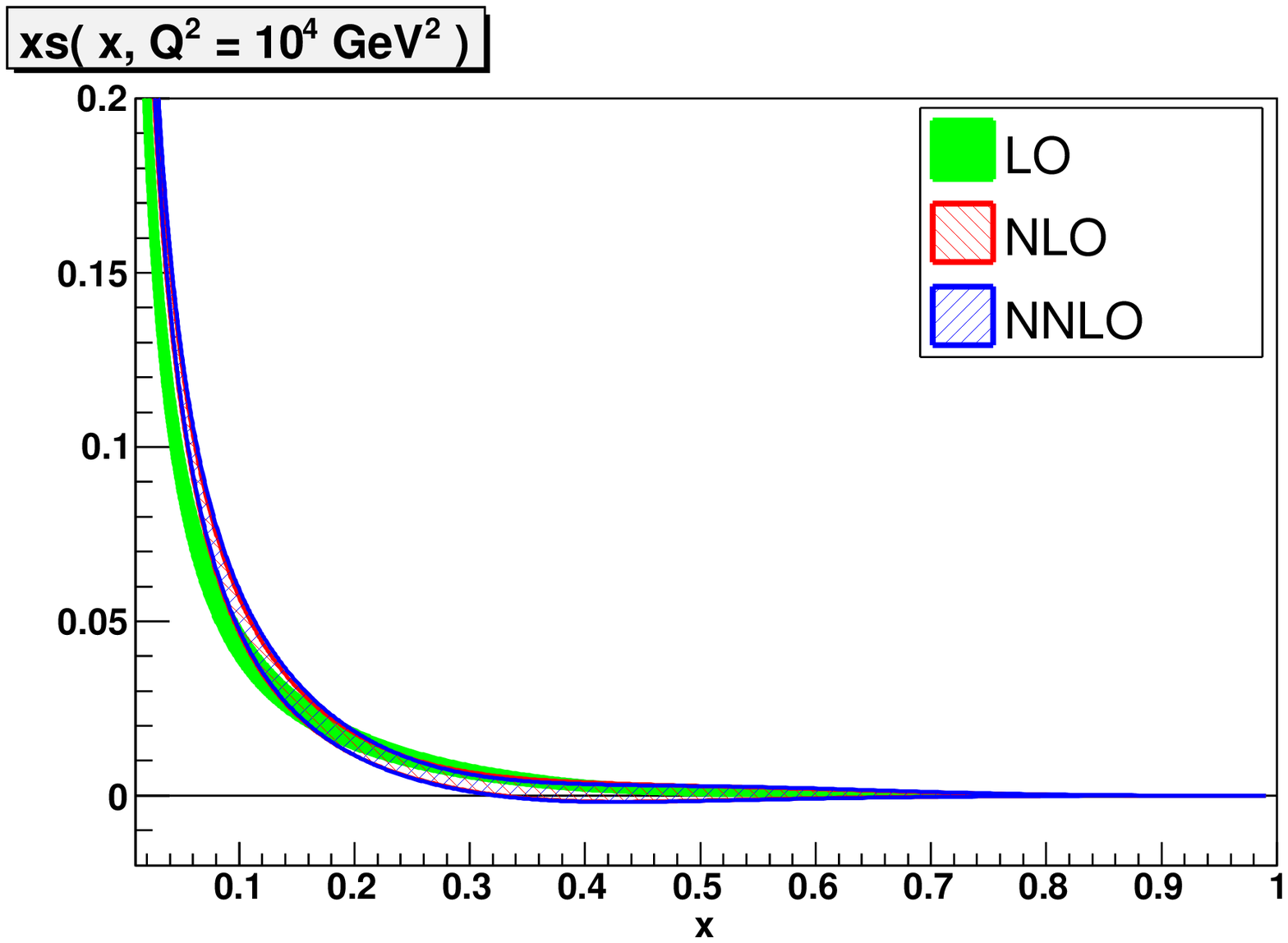}
\epsfig{width=0.49\textwidth,figure=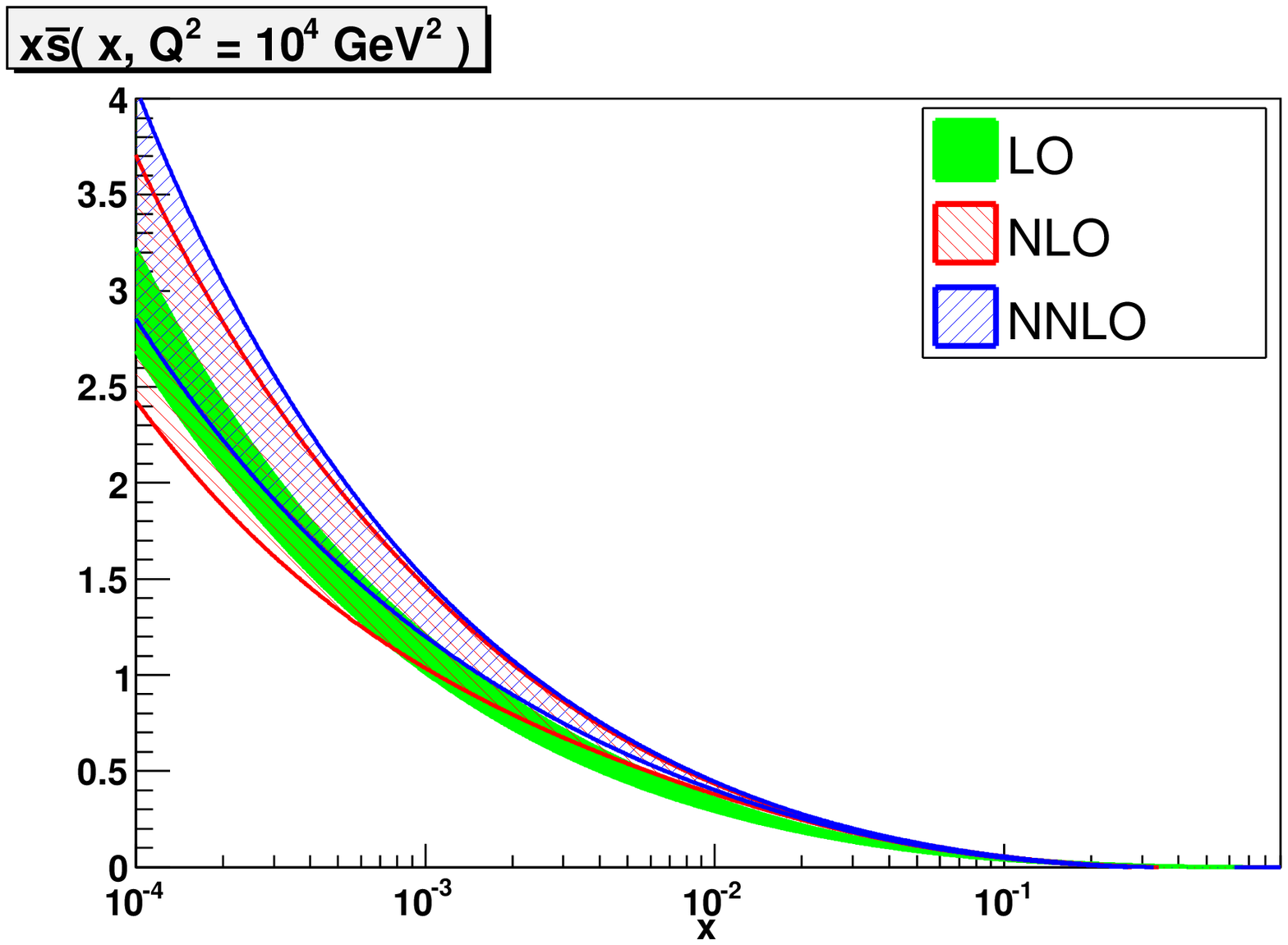}
\epsfig{width=0.49\textwidth,figure=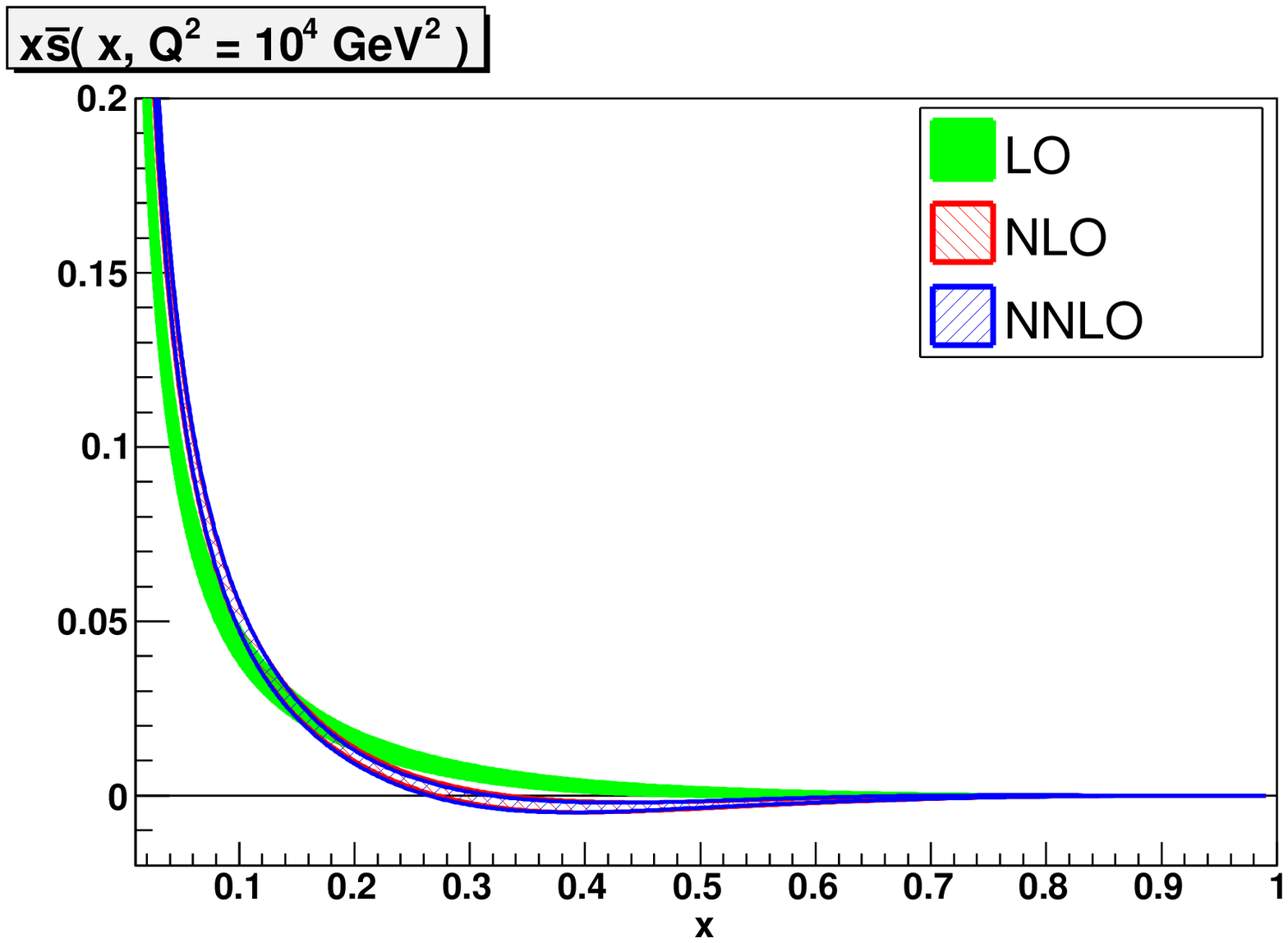}
\epsfig{width=0.49\textwidth,figure=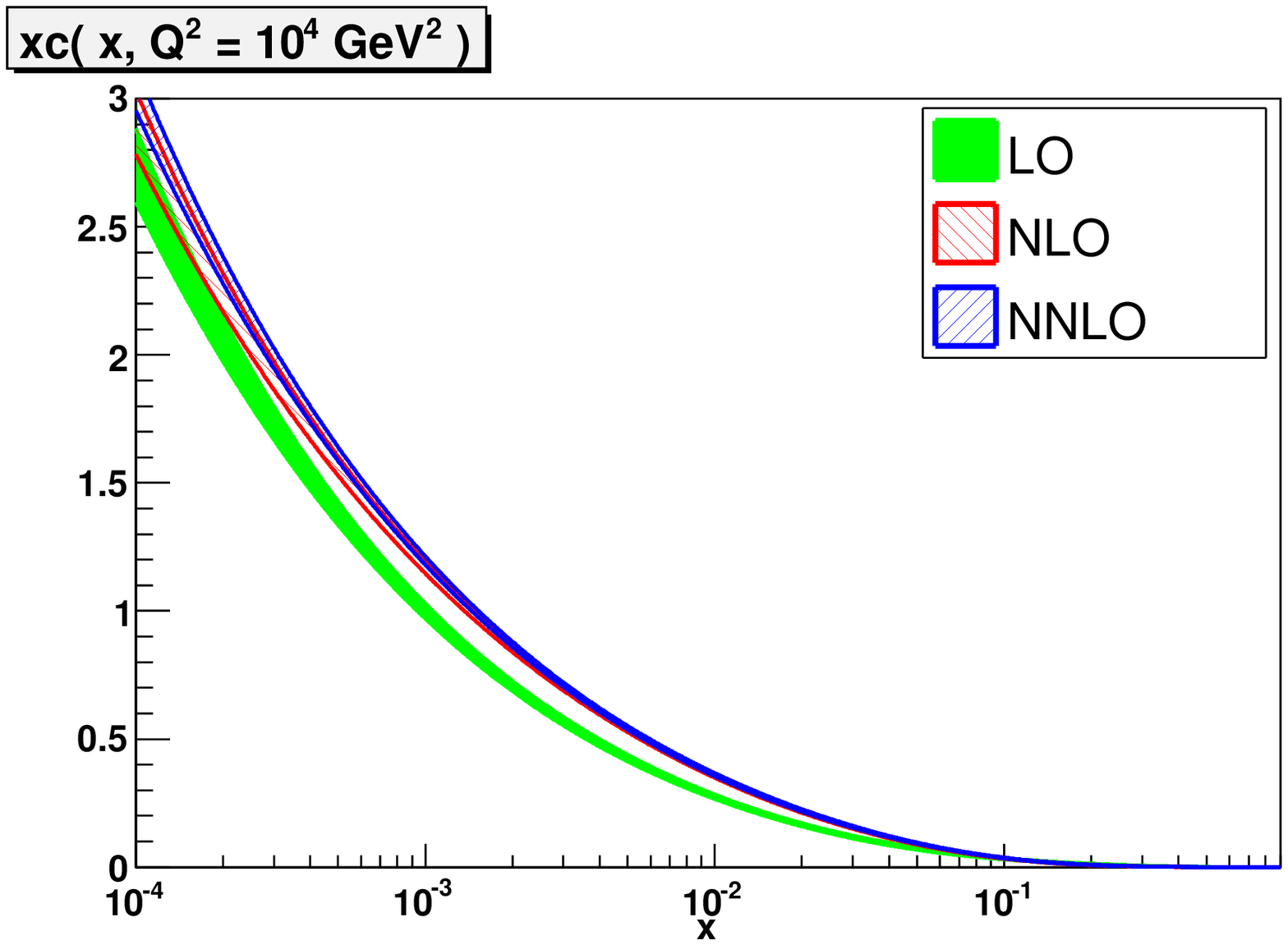}
\epsfig{width=0.49\textwidth,figure=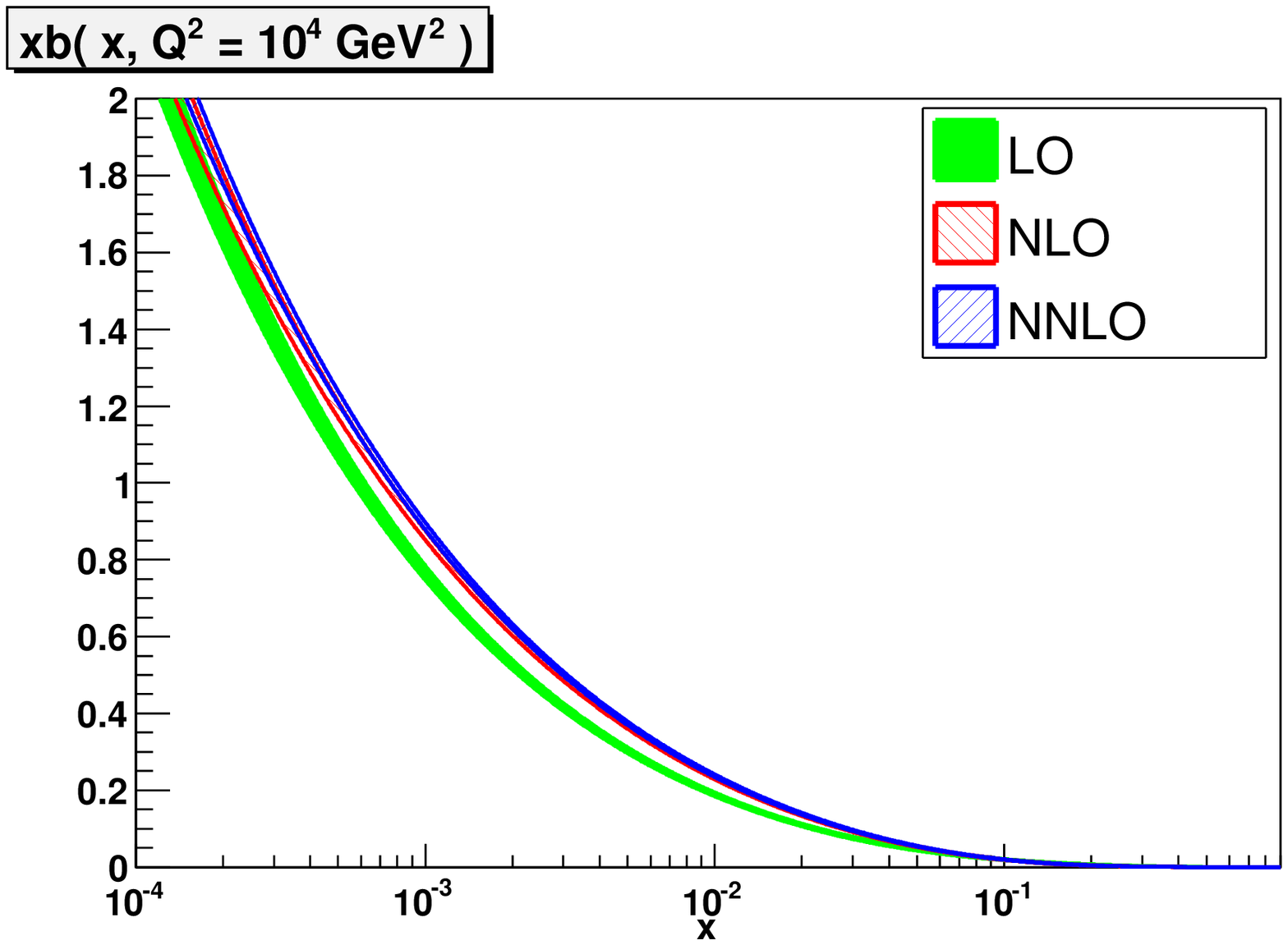}
\caption{\small Comparison of  NNPDF2.1 LO, NLO and NNLO   PDFs
in the flavour basis at $Q^2=10^4$~GeV$^2$: strange and heavy quarks.
 \label{fig:flavbasisPDFs-summary-2}} 
\end{center}
\end{figure}

We assess the perturbative stability of the PDF determination by
comparing the NNPDF2.1 PDFs as they go from LO to NNLO accuracy in
perturbative QCD.  The LO, NLO and NNLO NNPDF2.1 PDFs are compared in
Figs.~\ref{fig:singletPDFs-summary} and~\ref{fig:valencePDFs-summary}
at the starting scale $Q_0^2$=2~GeV$^2$ in the basis in which they are
independently parametrized by neural networks. All error bands shown
are defined as 68\% confidence levels, rather than as standard
deviations, so that possible deviations from gaussian behaviour are
accounted for.  In Figs.~\ref{fig:flavbasisPDFs-summary}
and~\ref{fig:flavbasisPDFs-summary-2} we provide a similar comparison
but this time at the scale $Q^2$= (100~GeV)$^2$ in the basis of
individual flavours.

The excellent convergence of the perturbative expansion within the kinematic region covered by the experimental data is clear from these
plots. In particular, even in the small $x$ and large $x$ region, where
we expect perturbation theory to become unstable and resummation to be
necessary~\cite{Laenen:2004pm,Forte:2009wh}, no evidence of
instability is seen in the PDFs, thus suggesting that resummation
corrections are smaller than current PDF uncertainties (at small
$x$, this is borne out by the dedicated study of Refs.~\cite{Caola:2009iy,Caola:2010cy}). 

It is also clear that the NNLO and NLO results for all PDFs almost always
agree within uncertainties. In particular, with one single exception,
at the starting scale
(Figs.~\ref{fig:singletPDFs-summary}-\ref{fig:valencePDFs-summary})
the NNLO central value is within (or just outside) the NLO uncertainty band, and
in fact it differs from the NLO central value by an amount which is usually
much smaller than the NLO uncertainty. The exception is the isospin
triplet distribution around the valence peak $x\sim0.3$, where the NLO
and NNLO bands overlap, but the NNLO central value is clearly outside the
NLO band. At a higher scale
(Figs.~\ref{fig:flavbasisPDFs-summary}-\ref{fig:flavbasisPDFs-summary-2})
the situation further improves, and the NLO and NNLO results become
almost indistinguishable, and only the small discrepancy in light
quark distributions for $x\gsim 10^{-3}$ already observed in
Fig.~\ref{fig:nnpdf21ratcomp} remains.

This leads to an important conclusion. At present, the PDF
uncertainties provided by NNPDF, and indeed all other PDF groups,
only reflect the data uncertainties: in particular 
they do not include the theoretical uncertainty due to higher perturbative orders,
which could be estimated by varying the renormalization and
factorization scale during the PDF fit. At NLO we can estimate the 
theoretical uncertainty by a direct comparison with the NNLO results. 
This comparison shows that at NLO (and beyond) it is at present
generally
a reasonable approximation to neglect the theoretical uncertainty, since it is
usually smaller than the PDF uncertainty coming from uncertainties in the data. 

On the other hand, at the starting scale the LO PDFs
differ by many standard deviations from NLO PDFs. One must
conclude that at LO the PDF uncertainty provided with NNPDF PDFs (as
well as with any other available LO set) is only a fraction of the
total uncertainty, the theoretical component here being the dominant
one. The situation improves somewhat at high scale 
(Figs.~\ref{fig:flavbasisPDFs-summary}-\ref{fig:flavbasisPDFs-summary-2}),
but the difference between LO and NLO remains large for the gluon.

\begin{figure}[t]
\begin{center}
\epsfig{width=0.60\textwidth,figure=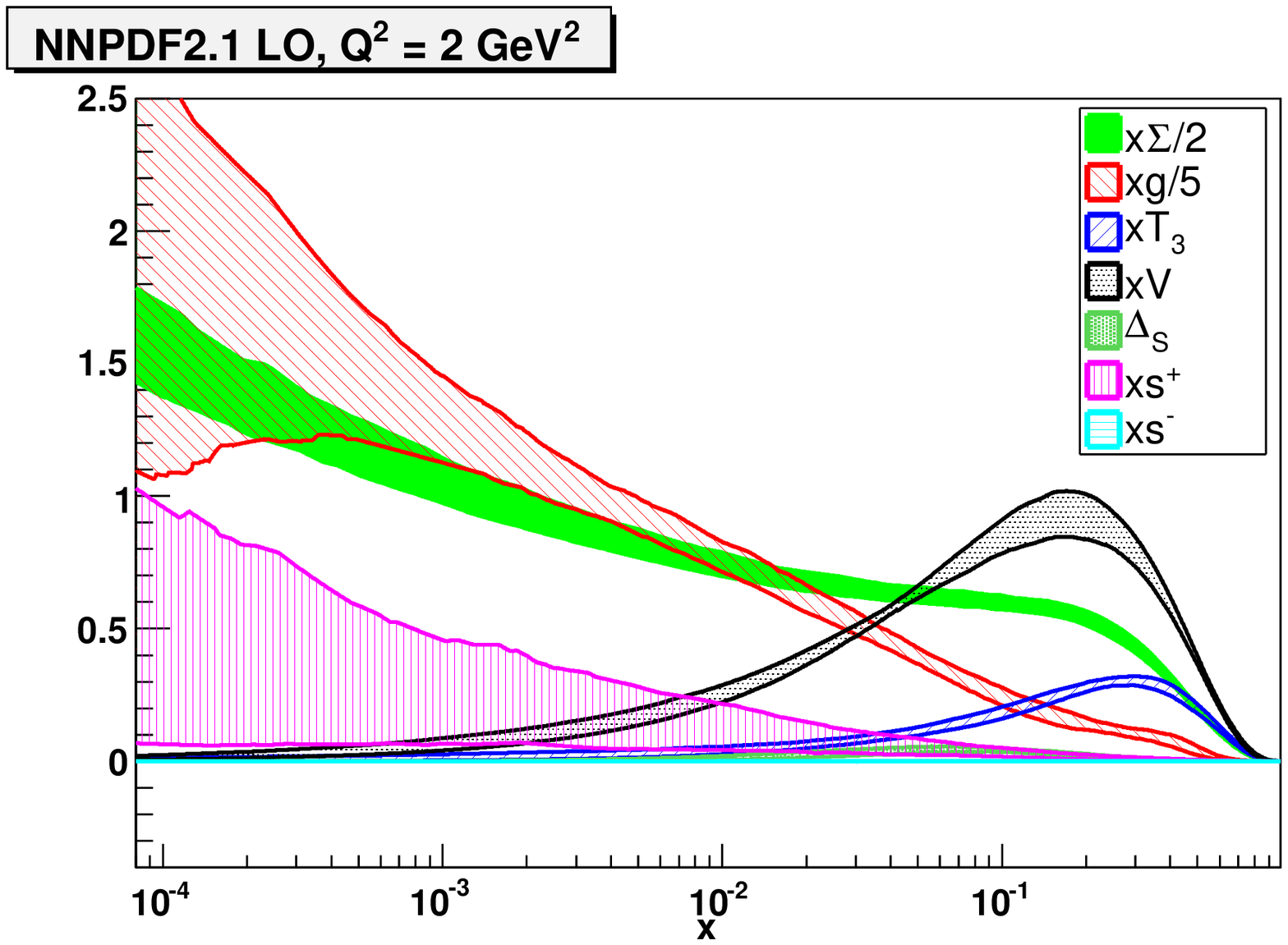}
\epsfig{width=0.60\textwidth,figure=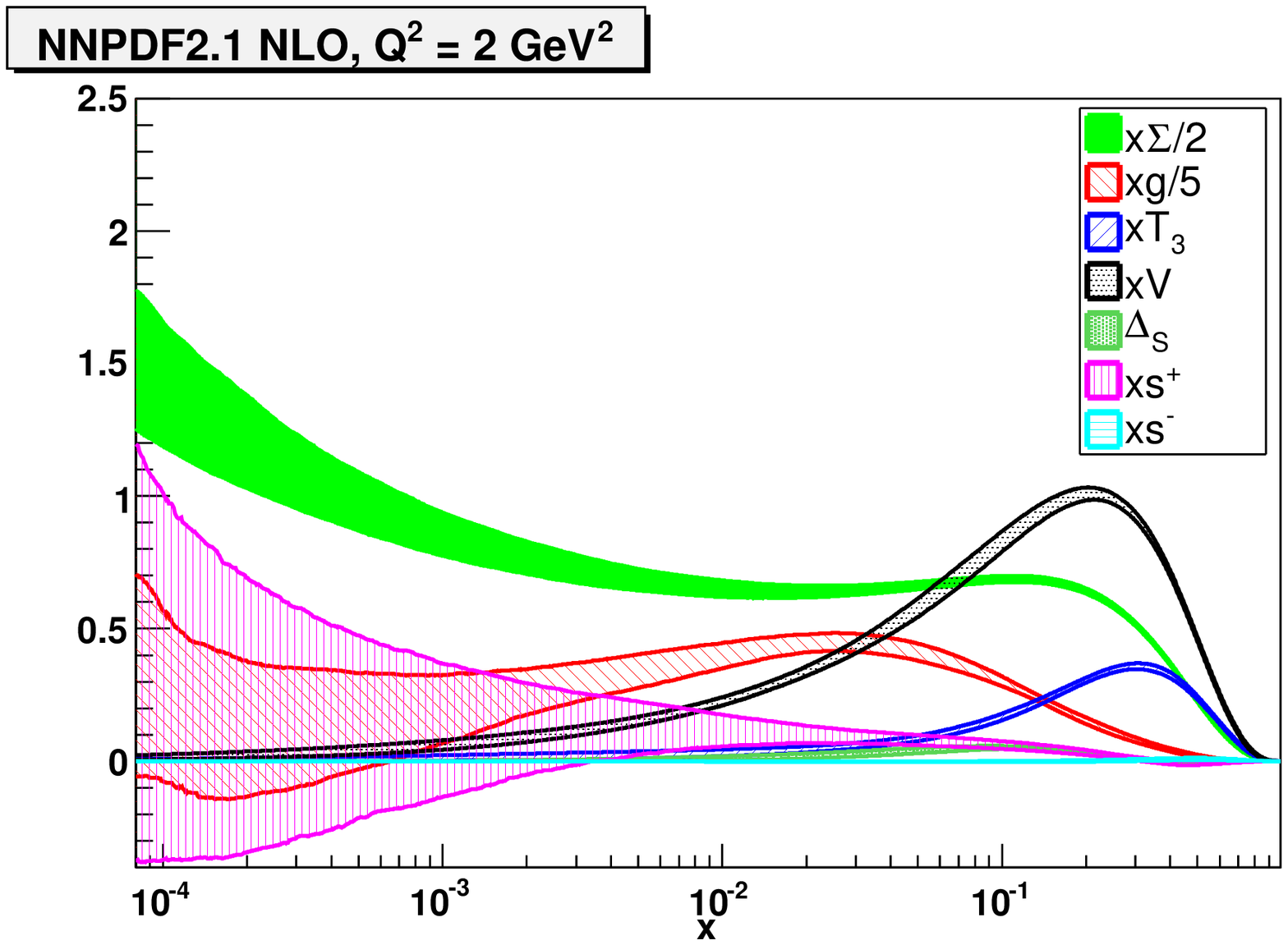}
\epsfig{width=0.60\textwidth,figure=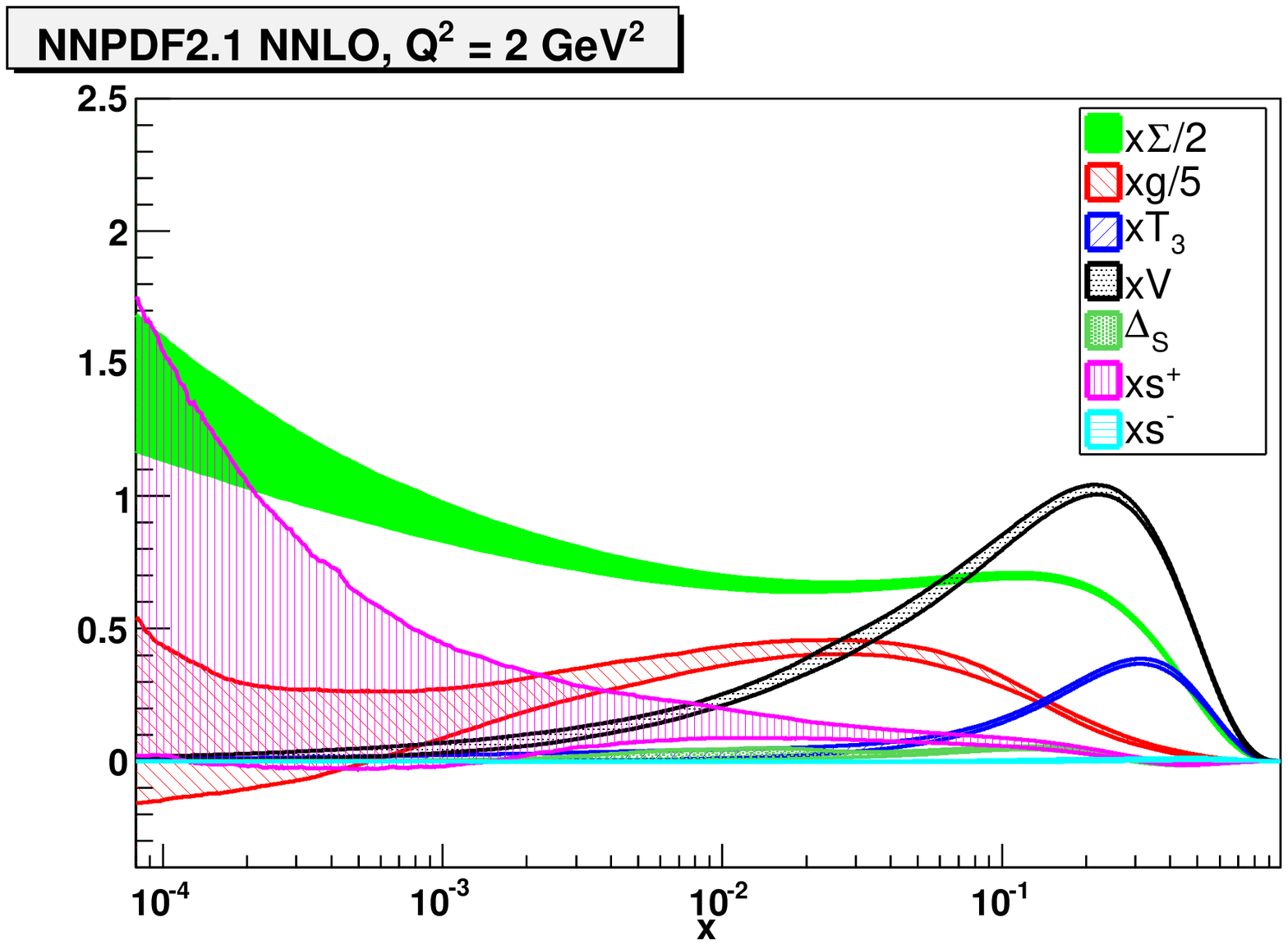}
\caption{\small Summary of the NNPDF2.1 LO (top),
NLO (center) and NNLO (bottom) PDF sets at $Q^2$=2~GeV$^2$.
All  uncertainty bands are defined as 68\% confidence levels.
 \label{fig:pdfplot-summary}} 
\end{center}
\end{figure}

\begin{figure}[t]
\begin{center}
\epsfig{width=0.60\textwidth,figure=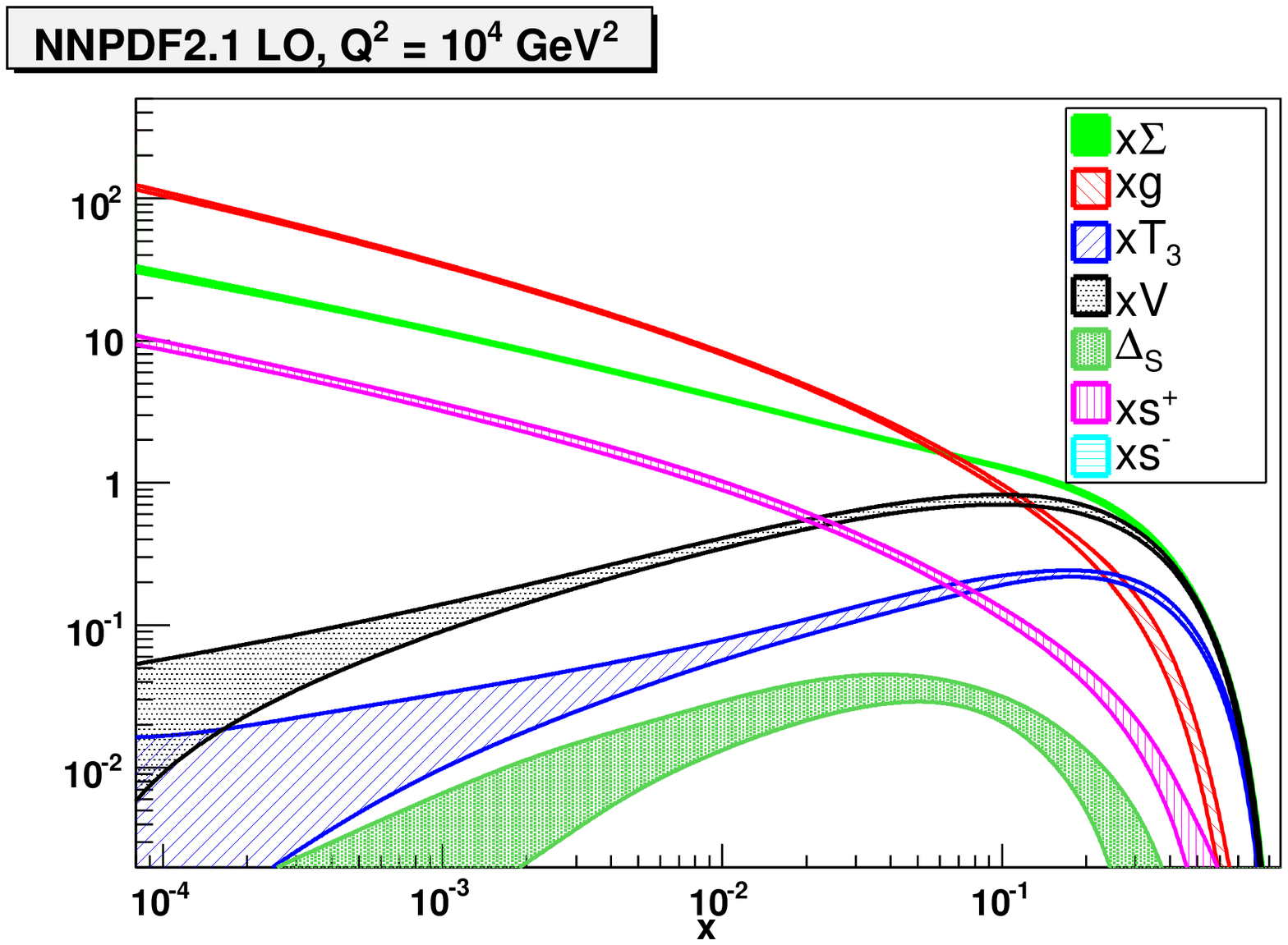}
\epsfig{width=0.60\textwidth,figure=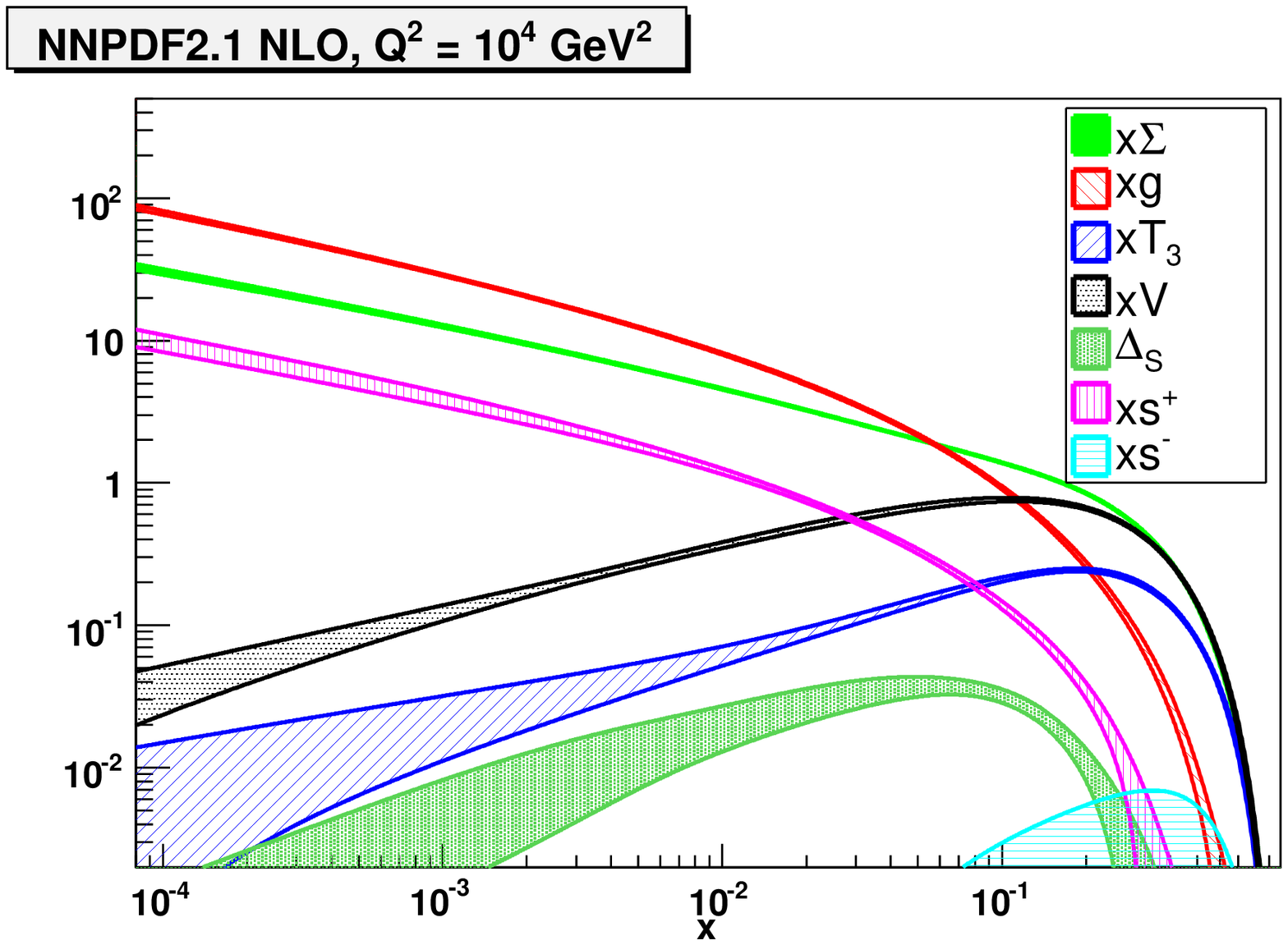}
\epsfig{width=0.60\textwidth,figure=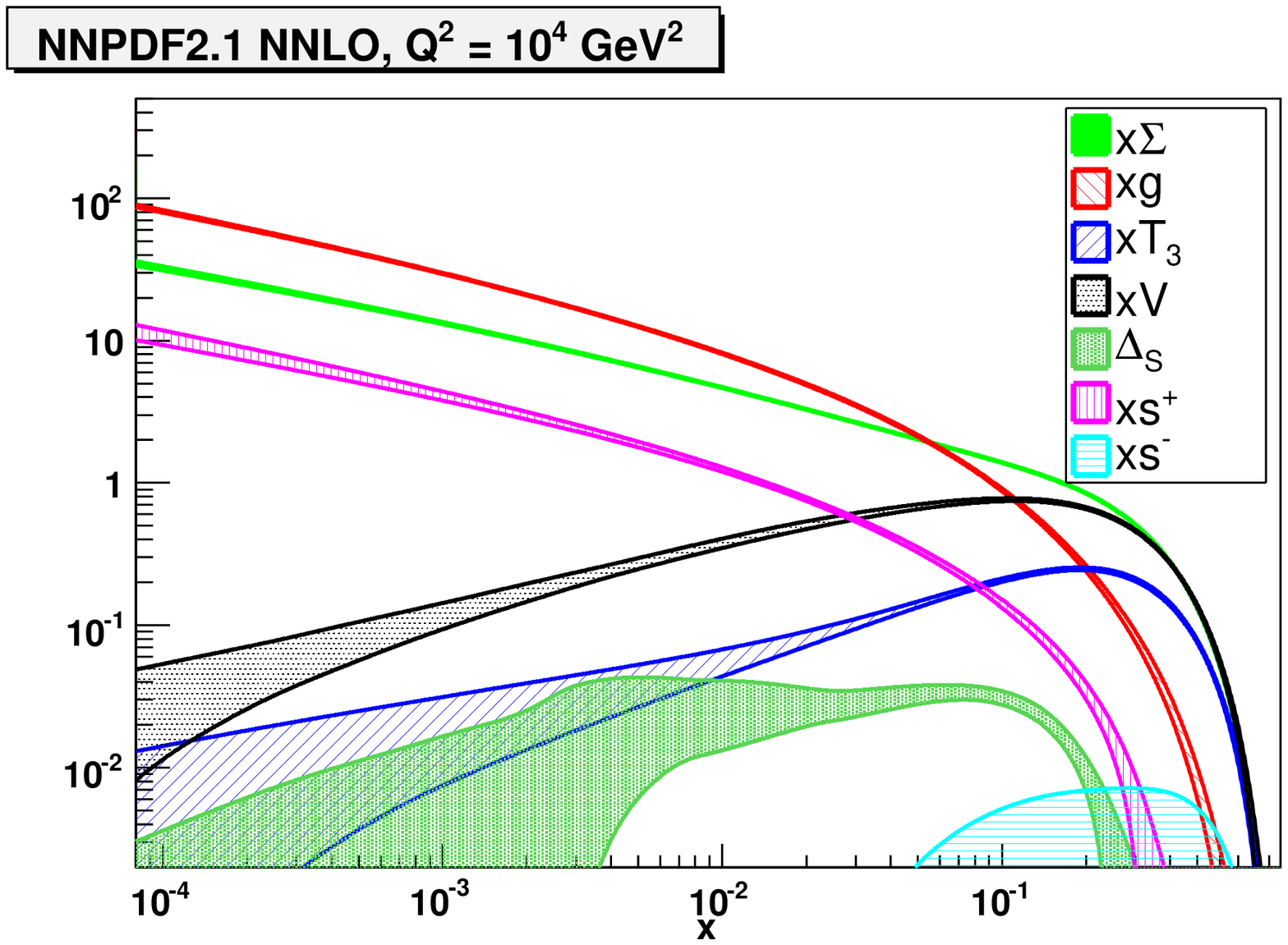}
\caption{\small Same as Fig.\ref{fig:pdfplot-summary} at $Q^2=10^4$~GeV$^2$.  Note the
logarithmic scale on the $y$ axis.
 \label{fig:pdfplot-summary-lhc}} 
\end{center}
\end{figure}

Finally, all seven independently parametrized LO, NLO and NNLO PDFs  
are collected in a single plot in
Fig.~\ref{fig:pdfplot-summary} at a low scale  $Q^2$=2~GeV$^2$
and in 
Fig.~\ref{fig:pdfplot-summary-lhc} at higher scale $Q^2=10^4$~GeV$^2$. These plots illustrate the relative
size of  individual PDFs. Note that at high scale the plot has a log
scale on the vertical axis because due to perturbative evolution the different PDFs
can differ by several orders of magnitude. 

\clearpage

\begin{figure}[ht]
\begin{center}
\epsfig{width=0.85\textwidth,figure=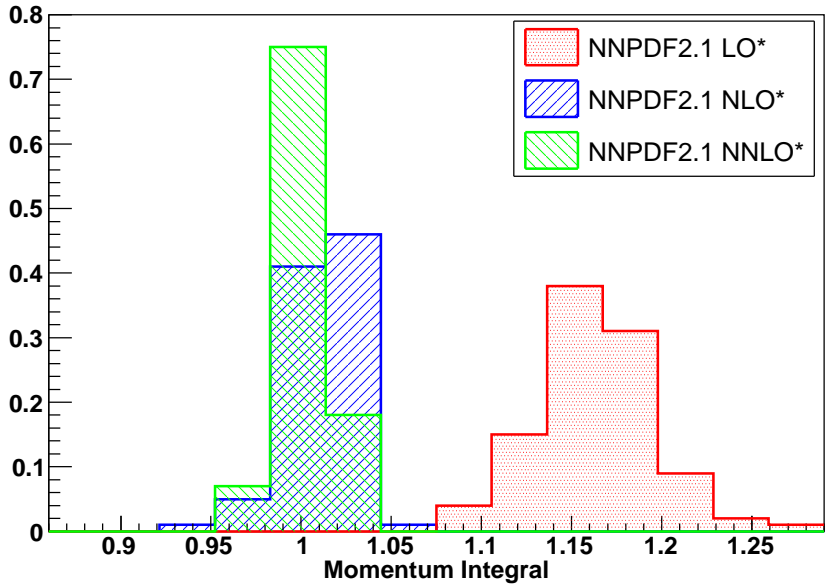}
\caption{\small Distribution of total momentum integrals
for the $N_{\rm rep}=100$ replicas in the
NNPDF2.1 LO*, NLO* and NNLO* PDF sets; $\alpha_s(M_z)=$0.119 in all cases.
 \label{fig:momsrthisto}} 
\end{center}
\end{figure}

\subsection{The momentum of quarks and gluons in the nucleon}

\label{sec:msr}

The value of the total momentum carried by quarks and gluons and its
dependence on the perturbative order provide a strong consistency
check of the perturbative QCD framework.
With the aim of testing this, 
we have performed NLO and NNLO PDF determinations in
which the momentum sum rule is relaxed, denoted as
NLO* and NNLO*, which  supplement the LO* fit
of Sect.~\ref{sec:lopdfs}. We take $\alpha_s(M_Z)=$0.119 at all
perturbative orders. 
In all cases, we find that
the fit quality is not changed in a significant way when 
relaxing the momentum sum rule. 

The momentum fraction carried by a parton distribution is
\be
\lc q \rc\lp Q^2\rp \equiv  \int_0^1 dx \, xq\lp x,Q^2\rp \ .
\ee
Using the LO*, NLO* and NNLO* PDF sets, we find that the total
momentum $\lc M \rc = \lc \Sigma \rc +\lc g \rc$ carried by partons is
\bea
\lc M \rc_{\rm LO} = 1.161\pm 0.032 \, , \nonumber \\
\lc M \rc_{\rm NLO} = 1.011\pm 0.018 \, , \label{eq:totmom} \\
\lc M \rc_{\rm NNLO} = 1.002\pm 0.014 \, . \nonumber
\eea
where the uncertainty is only from PDFs (and thus does not include any
theoretical uncertainty). 
The distributions of  total
momentum integrals over the 100 replicas for the NNPDF2.1 LO*, NLO*
and NNLO* sets is shown in Fig.~\ref{fig:momsrthisto}: they appear to
be Gaussian to a good approximation.

Estimating
the theoretical uncertainty as the difference
between results at two subsequent perturbative orders, we see that at LO the
theoretical uncertainty is dominant, as we already concluded from the
PDF plots
Figs.~\ref{fig:singletPDFs-summary}-\ref{fig:valencePDFs-summary}  in
Sect.~\ref{sec:pertstab}. The deviation of the LO momentum integral
from the QCD prediction is mostly driven by the gluon, which turns out
to be larger in the LO* set than in the default LO set with momentum
sum rule imposed. On the other hand already at NLO the
theoretical uncertainty is half of the PDF uncertainty, $\Delta^{\rm th}\lc
 M \rc_{\rm NLO}=0.01$, and thus at NNLO 
the theoretical uncertainty is likely
 to be negligible.

\begin{table}[t]
\centering
\begin{tabular}{|c||c|c|c|}
\hline
PDF combination &  LO* & NLO* & NNLO* \\
\hline
\hline
$\lc \Sigma +g\rc $ & $1.161\pm 0.032$  & $1.011 \pm 0.018$ 
& $1.002 \pm 0.014$ \\
\hline
\hline
\multicolumn{4}{|c|}{$Q^2_0=2$ GeV$^2$}\\
\hline
$\lc \Sigma \rc (Q_0^2)$ &  $0.550\pm 0.025$ &  $0.591\pm 0.010$ 
& $0.602 \pm 0.010$  \\
$\lc g\rc (Q^2_0)$  & $0.612\pm 0.028$  & $0.421\pm 0.021$
 & $0.400 \pm 0.018$ \\
$\lc \lp u+\bar{u}\rp\rc (Q^2_0) $  &  $0.346\pm 0.015$ &  $0.371\pm 0.005$ & 
$0.376 \pm 0.005$\\
$\lc \lp d+\bar{d}\rp\rc (Q^2_0) $  & $0.192\pm 0.011$  & $0.206\pm 0.005$& 
$0.209 \pm 0.003$\\
$\lc \lp s+\bar{s}\rp\rc(Q^2_0) $   & $0.012\pm 0.004$  & $0.014\pm 0.006$ & 
$0.017 \pm 0.006$\\
\hline
\multicolumn{4}{|c|}{$Q^2=10^4$ GeV$^2$}\\
\hline
$\lc \Sigma\rc (Q^2) $  & $0.546\pm 0.017$ &$0.528\pm 0.009$ & $0.527 \pm 0.005 $\\
$\lc g\rc (Q^2)$  &  $0.615\pm 0.020$ & $0.486\pm 0.018$ & $0.475 \pm 0.011$\\
$\lc \lp u+\bar{u}\rp\rc (Q^2)$  & $0.264\pm 0.009$ &  $0.256\pm 0.003$ & $0.255 \pm 0.002$ \\
$\lc \lp d+\bar{d}\rp\rc (Q^2)$  & $0.165\pm 0.007$ & $0.160\pm 0.002$ & 
$0.159 \pm 0.001$\\
$\lc \lp s+\bar{s}\rp\rc(Q^2)$  & $0.048 \pm 0.003$ & $0.047 \pm 0.004$ & 
$0.048 \pm 0.004$\\
$\lc \lp c+\bar{c}\rp\rc(Q^2)$  & $0.041 \pm 0.002$ & $0.039 \pm 0.002$ &
$0.039 \pm 0.001$ \\
$\lc \lp b+\bar{b}\rp\rc(Q^2)$  & $0.027 \pm 0.001$ &$0.025 \pm 0.001$ & 
$0.025 \pm 0.001$\\
\hline
\end{tabular}
\caption{\small Momentum fractions of  various PDF combinations
at low scale $Q_0^2=2$ GeV$^2$ and high scale $Q^2=10^4$~GeV$^2$  when
the momentum sum rule is not imposed (LO*, NLO* and NNLO* PDF sets).
 All results
are obtained with
$N_{\rm rep}=100$ replicas. 
\label{tab:momfrac}}
\end{table}

\begin{table}[t]
\centering
\begin{tabular}{|c|c|c|c|}
\hline
PDF combination & LO & NLO & NNLO  \\
\hline
\hline
$\lc \Sigma +g\rc $ & $1$  & $1$ 
& $1$ \\
\hline
\hline
\multicolumn{4}{|c|}{$Q^2_0=2$ GeV$^2$}\\
\hline
$\lc \Sigma\rc(Q_0^2)$ & $0.521 \pm 0.023$  & $0.590  \pm 0.009$ & $0.609  \pm  0.013$ \\
$\lc g \rc (Q^2_0)$ & $0.479  \pm 0.022$ & $0.411  \pm 0.009$ & $0.391  \pm 0.012$\\
$ \lc \lp u+\bar{u}\rp \rc(Q^2_0)$ & $0.328  \pm 0.012$ & $0.371  \pm 0.005$ &  $0.381  \pm 0.007$ \\
$\lc\lp d+\bar{d}\rp \rc (Q^2_0)$ & $0.181  \pm 0.010$ & $0.206  \pm 0.004$ & 
$0.211  \pm 0.005$\\
$\lc\lp s+\bar{s}\rp \rc (Q^2_0)$ & $0.012  \pm 0.005$  & $0.013  \pm 0.006$ &
$0.017  \pm 0.005$ \\
\hline
\multicolumn{4}{|c|}{$Q^2=10^4$ GeV$^2$}\\
\hline
$\lc\Sigma \rc (Q^2) $ & $0.492 \pm 0.010$ & $0.523 \pm 0.003$ & $0.529 \pm 0.004$ \\
$\lc g \rc (Q^2)$ & $0.509\pm 0.010$ & $0.477 \pm 0.003$& $0.471\pm 0.005$\\
$\lc \lp u+\bar{u}\rp \rc (Q^2)$ & $0.245 \pm 0.007$  & $0.255 \pm 0.003$ &
$0.257 \pm 0.004$ \\
$\lc \lp d+\bar{d}\rp \rc(Q^2)$ & $0.150 \pm 0.006$ & $0.159 \pm 0.002$ & 
$0.159 \pm 0.002$\\
$\lc \lp s+\bar{s}\rp \rc (Q^2)$ &  $0.041 \pm 0.003$ & $0.046 \pm 0.003 $  & 
$0.048 \pm 0.002$\\
$\lc \lp c+\bar{c}\rp \rc (Q^2)$ & $0.033 \pm 0.001$ & $0.0383 \pm 0.0004$ &
$0.0393 \pm 0.0006$ \\
$\lc \lp b+\bar{b}\rp \rc (Q^2)$ & $0.021 \pm 0.001$ & $0.0245 \pm 0.0002$ & 
$0.0249 \pm 0.0003$\\
\hline
\end{tabular}
\caption{\small Same as Table~\ref{tab:momfrac}, but when the momentum
  sum rule is imposed (LO, NLO and NNLO PDF sets).
\label{tab:momfrac2}}
\end{table}

\begin{figure}[t]
\begin{center}
\epsfig{width=0.9\textwidth,figure=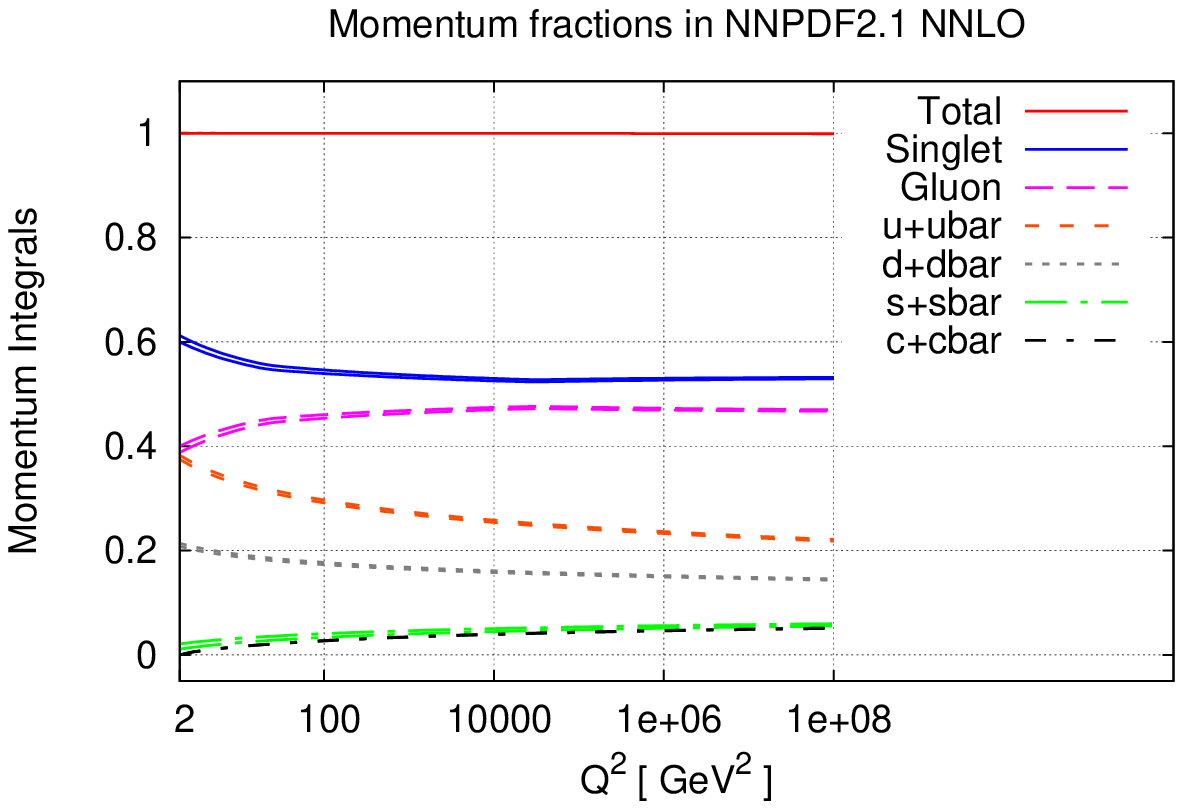}\\
\epsfig{width=0.9\textwidth,figure=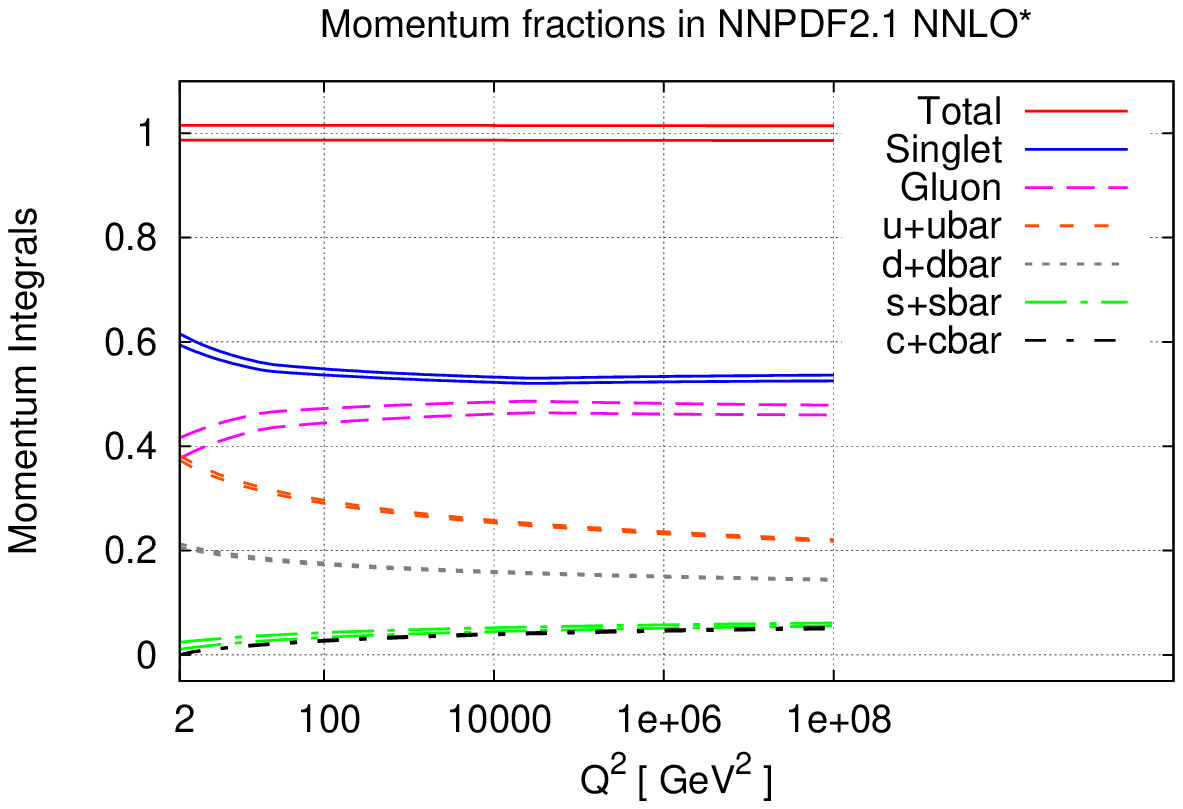}
\caption{\small Momentum fractions carried by various
combinations of quark and gluon 
distributions as a function of the scale
$Q^2$ at NNLO, when the momentum sum rule is (top) or is not (bottom)
imposed.
 \label{fig:momQ2}} 
\end{center}
\end{figure}

It is also interesting to determine the momentum fraction carried
by individual PDFs. These are tabulated
in Tables~\ref{tab:momfrac}-\ref{tab:momfrac2} 
at  a low scale
$Q^2_0=2$ GeV$^2$ and at a high scale $Q^2=10^4$~GeV$^2$, both before 
(Table~\ref{tab:momfrac}, * PDF sets) and after (Table~\ref{tab:momfrac2},
standard PDF sets) imposing the momentum sum rule.
They are also plotted as a function of scale in Fig.~\ref{fig:momQ2}.
We show the momentum
fractions of the light quarks, the gluon, and the total quark singlet
combination.

\begin{figure}[t]
\begin{center}
\epsfig{width=0.9\textwidth,figure=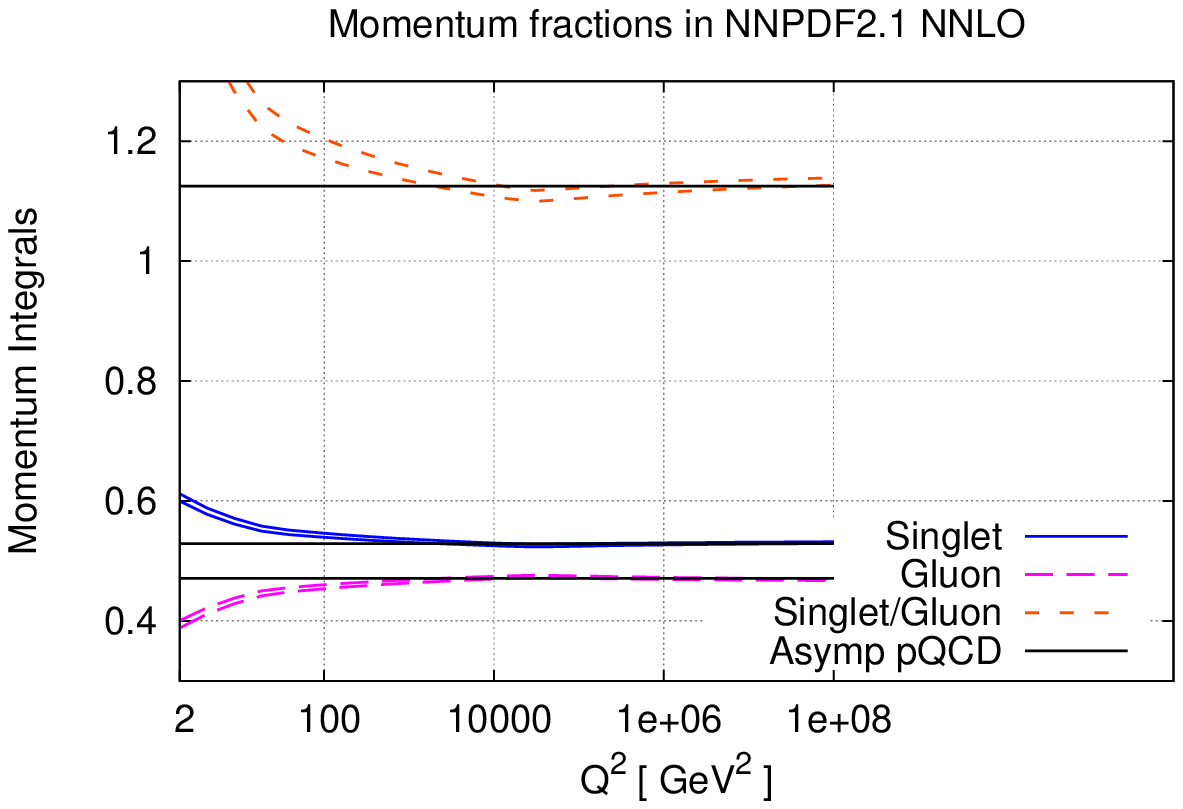}\\
\epsfig{width=0.9\textwidth,figure=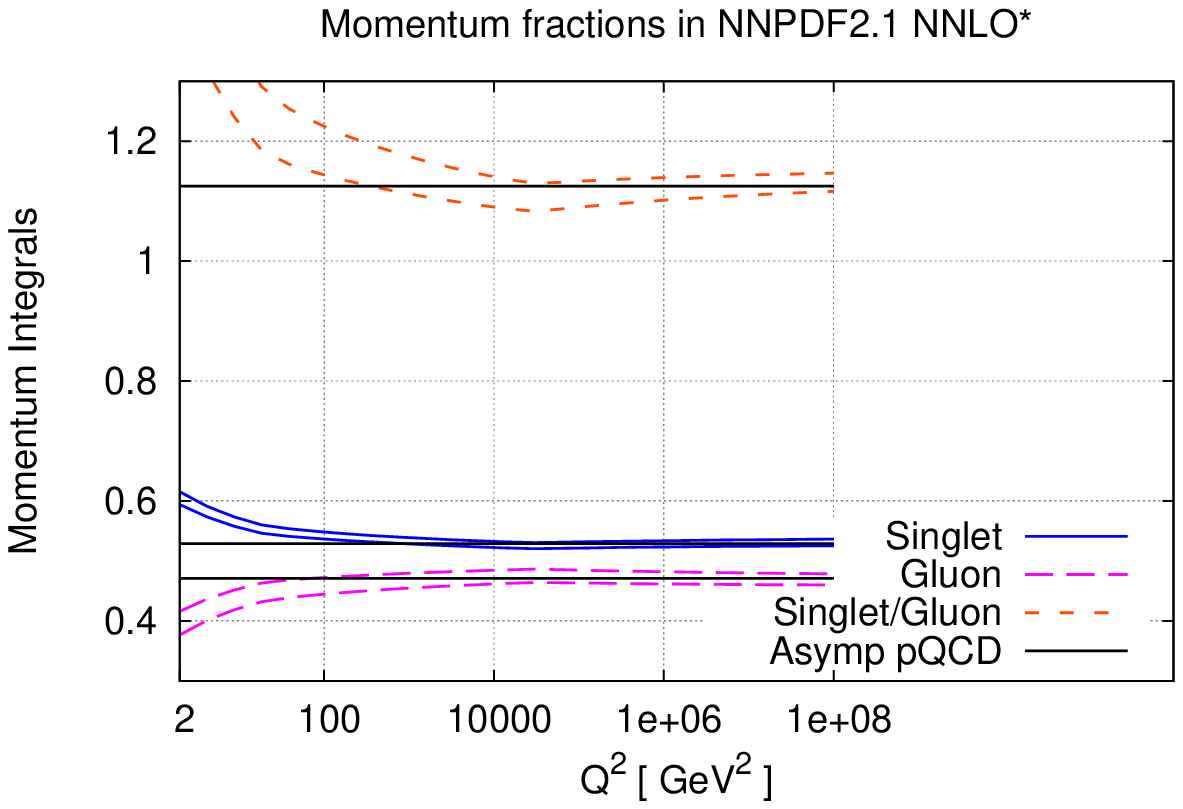}
\caption{\small Momentum fractions carried by quarks and
gluons, and their ratio, as a function of the scale
$Q^2$ at NNLO, when the momentum sum rule is (top) or is not (bottom)
imposed, compared to the asymptotic pQCD predictions,  Eq.~(\ref{eq:asympmom}).
 \label{fig:momQ2qcd}} 
\end{center}
\end{figure}

The asymptotic values of the momentum carried by the
total quark and gluon distributions are predicted in perturbative QCD
to be 
\bea
\label{eq:asympmom}
\lim_{Q^2\to\infty} [\Sigma](Q^2)=\frac{3 n_f}{16+3n_f}\approx0.5294;&\quad&
\lim_{Q^2\to\infty} [g](Q^2)=\frac{16}{16+3n_f}\approx0.4706 
\eea
(see e.g. Ref.~\cite{Ellis:1991qj}).  The results of
Tables~\ref{tab:momfrac}-\ref{tab:momfrac2} are in impressive
agreement with the QCD prediction Eq.~(\ref{eq:asympmom}). When the
momentum sum rule is imposed (Table~\ref{tab:momfrac2}) the accuracy
of the determination of each momentum component improves, and the
agreement with the QCD prediction Eq.~(\ref{eq:asympmom}) improves
accordingly. 
In Fig.~\ref{fig:momQ2qcd} we compare for the NNPDF2.1 NNLO and
NNLO* fits the gluon and singlet momentum fraction and their
ratio with the corresponding asymptotic values predicted by
pQCD. This confirms the excellent agreement, both
with and without the momentum sum rule imposed.

It is interesting to observe that even though the
limits Eq.~(\ref{eq:asympmom}) hold at any perturbative order,
because of asymptotic freedom, the results of
Tables~\ref{tab:momfrac}-\ref{tab:momfrac2} agree better with the limits as the
perturbative order increases. This is due to the fact that the
uncertainties given in the tables do not include the theoretical
uncertainty, which decreases as the perturbative order
increase. Indeed, comparison of central values of the various momentum
fractions at subsequent perturbative orders shows that at NLO this
uncertainty is small, but not entirely negligible, as already
concluded in Sect.~\ref{sec:pdfcomp}, while at NNLO it is likely much 
less than the PDF uncertainties.

%% file: sec-pheno.tex
\section{Phenomenological implications}
\label{sec:pheno}

We will now present the parton luminosities determined from NNPDF2.1 NNLO
PDFs, and use them to compute several LHC standard cross-sections. As we
shall see, the first LHC data already have discriminating power
between existing PDF sets. They are thus likely to lead to a
substantial improvement in PDF accuracy in the near future.

\subsection{Parton luminosities}

At a hadron collider, all factorized observables depend on parton
distributions through a parton luminosity, which, following Ref.~\cite{Campbell:2006wx}, we define 
 as
\be
\Phi_{ij}\lp M_X^2\rp = \frac{1}{s}\int_{\tau}^1
\frac{dx_1}{x_1} f_i\lp x_1,M_X^2\rp f_j\lp \tau/x_1,M_X^2\rp \ ,
\label{eq:lumdef}
\ee
where $f_i(x,M^2)$ is a PDF and $\tau \equiv M_X^2/s$.
Therefore, a good deal of information on the dependence of
hadron-level cross-sections on PDFs can be gathered by simply looking
at the luminosities which correspond to individual parton subprocesses.
We consider in particular the gluon-gluon luminosity, 
the total quark-gluon and quark-antiquark
luminosities defined as
\be
\Phi_{qg}\equiv\sum_{i=1}^{n_f}
\Phi_{q_ig};\quad\Phi_{q\bar{q}}\equiv\sum_{i=1}^{n_f} \Phi_{q_i\bar q_i},
\label{eq:qqqgdef}
\ee
and the charm and beauty quark-antiquark luminosities.

\begin{figure}[ht!]
\centering
\epsfig{width=0.49\textwidth,figure=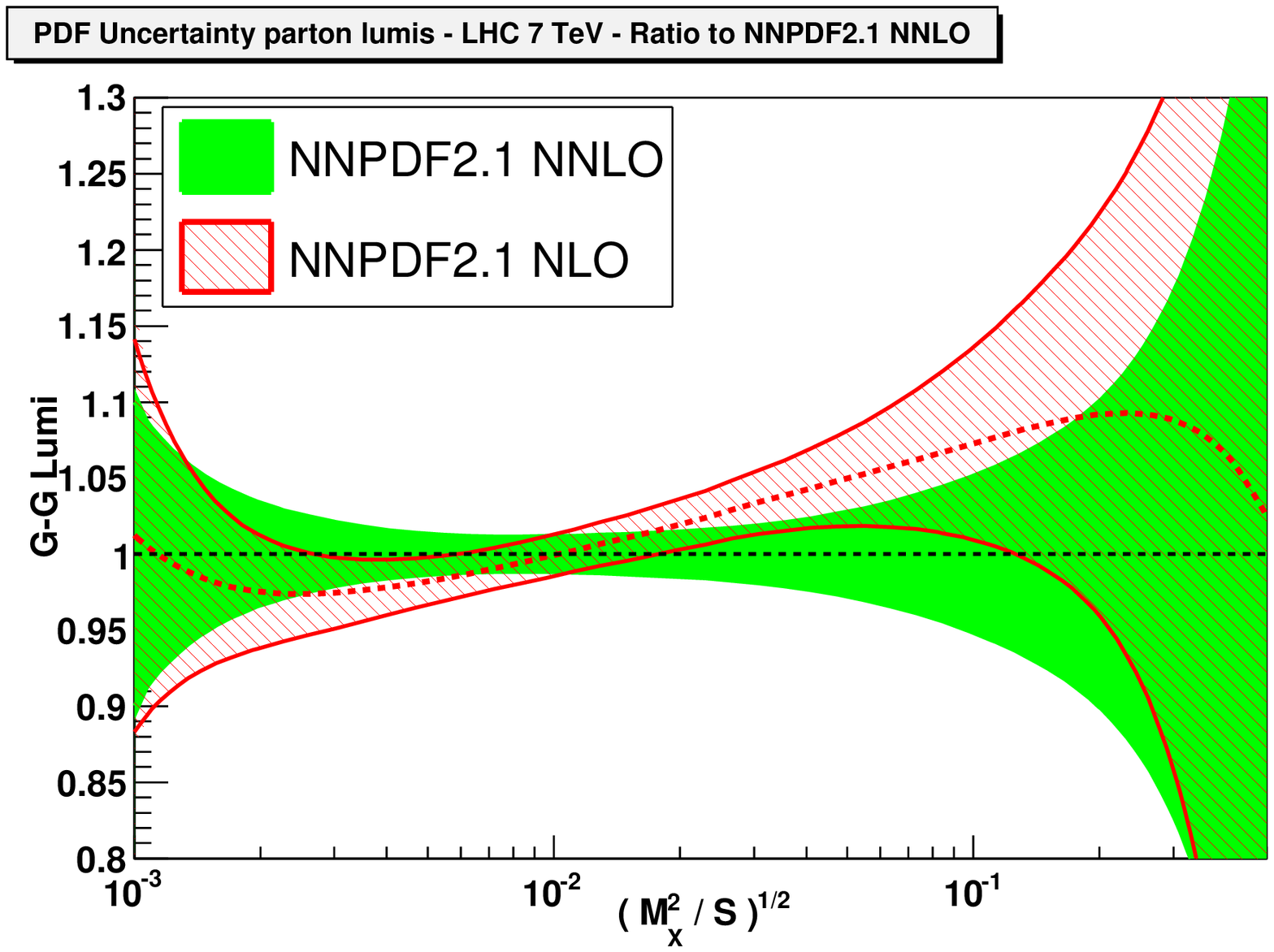}
\epsfig{width=0.49\textwidth,figure=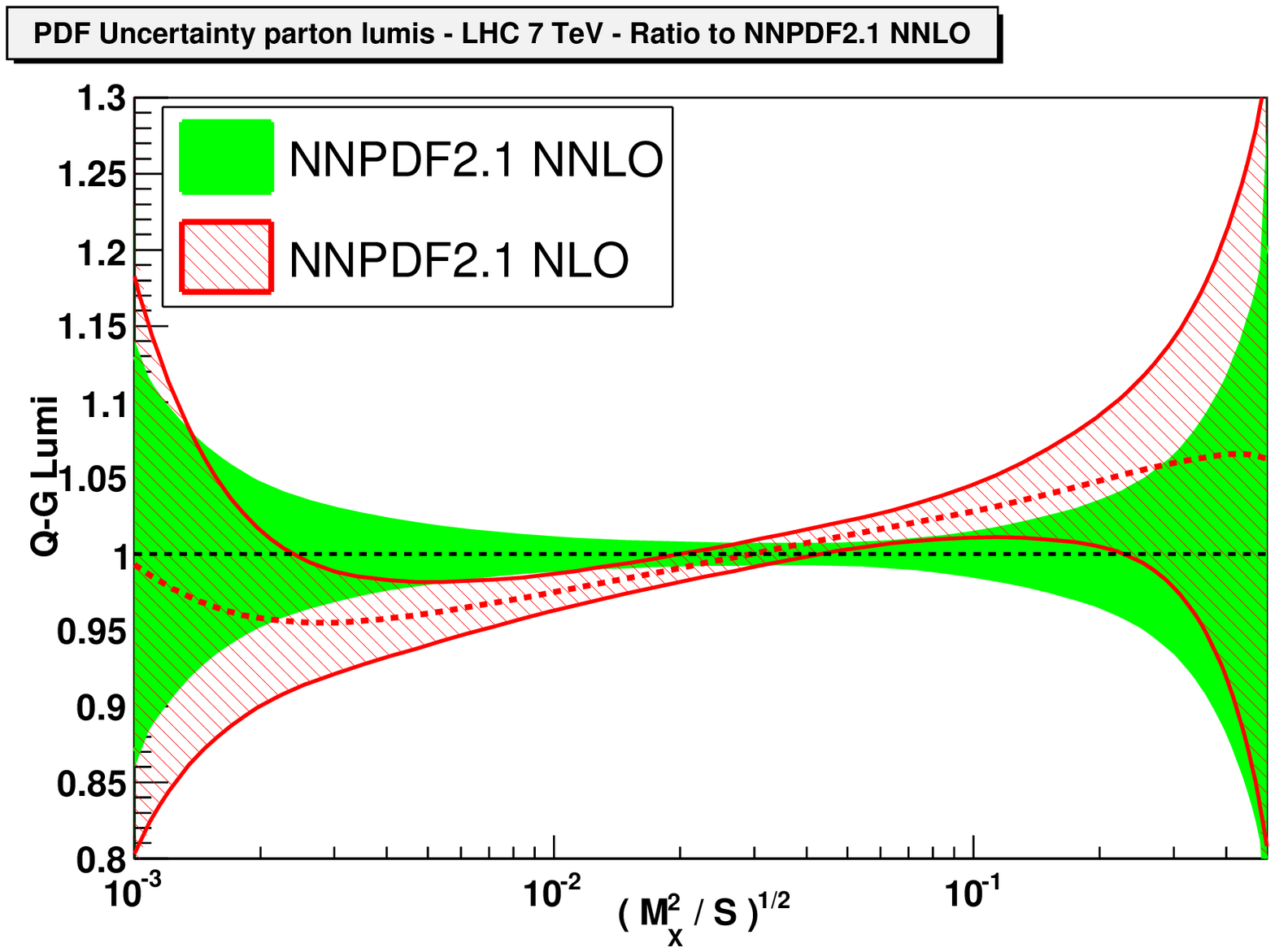}\\
\epsfig{width=0.49\textwidth,figure=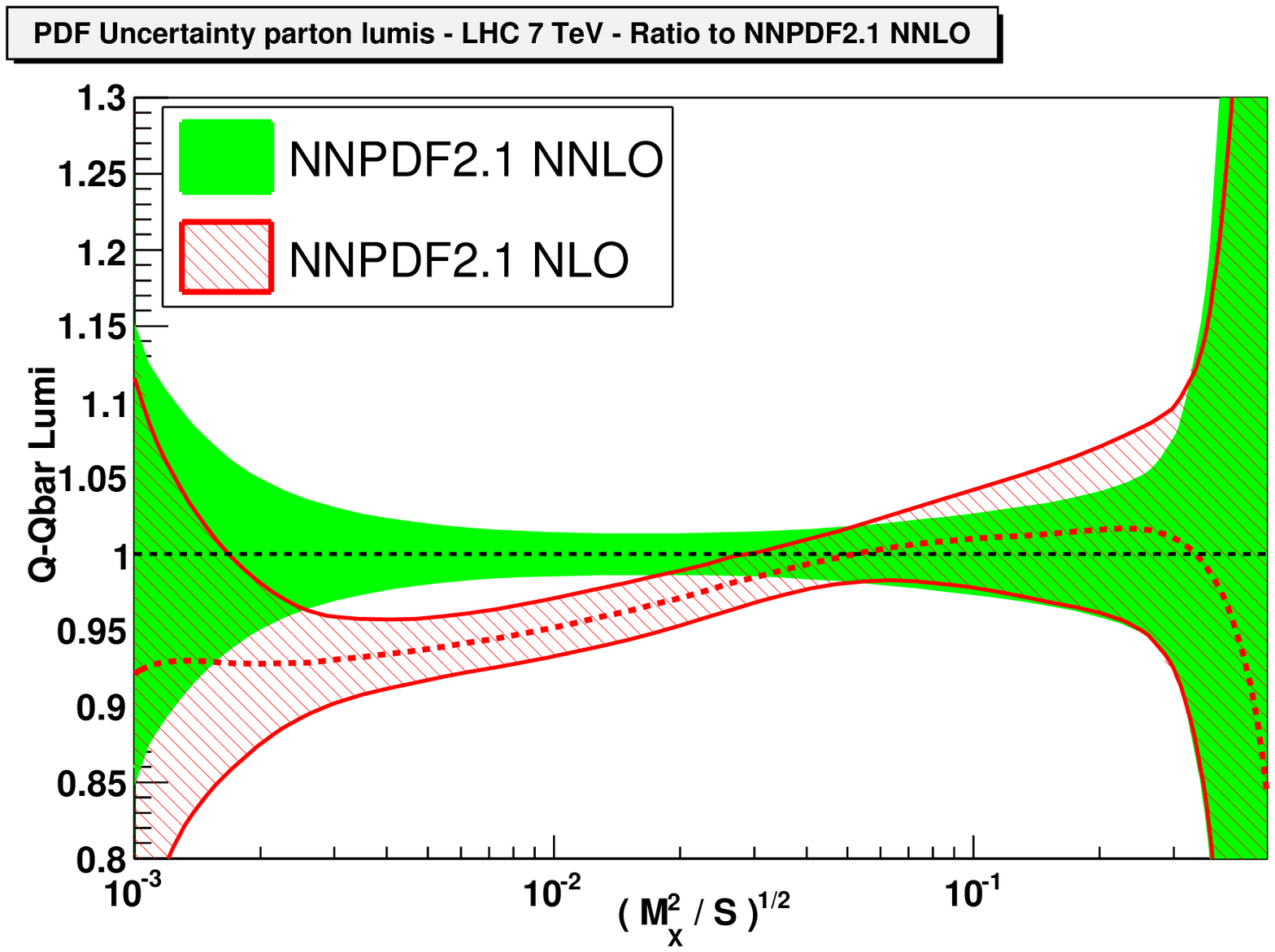}
\epsfig{width=0.49\textwidth,figure=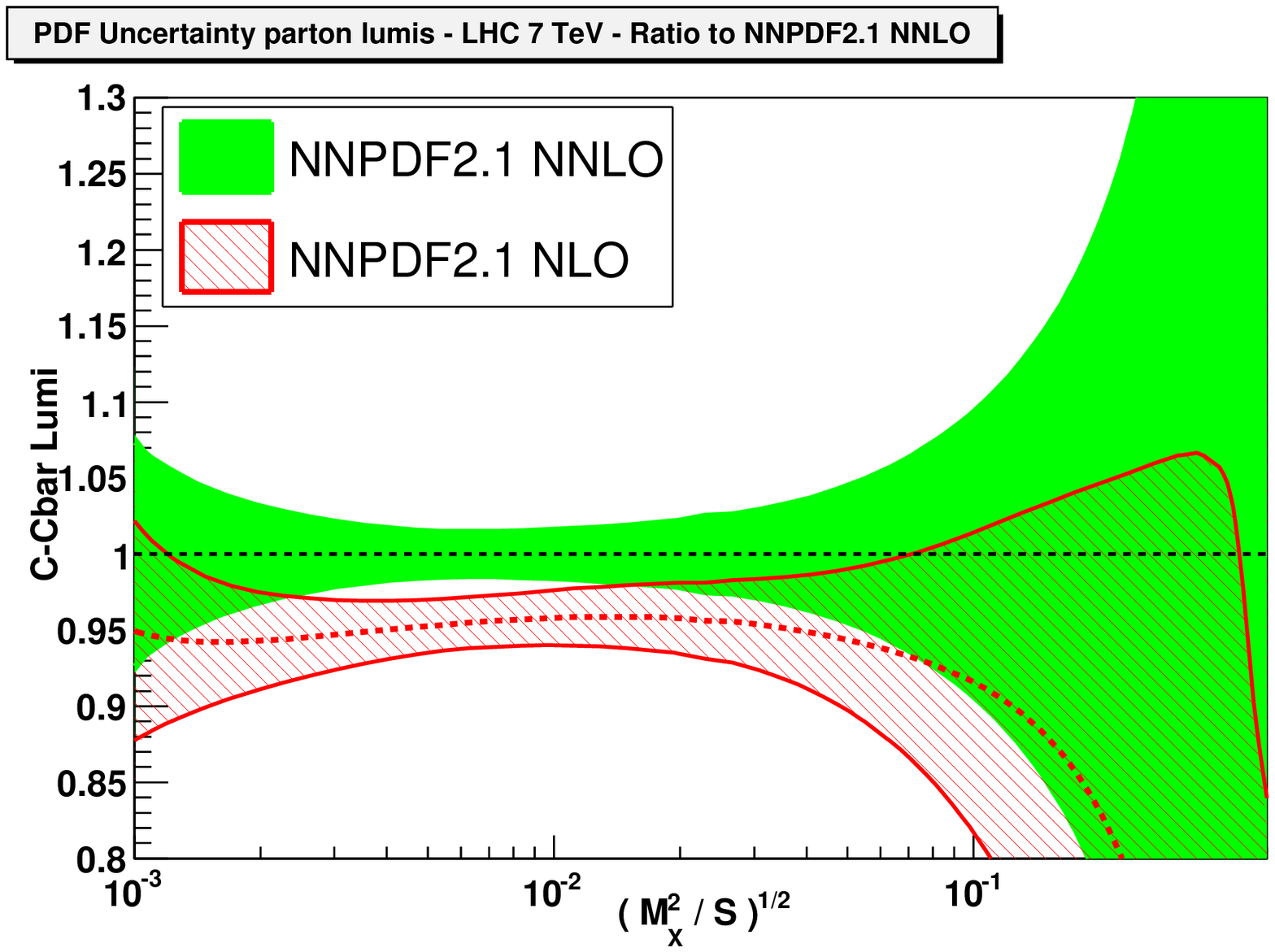}\\
\epsfig{width=0.49\textwidth,figure=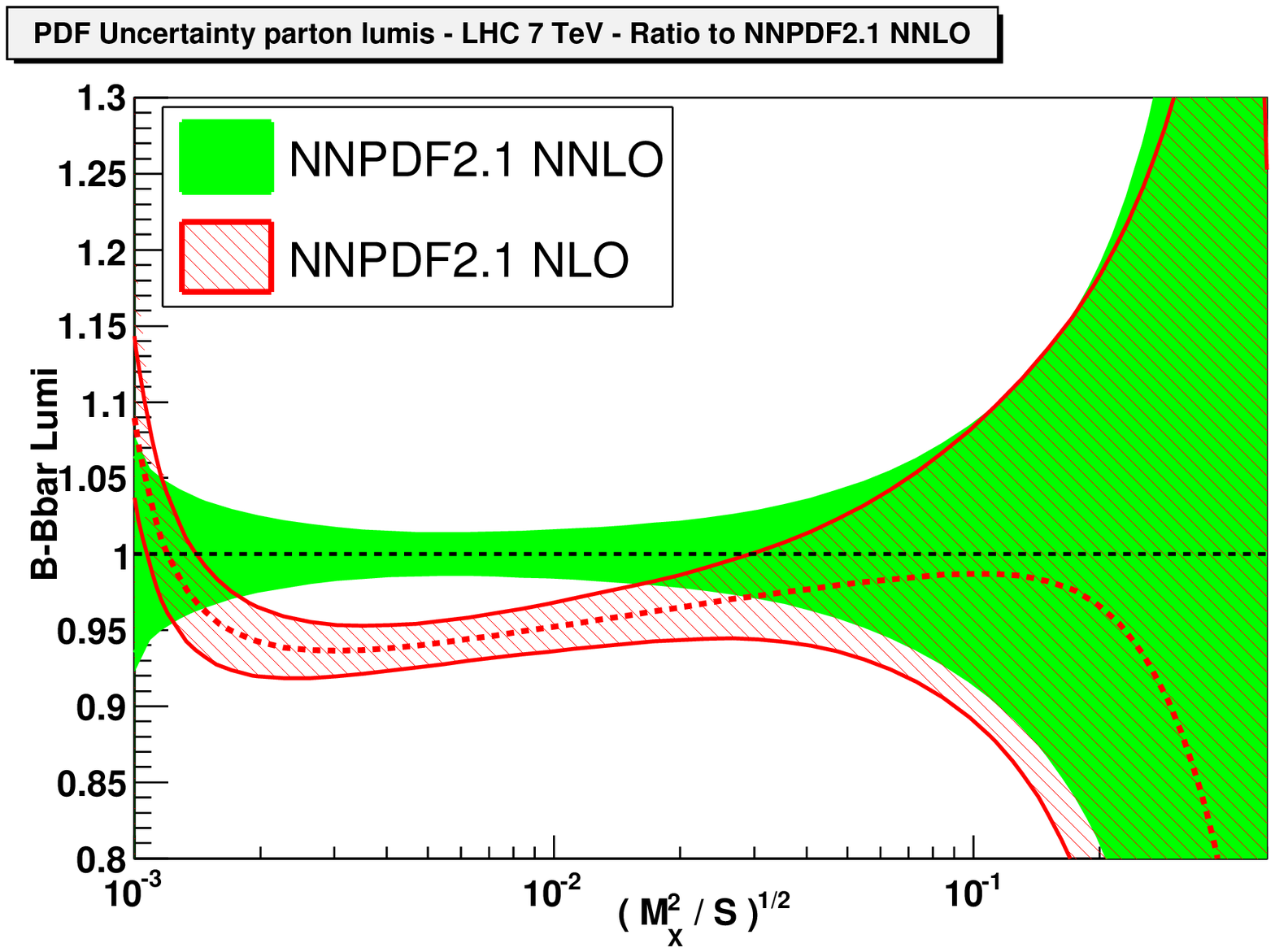}
\epsfig{width=0.49\textwidth,figure=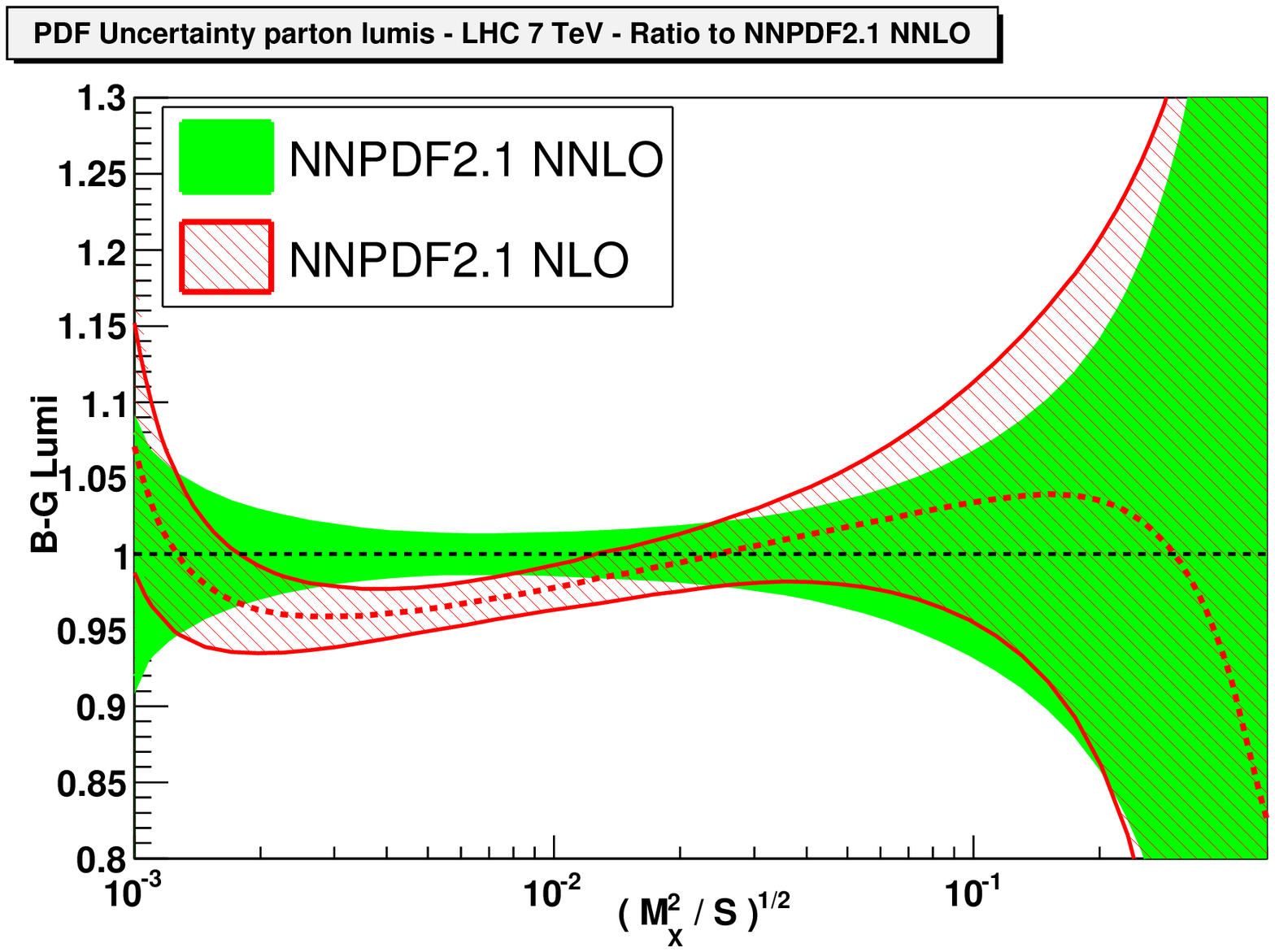}
\caption{\small Comparison of the parton luminosities
  Eqs.~(\ref{eq:lumdef}-\ref{eq:qqqgdef}) for LHC at 7~TeV, computed using the
NNPDF2.1 NLO and NNLO PDFs, using $N_{\rm rep}=100$ replicas
from both sets. 
From left to right we show $\Phi_{gg}$,  $\Phi_{qg}$, (top)
 $\Phi_{q\bar{q}}$,  $\Phi_{c\bar{c}}$, (middle)  $\Phi_{b\bar{b}}$, 
$\Phi_{bg}$ (bottom).
 All luminosities
are plotted as ratios to the NNPDF2.1 NNLO central value. 
All uncertainties shown are one sigma.
\label{fig_fluxes}}
\end{figure}

The luminosities computed from the NLO and NNLO NNPDF2.1 PDF sets are
compared in Fig.~\ref{fig_fluxes}, all  normalized to the NNPDF2.1 NNLO
central value. The compatibility, and thus the perturbative
stability,
is good for of all luminosities, 
as expected from the PDF comparison of
Figs.~\ref{fig:singletPDFs}-\ref{fig:dist_20_21}. In particular, the 
gluon-gluon luminosity, which is relevant for Higgs production at the
LHC, is quite stable in the region which corresponds to standard Higgs
production, though for larger invariant masses the
NNLO luminosity becomes smaller. Non-negligible differences are seen for
the quark-antiquark luminosity, which is significantly larger at NNLO
in the region relevant
for $W$ and $Z$ production, as a consequence of the discrepancy in
light quark distributions already noted in
Fig.~\ref{fig:nnpdf21ratcomp}, which gets squared at the level of luminosities. Similar but somewhat smaller
differences are seen in the quark-gluon channel.
The heavy quark PDFs follow the behaviour of the gluon, from which they
are generated dynamically via perturbative evolution. 
Note that the masses of the heavy quarks
$m_c$ and $m_b$ are the same in the NLO and NNLO analyses.

\begin{figure}[ht!]
\centering
\epsfig{width=0.49\textwidth,figure=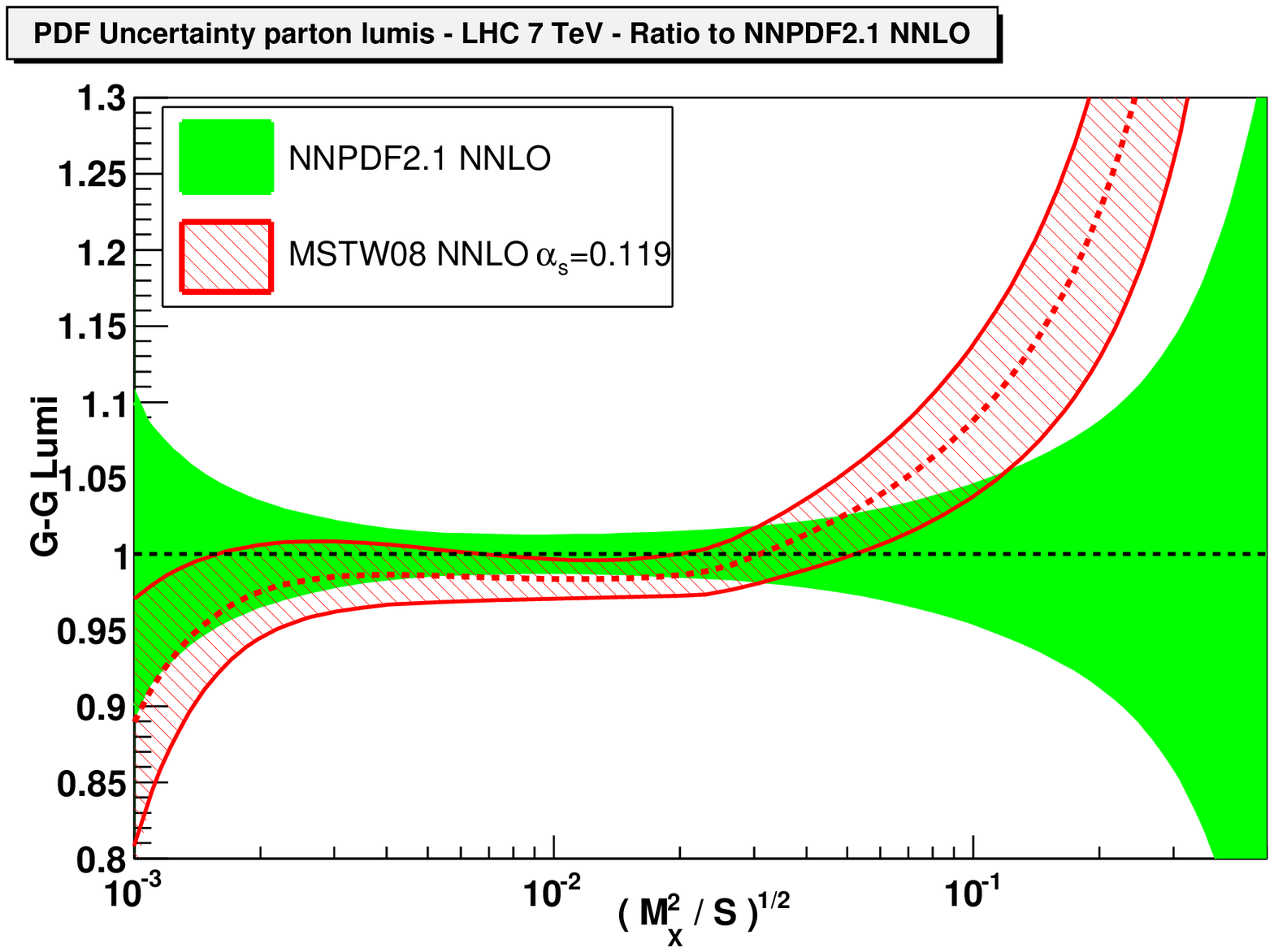}
\epsfig{width=0.49\textwidth,figure=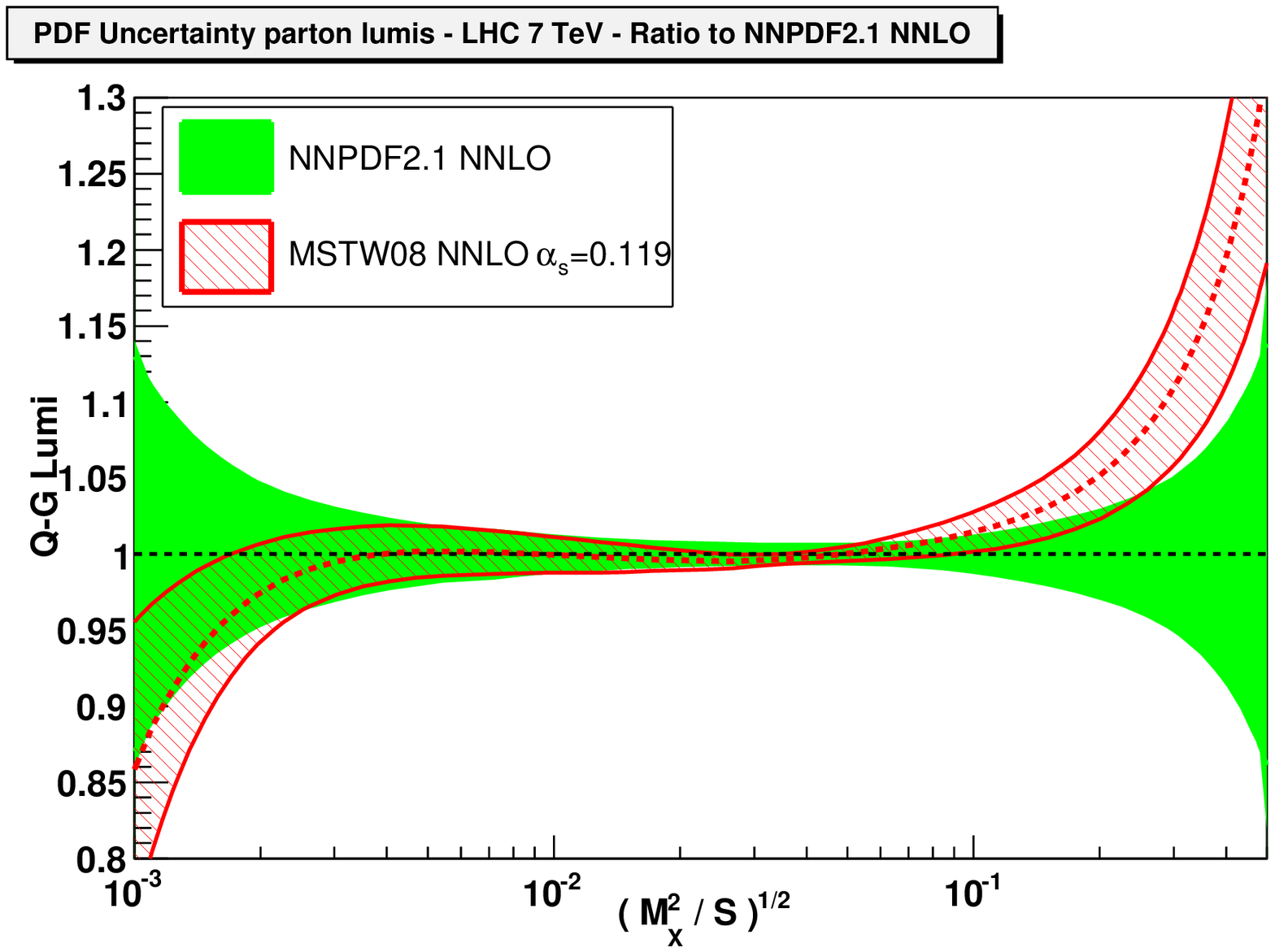}\\
\epsfig{width=0.49\textwidth,figure=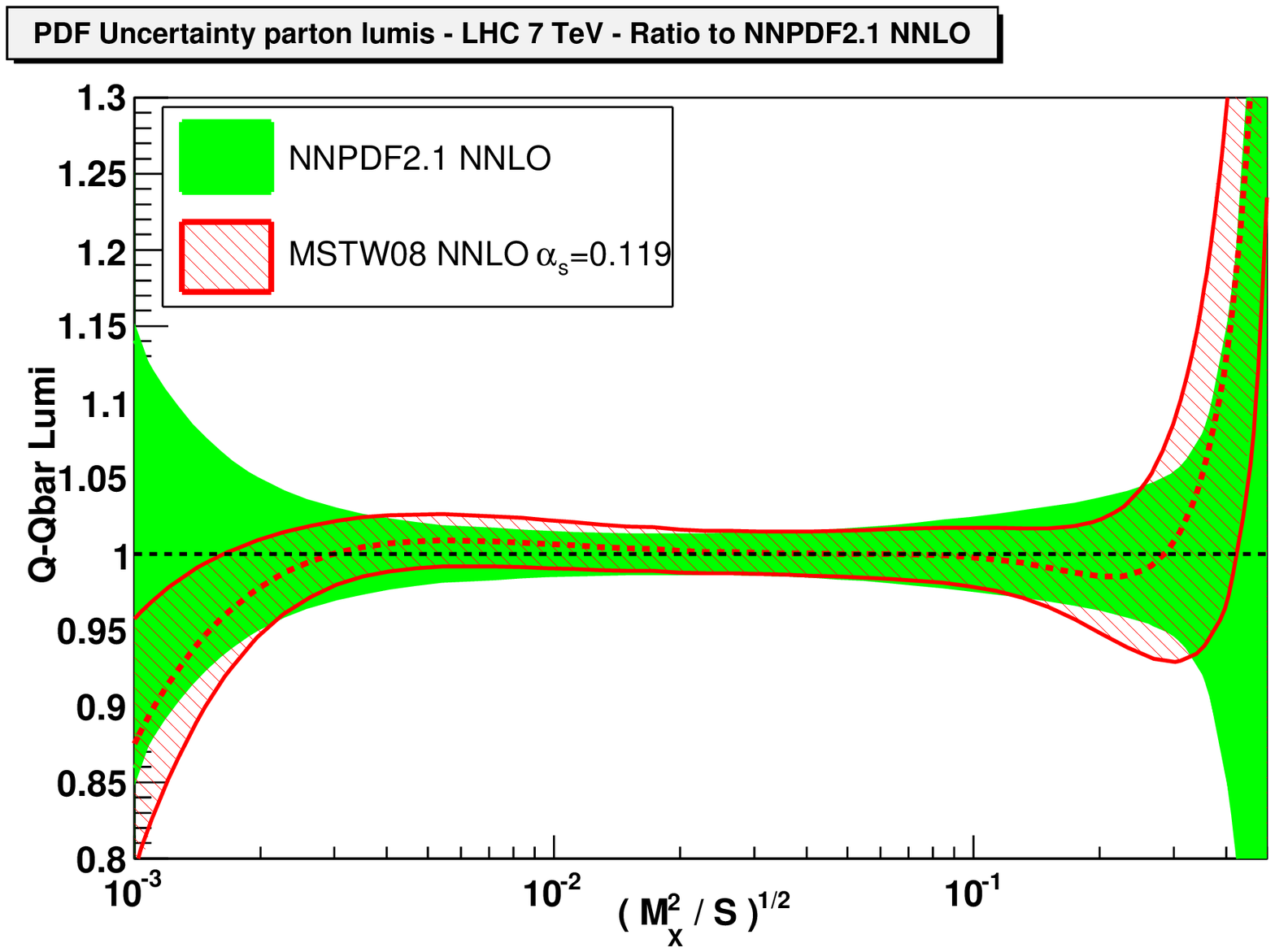}
\epsfig{width=0.49\textwidth,figure=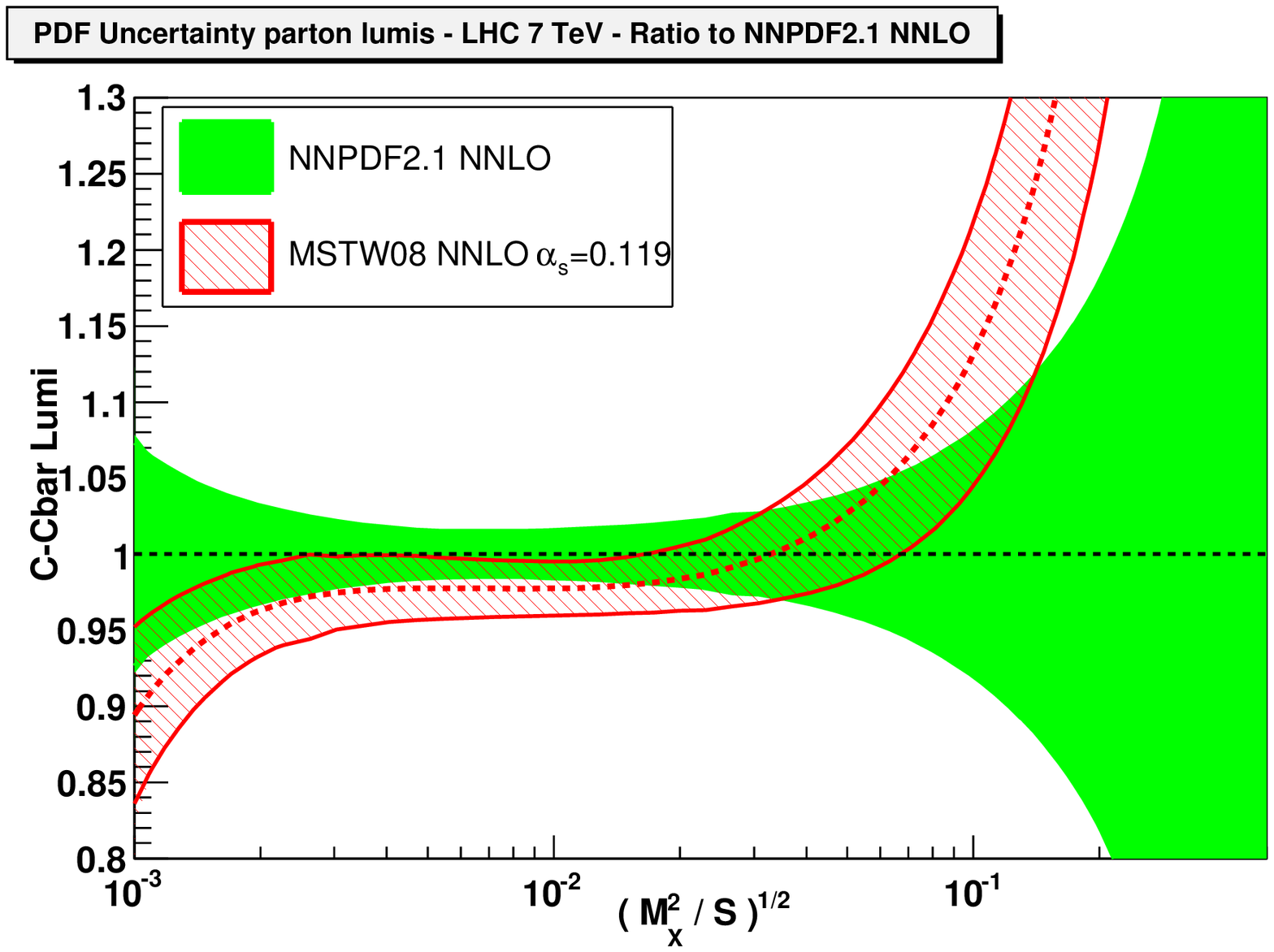}\\
\epsfig{width=0.49\textwidth,figure=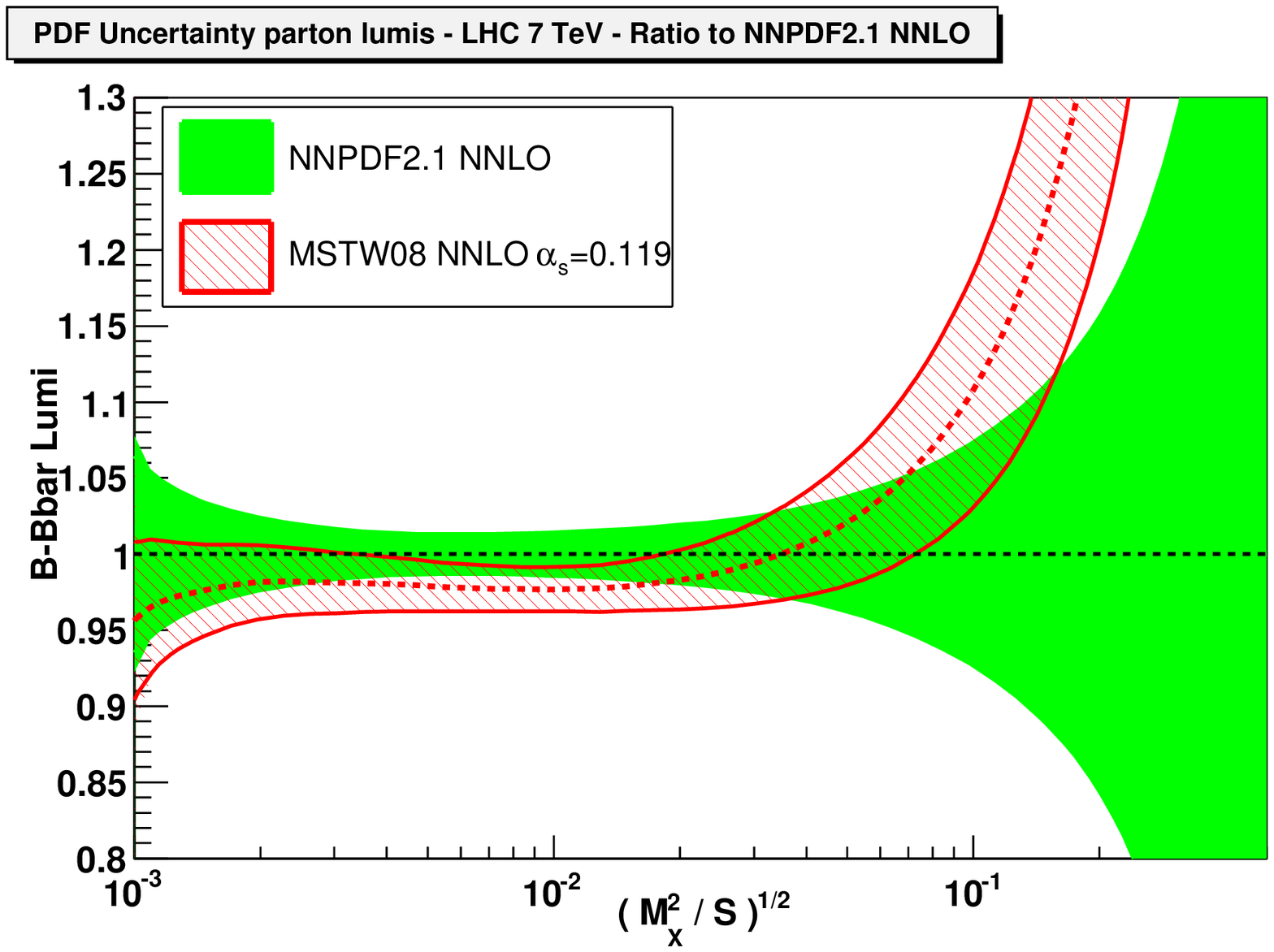}
\epsfig{width=0.49\textwidth,figure=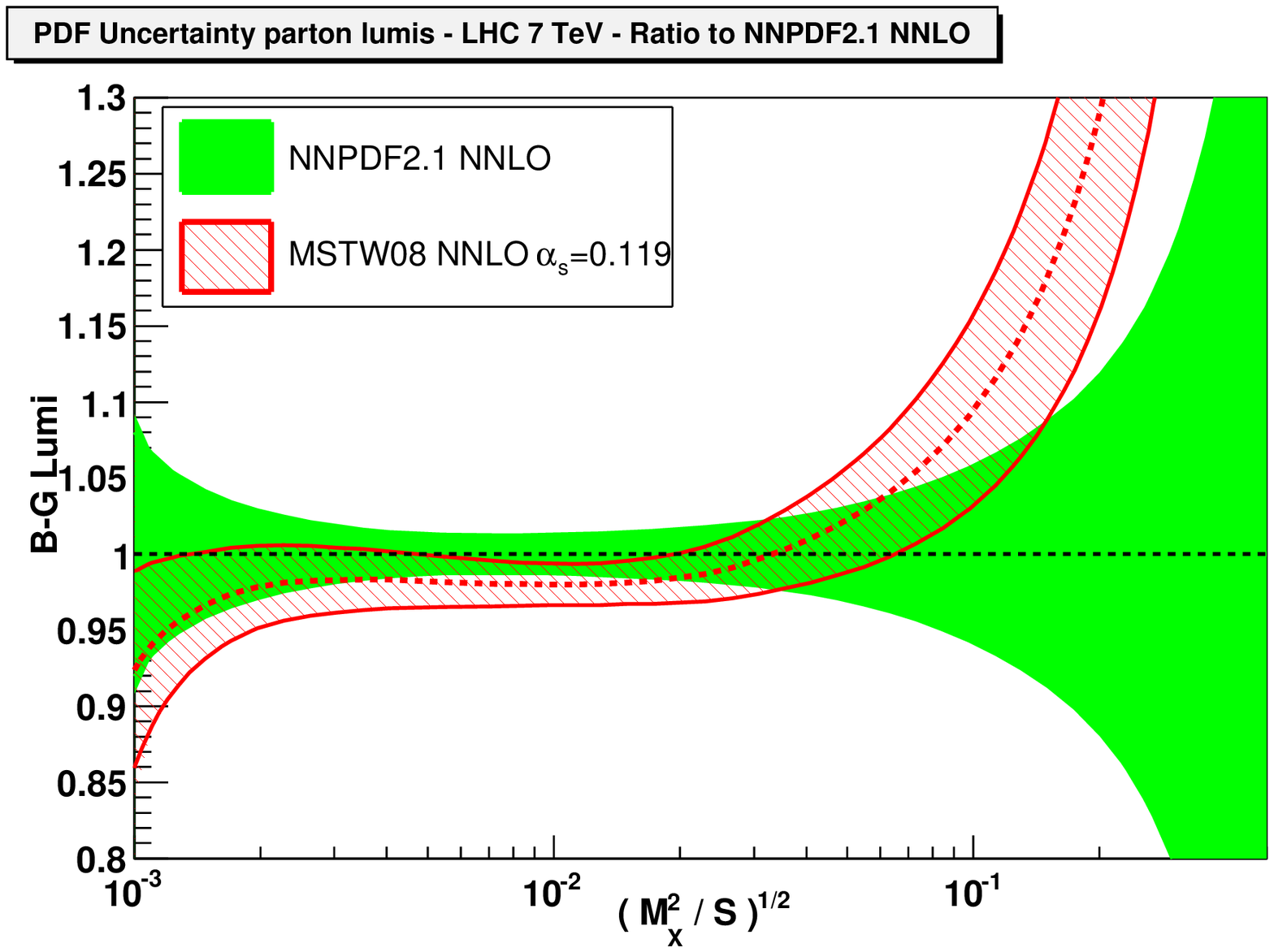}
\caption{\small Same as Fig.~\ref{fig_fluxes}, but for NNPDF2.1 NNLO 
and MSTW2008 NNLO PDFs. For both sets  PDFs corresponding to the same value
$\alpha_s=0.119$ have
been used for consistency.
\label{fig_fluxes-mstw}}
\end{figure}

In Fig.~\ref{fig_fluxes-mstw} the NNPDF2.1 NNLO luminosities are
compared to MSTW08 NNLO (at a common value of 
$\alpha_s=0.119$), all
plotted as ratios to the NNPDF2.1 NNLO central value. The agreement is
generally quite good in the region of $M_X^2/s$ which corresponds to typical
electroweak final state masses $M_X$ at the LHC, 
but it deteriorates for very low and
especially very high $M_X$. This is mostly a consequence of the
strikingly different behaviour of the singlet and especially the gluon
distribution
at small $x$ seen in Fig.~\ref{fig:singletPDFs-lhapdf}, related to the
unstable behaviour of the MSTW08 NNLO gluon.

\subsection{Predictions for LHC observables}
\label{sec:lhcimplications}

NNLO computations are at the
current frontier of perturbative QCD, and are thus available for only a
small number of processes. We will now present NNLO results for
the total cross-section
for Higgs and  weak vector boson production. We will 
consider also the approximate NNLO computation of the $t\bar{t}$
production cross-section. 

When comparing predictions for physical observables with the aim of
understanding their dependence on PDFs and their associated uncertainty it is
important not to mix the uncertainties due to PDFs with uncertainties due to 
the choice of external parameters~\cite{Watt:2011kp}. 
In particular, collider
observables have a nontrivial dependence on the value of the strong
coupling, both through the hard matrix elements and due to the
correlation of PDFs with the value of $\alpha_s$~\cite{Pumplin:2005rh,Martin:2009bu,LHas}, which is
particularly significant for the gluon distribution. As a consequence, as already
discussed in Sect.~\ref{sec:results}, the only NNLO PDF set with
which a detailed quantitative comparison is currently possible is MSTW08. For
the sake of illustration, however, we will also present comparisons
with the ABKM09 NNLO set, even though it should be kept in mind that
these PDFs are provided for $\alpha_s(M_Z)=0.1135\pm0.0014$, and their
uncertainties always include also the uncertainty due to the variation
of $\alpha_s$ in this range.

\begin{figure}[t]
\centering
\epsfig{width=0.7\textwidth,figure=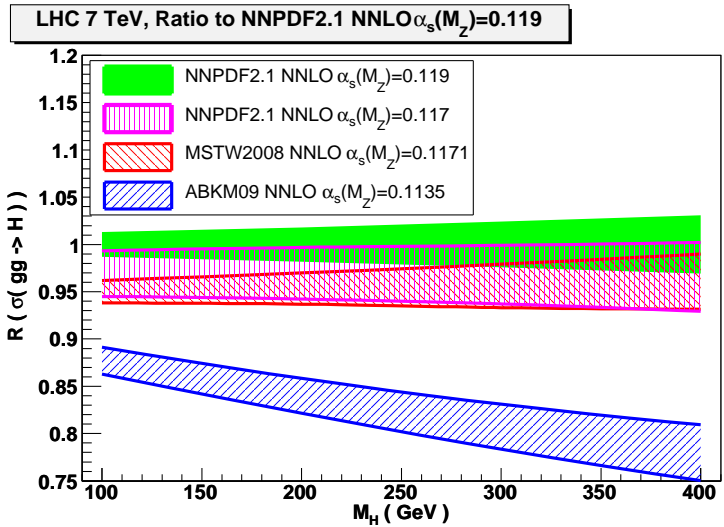}
\caption{\small The total cross-section for Higgs boson
production via gluon fusion at NNLO as a function
of $m_H$. Results are shown for  NNPDF2.1 
with $\alpha_s(M_Z)=0.119$ and $\alpha_s(M_Z)=0.117$,
 MSTW08 with  $\alpha_s(M_Z)=0.1171$, and ABKM09 with
 $\alpha_s(M_Z)=0.1135\pm0.0014$, all displayed as ratios to the
central NNPDF2.1 curve.  The NNPDF result is obtained using $N_{\rm
  rep}=100$  replicas. All uncertainties shown are one sigma; for
NNPDF and MSTW they are pure PDF uncertainties, while for ABKM they
also include the $\alpha_S$ uncertainty corresponding to their given range.
\label{fig:higgs}}
\end{figure}

We consider first the total inclusive cross-section for Higgs
production from gluon-gluon fusion,  whose
dependence on PDFs has attracted
considerable attention recently, in view of claims in the literature
(see Ref.~\cite{Baglio:2011hc} and references therein) that the
recommended~\cite{Dittmaier:2011ti} determination of
PDF
uncertainties through  the so-called PDF4LHC 
prescription~\cite{Botje:2011sn}  might be substantially
underestimated (see Refs.~\cite{Ball:2011we} and especially
Ref.~\cite{Thorne:2011kq}  for a
thorough discussion of this issue).
We have computed the NNLO Higgs production cross-section
in the gluon fusion channel using the code of 
Refs.~\cite{Bonciani:2007ex,Aglietti:2006tp}. Results are shown
in Fig.~\ref{fig:higgs}
 as a function of the Higgs mass $m_H$, determined using
the NNLO
NNPDF2.1, MSTW08 and ABKM09 PDF sets. The predictions for each
group are shown at the respective default value of $\alpha_s$, namely
$\alpha_s(M_Z)=0.119$ for NNPDF, $\alpha_s(M_Z)=0.1171$ for MSTW08 and
$\alpha_s(M_Z)=0.1135\pm0.0014$ for ABKM09; for NNPDF the prediction for 
$\alpha_s(M_Z)=0.117$ is also shown in order to allow for a direct
comparison with MSTW08 (NNPDF2.1 NNLO predictions for more values of
$\alpha_s$ are shown and discussed in Fig.~\ref{fig:higgs-alphas} of
Sect.~\ref{sec:alphas} below).  
All uncertainties shown are
one sigma PDF uncertainties only 
(but in the case of ABKM, as mentioned, they include also the
uncertainty on $\alpha_s$).
The NNPDF and MSTW results are in excellent agreement, provided the same value
of $\alpha_s$ is used. On the other hand, the ABKM result disagrees by many
standard deviations: even though a sizable fraction of this disagreement is
due to the different value of $\alpha_s$, it would persist even when
the same value of $\alpha_s$ is adopted (see
Fig.~\ref{fig:higgs-alphas} below). 

\begin{figure}[t]
\centering
\epsfig{width=0.49\textwidth,figure=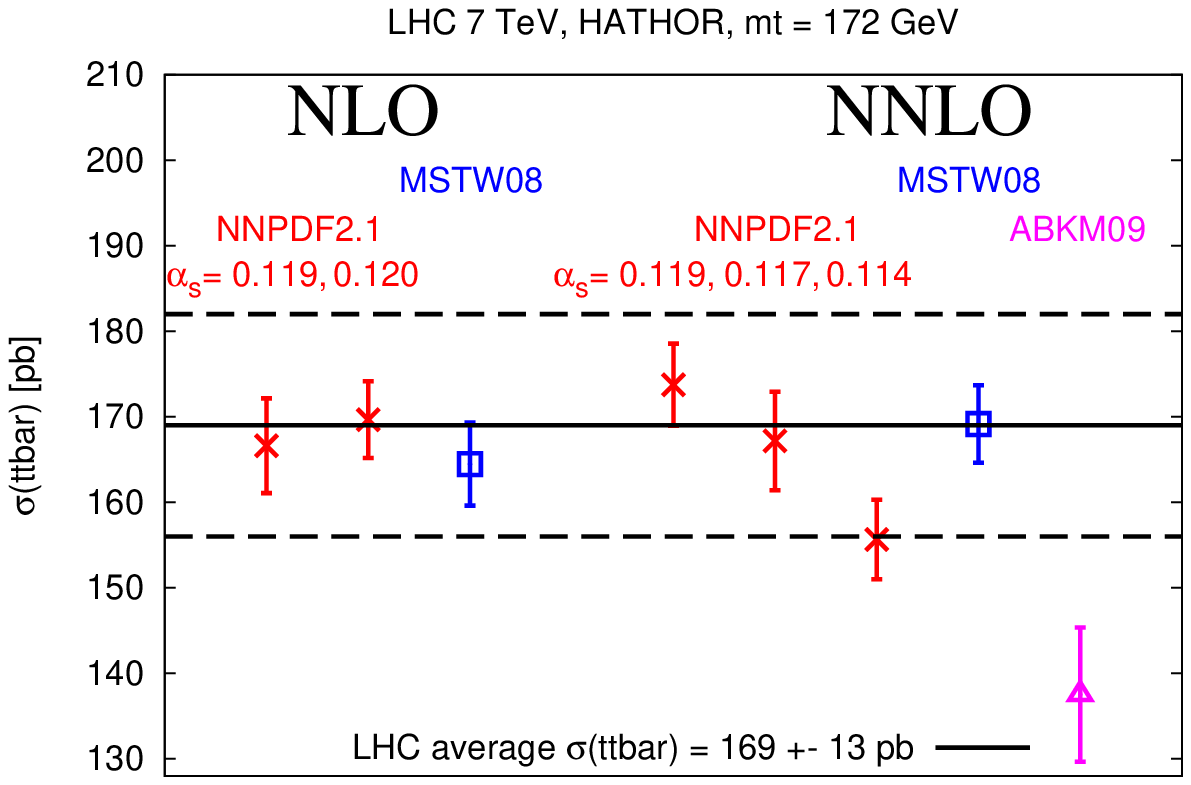}
\epsfig{width=0.49\textwidth,figure=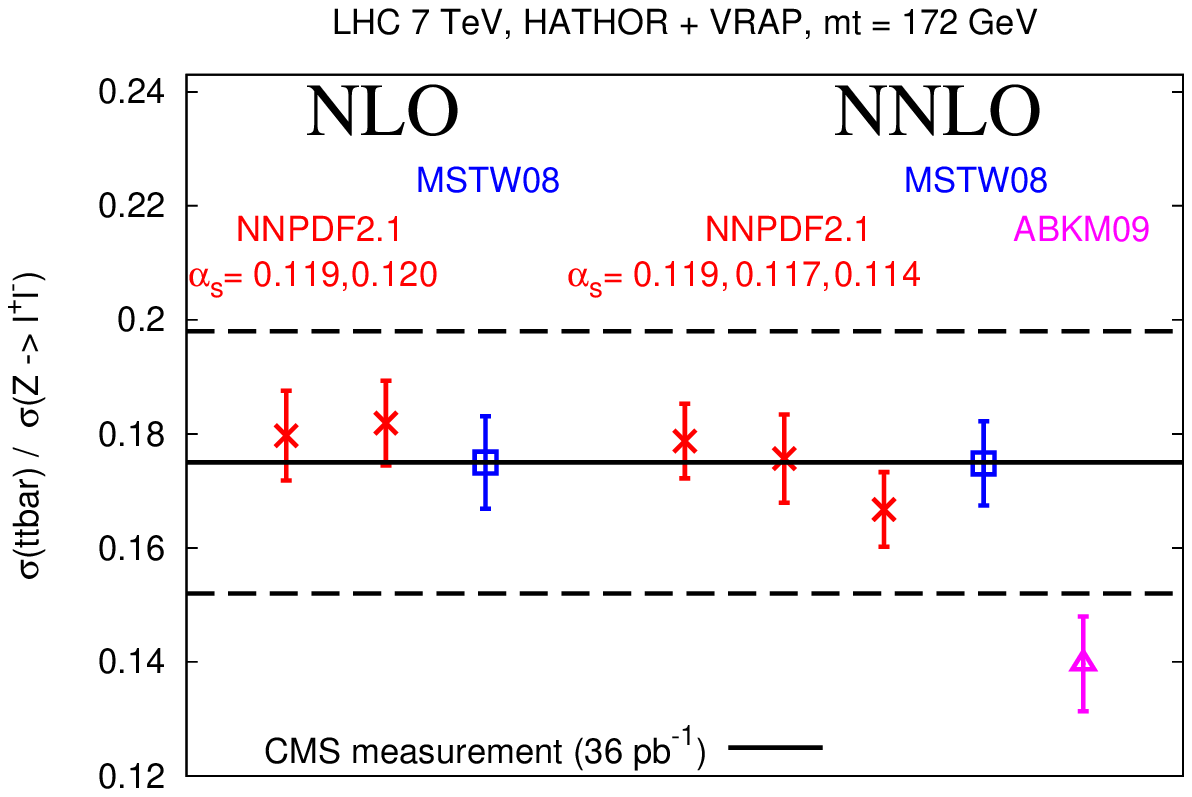}
\caption{\small The total cross-section for $t\bar{t}$
production at the LHC 7 TeV computed using {\tt HATHOR} (left) and its ratio
to the total $Z$ production cross-section computed using {\tt VRAP}
(right). 
Results are shown for  NNPDF2.1 
with $\alpha_s(M_Z)=0.119$ (NLO and NNLO), $\alpha_s(M_Z)=0.120$ (NLO)
and $\alpha_s(M_Z)=0.114,0.117$ (NNLO),
 MSTW08 with  $\alpha_s(M_Z)=0.1202$ (NLO) and
$\alpha_s(M_Z)=0.1171$ (NNLO), and ABKM09 with
 $\alpha_s(M_Z)=0.1135\pm0.0014$ (NNLO). 
The NNPDF results are obtained using $N_{\rm
  rep}=100$  replicas. All uncertainties shown are one sigma; for
NNPDF and MSTW they are pure PDF uncertainties, while for ABKM they
also include the $\alpha_S$ uncertainty corresponding to the given range.
The band corresponds to the combination of the most recent CMS
and ATLAS measurements (left, see text) and to the CMS
measurement~\cite{Chatrchyan:2011nb} (right).
\label{fig:tt-hathor-lhc}}
\end{figure}

Another LHC process which is very sensitive to QCD dynamics  is top
production. Full NNLO corrections for this process are not
available, however, approximate NNLO expressions  based on threshold 
resummation have been constructed~\cite{Moch:2008qy} and implemented
in the public {\tt HATHOR} code~\cite{Aliev:2010zk}.
In Fig.~\ref{fig:tt-hathor-lhc} we show
the prediction for the total $t\bar{t}$ cross-section determined at
NLO and NNLO at the LHC~7~TeV with  $m_t=$172 GeV (pole mass). Theoretical
predictions  are shown for NNPDF2.1
at NLO and NNLO using the corresponding  parton sets with 
$\alpha_s(M_Z)=0.119$; they are compared to the NLO and NNLO MSTW08
predictions which have $\alpha_s(M_Z)=0.1202$ at NLO and
$\alpha_s(M_Z)=0.1171$ at NNLO, and to the ABKM NNLO prediction 
which has $\alpha_s(M_Z)=0.1135\pm0.0014$,
by also displaying the NNPDF results
with $\alpha_s(M_Z)=0.120$ (NLO)
and $\alpha_s(M_Z)=0.114,0.117$ (NNLO). These theoretical
predictions can be compared to the 
average of the recent measurements from 
CMS~\cite{Chatrchyan:2011nb,CMS-PAS-TOP-11-001}, $\sigma\lp t\bar{t}\rp=158
\pm 19$ pb, and
ATLAS~\cite{ATLAS-CONF-2011-040}, $\sigma\lp t\bar{t}\rp=180
\pm 19$ pb.
Averaging the most accurate results, which have been
obtained with a luminosity of $\sim 36$ pb$^{-1}$, and assuming
that the two measurements are independent, yields
$\sigma \lp t\bar{t} \rp=169 \pm 13 $~pb (shown in
Fig.~\ref{fig:tt-hathor-lhc} as a dashed band).

Figure~\ref{fig:tt-hathor-lhc} shows that the NNPDF2.1 
and MSTW08 predictions are in good agreement both at NLO and NNLO:
however, once again, it is important that a common value of $\alpha_s$
be used. Also, of course, one should remember that the uncertainties shown in  
Fig.~\ref{fig:tt-hathor-lhc} 
are only PDF uncertainties. In particular theoretical  
uncertainties, such as may be estimated by scale variation, and  
uncertainties due to the dependence on the top mass, are not shown and  
might also be significant. Again, the ABKM09 prediction is  
significantly lower, and the disagreement persists even when a common  
value of $\alpha_s$ is adopted: already the LHC data are starting to  
discriminate between PDF sets.

\begin{figure}[t]
\begin{center}
\epsfig{width=0.49\textwidth,figure=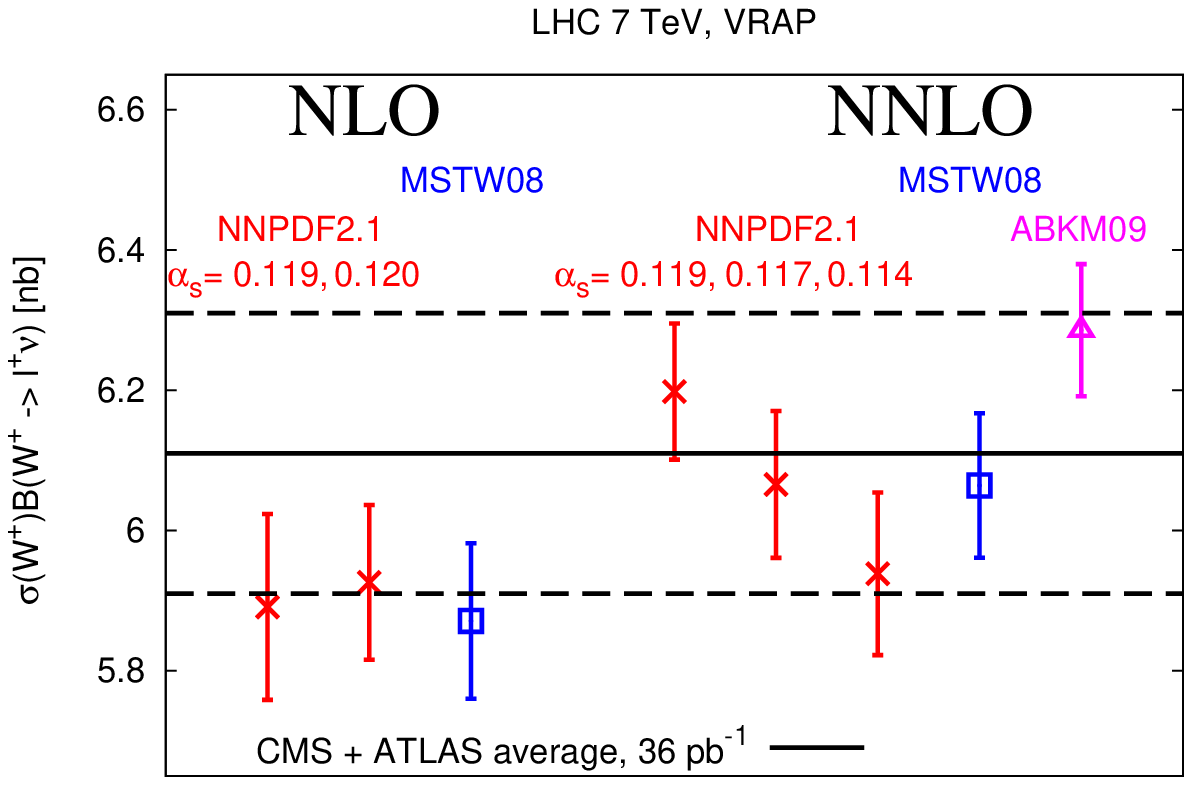}
\epsfig{width=0.49\textwidth,figure=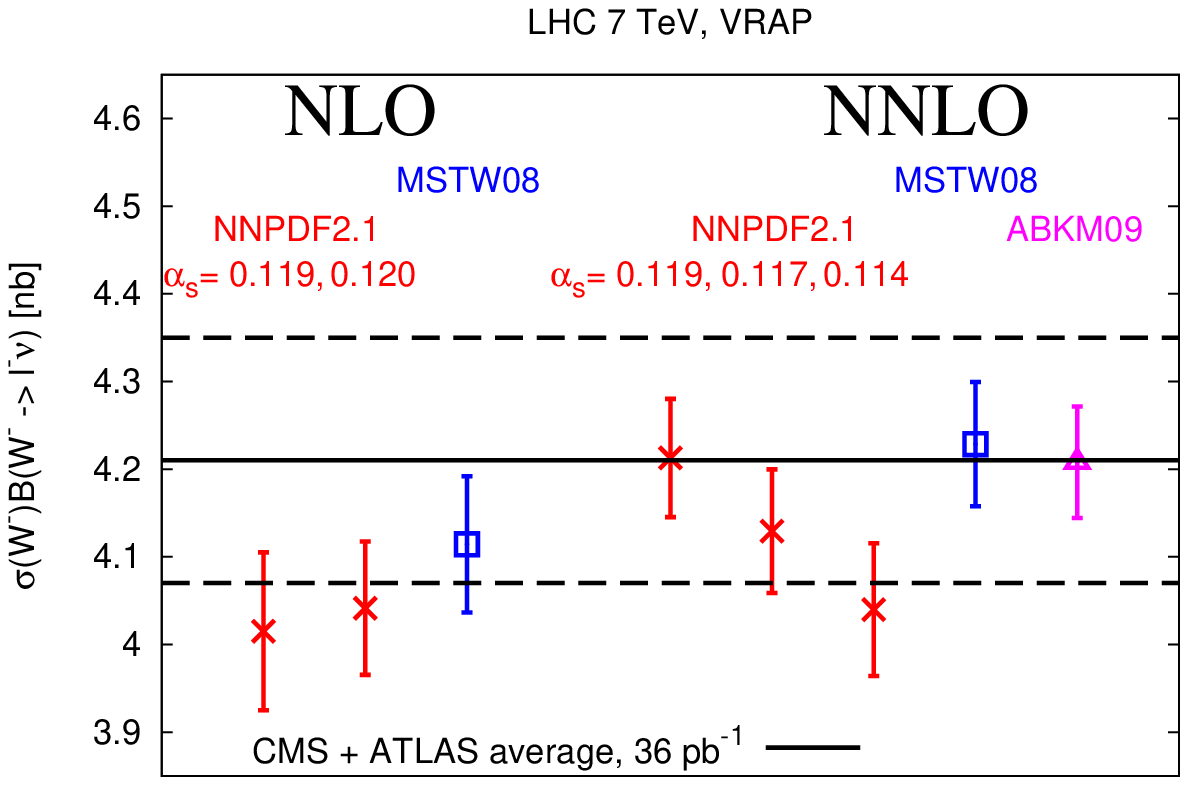}
\epsfig{width=0.49\textwidth,figure=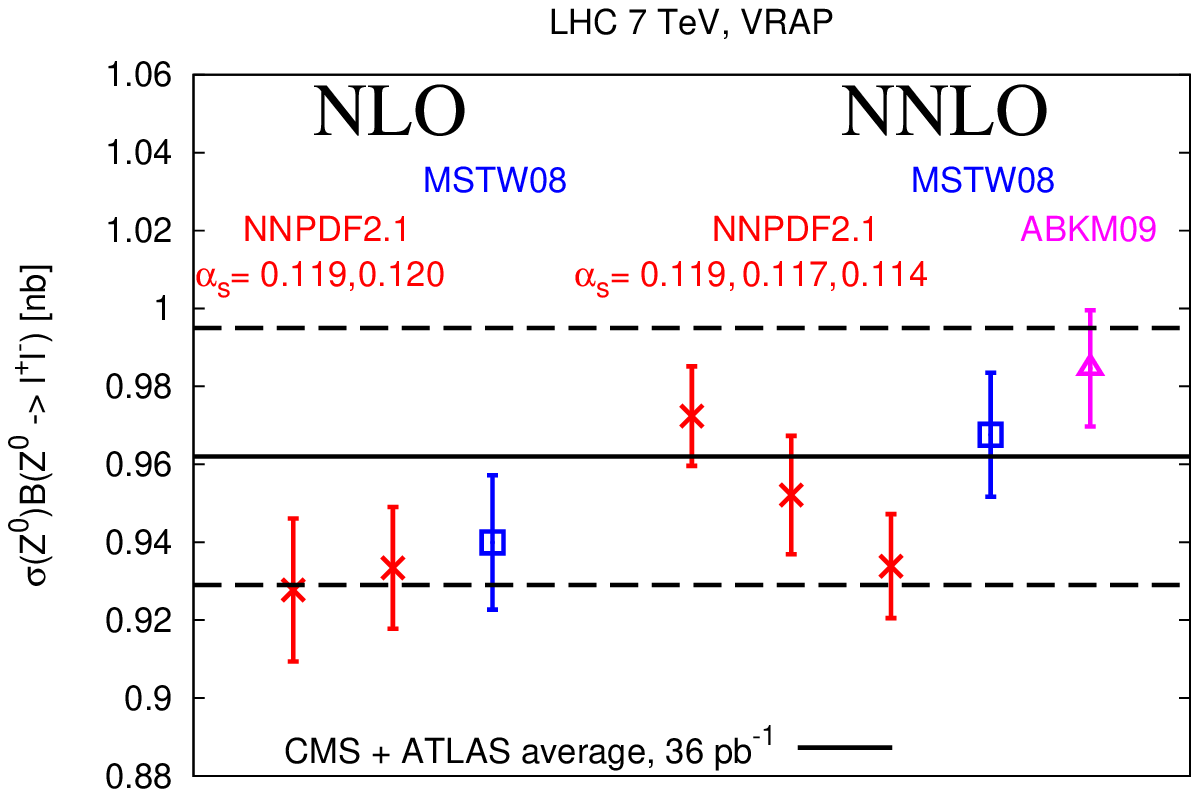}
\end{center}\begin{center}
\epsfig{width=0.49\textwidth,figure=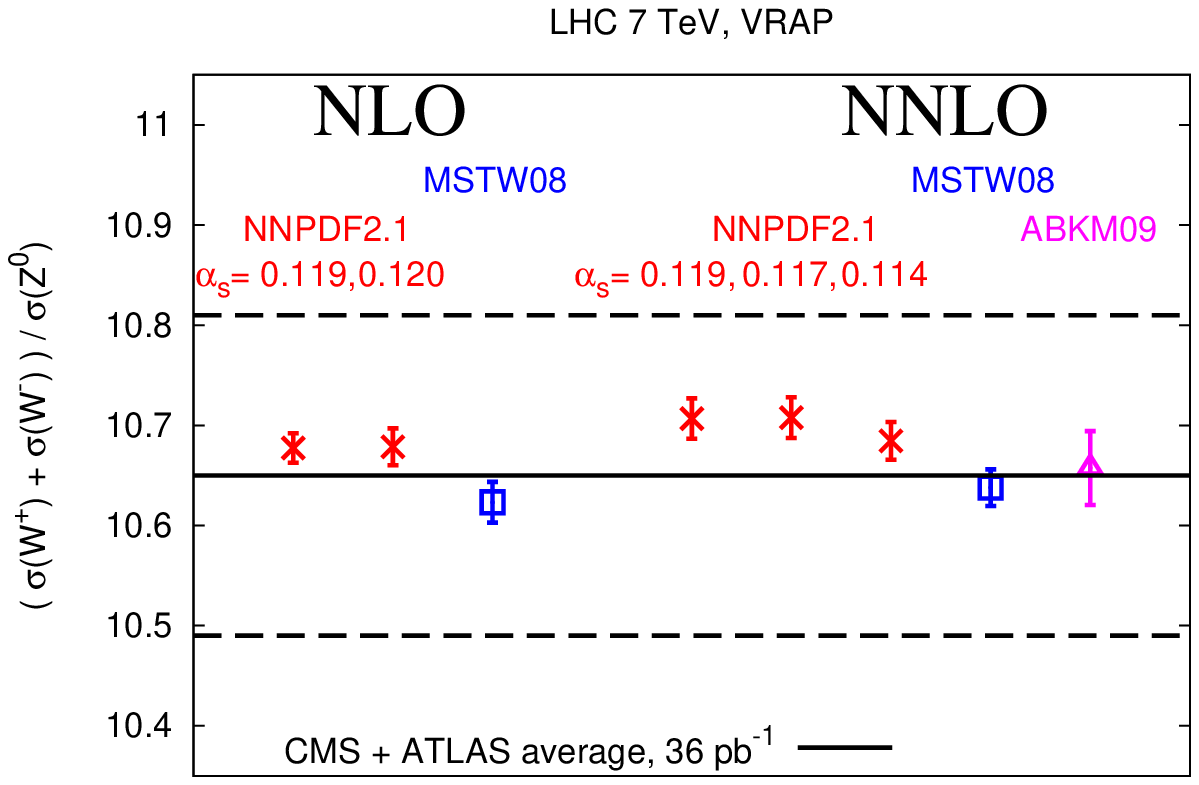}
\epsfig{width=0.49\textwidth,figure=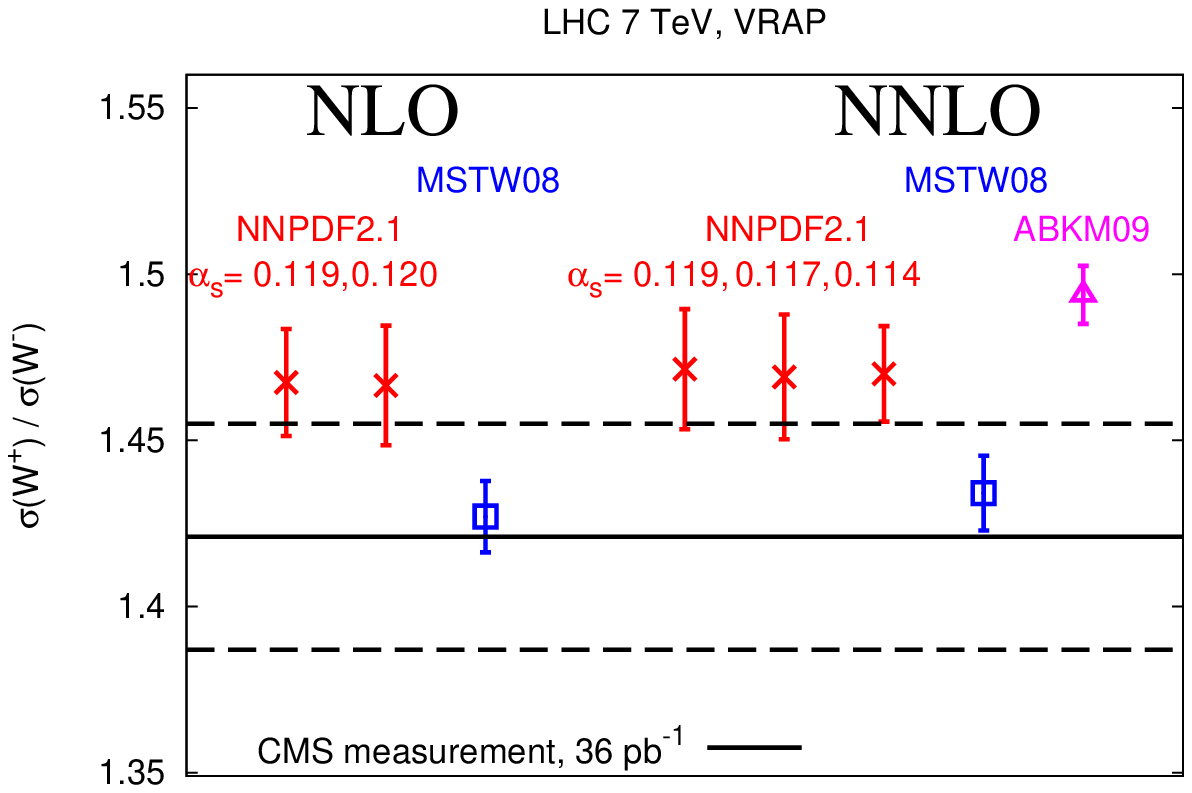}
\end{center}
\caption{\small The total cross-sections for $W^+$, $W^-$ and
$Z^0$ production at the LHC 7 TeV and their
ratios $\sigma(W^+ +W^-)/\sigma(Z^0)$ and
$\sigma(W^+)/\sigma(W^-)$. Predictions are shown for the same PDF sets
and values of $\alpha_s$ shown in Fig.~\ref{fig:tt-hathor-lhc}, and
compared to the ATLAS and CMS data summarized in Table~\ref{tab:ewk},
shown as a dashed band.
Uncertainties shown are one sigma for absolute
cross-sections, and 68\% confidence levels for cross-section ratios.
\label{fig:vrap-lhc}}
\end{figure}

A closely related recent measurement by CMS~\cite{Chatrchyan:2011nb}
is the ratio of $t\bar{t}$ and $Z$ cross-sections. We have computed
predictions for this observable using  
the {\tt VRAP} code~\cite{Anastasiou:2003ds} together with {\tt
  HATHOR}, for the same PDF sets and settings. At NNLO this ratio 
is only weakly dependent on the value of $\alpha_s$. 
 Results are also shown in
Fig.~\ref{fig:tt-hathor-lhc} and compared to the CMS measurement,
again shown as a dashed band.  The conclusions are similar.

Finally, we consider the total cross-sections for
electroweak gauge boson production and their ratios at the LHC.
We have computed these 
observables with the {\tt VRAP} code~\cite{Anastasiou:2003ds},
within the narrow-width approximation (including the $\gamma^*$
contribution to gauge boson production). Results are
collected in Fig.~\ref{fig:vrap-lhc}. As for the case of top production, we
show NNPDF2.1 at NLO and NNLO, using our preferred value $\alpha_s=0.119$, and 
also compare NNPDF2.1 with
MSTW08 at NLO and NNLO, and ABKM09 at NNLO, using their preferred
values of $\alpha_s$. 
Again, we have averaged
the most updated CMS~\cite{CMS-EW} and ATLAS~\cite{ATLAS-EW} results for these
observables, assuming they are uncorrelated. The individual
ATLAS and CMS results, corresponding
to an integrated luminosity of $36$ pb$^{-1}$, together
with their average are summarized
in Table~\ref{tab:ewk}. 
The results are shown as dashed bands on the plots. Note that for
cross-section ratios the uncertainty shown is computed as a 68\%
confidence 
level, because we have verified that  the distribution of results can
be markedly non-gaussian.

\begin{table}
\small
\begin{center}
\begin{tabular}{|c|c|c||c|}
\hline
 & ATLAS & CMS & Average \\
\hline
\hline
$\sigma\lp W^+\rp B\lp l^+\nu\rp$ (nb) & $6.26 \pm 0.32$  &  $6.02\pm 0.26$ &  $6.11\pm 0.20$ \\
\hline
$\sigma\lp W^-\rp B\lp l^-\nu\rp$ (nb) &  $4.15\pm 0.21$ & $4.26\pm 0.19$  &  $4.21\pm 0.14$ \\
\hline
$\sigma\lp Z^0\rp B\lp l^+l^-\rp$ (nb) & $0.945\pm 0.051$ & $0.975\pm 0.044$  &  $0.962\pm 0.033$\\
\hline
$\sigma\lp W^++W^-\rp B\lp l\nu\rp/\sigma\lp Z^0\rp B\lp  l^+l^-\rp$  & $10.91\pm 0.28$  &  $10.54\pm 0.19$&  $10.65\pm 0.16$\\
\hline
$\sigma\lp W^+\rp/\sigma\lp W^-\rp$  & -  & $1.421\pm 0.034$  & $1.421\pm 0.034$  \\
\hline
\end{tabular}
\end{center}
\caption{\small Recent results from CMS at ATLAS for the
total cross-sections for $W^+$, $W^-$ and $Z^0$ production and their ratios,
obtained with an integrated luminosity of $\sim 36$ pb$^{-1}$, together
with their average. The average has been obtained assuming the two measurements
to be completely uncorrelated. \label{tab:ewk}}
\end{table}

The dependence on $\alpha_s$ is weaker for these processes, which are
quark-dominated,   independent of the strong coupling at Born level,
and affected by smaller NNLO corrections. However, the dependence of
the predictions on the perturbative order is still not negligible. Differences between PDF
sets are also less significant, except for the $W^+/W^-$ cross-section
ratio which is a very sensitive probe of the quark flavor
decomposition. The general conclusions are similar to those from 
top production: the LHC data (in particular the $W$ cross-section ratio)
already show some discrimination between PDFs. Moreover, for some of
these standard candle observables the LHC data may 
be able to discriminate between NLO and NNLO.

As a general conclusion of this first pehnomenological study, we note
that the
pattern of comparison between PDF sets is essentially unchanged when
going from NLO to NNLO. Therefore, the arguments supporting the
PDF4LHC recommendation~\cite{Botje:2011sn} (and specifically its use
in the computation of Higgs exclusion limits~\cite{Dittmaier:2011ti}) 
apply equally at NLO and NNLO.

%% file: sec-modfits.tex
\section{Accuracy of the NNLO PDF determination}
\label{sec:modfits}

The NNLO PDF determination is based on the most accurate available
theory. It is therefore worth discussing the dependence of
results on the  main sources of uncertainty. As shown in detail in our
previous studies Ref.~\cite{Ball:2010de,phystat}, the main factor
which drives PDF uncertainties is the underlying dataset. Also
important are the
values of the QCD parameters used in the PDF extraction, primarily the
value of the strong coupling $\alpha_s$, but also of the quark
masses: their impact is sometimes comparable to that of the data, in 
that the uncertainty due to their variation 
is  similar to the PDF uncertainty when they are kept
fixed~\cite{LHas,Demartin:2010er,Ball:2011mu,LHhq}. Here we 
study both aspects: first, the dependence of PDFs on
parameters (specifically $\alpha_s$), by repeating the NNPDF2.1 NNLO
determination as the underlying parameters are varied, 
and then the dependence  on the size of the dataset by
repeating the NNPDF2.1 NNLO determination with various subsets of the
global dataset.
Theoretical uncertainties will not be studied here: as we argued in
Sect.~\ref{sec:pertstab}, at NNLO the uncertainty related to higher order
corrections (as might be estimated by renormalization and
factorization scale variation) is usually subdominant, as are those
related to the treatment of heavy quarks \cite{LHhq}.

\begin{figure}[t]
  \centering
  \epsfig{width=0.48\textwidth,figure=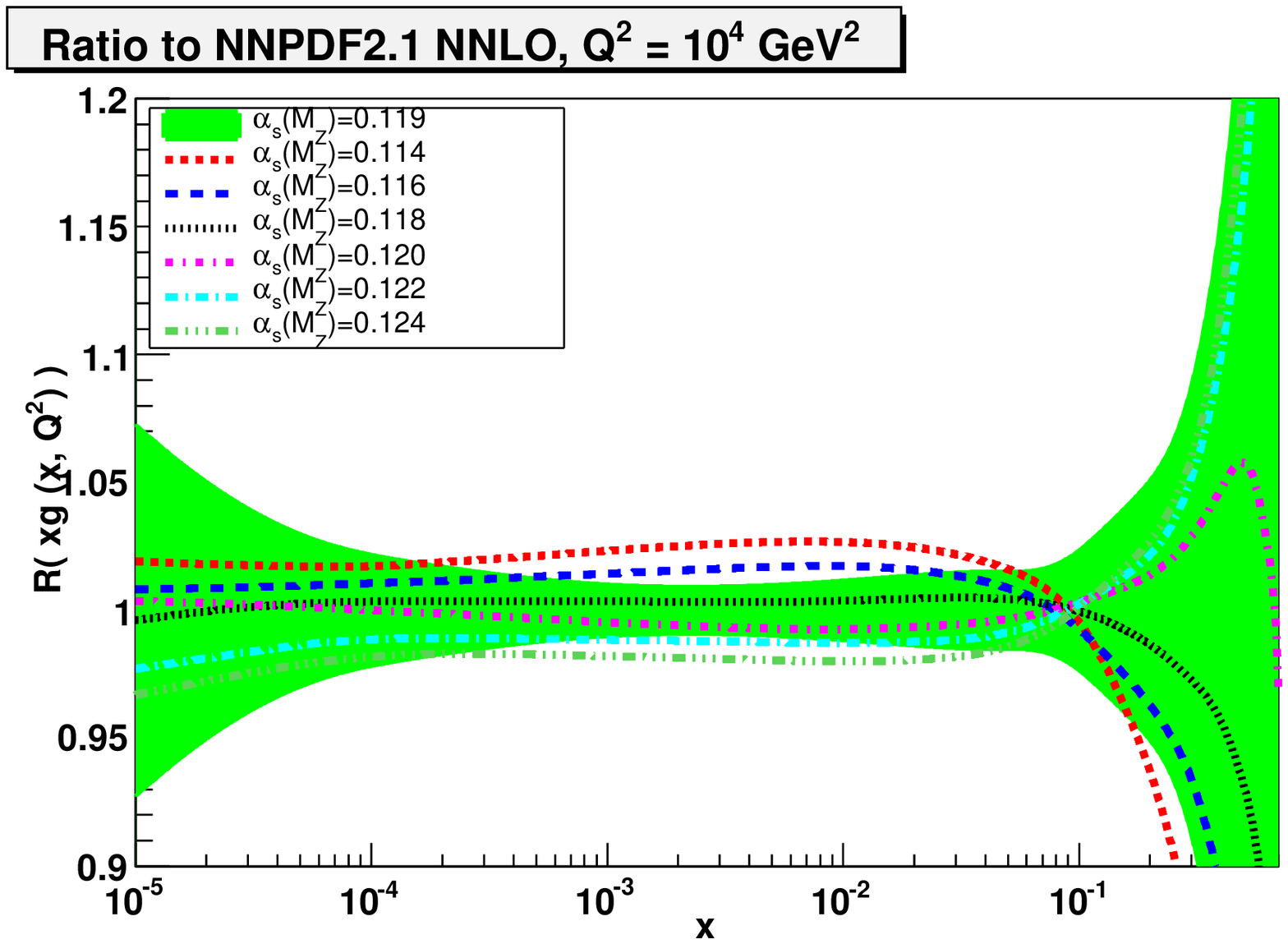}
  \epsfig{width=0.48\textwidth,figure=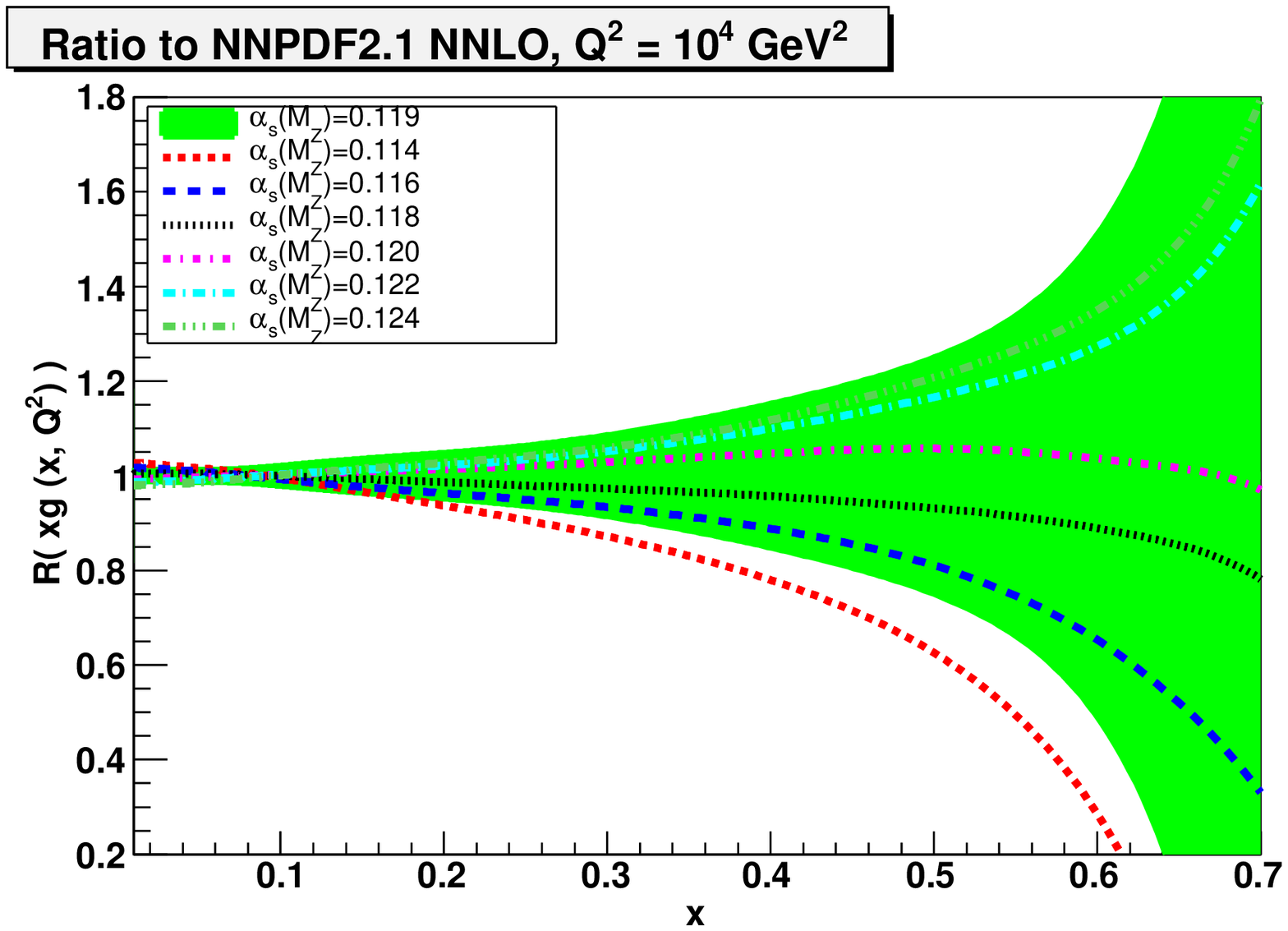}
  \epsfig{width=0.48\textwidth,figure=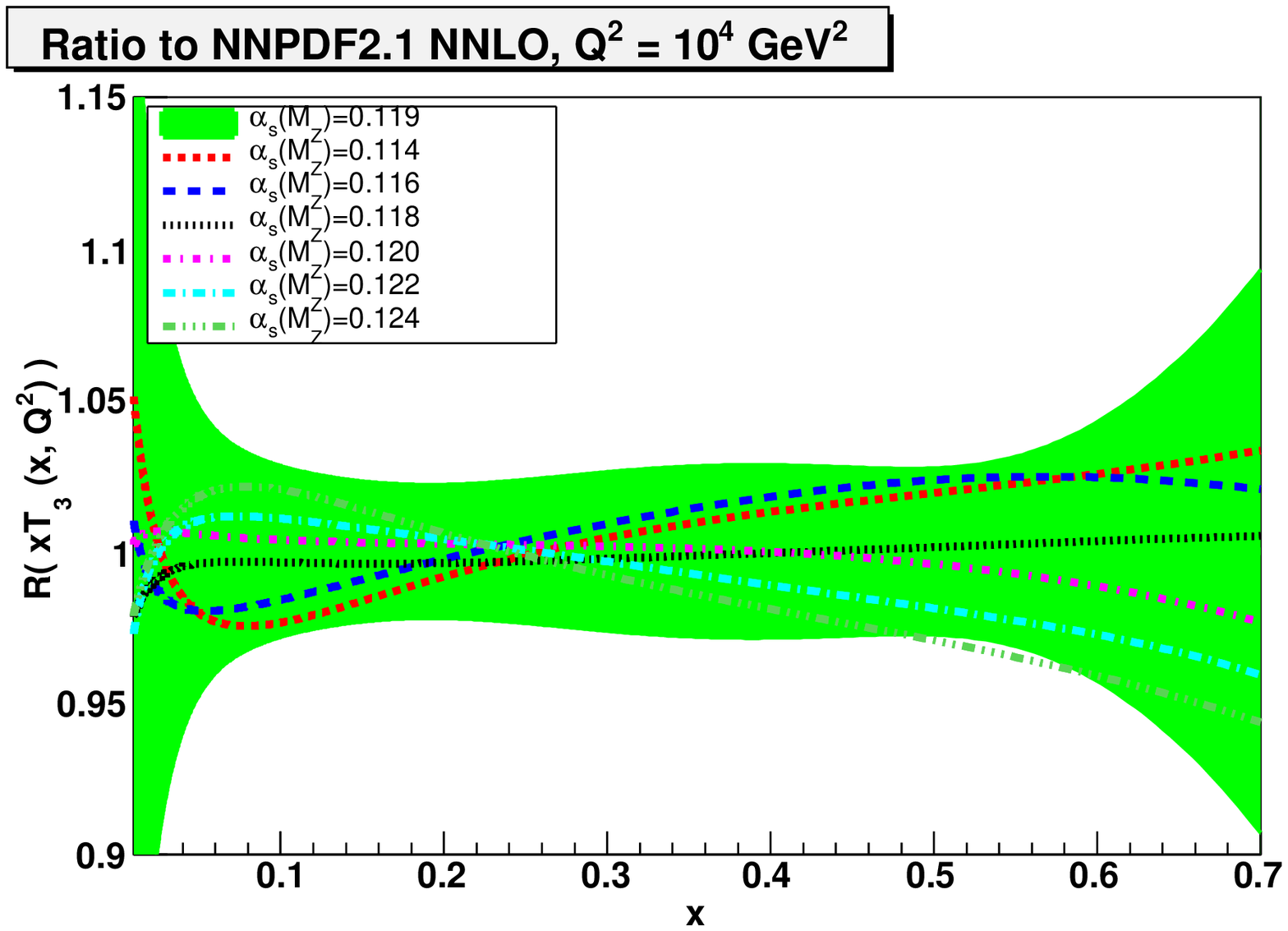}
  \epsfig{width=0.48\textwidth,figure=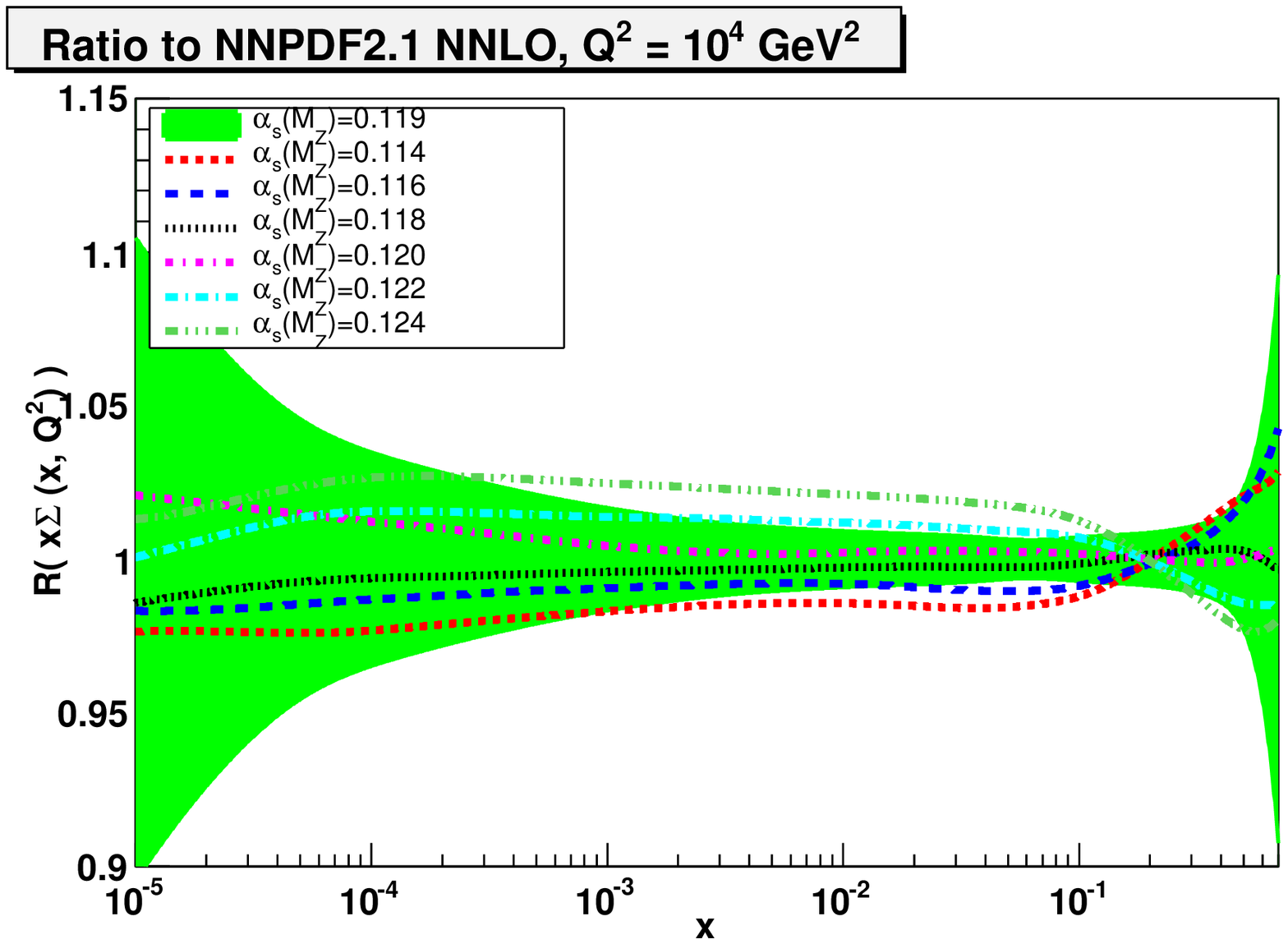}
  \caption{\small Comparison of NNPDF2.1 NNLO sets with different values
    of the strong coupling, shown as ratios to  reference
    set with $\alpha_s=0.119$ at $Q^2=10^4$~GeV$^2$: 
gluon at small and 
    large $x$ (top), triplet at large $x$ and 
singlet at small $x$
(bottom). To improve readability  PDF uncertainties are only shown for the  
$\alpha_s=0.119$ set.
    \label{fig:PDFas}}
\end{figure}

\subsection{Dependence on $\alpha_s\lp M_Z\rp$}
\label{sec:alphas}

In order to determine the dependence of PDFs on
the value of $\alpha_s$ we have repeated the NNPDF2.1 NNLO
determination as $\alpha_s$ is varied: we provide sets with 
 $\alpha_s\lp M_Z\rp$ in the range from 0.114 to 0.124 in
steps of 0.001, each including $N_{\rm rep}=100$ PDF replicas.
Results for the gluon, the quark singlet and isospin triplet are displayed 
in Fig.~\ref{fig:PDFas}, where the ratio of the central PDFs for each
value of $\alpha_s$ to the default  $\alpha_s\lp
M_Z\rp=0.119$ set is shown, and compared to the PDF uncertainty on the
central set. Results are qualitatively similar to their NLO
counterparts of Ref.~\cite{LHas}: the PDF which is most sensitive to
the choice of value of $\alpha_s$ is the gluon, which is mostly
determined by scaling violations.

\begin{figure}[t]
\centering
\epsfig{width=0.70\textwidth,figure=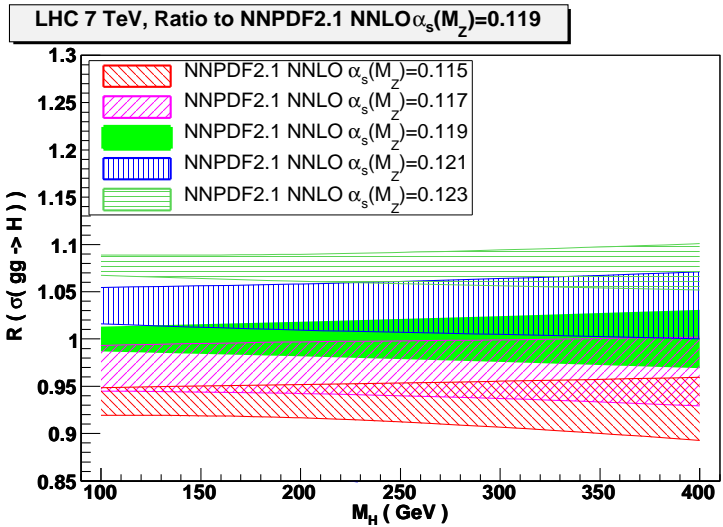}
\caption{\small Same as Fig.~\ref{fig:higgs}, but for
NNPDF2.1 NNLO sets with
different values of
$\alpha_s(M_Z)$.
Results are shown as ratios to the NNPDF2.1 NNLO
reference with $\alpha_s(M_Z)=0.119$.
\label{fig:higgs-alphas}}
\end{figure}

As an illustration of the phenomenological impact of the choice of $\alpha_s$,
in Fig.~\ref{fig:higgs-alphas} we present results for the NNLO
Higgs cross-section from gluon-gluon fusion (same as
Fig.~\ref{fig:higgs})  computed using these PDF sets with varying
$\alpha_s$, again normalized to the central prediction. The dependence
on $\alpha_s$ of this observable is quite strong, both because it is
quadratic in the gluon, and also because it starts at ${\mathcal
  O}(\alpha_s^2)$, and it undergoes NLO corrections which are as
large as the LO contribution, and NNLO corrections which are about
half of the LO
(see
Ref.~\cite{Demartin:2010er,Thorne:2011kq} 
for detailed studies of the relative size
of PDF and $\alpha_s$ uncertainties on this process).

Finally, in order to study the dependence on the value of the heavy
quark masses, we have also produced PDF sets with
$m_c=1.5,~1.6$~and~$1.7$~GeV (in addition to the default $m_c=\sqrt{2}$~GeV),
and  with $m_b=4.25,~4.5,~5.0$~and~$5.25$~GeV
(in addition to the default $m_b=4.75$~GeV). The dependence of PDFs on
the heavy quark masses is similar to that observed at NLO and
discussed in Ref.~\cite{Ball:2011mu}.


\subsection{Dependence on the dataset}
\label{sect:reddataset}

We now turn to the dependence of PDFs and their uncertainties 
on the underlying dataset.
This  dependence is relevant in that one may think that
even though a wider dataset always carries more information, smaller
datasets may be more consistent (from either the experimental or theoretical 
point of view), and thus perhaps more reliable. It is thus important to 
understand what is the price that one pays for this putative improved consistency. 

To this
purpose, we have constructed four new PDF sets, each based on a subset
of the full NNPDF2.1 NNLO dataset, but using exactly the same
methodology. The possibility of obtaining reliable PDFs from datasets
of widely varying size (more than a factor three, in our case) without
having to modify any aspect of the methodology (and in particular
without having to change the parametrization, see
Ref.~\cite{Rojo:2008ke}) is an advantage of the NNPDF approach, since it 
allows a meaningful comparison of uncertainties. Each of these
PDF sets  is made available through the standard LHAPDF interface, and each
so far as it goes is as good as the default one, the only difference between them 
being the smaller amount of experimental information that goes into them.

{
\begin{table}[t]
\centering
\small
\begin{tabular}{|c||c|c|c|c|c|}
\hline 
Experiment    & Global & HERA--only & DIS--only & DIS+DY    & Collider--only \\
\hline
\hline
$N_{\rm dat}$ & 3357  & 834 &2783 & 3171& 1090\\

\hline 
\hline 
Total   & 1.16  & 1.07 & 1.15& 1.18  & 1.02 \\
\hline 
\hline 
NMC-pd    & 0.93 &  [13.15] & 0.88 & 0.94  & [3.43] \\
\hline
NMC       & 1.63  & [1.91] & 1.69 & 1.69 &  [2.06]\\
\hline
SLAC      & 1.01 &  [3.17] & 0.97 & 1.03 &  [1.23]\\
\hline
BCDMS &  1.32 & [2.15] & 1.28 & 1.30  &  [2.22] \\
\hline
HERAI-AV & 1.10 & 1.05 & 1.09 &1.09  & 1.06  \\
\hline
CHORUS   & 1.12& [2.63] & 1.08 & 1.13 & [1.74] \\
\hline
FLH108   & 1.26 & 1.32 & 1.27 & 1.26 & 1.26 \\
\hline
NTVDMN   & 0.49 & [60.51] & 0.45 & 0.54  & [23.02] \\
\hline
ZEUS-H2  & 1.31 &  1.21 & 1.26 & 1.28 & 1.30  \\
\hline
ZEUSF2C & 0.88  &  0.77& 0.86 & 0.88& 0.75  \\
\hline
H1F2C  &  1.46 & 1.30& 1.47 & 1.50  & 1.24 \\
\hline
DYE605  & 0.81  & [9.06] & [6.86] & 0.82&  [1.34] \\
\hline
DYE866  & 1.32 & [12.41] & [2.70] & 1.32  & [5.76] \\
\hline
CDFWASY & 1.65 & [7.71] & [13.94] & 1.64  & 1.07 \\
\hline
CDFZRAP &  2.12 & [3.74] & [2.15] & 1.91& 1.22 \\
\hline
D0ZRAP   & 0.67  & [1.11]  & [0.67]  & 0.65& 0.61 \\
\hline
CDFR2KT  & 0.74 & [1.15] & [0.99]  & [1.25] &  0.64 \\
\hline
D0R2CON  & 0.82 & [1.28] & [0.88] & [1.03]  & 0.83 \\
\hline
\end{tabular}
\caption{\small Quality of the fit for
  NNLO PDF sets based on datasets of varying size. The total number of data
  points is given in the first row, followed by the $\chi^2$
  normalized to the number of data points both for the total fitted
  set and for each of the individual experiments. The $\chi^2$ values
  for experiments which are not fitted are also shown in square brackets. 
\label{tab:redfitschi2}}
\end{table}
}

\begin{figure}[t]
\begin{center}
\epsfig{width=0.99\textwidth,figure=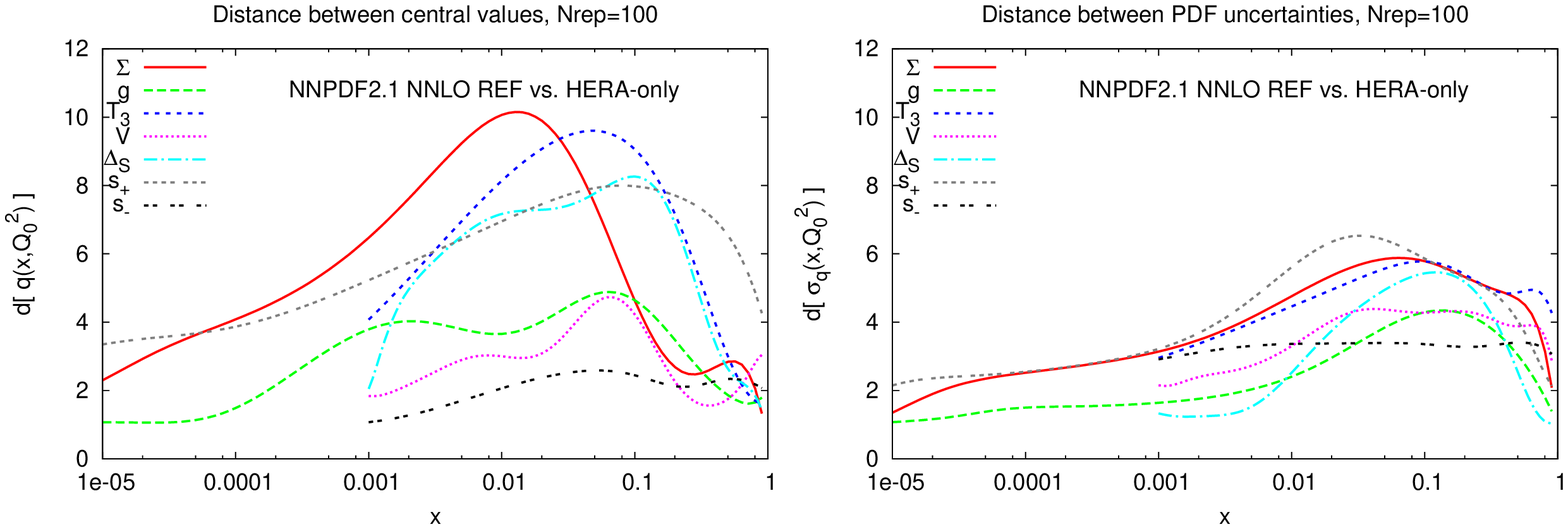}
\caption{\small Distances between central values (left) and
  uncertainties (right) for PDFs in the 
HERA-only and default NNPDF2.1 NNLO fits. All distances are computed from
sets of $N_{\rm rep}=100$ replicas. \label{fig:dist_nnlo-heraonly}}
\end{center}
\end{figure}
\begin{figure}[t]
\begin{center}
\epsfig{width=0.32\textwidth,figure=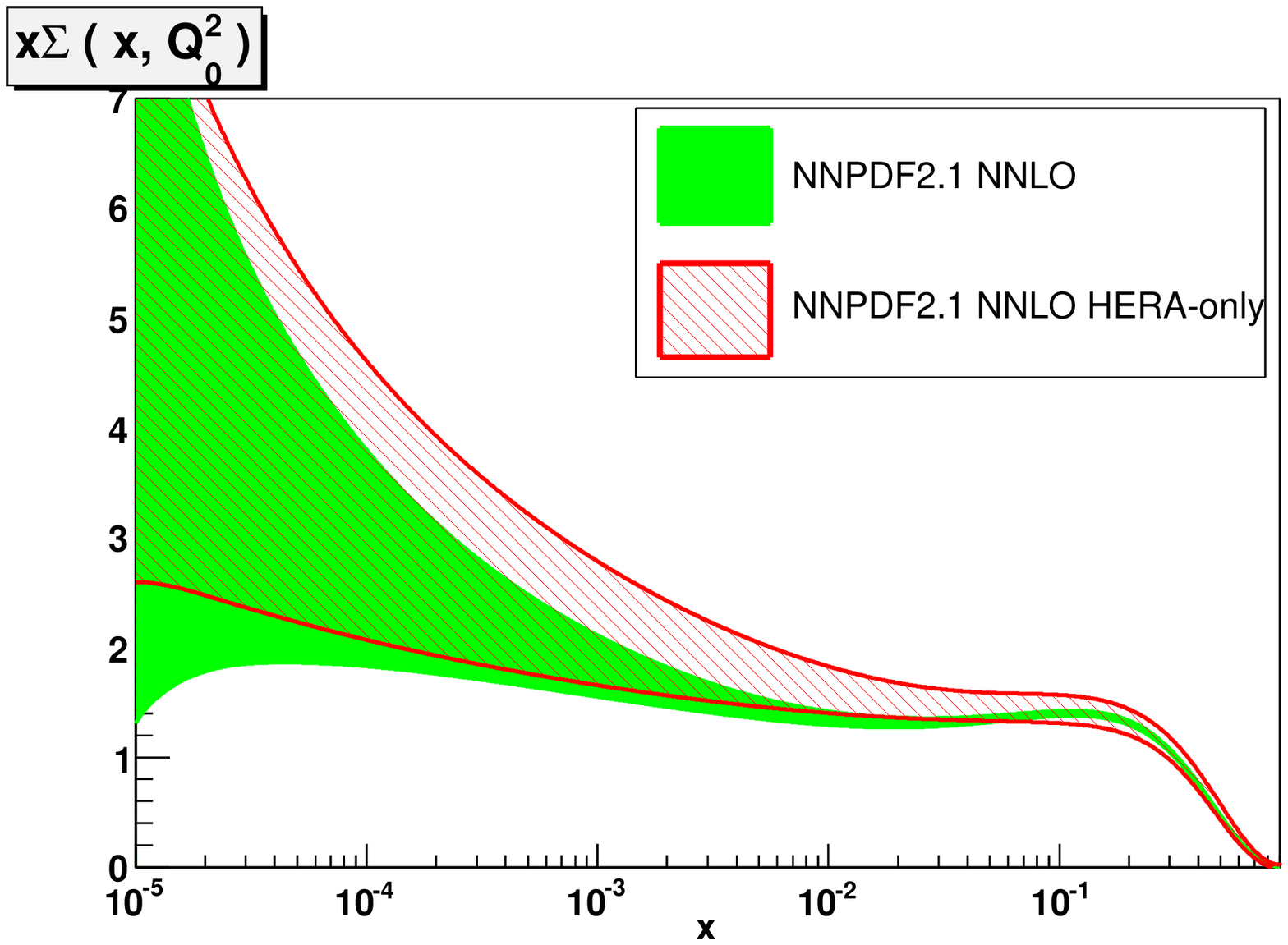}
\epsfig{width=0.32\textwidth,figure=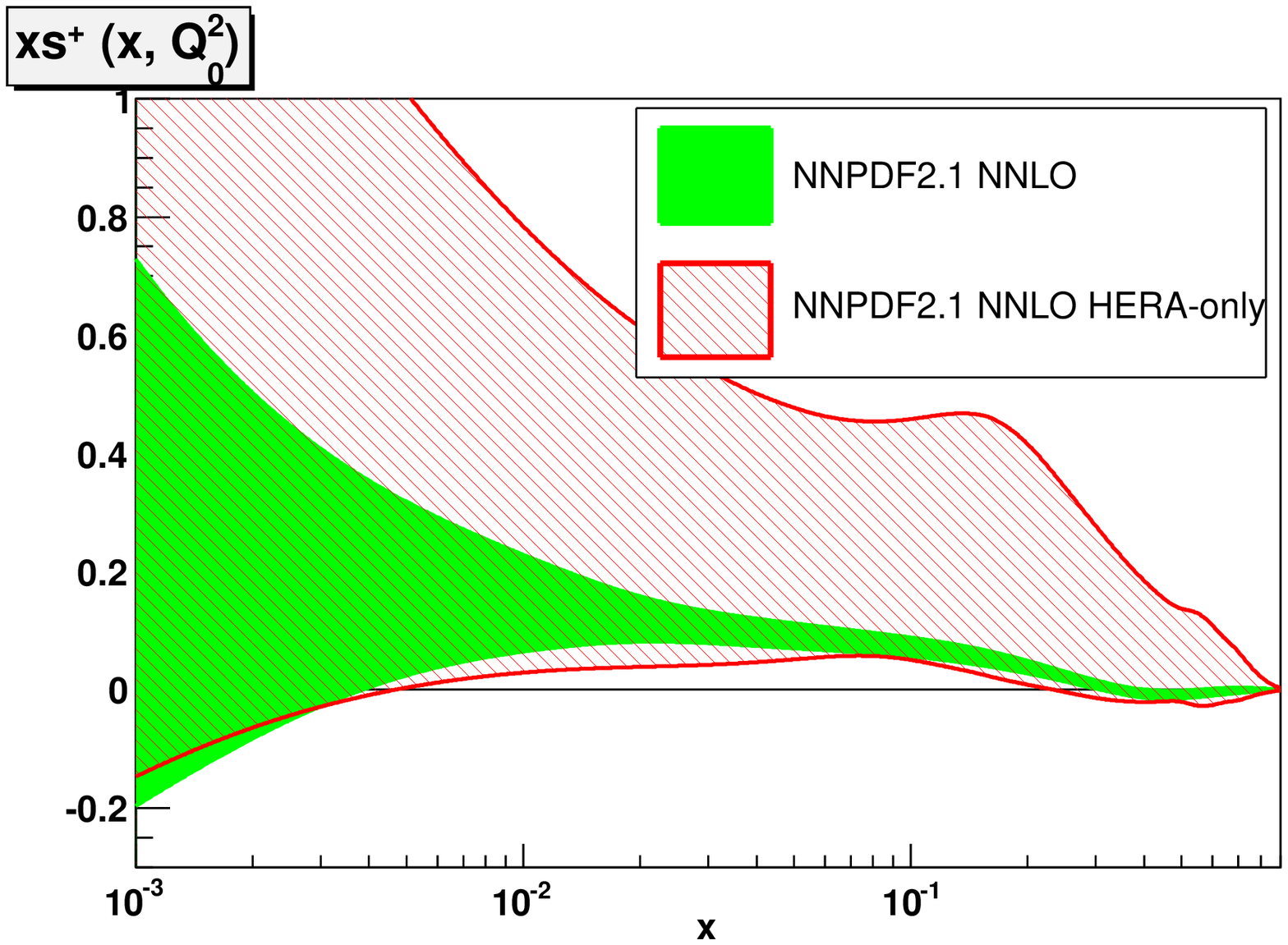}
\epsfig{width=0.32\textwidth,figure=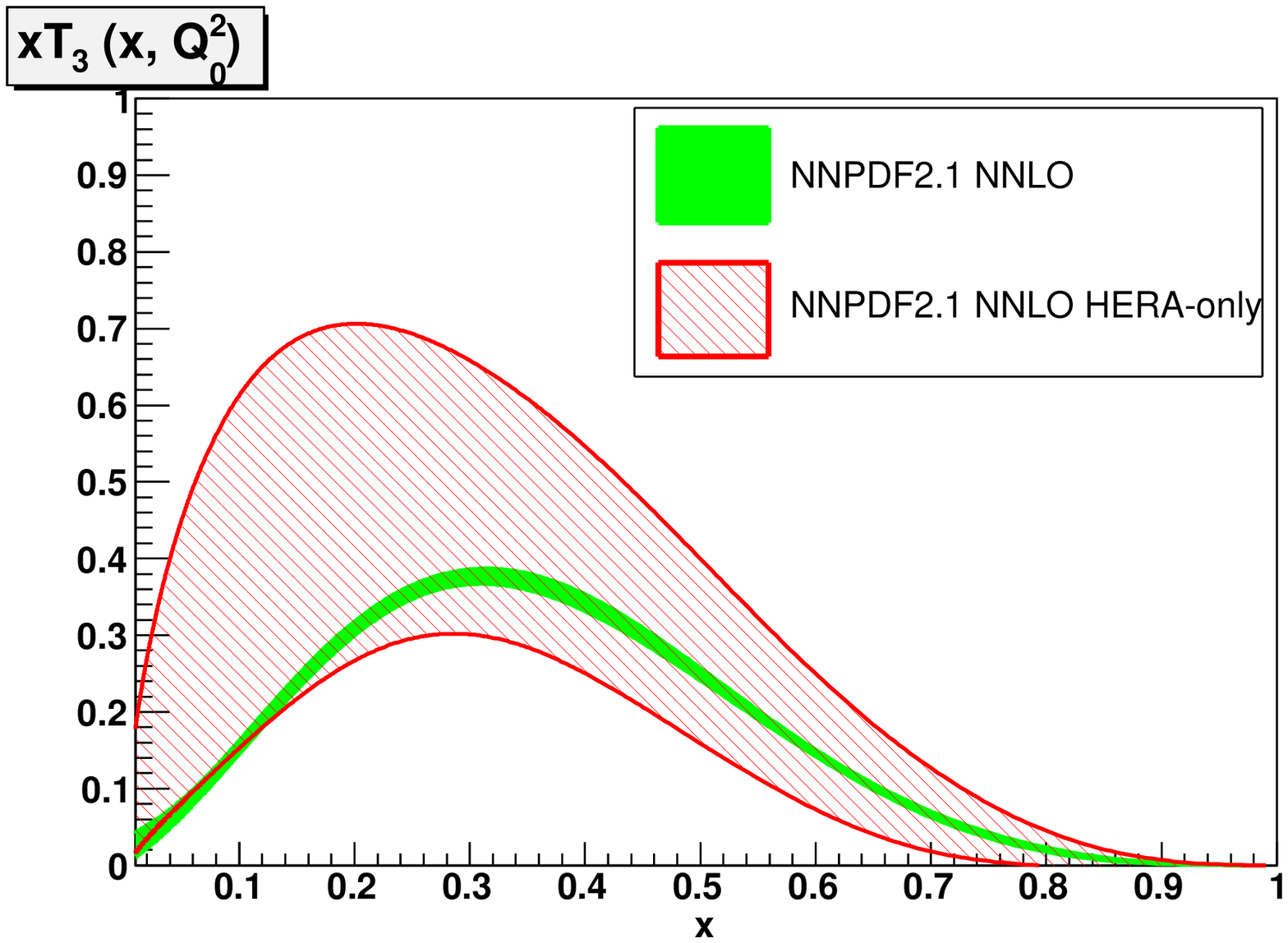}
\caption{\small Comparison of  NNPDF2.1 NNLO  singlet, total
strangeness and
isotriplet PDFs in the global and in the HERA-only fits.
 \label{fig:PDFs-red-global-heraonly}} 
\end{center}
\end{figure}

In particular the fits that we discuss in this section are,
in order of increasing complexity:
\begin{itemize}
\item HERA data only. This determination exploits a
  maximally consistent set of data: the combined HERA-I inclusive
data, the H1 and ZEUS $F_2^c$ data and the ZEUS HERA-II data. PDFs
based on this dataset have also been determined and published by
the HERAPDF group~\cite{H1:2009wt}.
\item Deep-inelastic scattering (DIS) only. This determination does not include hadron-hadron
  data, which one may perhaps consider theoretically or experimentally
  less clean than lepton-hadron data. 
\item Deep-inelastic scattering and Drell-Yan (DIS+DY) 
only. This determination
  is in principle the only truly NNLO one, as it excludes jet data,
  for which only approximate NNLO matrix elements are known. In order
  to better understand the relevance of approximate NNLO jet
  corrections, we will also construct a set which only  differs from
  the default one by setting to zero the approximate NNLO terms in jet matrix elements. 
\item Lepton and hadron collider data only. This determination
  excludes fixed-target data, which are less clean, both
  because of the lower energy and because a sizable fraction of them
  (in particular, all the neutrino DIS data) are obtained using nuclear targets.
  This determination is of greater complexity than DIS+DY, 
  despite having a smaller
  number of datapoints, because it also includes jet data.
\end{itemize}
In each case, a set of $N_{\rm rep}=100$ PDF replicas has been
constructed.
The NLO counterparts of the DIS and DIS+DY PDF determinations were discussed in
Ref.~\cite{Ball:2010de} and are available from LHAPDF both for
NNPDF2.0 and NNPDF2.1; the HERA-only NLO PDFs were briefly discussed
in Ref.~\cite{Lionetti:2011pw}; collider-only PDFs are presented here
for the first time.

The  total $\chi^2$ and that
of the individual experiments for each of these PDF fits are shown in
Table~\ref{tab:redfitschi2}, along with the number of data in each
reduced dataset, and compared to those of the default NNPDF2.1 fit.
In each case, the
$\chi^2$ of  experiments not included in the corresponding
fit is also shown, in square brackets. 

We now turn to a discussion of these fits, and in particular a
comparison between each of them, the default NNPDF2.1, and the fit
with immediately greater complexity, in order to assess the impact of
individual data. These comparisons are based on the computation of
distances
(defined as in Appendix~A of Ref.~\cite{Ball:2010de}) between PDFs in
the two sets which are being compared, bearing in mind that $d\sim1$
means statistical equivalence (the data added have no impact on the
given PDF), while
$1<d\lsim 7$ (for $N_{\rm rep}=100$ replicas) 
means statistical inequivalence but compatibility at the
one sigma level (the data added do have an impact, but only
at the one sigma level). Some of the pairs of
PDFs with the largest distances will also be compared directly.

The HERA-only PDF set is subject to the limitation that charged-current
DIS data are enough to determine at most four independent linear
combinations of quark PDFs (see e.g. Ref.~\cite{Forte:2010dt}), hence
strangeness in this set is entirely uncertain --- in the
HERAPDF~\cite{H1:2009wt} set an independent parametrization is only
provided for the combinations  $d+s$ and $\bar d+\bar s$, but not
separately for strangeness. 
Indeed, for this fit  the quality of the fit to NuTeV
dimuon data is extremely poor. This  set also provides a poor
description of all datasets which are sensitive to the singlet-triplet
separation (such as fixed-target DIS and DY data) to the light sea
decomposition (such as $W$ production data) and, to 
 a lesser
extent, the valence-sea separation (such as neutrino data). The
description of the jet data is also of marginal quality.

\begin{figure}[t]
\begin{center}
\epsfig{width=0.99\textwidth,figure=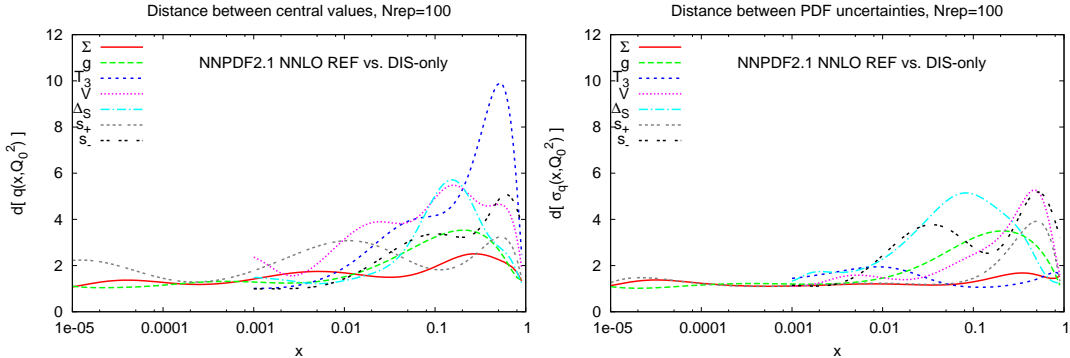}
\caption{\small 
Distances between central values (left) and
  uncertainties (right) for PDFs in the 
DIS-only and default NNPDF2.1 NNLO fits. All distances are computed from
sets of $N_{\rm rep}=100$ replicas.\label{fig:dist_nnlo-dis}}
\end{center}
\end{figure}
\begin{figure}[t]
\begin{center}
\epsfig{width=0.99\textwidth,figure=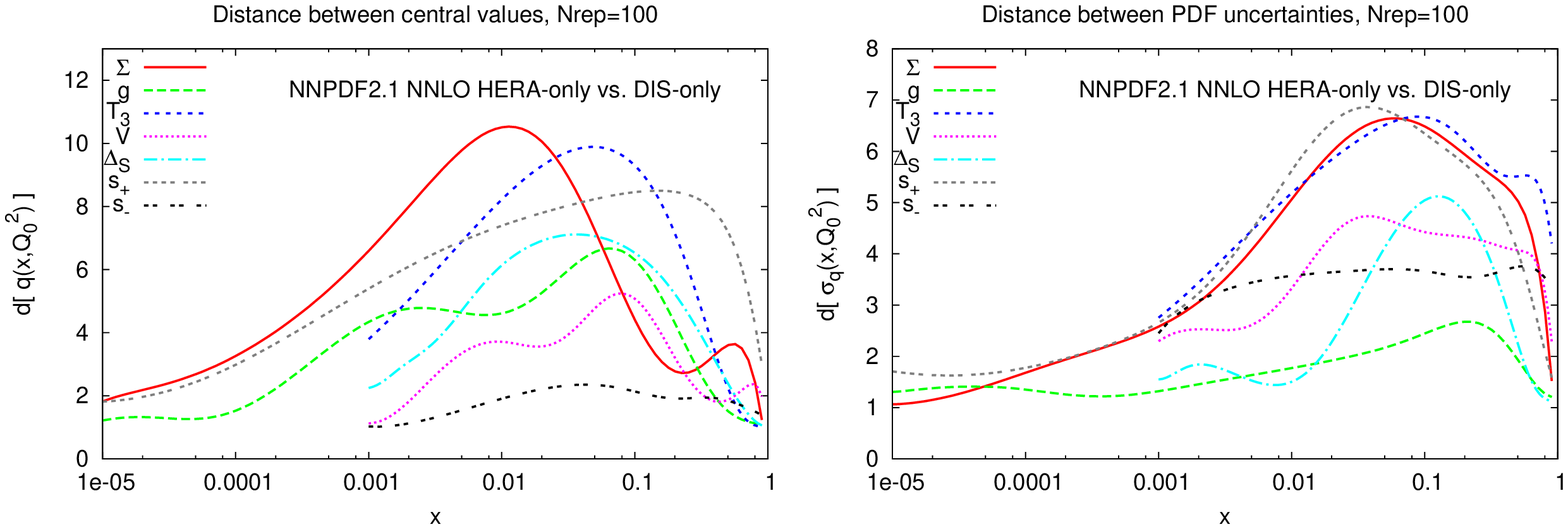}
\caption{\small 
Distances between central values (left) and
  uncertainties (right) for PDFs in the HERA-only and
DIS-only NNPDF2.1 NNLO fits. All distances are computed from
sets of $N_{\rm rep}=100$ replicas. \label{fig:dist_nnlo-heraonly-dis}}
\end{center}
\end{figure}
\begin{figure}[t]
\begin{center}
\epsfig{width=0.32\textwidth,figure=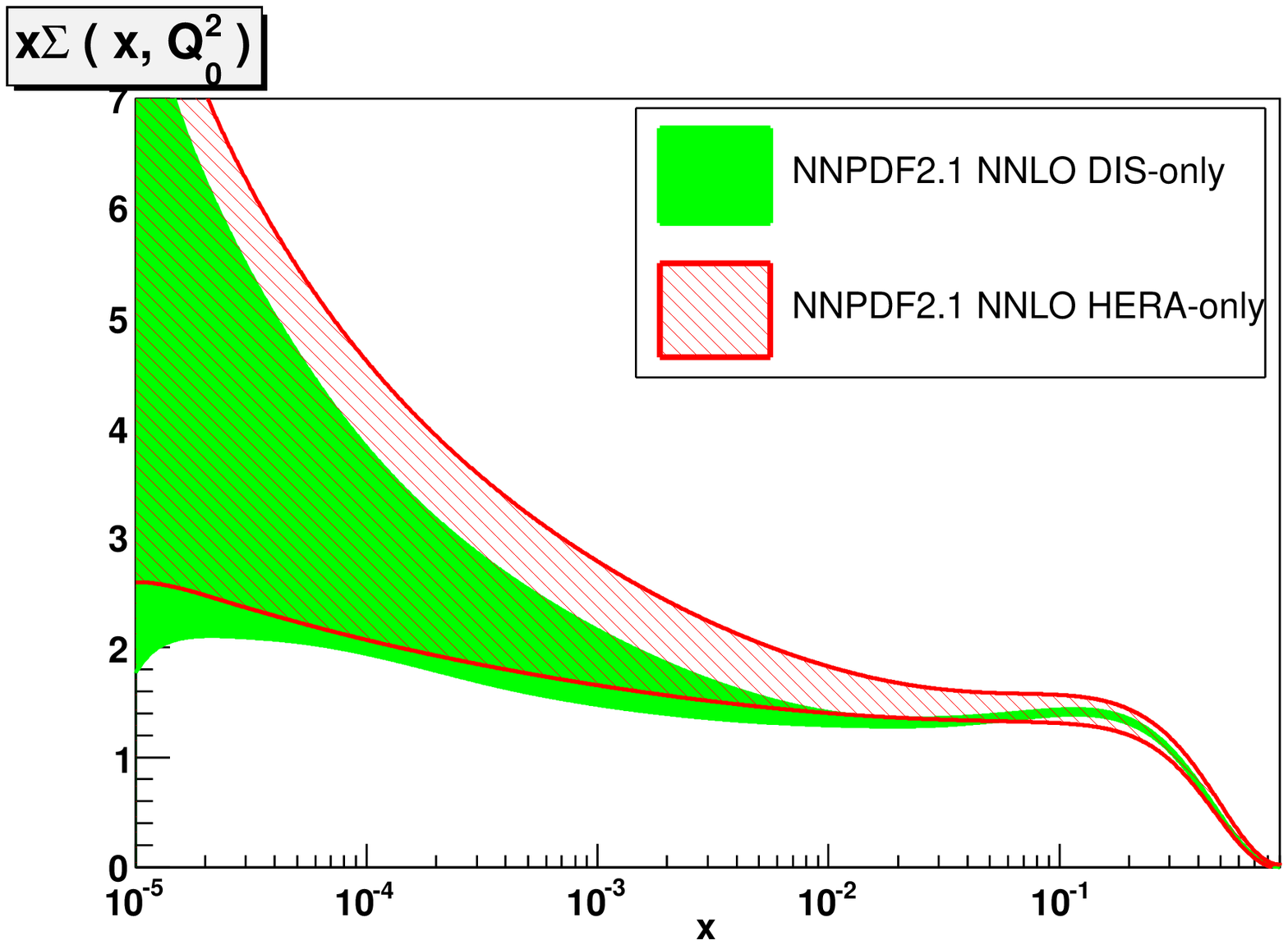}
\epsfig{width=0.32\textwidth,figure=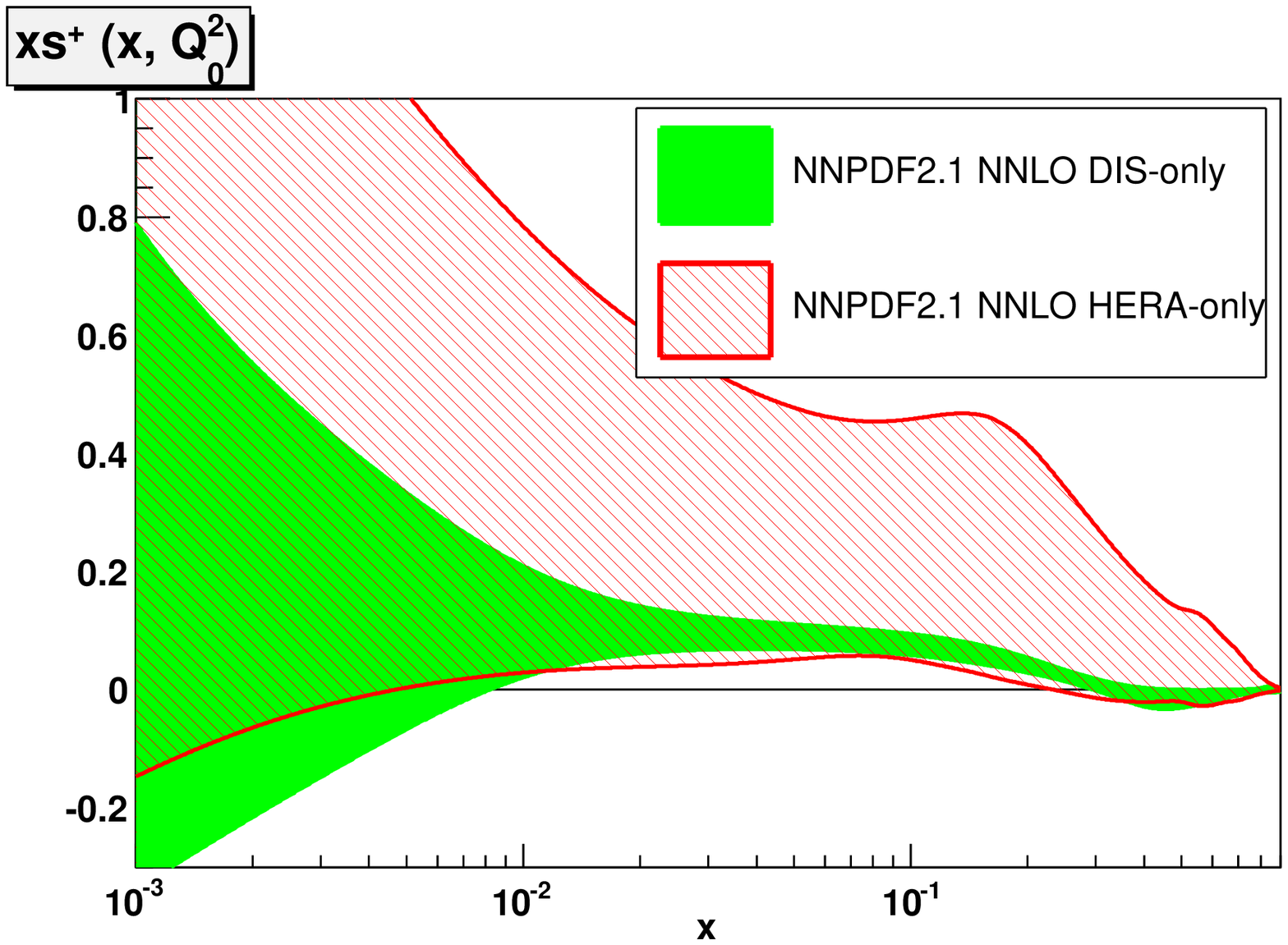}
\epsfig{width=0.32\textwidth,figure=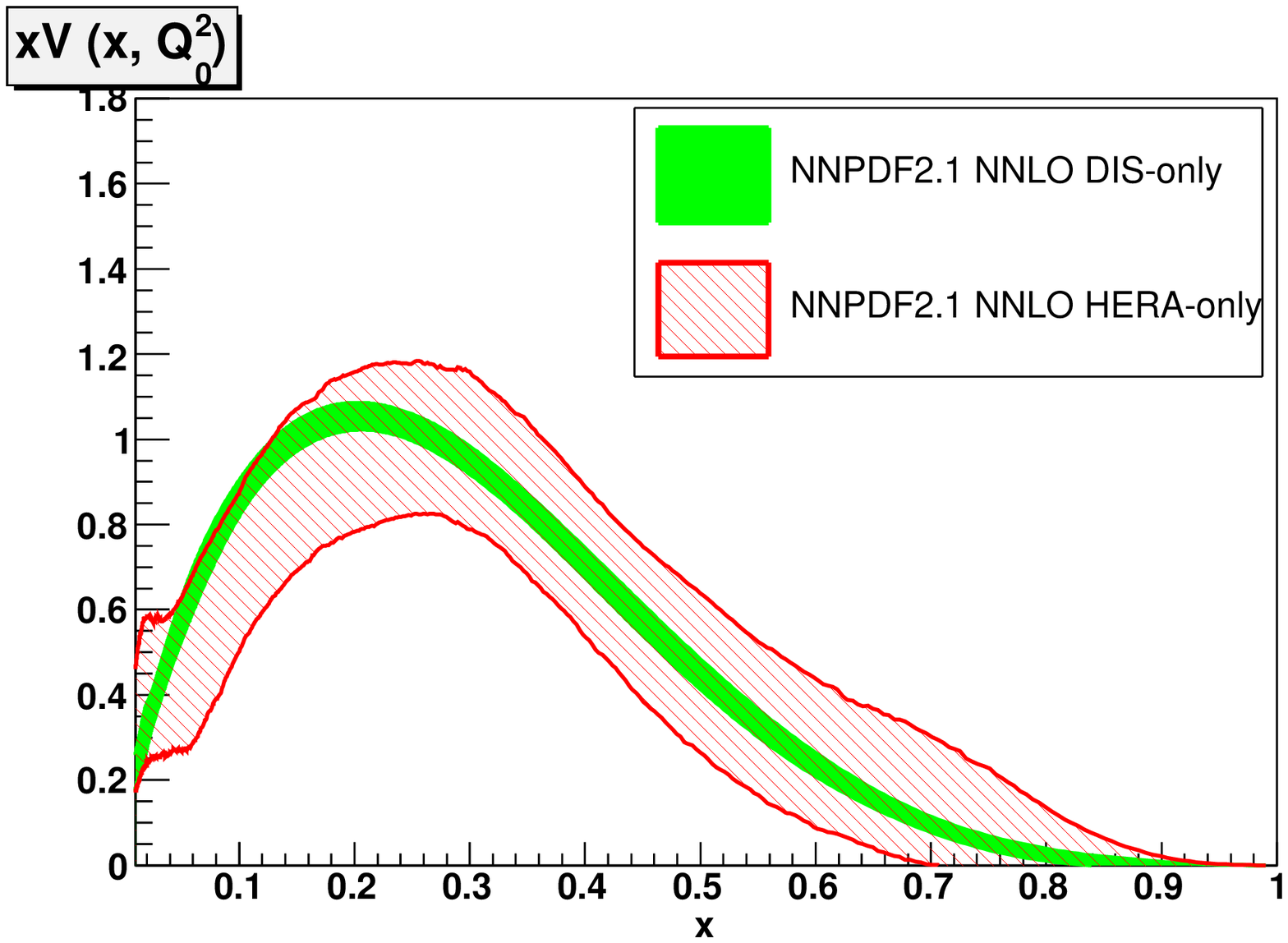}
\caption{\small Comparison of  NNPDF2.1 NNLO  singlet, total
strangeness and
total valence PDFs in the DIS-only and in the HERA-only fits.
 \label{fig:PDFs-red-heraonly-dis}} 
\end{center}
\end{figure}

The distances between
uncertainties (see Fig.~\ref{fig:dist_nnlo-heraonly}) 
for this fit and the default are largest for the
triplet, strangeness, sea asymmetry, and valence. The distance between
central values remains large for triplet and strangeness, but it is
yet larger for the singlet and gluon: the large shifts in strangeness and
triplet are accompanied by a corresponding increase in their uncertainty,
while the increase in the uncertainty of the singlet and gluon is more
moderate, so the change in central value ends up being  statistically more
significant, as can be clearly seen in the direct PDF comparison of
Fig.~\ref{fig:PDFs-red-global-heraonly}. All PDFs remain compatible with those of the global fit
at the level of 90\%\ confidence level, but the deviation of singlet and gluon is
greater than one sigma. Phenomenology based on this PDF set is
necessarily very uncertain in any contribution which depends on
strangeness, as cross-sections 
using HERA-only PDF sets have a theoretical 
uncertainty of order of the size of the
strange contribution, which for instance for $W$ and $Z$ production at
the LHC is of order 15-25\% of the total cross-section~\cite{Bozzi:2011ww}.
This uncertainty  cannot be reduced
regardless of the accuracy of the HERA data that go into the PDF determination.

The most severe problems of the HERA-only fit disappear in the DIS-only
fit, which thanks to the presence of data with neutrino beams or
deuterium targets, achieves
reasonably accurate
flavor decomposition: indeed, distances between strangeness, valence
and light sea asymmetry determined in this fit and those of the global
fit, shown in Fig.~\ref{fig:dist_nnlo-dis}, are rather smaller than
one sigma (though uncertainties are still significantly larger), 
and the singlet and gluon are now in near-perfect
agreement with those of the global fit, with only the gluon showing a
deviation at the half-sigma level in the large $x$ region. The
improvement in accuracy in the flavour decomposition over the
HERA-only fit is clear both in the distances between the uncertainties
in these two fits (see Fig.~\ref{fig:dist_nnlo-heraonly-dis}), and
directly comparing PDFs for which the improvement is most dramatic: singlet,
strangeness, and valence (Fig.~\ref{fig:PDFs-red-heraonly-dis}). 
The triplet distribution also shows a significant decrease in
uncertainty, but around the valence peak 
it only agrees with that of the global fit at the
90\%\ confidence level: this may suggest some tension
between deuterium DIS and hadron collider data (as has been
discussed elsewhere~\cite{Ball:2010gb,Lai:2010vv}), though it could
also be a statistical fluctuation.

As a consequence of all this, the
quality of the agreement of the DIS-only PDFs with all Drell-Yan data
shown in Table~\ref{tab:redfitschi2}
(specifically  $W$ and $Z$ production data) remains poor (at the level of many
units of $\chi^2$), thus showing that a
DIS-only PDF fit is not adequate for precision hadron collider
phenomenology.
On the other hand, it is interesting to observe that the gluon
distribution determined by DIS scaling violations is in good agreement
with that of the global fit, even in the large $x$ region where jet
data have an impact on its uncertainty. Even though the gluon
uncertainty at large $x$ is rather smaller in the global fit, the
central value agrees at the half sigma level, and indeed the
quality of the fit to jet data provided by the DIS-only fit is quite
good, though of course not as good as when these data are also fitted.
It is interesting to observe that other PDF fits based on reduced
datasets (such as HERAPDF or ABKM) 
do not seem~\cite{Thorne:2011kq} to provide an equally good fit to jet
data, presumably because of their less flexible PDF parametrization.

\begin{figure}[t]
\begin{center}
\epsfig{width=0.99\textwidth,figure=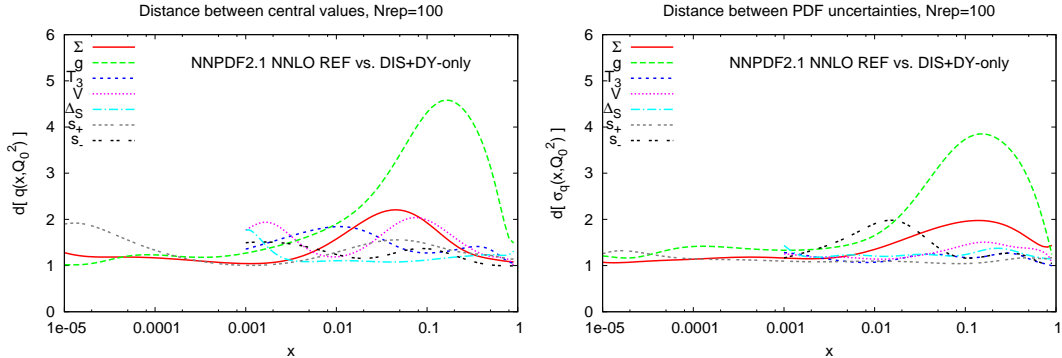}
\caption{\small 
Distances between central values (left) and
  uncertainties (right) for PDFs in the DIS+DY and
default NNPDF2.1 NNLO fits. All distances are computed from
sets of $N_{\rm rep}=100$ replicas. \label{fig:dist_nnlo-dis+dy}}
\end{center}
\end{figure}

\begin{figure}[t]
\begin{center}
\epsfig{width=0.99\textwidth,figure=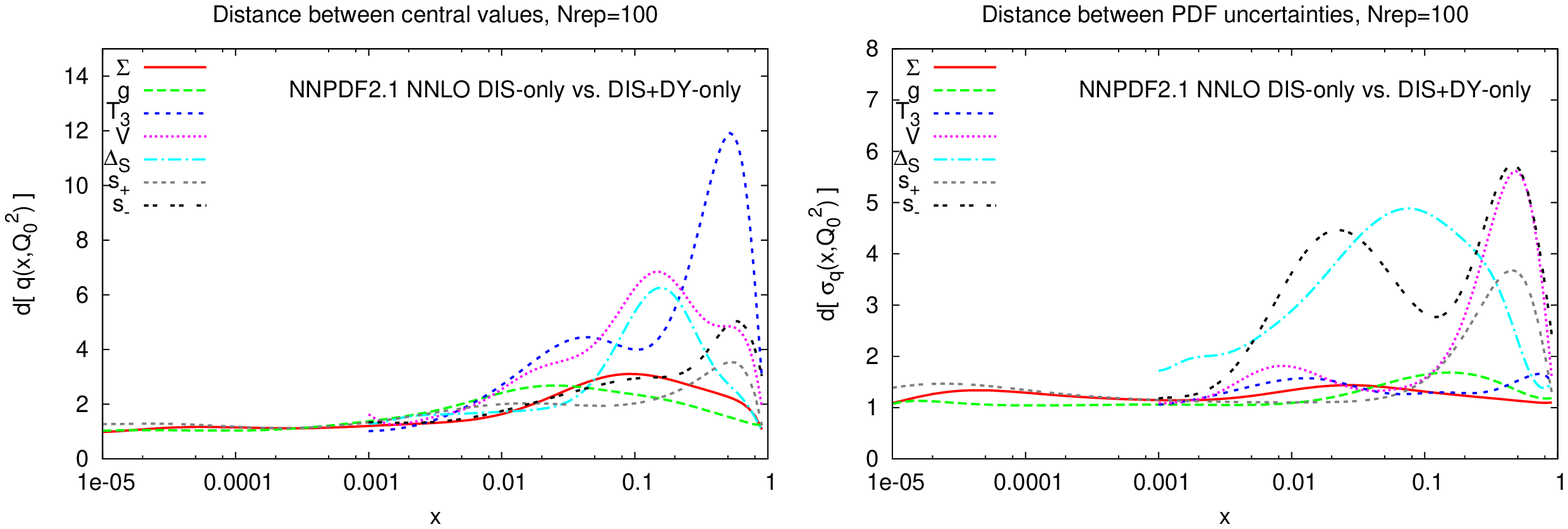}
\caption{\small 
Distances between central values (left) and
  uncertainties (right) for PDFs in the DIS-only and
DIS+DY  NNPDF2.1 NNLO fits. All distances are computed from
sets of $N_{\rm rep}=100$ replicas. \label{fig:dist_nnlo-dis-dis+dy}}
\end{center}
\end{figure}

\begin{figure}[t]
\begin{center}
\epsfig{width=0.32\textwidth,figure=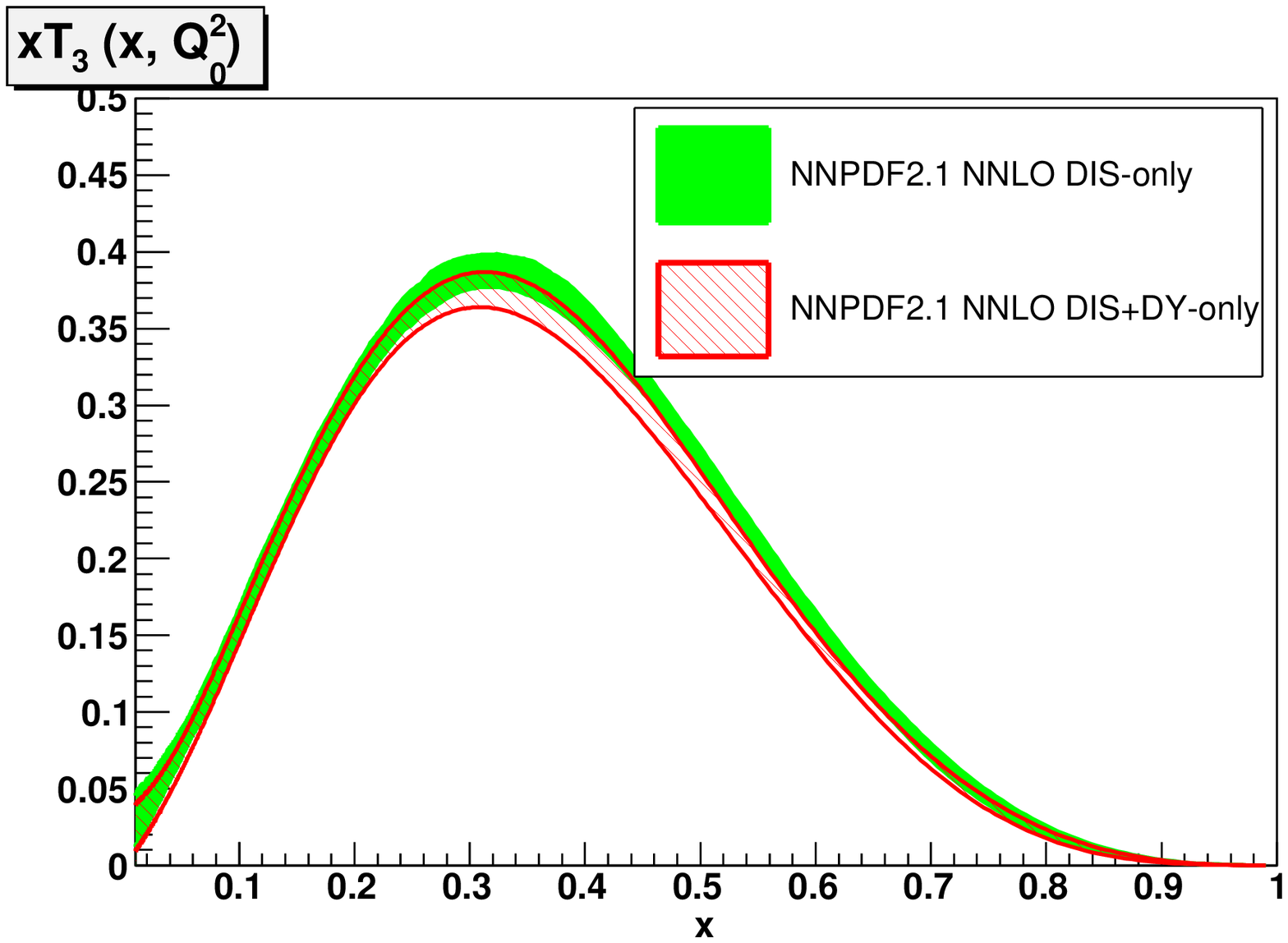}
\epsfig{width=0.32\textwidth,figure=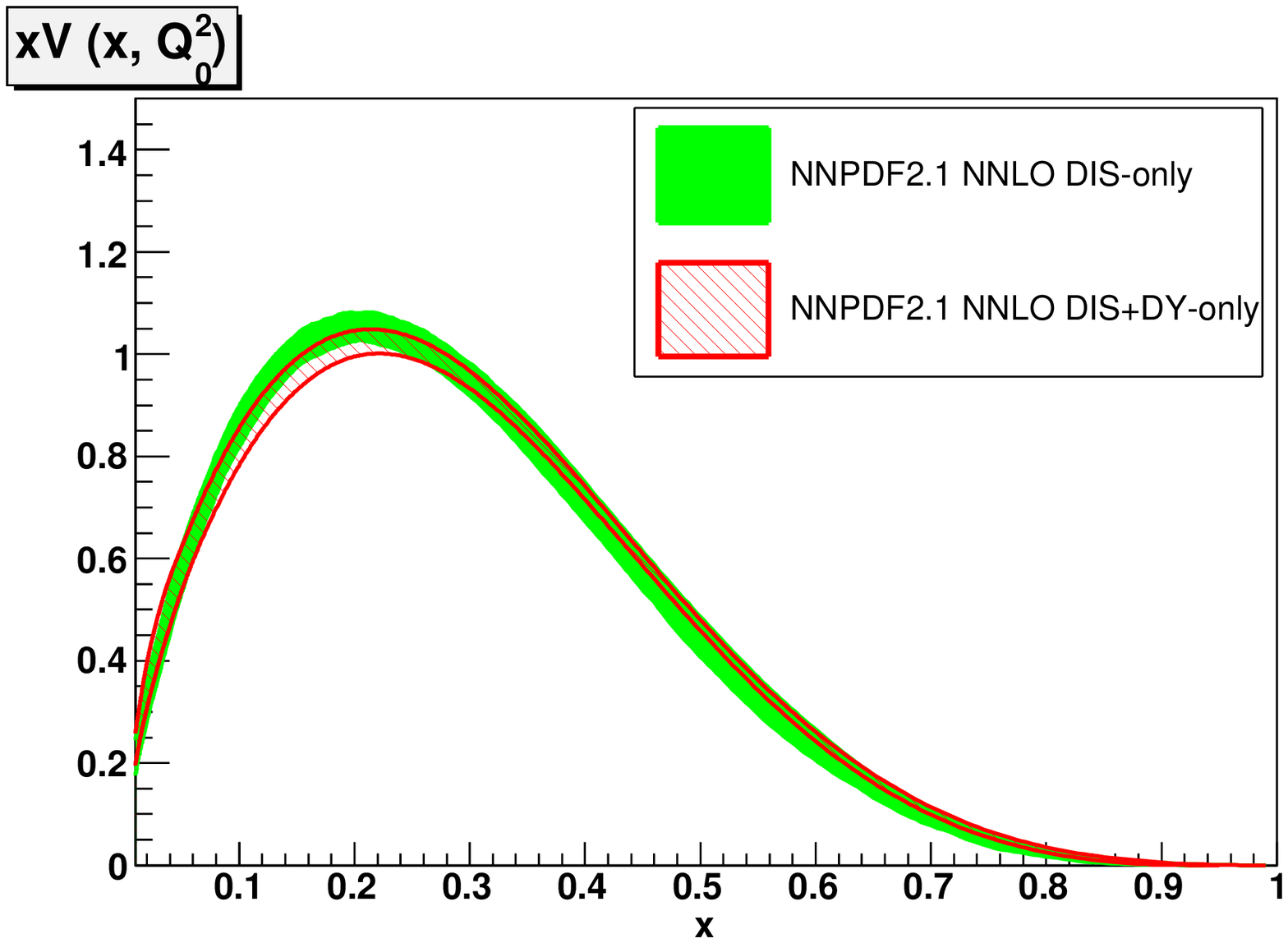}
\epsfig{width=0.32\textwidth,figure=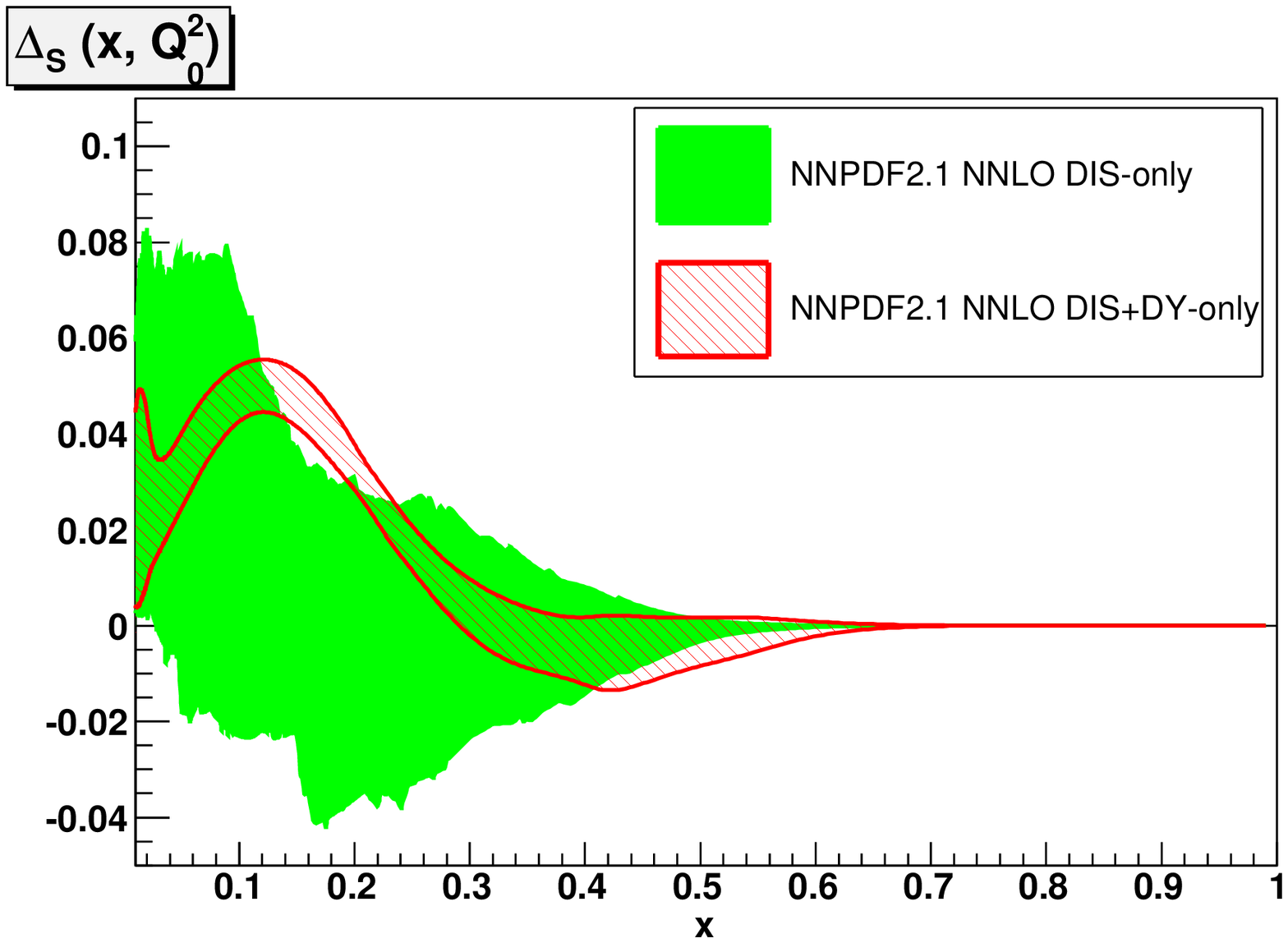}
\caption{\small Comparison of  NNPDF2.1 NNLO  
isotriplet, total valence and sea asymmetry PDFs in the 
DIS-only and in the DIS+DY fits.
 \label{fig:PDFs-red-dis-dis+dy}} 
\end{center}
\end{figure}

Based on this, we expect the DIS+DY fit to be quite close to the
default global fit. This expectation is borne out by the plot of the
distance between these two fits, Fig.~\ref{fig:dist_nnlo-dis+dy}. In
fact, almost all PDFs in this pair of  fits are statistically
equivalent. In comparison to the DIS-only fit 
(Fig.~\ref{fig:dist_nnlo-dis-dis+dy}), the addition of 
the Drell-Yan data has a significant impact on all individual quark
flavours, reducing the uncertainties of the flavour
decomposition, as can also be seen by directly comparing PDFs 
(Fig.~\ref{fig:PDFs-red-dis-dis+dy}). 

\begin{figure}[t]
\begin{center}
\epsfig{width=0.99\textwidth,figure=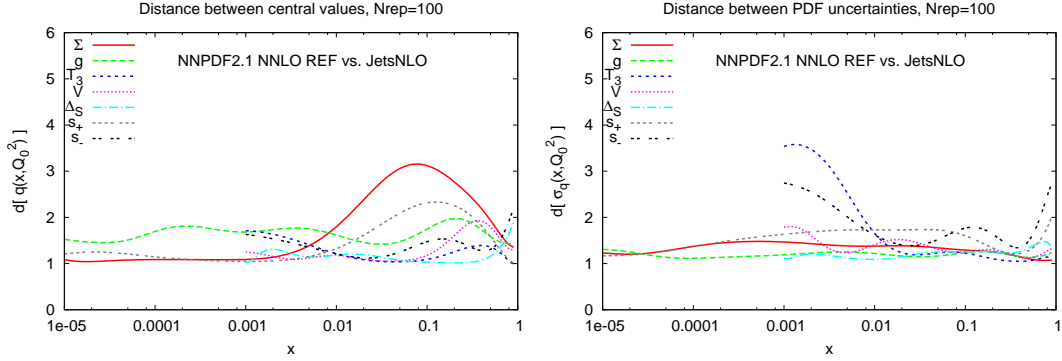}
\caption{\small
Distances between central values (left) and
  uncertainties (right) for PDFs 
in the default
 NNPDF2.1 NNLO fit when approximate NNLO jet matrix elements are
 switched off. All distances are computed from
sets of $N_{\rm rep}=100$ replicas. 
\label{fig:dist_nnlo-jetsnlo}}
\end{center}
\end{figure}

The only significant impact of jet data is
on the gluon at large $x$ whose central value shifts by
about half a sigma with a corresponding reduction in uncertainty when
going from the DIS+DY to the global fit, also leading to a slight
shift of the singlet and valence distributions.

Because of all this, one may wonder whether it might be advantageous to use
the DIS+DY fit as the default global NNLO fit, given that the NNLO jet matrix
elements are only approximate. To address this issue, we have 
produced a NNLO fit in which the jet matrix elements are computed at NLO,
everything else being as in the NNLO global fit. Distances between
this PDF set and the default are displayed in
Fig.~\ref{fig:dist_nnlo-jetsnlo}. These distances are all smaller than
the distances  of Fig.~\ref{fig:dist_nnlo-dis+dy} between the
reference and DIS+DY fits. Hence, the impact of NNLO corrections to
the matrix elements is negligible in comparison to the impact of the
jet data (however moderate this be). If one reasonably assumes that
the difference between exact and approximate NNLO matrix elements is
smaller, or at least not much larger, than the distance between NLO
and approximate NNLO, one must conclude that there is no advantage in
downgrading to the DIS+DY determination. Thus, the fit with jets at
approximate NNLO (i.e. the NNPDF2.1 NNLO default) provides the most
accurate PDF determination.

\begin{figure}[t]
\begin{center}
\epsfig{width=0.99\textwidth,figure=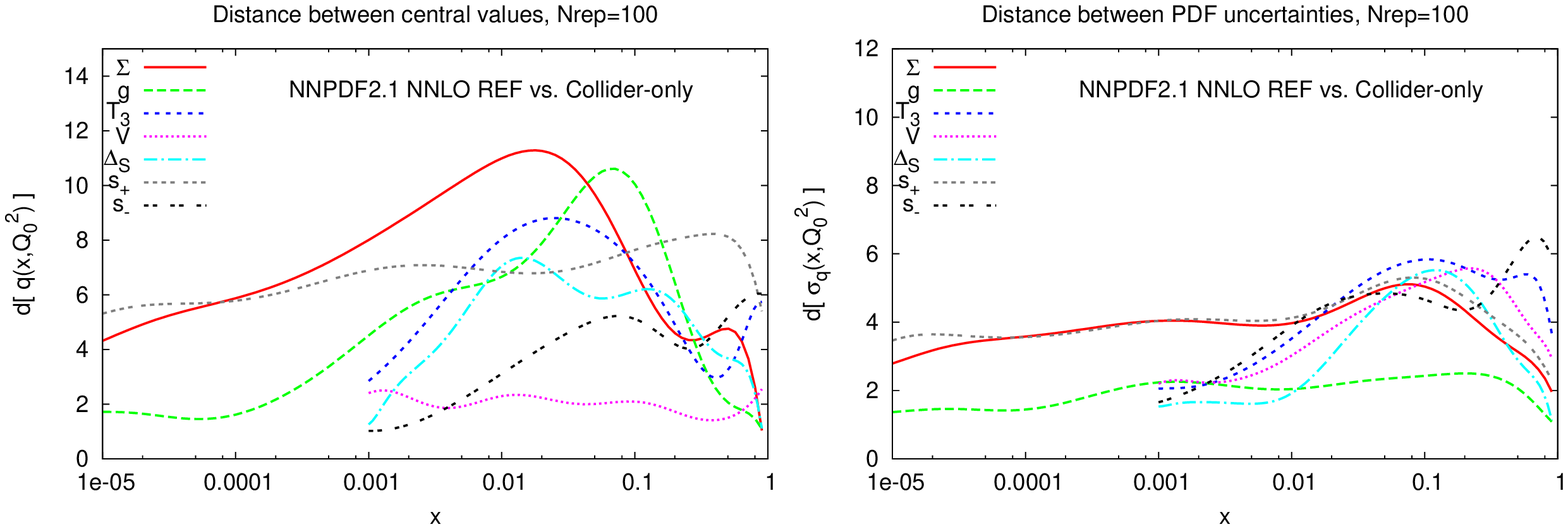}
\caption{\small 
Distances between central values (left) and
  uncertainties (right) for PDFs in the collider-only and
default NNPDF2.1 NNLO fits. All distances are computed from
sets of $N_{\rm rep}=100$ replicas.\label{fig:dist_nnlo-collider}}
\end{center}
\end{figure}

\begin{figure}[t]
\begin{center}
\epsfig{width=0.32\textwidth,figure=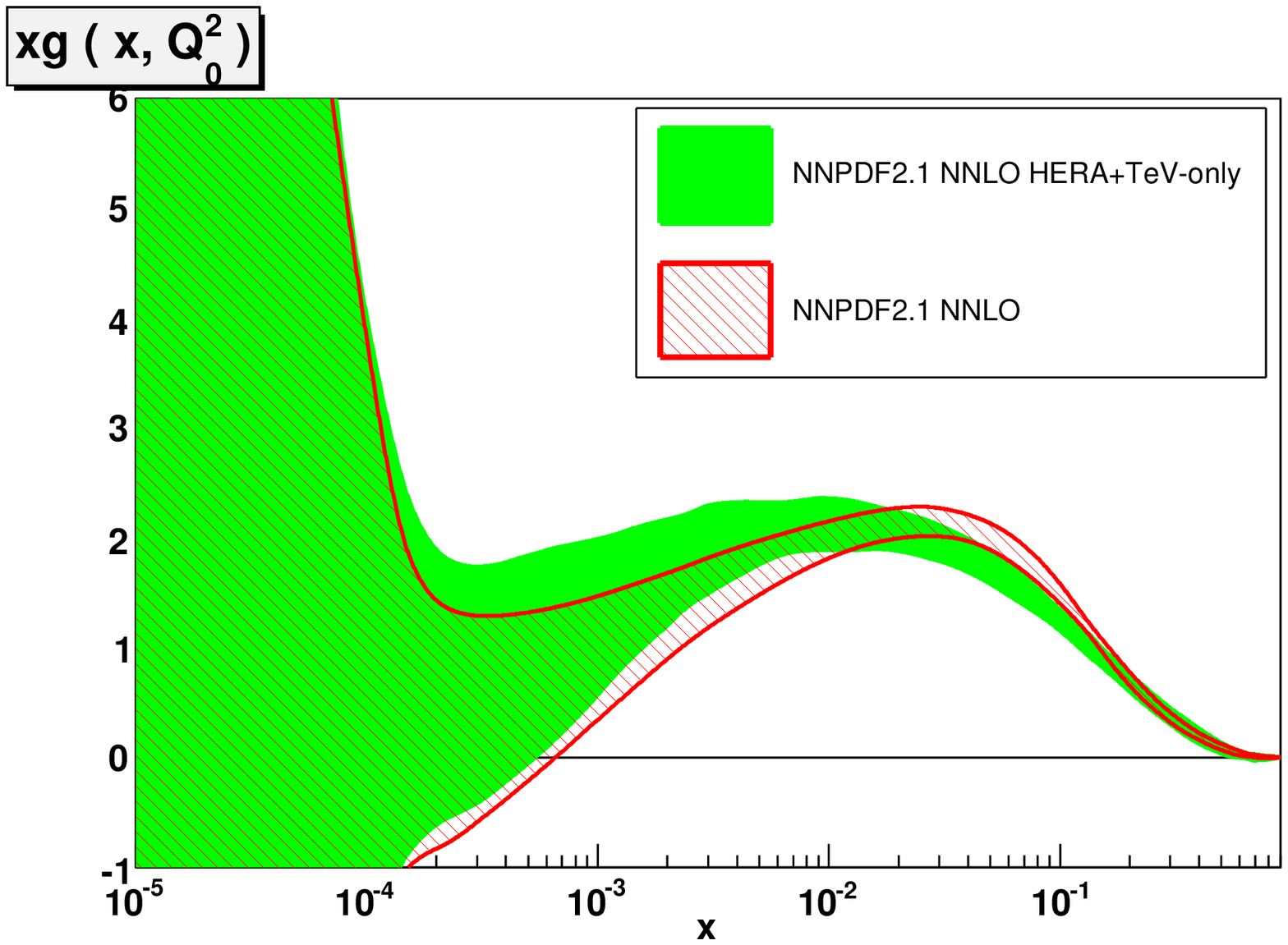}
\epsfig{width=0.32\textwidth,figure=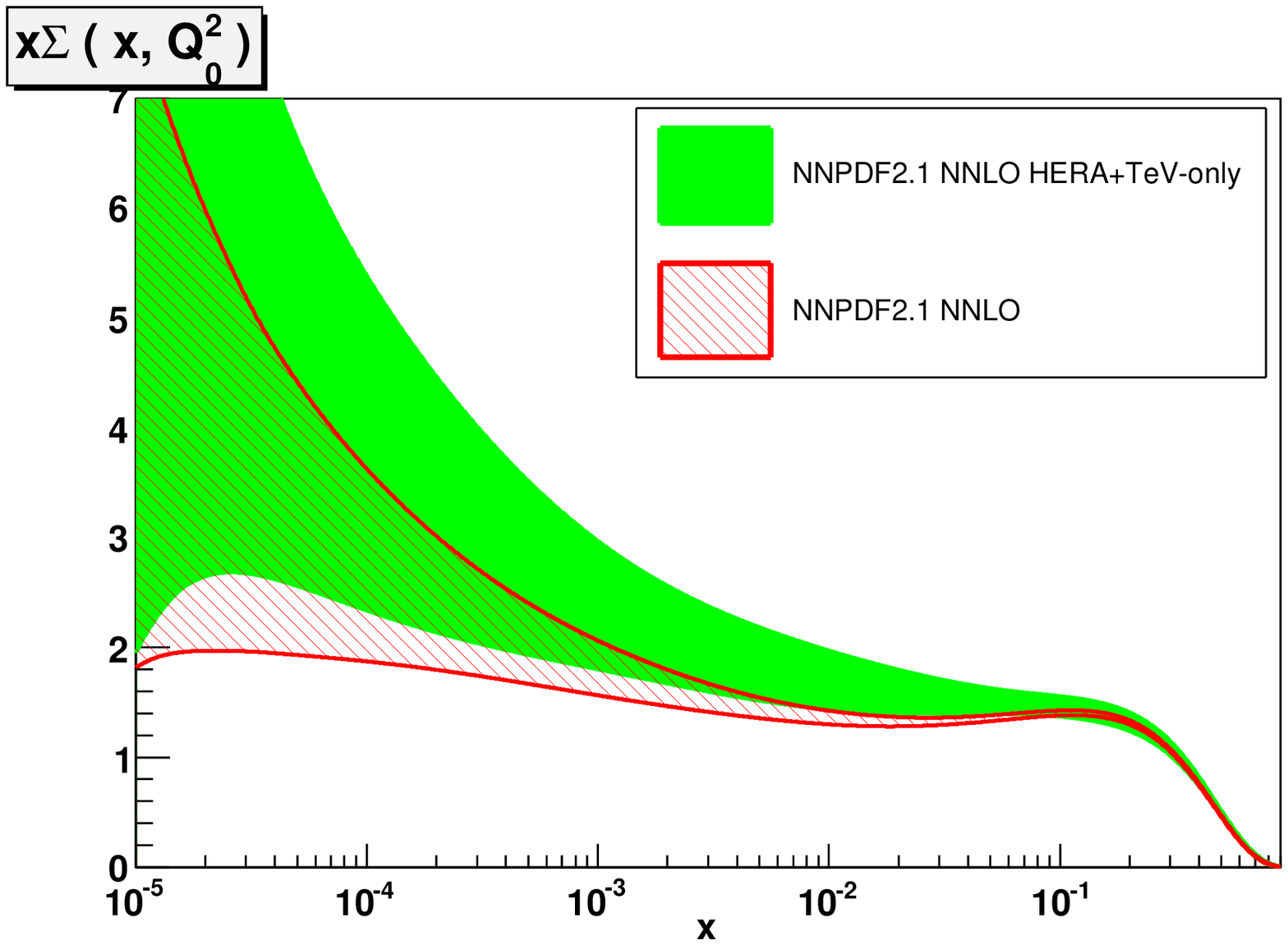}
\epsfig{width=0.32\textwidth,figure=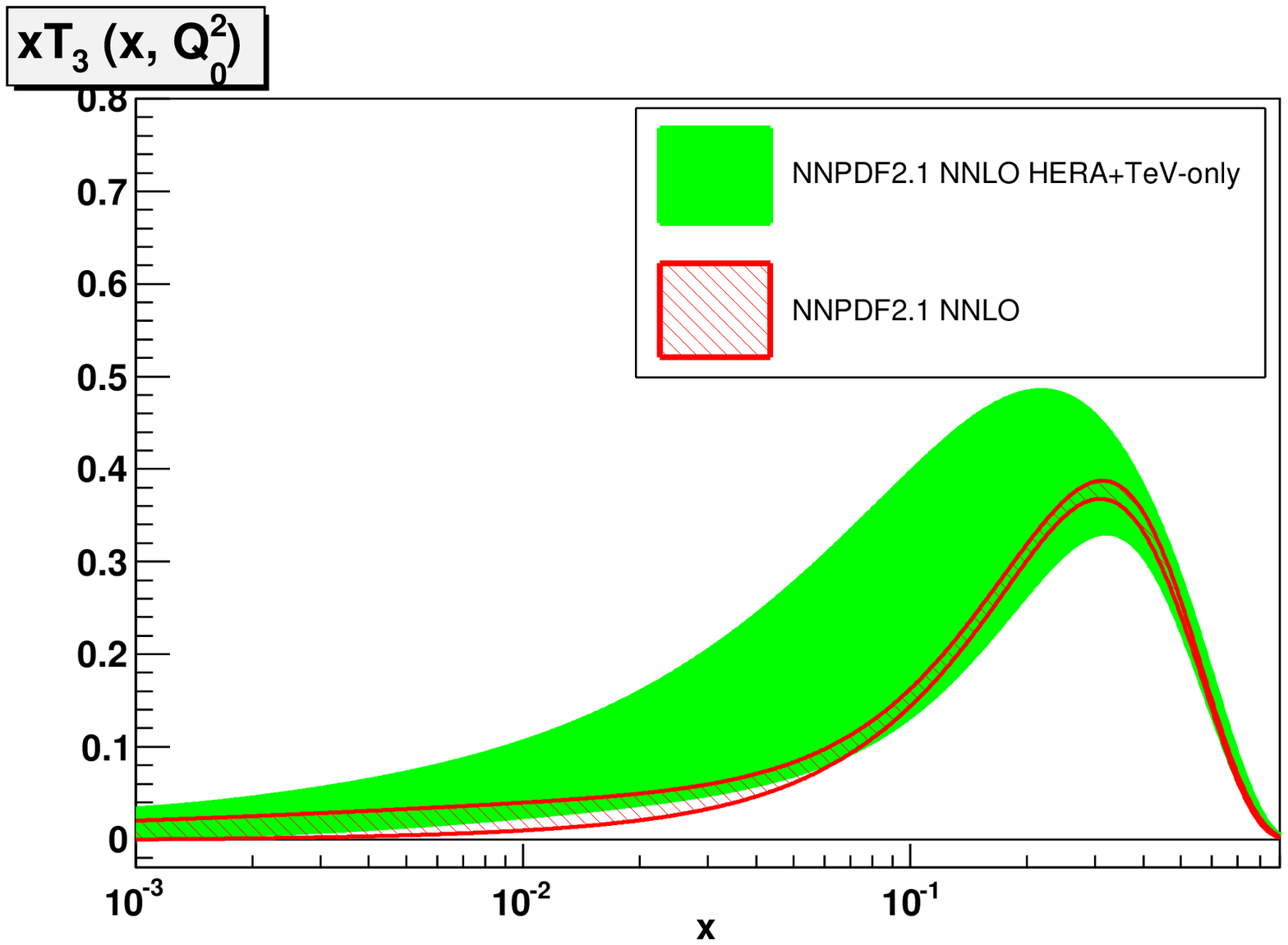}
\caption{\small Comparison of  NNPDF2.1 NNLO  
gluon, singlet and triplet PDFs in the 
 collider-only and reference fits.
 \label{fig:PDFs-red-nnpdf21nnlo-collider}} 
\end{center}
\end{figure}

The last PDF determination that we consider is 
based on a fit to collider data only. To be precise,
we use the same datasets as in the HERA--only fits supplemented by the
Tevatron weak-boson production and inclusive jet
production data. The quality of the global fit in this case, with
$\chi^2 = 1.02$, is significantly better than that of the default global fit:
this supports the idea that a collider-only dataset  
might be more consistent than one which also includes fixed-target data.

However, the distances between the collider-only and global PDFs
(Fig.~\ref{fig:dist_nnlo-collider}) show that they are far from being
equivalent: in fact almost all the PDFs show a shift at the one sigma
level, with uncertainties significantly larger in the collider-only
fit.  Correspondingly, the quality of the fit of the collider-only
PDFs to the fixed target data is generally poor, especially for
deuterium data (such as NMC-pd or DYE866), which control the up-down
separation, and the dimuon data (which control strangeness). These
increased uncertainties end up contaminating also the singlet and
gluon PDFs, as shown by the distance computation and by direct PDF
comparison (Fig.~\ref{fig:PDFs-red-nnpdf21nnlo-collider}).
Hence, we must conclude that a collider-only fit at present does not
provide competitive accuracy for phenomenology due to its extremely poor
determination of strangeness and light quark asymmetry, 
and rather poor determination of valence, triplet and even singlet
PDFs. This situation might improve with forthcoming HERA-II and LHC data, 
so the collider-only fit remains an appealing possibility for the future.

%% file: sec-conclusions.tex

\section{Conclusions}
\label{sec:conclusions}

The family of  NNPDF2.1 PDF determinations --- including the NLO set 
of Ref.~\cite{Ball:2011mu} and the NNLO and LO sets presented here ---
provides arguably the most accurate available PDF determination: it is based on
a wider dataset than ABKM09~\cite{Alekhin:2009ni} or HERAPDF~\cite{H1:2009wt}, unlike
CT10~\cite{Lai:2010vv} it also includes NNLO PDFs, and it is based on
a more up-to-date dataset and treatment of uncertainties than
MSTW08~\cite{Martin:2009iq} (in particular it uses the combined HERA-I
data, and correlated uncertainties for all the experiments which supply them).  
Of course, this is in part due to it being the most recent PDF set, and
updates from the other PDF fitting groups are to be expected in the future. 
However, the NNPDF set is unique in that it is based on a methodology which has
been demonstrated~\cite{phystat,Ball:2010gb} to lead to PDFs and
uncertainties which behave in a statistically consistent way and
minimize  
parametrization bias.

In Ref.~\cite{Forte:2010dt} it was proposed that an ideal parton set
should satisfy a set of general requirements: NNPDF2.1 now satisfies essentially
all of them. Its only shortcoming is that, like all other parton sets to date, it
does not include an estimate of the uncertainty related to the
truncation of the perturbative expansion (as given for instance by
performing renormalization and factorization scale variation). However
this is only a problem at NNLO, since at
LO and NLO the availability of a higher-order set already
provides this information. Furthermore there is every indication that at NNLO
the theoretical uncertainty is small compared to the data uncertainties.

With the most relevant methodological and conceptual issues now
solved, the emphasis now shifts to making an optimal use of the data
which have started to flow so copiously from the LHC experiments.

\bigskip
\bigskip
\begin{center}
\rule{5cm}{.1pt}
\end{center}
\bigskip
 \bigskip

All the NNPDF2.1 NNLO and LO PDF sets that have been discussed in this work
are available from the NNPDF web site,
\begin{center}
{\bf \url{http://sophia.ecm.ub.es/nnpdf}~}
\end{center}
and through the LHAPDF interface~\cite{Bourilkov:2006cj} from v.5.8.6-beta onwards. The older NNPDF2.1 NLO PDF sets (called simply NNPDF2.1 on LHAPDF), 
described in detail in Ref.~\cite{Ball:2011mu}, remain unchanged.

Specifically, the new PDF sets that have been produced in the present analysis and are available in LHAPDF are the following:

\begin{itemize}

\item The reference NNPDF2.1 NNLO sets, sets of $N_{\rm rep}=100$ and 
1000 replicas:\\
{\tt NNPDF21\_nnlo\_100.LHgrid} and {\tt NNPDF21\_nnlo\_1000.LHgrid}.

\item NNPDF2.1 NNLO sets of $N_{\rm rep}=100$ replicas  with $\alpha_s$ varied 
from 0.114 to 0.124 with steps of $\delta\alpha_s=0.001$:\\ 
{\tt NNPDF21\_nnlo\_as\_0114\_100.LHgrid},
$\ldots$,  {\tt NNPDF21\_nnlo\_as\_0124\_100.LHgrid}.

\item NNPDF2.1 NNLO sets of $N_{\rm rep}=100$ replicas  with $m_c$ varied:\\ 
{\tt NNPDF21\_nnlo\_mc\_15\_100.LHgrid},
{\tt NNPDF21\_nnlo\_mc\_16\_100.LHgrid},\\
{\tt NNPDF21\_nnlo\_mc\_17\_100.LHgrid}.

\item NNPDF2.1 NNLO sets of $N_{\rm rep}=100$ replicas  with $m_b$ varied:\\ 
{\tt NNPDF21\_nnlo\_mb\_425\_100.LHgrid},
{\tt NNPDF21\_nnlo\_mb\_45\_100.LHgrid},\\
{\tt NNPDF21\_nnlo\_mb\_50\_100.LHgrid},
{\tt NNPDF21\_nnlo\_mb\_525\_100.LHgrid}.

\item The NNPDF2.1 NNLO PDF sets based on reduced datasets:\\ 
{\tt NNPDF21\_dis\_100.LHgrid}, {\tt NNPDF21\_dis+dy\_100.LHgrid},\\ 
{\tt NNPDF21\_heraonly\_100.LHgrid}, {\tt NNPDF21\_collider\_100.LHgrid}.

\item The reference NNPDF2.1 LO sets, sets of $N_{\rm rep}=100$,
for alphas=0.119 and 0.130:\\
{\tt NNPDF21\_lo\_as\_0119\_100.LHgrid} and {\tt NNPDF21\_lo\_as\_0130\_100.LHgrid}.

\item The NNPDF2.1 LO* sets, without the momentum sum rule
imposed, sets of $N_{\rm rep}=100$,
for alphas=0.119 and 0.130:\\
{\tt NNPDF21\_lostar\_as\_0119\_100.LHgrid} and {\tt NNPDF21\_lostar\_as\_0130\_100.LHgrid}.

\end{itemize}

\bigskip
\bigskip
\begin{center}
\rule{5cm}{.1pt}
\end{center}
\bigskip
\bigskip

{\bf\noindent  Acknowledgments \\}

We are grateful to Lance Dixon for correspondence on the {\tt VRAP}
code, to Frank Petriello  for help with {\tt DYrap} and to 
Marco Bonvini for help with the treatment of fixed-target Drell-Yan
data. We thank Kevin Einsweiler,  Ronan McNulty, Frank--Peter Schilling
and Michael Schmitt
for assistance with LHC data.
We also acknowledge discussions with A.~Accardi, J.~Andersen, P.~Nadolsky, F.~Olness, 
R.~Thorne, A.~Vicini and G.~Watt. We afe grateful to G.~Watt for
pointing out an error in Fig.~\ref{fig_fluxes-mstw} and to G.~Salam
for comments on the preprint version of this paper. 
M.U. is supported by the Bundesministerium f\"ur Bildung and Forschung (BmBF) of the Federal 
Republic of Germany (project code 05H09PAE).
This work was 
partly supported by the Spanish
MEC FIS2007-60350 grant. 
We would like to acknowledge the use of the computing resources provided 
by the Black Forest Grid Initiative in Freiburg and by the Edinburgh Compute 
and Data Facility (ECDF) (http://www.ecdf.ed.ac.uk/). The ECDF is partially 
supported by the eDIKT initiative (http://www.edikt.org.uk).



%% file: sec-massive-nc-mellin.tex
\section{Heavy quark coefficient functions to $\mathcal{O}\lp\alpha_s^2\rp$ in  Mellin space}
\label{sec:massive-nc-comp}

In this Appendix we provide expressions for the Mellin transforms
of the
$\mathcal{O}\lp \alpha_s^2\rp$ massive heavy quark coefficient
functions in the  $Q^2\to\infty$ limit. 
These asymptotic coefficient functions were first computed
long in Ref.~\cite{Buza:1995ie} in $x$--space. Their Mellin transforms
have been given in Refs.~\cite{Blumlein:2006mh,Bierenbaum:2007qe}, and
will be rederived here in a form which is suitable for our purposes.

In order to perform 
the Mellin transform of the $x$--space FFN heavy quark coefficient
functions of Ref.~\cite{Buza:1995ie} it is convenient to rewrite them
in terms of independent Mellin integrals, which we can then tabulate.

Following the notation introduced in Ref.~\cite{Buza:1995ie}, we first
consider the gluon coefficient function for $F_L$. The corresponding
Mellin transform can be written as
\begin{equation}
H_{L,g}^{(2)}\left(N,\frac{Q^2}{m^2},\frac{\mu^2}{m^2}\right) = 4T_f\left[C_{L,g}^{\rm const}(N)+C_{L,g}^{Q}(N)\ln\frac{Q^2}{m^2}-C_{L,g}^{\mu}(N)\ln\frac{\mu^2}{m^2}\right].
\end{equation}
The coefficient function has been separated into three terms:
a $Q^2$--independent term  $C^{\rm const}(N)$,
a collinear log term $C^{Q}(N)$, and a scale variation term
$C^{\mu}(N)$. A similar
decomposition will be performed for all coefficient functions. 

The individual terms are:
\begin{subequations}
\begin{equation}
\begin{array}{rcl}
C_{L,g}^{\rm const} &=C_F&\displaystyle \big[ \smallfrac{16}{15}  A_{10}^{(-2)} - \smallfrac{16}{3}  A_{10}^{(1)} + \smallfrac{32}{5}  A_{10}^{(3)} + 8  A_{8}^{(1)} + 8  A_{6}^{(1)} - \smallfrac{16}{3}  A_{3}^{(1)} - \smallfrac{16}{5}  A_{3}^{(3)} \\
                &   &\displaystyle -  \zeta(2)\big( \smallfrac{16}{3}  A_{1}^{(1)} - \smallfrac{32}{5}  A_{1}^{(3)} \big)  + 4  A_{4}^{(0)} + 12  A_{4}^{(1)} - 16  A_{4}^{(2)} - \smallfrac{16}{15}  A_{2}^{(-1)} \\
                &   &\displaystyle  - \smallfrac{52}{15}  A_{2}^{(0)} - \smallfrac{104}{5}  A_{2}^{(1)} + \smallfrac{48}{5}  A_{2}^{(2)} + \smallfrac{16}{15}  A_{1}^{(-1)} - \smallfrac{64}{15}  A_{1}^{(0)} - \smallfrac{152}{5}  A_{1}^{(1)} + \smallfrac{168}{5}  A_{1}^{(2)} \big]\\
                &+C_A&\displaystyle\big[ 16  A_{10}^{(1)} + 16  A_{10}^{(2)} - 32  A_{8}^{(1)} + 16  \zeta(2)  A_{1}^{(2)} - 48  A_{6}^{(1)} + 16  A_{6}^{(2)}  \\
                &    &\displaystyle  + 8  A_{5}^{(1)} - 8  A_{5}^{(2)} + 24  A_{3}^{(1)} + \smallfrac{8}{3}  A_{4}^{(-1)} - 8  A_{4}^{(0)} - 72  A_{4}^{(1)} + \smallfrac{232}{3}  A_{4}^{(2)} \\
                &    &\displaystyle  + 8  A_{2}^{(0)} + 64  A_{2}^{(1)} - 104  A_{2}^{(2)} - \smallfrac{8}{9}  A_{1}^{(-1)} + \smallfrac{8}{3}  A_{1}^{(0)} + \smallfrac{136}{3}  A_{1}^{(1)}  - \smallfrac{424}{9}  A_{1}^{(2)} \big],
\end{array}
\end{equation}
\begin{equation}
\begin{array}{rl}
C_{L,g}^{Q} & \displaystyle = C_F [ 8  A_{2}^{(1)} + 4  A_{1}^{(0)} + 4  A_{1}^{(1)} - 8  A_{1}^{(2)} ]\\
           & \displaystyle + C_A  \big[ 16  A_{4}^{(1)} - 16  A_{4}^{(2)} - 32  A_{2}^{(1)} + \smallfrac{8}{3}  A_{1}^{(-1)}- 8  A_{1}^{(0)}
                 - 40  A_{1}^{(1)} + \smallfrac{136}{3}  A_{1}^{(2)} \big],
\end{array}
\end{equation}
\begin{equation}
C_{L,g}^{\mu}= C_A  \big[ 16  A_{4}^{(1)} - 16  A_{4}^{(2)} - 32  A_{2}^{(1)} + \smallfrac{8}{3}  A_{1}^{(-1)}- 8  A_{1}^{(0)}
                 - 40  A_{1}^{(1)} + \smallfrac{136}{3}  A_{1}^{(2)} \big],
\end{equation}
\end{subequations}
where the $A_N^{(l)}$ are the independent Mellin integrals 
\begin{equation}
\label{eq:acoeff}
A^{(l)}_n \equiv A_n(N+l).
\end{equation}
Those for which we will use closed-form analytic expressions
are collected in
Tables~\ref{mellintranscc} and~\ref{mellintranscc1}.  Some of these
Mellin transforms were already computed in Ref.~\cite{Blumlein:1998if}
and are repeated here for completeness. The remainder, for which we
will use nymerical approximations, are evaluated at the end of 
this appendix.

The quark coefficient function for $F_L$ can be similarly written as
\begin{equation}
H_{L,q}^{(2)}\left(N,\frac{Q^2}{m^2},\frac{\mu^2}{m^2}\right) =
4T_f\left[C_{L,q}^{\rm
    const}(N)+C_{L,q}^{Q}(N)\ln\frac{Q^2}{m^2}-C_{L,q}^{\mu}(N)\ln\frac{\mu^2}{m^2}\right], 
\end{equation}
where
\begin{subequations}
\begin{equation}
\begin{array}{rcl}
C_{L,q}^{\rm const} &=C_F& \displaystyle\big[ - 8  A_{8}^{(1)} - 8  A_{6}^{(1)} + 8  A_{3}^{(1)}  + \smallfrac{8}{3}  A_{4}^{(-1)} - 8  A_{4}^{(0)}  + \smallfrac{16}{3}  A_{4}^{(2)} + 8  A_{2}^{(0)}\\
              &    &\displaystyle - 8  A_{2}^{(1)} - 16  A_{2}^{(2)} - \smallfrac{8}{9}  A_{1}^{(-1)} + \smallfrac{8}{3}  A_{1}^{(0)} - \smallfrac{32}{3}  A_{1}^{(1)} + \smallfrac{80}{9}  A_{1}^{(2)}\big],
\end{array}
\end{equation}
\begin{equation}
C_{L,q}^{Q}= C_{L,q}^{\mu}= C_F \big[ - 8  A_{2}^{(1)} + \smallfrac{8}{3}  A_{1}^{(-1)} - 8  A_{1}^{(0)} + \smallfrac{16}{3}  A_{1}^{(2)}  \big].
\end{equation}
\end{subequations}

Finally, the gluon radiation coefficient function for $F_L$ can be written as
\begin{equation}
H_{L,GR}^{(2)}\left(N,\frac{Q^2}{m^2},\frac{\mu^2}{m^2}\right) = 4T_f\left[C_{L,GR}^{\rm const}(N)+C_{L,GR}^{Q}(N)\ln\frac{Q^2}{m^2}\right],
\end{equation}
where
\begin{subequations}
\begin{equation}
C_{L,GR}^{\rm const} =C_F\big[ \smallfrac{4}{3} ( A_{4}^{(1)} - 2  A_{2}^{(1)}+  A_{1}^{(0)}) - \smallfrac{50}{9}  A_{1}^{(1)} \big],
\end{equation}
\begin{equation}
C_{L,GR}^{Q}= C_F \smallfrac{4}{3} A_{1}^{(1)}.
\end{equation}
\end{subequations}

Let us now turn to the $F_2$ heavy quark coefficient functions.
In comparison  to the longitudinal structure function,
there are extra pieces $C^{2Q}(N)$ and $C^{\mu Q}(N)$ arising from the double
collinear logarithm.
The gluon coefficient function for $F_2$ can be written as
\begin{equation}
\begin{array}{rl}
\displaystyle H_{2,g}^{(2)}\left(N,\frac{Q^2}{m^2},\frac{\mu^2}{m^2}\right) & \displaystyle = 4T_f\bigg[C_{2,g}^{\rm const}(N)+C_{2,g}^{2Q}(N)\ln^2\frac{Q^2}{m^2}+C_{2,g}^{Q}(N)\ln\frac{Q^2}{m^2}\\
& \displaystyle \qquad\qquad -C_{2,g}^{\mu}(N)\ln\frac{\mu^2}{m^2}
-C_{2,g}^{\mu Q}(N)\ln\frac{\mu^2}{m^2}\ln\frac{Q^2}{m^2}\bigg],
\end{array}
\end{equation}
where
\begin{subequations}
\begin{equation}
\nonumber
\begin{array}{rcl}
C_{2,g}^{\rm const} & = C_F &  \displaystyle \big\{ 4 \big[ - 2  B_{16}^{+} - 2 \zeta(2) B_{17}^{+} + B_{21}^{+} - \smallfrac{3}{2} B_{8}^{+}\big] \\
              &      & \displaystyle + 2\big[ - 4 B_{15}^{-}  - 2 B_{22}^{-}  + 12  B_{19}^{-}  +  B_{18}^{-}   - \smallfrac{5}{2} B_{14}^{-}  + 2 B_{12}^{-}  + B_{13}^{-} \\
              &      & \displaystyle 
 - 4  B_{20}^{-}   + 4  \zeta(2)  B_{2}^{-}  - \smallfrac{1}{3}  B_{11}^{-} + 3  B_{6}^{-}  \big] + 2 \big( 4  A_{15}^{(2)} - 2  A_{22}^{(2)} + A_{18}^{(2)} \\
              &      & \displaystyle - \smallfrac{7}{2}  A_{14}^{(2)}  + 4  A_{12}^{(2)} - A_{13}^{(2)} - 4 \zeta(2)  A_{4}^{(2)} + 4  A_{23}^{(2)} \\
              &      & \displaystyle + 4 \zeta(2)  A_{2}^{(2)} - A_{11}^{(2)} + 11  A_{8}^{(2)} \big) + 32  A_{19}^{(1)} + \zeta(3) ( 28  A_{1}^{(0)} \\
              &      & \displaystyle - 24  A_{1}^{(1)} + 48  A_{1}^{(2)} ) - 28  A_{6}^{(1)} + 36  A_{6}^{(2)} + \smallfrac{4}{15}  A_{10}^{(-2)} + 24  A_{10}^{(0)} \\
              &      & \displaystyle + \smallfrac{32}{3}  A_{10}^{(1)} + \smallfrac{48}{5}  A_{10}^{(3)} + \zeta(2) \big( 12  A_{1}^{(0)} - \smallfrac{52}{3}  A_{1}^{(1)}+ 26  A_{1}^{(2)} + \smallfrac{48}{5}  A_{1}^{(3)} \big)  \\
              &      & \displaystyle  - \smallfrac{11}{2}  A_{5}^{(0)} + 22  A_{5}^{(1)}  - 21  A_{5}^{(2)} - A_{3}^{(0)} + \smallfrac{2}{3}  A_{3}^{(1)} - 13  A_{3}^{(2)} - \smallfrac{24}{5}  A_{3}^{(3)} \\
              &      & \displaystyle + 7  A_{4}^{(0)} - 33  A_{4}^{(1)} + 24  A_{4}^{(2)} - \smallfrac{4}{15}  A_{2}^{(-1)} - \smallfrac{178}{15}  A_{2}^{(0)} + \smallfrac{34}{5}  A_{2}^{(1)} \\
              &      & \displaystyle - \smallfrac{168}{5}  A_{2}^{(2)} + \smallfrac{4}{15}  A_{1}^{(-1)} - \smallfrac{226}{15}  A_{1}^{(0)} + \smallfrac{17}{5}  A_{1}^{(1)} + \smallfrac{82}{5}  A_{1}^{(2)} \big\}\\
              & + C_A &\displaystyle \big\{ 4 \big[ C_{24}  - C_{25} - C_{26}+ \smallfrac{3}{4} C_{21}  + \smallfrac{3}{2} C_{20} \big]  \\
              &      & \displaystyle + 2 \big[ 5  ( A_{15}^{(0)} + 2  A_{15}^{(1)} )+ ( A_{16}^{(0)} + 2  A_{16}^{(1)} ) - 3  ( A_{19}^{(0)} + 2  A_{19}^{(1)} ) + \zeta(2)  ( A_{17}^{(0)} + 2  A_{17}^{(1)} ) \big] \\
              &      & \displaystyle + 8  ( 2  A_{15}^{(1)} + A_{23}^{(1)} ) - 16  A_{22}^{(1)} + 2  A_{21}^{(2)} - 4  A_{20}^{(2)} + 12  A_{14}^{(1)} - 4  A_{14}^{(2)} \\
              &      & \displaystyle - 2  A_{12}^{(0)} - 16  A_{12}^{(1)} + 4  A_{12}^{(2)} + 4  A_{13}^{(0)} + 16  A_{13}^{(1)} - \zeta(2)  ( 10  A_{4}^{(0)} - 12  A_{4}^{(1)} \\
              &      & \displaystyle + 16  A_{4}^{(2)} ) + \smallfrac{4}{3}  A_{11}^{(0)} + 4  A_{11}^{(1)} - \zeta(2)  ( 4  A_{2}^{(0)} + 40  A_{2}^{(1)} - 8  A_{2}^{(2)} ) \\
              &      & \displaystyle - \zeta(3)  ( 3  A_{1}^{(0)} + 14  A_{1}^{(1)} + 2  A_{1}^{(2)} ) - \smallfrac{8}{3}  A_{10}^{(-1)} - 12  A_{10}^{(0)} + 4  A_{10}^{(1)} \\
              &      & \displaystyle + \smallfrac{52}{3}  A_{10}^{(2)} + \smallfrac{16}{3}  A_{8}^{(-1)}  + 5  A_{8}^{(0)} - 16  A_{8}^{(1)} + \smallfrac{20}{3}  A_{8}^{(2)} \\
              &      & \displaystyle - \zeta(2)  \big( 8  A_{1}^{(-1)} + A_{1}^{(0)} + 52  A_{1}^{(1)} - \smallfrac{199}{3}  A_{1}^{(2)} \big) + 4  A_{6}^{(0)} \\
              &      & \displaystyle - 72  A_{6}^{(1)} + 73  A_{6}^{(2)} + \smallfrac{4}{3}  A_{5}^{(-1)} - \smallfrac{3}{2}  A_{5}^{(0)} + 16  A_{5}^{(1)} - \smallfrac{107}{6}  A_{5}^{(2)} \\
              &      & \displaystyle + 46  A_{3}^{(1)} - \smallfrac{57}{2}  A_{3}^{(2)} + \smallfrac{52}{9}  A_{4}^{(-1)} - \smallfrac{28}{3}  A_{4}^{(0)}  - \smallfrac{215}{3}  A_{4}^{(1)} + \smallfrac{749}{9}  A_{4}^{(2)} + \smallfrac{73}{3}  A_{2}^{(0)}  \\
              &      & \displaystyle + 83  A_{2}^{(1)}- \smallfrac{1445}{9}  A_{2}^{(2)} + \smallfrac{20}{9}  A_{1}^{(-1)} + \smallfrac{233}{18}  A_{1}^{(0)} + \smallfrac{65}{9}  A_{1}^{(1)} - \smallfrac{439}{18}  A_{1}^{(2)} \big\},
\end{array}
\end{equation}
\begin{equation}
\begin{array}{rcl}
C_{2,g}^{2Q} &=C_F&\displaystyle \big[ 2  A_{4}^{(0)} - 4  A_{4}^{(1)} + 4  A_{4}^{(2)} - A_{2}^{(0)} + 2  A_{2}^{(1)} - 4  A_{2}^{(2)} - \smallfrac12 A_{1}^{(0)}  + 2  A_{1}^{(1)} \big] \\
              &+C_A&\displaystyle \big[ 2  A_{4}^{(0)} - 4  A_{4}^{(1)} + 4  A_{4}^{(2)} + 2  A_{2}^{(0)} + 8  A_{2}^{(1)}   + \smallfrac{4}{3}  A_{1}^{(-1)}  + A_{1}^{(0)} + 8  A_{1}^{(1)}\\
              &    &\displaystyle  - \smallfrac{31}{3}  A_{1}^{(2)} \big],
\end{array}
\end{equation}
\begin{equation}
\begin{array}{rcl}
C_{2,g}^{Q} & = C_F & \displaystyle [ 2  A_{8}^{(0)} - 4  A_{8}^{(1)} - ( 8  A_{1}^{(0)} - 16  A_{1}^{(1)} + 16  A_{1}^{(2)} )  \zeta(2) - 6  A_{6}^{(0)}  \\
          &       & \displaystyle  + 12  A_{6}^{(1)} - 16  A_{6}^{(2)} + 4  A_{5}^{(0)} - 8  A_{5}^{(1)} + 8  A_{5}^{(2)} + 2  A_{3}^{(0)} - 4  A_{3}^{(1)}  \\
          &       & \displaystyle  + 8  A_{3}^{(2)}  - 7  A_{4}^{(0)} + 24  A_{4}^{(1)} - 20  A_{4}^{(2)} + 2  A_{2}^{(0)} - 12  A_{2}^{(1)} + 20  A_{2}^{(2)} \\
          &       & \displaystyle + 9  A_{1}^{(0)} - 17  A_{1}^{(1)} + 4  A_{1}^{(2)} ] \\
          & + C_A &  \displaystyle \big[ - 4  A_{10}^{(0)} - 8  A_{10}^{(1)} - 8 A_{10}^{(2)} + 4  A_{8}^{(0)} + 16  A_{8}^{(1)} \\ 
          &       & \displaystyle  - ( 4  A_{1}^{(0)} + 8  A_{1}^{(2)} )  \zeta(2) + 24  A_{6}^{(1)} - 8  A_{6}^{(2)} + 2  A_{5}^{(0)} - 4  A_{5}^{(1)} \\
          &       & \displaystyle + 4  A_{5}^{(2)} - 4  A_{3}^{(0)} - 12  A_{3}^{(1)} + \smallfrac{8}{3}  A_{4}^{(-1)} - 2  A_{4}^{(0)} + 40  A_{4}^{(1)} - \smallfrac{134}{3}  A_{4}^{(2)}\\
          &       & \displaystyle  - 48  A_{2}^{(1)} + 50  A_{2}^{(2)}  + \smallfrac{52}{9}  A_{1}^{(-1)} - \smallfrac{55}{3}  A_{1}^{(0)}   - \smallfrac{92}{3}  A_{1}^{(1)} + \smallfrac{407}{9}  A_{1}^{(2)} \big],
\end{array}
\end{equation}
\begin{equation}
\begin{array}{rcl}
C_{2,g}^{\mu} & = C_A & \displaystyle \big\{4  A_{8}^{(0)} + 16  A_{8}^{(1)} - ( 4  A_{1}^{(0)} - 8  A_{1}^{(1)} + 8  A_{1}^{(2)} )  \zeta(2) + 24  A_{6}^{(1)}  \\
            &       & \displaystyle  - 8  A_{6}^{(2)} + 4  A_{5}^{(0)} - 8  A_{5}^{(1)} + 8  A_{5}^{(2)} - 2  A_{3}^{(0)} - 8  A_{3}^{(1)} + \smallfrac{8}{3}  A_{4}^{(-1)} \\
            &       & \displaystyle  - 2  A_{4}^{(0)}+ 48  A_{4}^{(1)} - \smallfrac{158}{3}  A_{4}^{(2)} - 2  A_{2}^{(0)} - 64  A_{2}^{(1)} \\
            &       & \displaystyle + \smallfrac{62}{3}  A_{2}^{(2)} + \smallfrac{4}{3}  A_{1}^{(-1)} - \smallfrac{43}{3}  A_{1}^{(0)} - \smallfrac{242}{3}  A_{1}^{(1)} + \smallfrac{281}{3}  A_{1}^{(2)} \big\},
\end{array}
\end{equation}
\begin{equation}
\begin{array}{rcl}
C_{2,g}^{\mu Q} & = C_A & \displaystyle\big[ 4  A_{4}^{(0)} - 8  A_{4}^{(1)} + 8  A_{4}^{(2)} + 4  A_{2}^{(0)} + 16  A_{2}^{(1)} + \smallfrac{8}{3}  A_{1}^{(-1)} + 2  A_{1}^{(0)} \\
            &       & \displaystyle + 16  A_{1}^{(1)} - \smallfrac{62}{3}  A_{1}^{(2)} \big],
\end{array}
\end{equation}
\end{subequations}
where
\begin{equation}
\label{eq:bcoeff}
B^{\pm}_n \equiv A_n^{(0)}\pm 2A_n^{(1)}+A_n^{(2)},\qquad C_n \equiv A_n^{(0)}+2A_n^{(1)}+2A_n^{(2)}.
\end{equation}

The quark coefficient function for $F_2$ can be written similarly as
\begin{equation}
\begin{array}{rl}
\displaystyle H_{2,q}^{(2)}\left(N,\frac{Q^2}{m^2},\frac{\mu^2}{m^2}\right) &\displaystyle = 4T_f\bigg[C_{2,q}^{\rm const}(N)+C_{2,q}^{2Q}(N)\ln^2\frac{Q^2}{m^2}+C_{2,q}^{Q}(N)\ln\frac{Q^2}{m^2}\\
&\displaystyle \qquad - C_{2,q}^{\mu}(N)\ln\frac{\mu^2}{m^2}
- C_{2,q}^{\mu Q}(N)\ln\frac{\mu^2}{m^2}\ln\frac{Q^2}{m^2}\bigg],
\end{array}
\end{equation}
where
\begin{subequations}
\begin{equation}
\begin{array}{rcl}
C_{2,q}^{\rm const} & = C_F & \displaystyle \big[ 8  ( A_{15}^{(0)} + A_{15}^{(1)} ) + 4  ( A_{13}^{(0)} + A_{13}^{(1)} ) + 2  ( A_{14}^{(0)} + A_{14}^{(1)} ) - 4  ( A_{12}^{(0)} + A_{12}^{(1)} )\\
             &       & \displaystyle  - 8  \zeta(2) ( A_{2}^{(0)} + A_{2}^{(1)} ) + \smallfrac{4}{3}  ( A_{11}^{(0)} + A_{11}^{(1)} ) - \smallfrac{8}{3}  A_{10}^{(-1)} - 8  A_{10}^{(0)} - 8  A_{10}^{(1)}\\
             &       & \displaystyle  - \smallfrac{8}{3}  A_{10}^{(2)} + \smallfrac{16} {3} A_{8}^{(-1)} + 4  A_{8}^{(0)} - 4  A_{8}^{(1)} + \smallfrac{8}{3}  A_{8}^{(2)} - \zeta(2) \big( 8  A_{1}^{(-1)} + 4  A_{1}^{(0)} \\
             &       & \displaystyle  + 4  A_{1}^{(1)} - \smallfrac{16}{3}  A_{1}^{(2)} \big) + 8  A_{6}^{(2)} + \smallfrac{4}{3}  A_{5}^{(-1)} + A_{5}^{(0)} - A_{5}^{(1)} - \smallfrac{4}{3}  A_{5}^{(2)} \\
             &      & \displaystyle + 10  A_{3}^{(1)} - 4  A_{3}^{(2)} + \smallfrac{52}{9}  A_{4}^{(-1)} - \smallfrac{52}{3}  A_{4}^{(0)} + \smallfrac{40}{3}  A_{4}^{(1)}  - \smallfrac{16}{9}  A_{4}^{(2)} + \smallfrac{70}{3}  A_{2}^{(0)} \\
             &      & \displaystyle - 22  A_{2}^{(1)} - \smallfrac{176}{9}  A_{2}^{(2)} + \smallfrac{20}{9}  A_{1}^{(-1)} + \smallfrac{76}{9}  A_{1}^{(0)} - \smallfrac{304}{9}  A_{1}^{(1)} + \smallfrac{208}{9}  A_{1}^{(2)} \big],
\end{array}
\end{equation}
\begin{equation}
\begin{array}{rcl}
C_{2,q}^{2Q} & = C_F & \displaystyle \big[ 2  A_{2}^{(0)} + 2  A_{2}^{(1)} + \smallfrac{4}{3}  A_{1}^{(-1)} + A_{1}^{(0)} - A_{1}^{(1)} - \smallfrac{4}{3}  A_{1}^{(2)} \big],\qquad\qquad
\end{array}
\end{equation}
\begin{equation}
\begin{array}{rcl}
C_{2,q}^{Q} & = C_F & \displaystyle \big[ 4  A_{8}^{(0)} + 4  A_{8}^{(1)} + 4  A_{6}^{(0)} + 4  A_{6}^{(1)} - 4  A_{3}^{(0)} - 4  A_{3}^{(1)} + \smallfrac{8}{3}  A_{4}^{(-1)} \\
          &       & \displaystyle + 2  A_{4}^{(0)} - 2  A_{4}^{(1)} - \smallfrac{8}{3}  A_{4}^{(2)} + 8  A_{2}^{(2)} + \smallfrac{52}{9}  A_{1}^{(-1)} - \smallfrac{52}{3}  A_{1}^{(0)} \\
          &       & \displaystyle + \smallfrac{40}{3}  A_{1}^{(1)} - \smallfrac{16}{9}  A_{1}^{(2)} \big],
\end{array}
\end{equation}
\begin{equation}
\begin{array}{rcl}
C_{2,q}^{\mu} & = C_F & \displaystyle \big\{4  A_{8}^{(0)} + 4  A_{8}^{(1)} + 4  A_{6}^{(0)} + 4  A_{6}^{(1)} - 2  A_{3}^{(0)} - 2  A_{3}^{(1)} + \smallfrac{8}{3}  A_{4}^{(-1)} \\
            &       & \displaystyle  + 2  A_{4}^{(0)} - 2  A_{4}^{(1)} - \smallfrac{8}{3}  A_{4}^{(2)} - 2  A_{2}^{(0)} - 10  A_{2}^{(1)} + \smallfrac{8}{3}  A_{2}^{(2)} + \smallfrac{4}{3}  A_{1}^{(-1)} \\
            &       & \displaystyle - \smallfrac{40}{3}  A_{1}^{(0)} + \smallfrac{4}{3}  A_{1}^{(1)} + \smallfrac{32}{3}  A_{1}^{(2)} \big\},
\end{array}
\end{equation}
\begin{equation}
\begin{array}{rcl}
C_{2,q}^{\mu Q} & = C_F & \displaystyle \big[ 4  A_{2}^{(0)} + 4  A_{2}^{(1)} + \smallfrac{8}{3}  A_{1}^{(-1)} + 2  A_{1}^{(0)} - 2  A_{1}^{(1)} - \smallfrac{8}{3}  A_{1}^{(2)} \big].
\end{array}
\end{equation}\end{subequations}

The gluon radiation coefficient function for $F_2$ can be written as
\begin{equation}
H_{2,GR}^{(2)}\left(N,\frac{Q^2}{m^2},\frac{\mu^2}{m^2}\right) = 4T_f\left[C_{2,GR}^{\rm const}(N)+C_{2,GR}^{2Q}(N)\ln^2\frac{Q^2}{m^2}+C_{2,GR}^{Q}(N)\ln\frac{Q^2}{m^2}\right],
\end{equation}
where
\begin{subequations}
\begin{equation}
\begin{array}{rcl}
C_{2,GR}^{\rm const} & = C_F & \displaystyle \big[ - \smallfrac{2}{3} ( A_{33}^{(0)} + A_{33}^{(2)} ) - \smallfrac{2 \zeta(2)}{3} ( 2  A_{27}^{(0)} - A_{1}^{(0)} - A_{1}^{(1)} ) - \smallfrac{4}{3} ( A_{32}^{(0)} + A_{32}^{(2)} )\\
               &      & \displaystyle  + \smallfrac{1}{3} ( 2  A_{30}^{(0)} - A_{5}^{(0)} - A_{5}^{(1)} ) + A_{31}^{(0)} + A_{31}^{(2)}- \smallfrac{29}{18}  ( 2  A_{28}^{(0)} - A_{4}^{(0)}- A_{4}^{(1)} )\\
               &      & \displaystyle   + \smallfrac{67}{18}  ( A_{29}^{(0)} + A_{29}^{(2)} ) + \smallfrac{359}{108} ( 2  A_{27}^{(0)}  - A_{1}^{(0)} - A_{1}^{(1)} ) + \smallfrac{1}{6} A_{4}^{(0)} + \smallfrac{13}{6}  A_{4}^{(1)}\\
               &      & \displaystyle  - \smallfrac{1}{2} A_{2}^{(0)} - \smallfrac{23}{6}  A_{2}^{(1)}  + \smallfrac{29}{36}  A_{1}^{(0)} - \smallfrac{295}{36} A_{1}^{(1)} + \smallfrac{134 \zeta(2)}{18} + \smallfrac{265}{36} \big],
\end{array}
\end{equation}
\begin{equation}
C_{2,GR}^{2Q}= C_F \big[ \smallfrac{1}{3} ( 2  A_{27}^{(0)} - A_{1}^{(0)} - A_{1}^{(1)} ) + \smallfrac{1}{2} \big],
\end{equation}
\begin{equation}
\begin{array}{rcl}
C_{2,GR}^{Q} & = C_F & \displaystyle \big[ \smallfrac{2}{3}  ( 2  A_{28}^{(0)} - A_{4}^{(0)} - A_{4}^{(1)} )- \smallfrac{29}{18}  ( 2  A_{27}^{(0)} - A_{1}^{(0)} - A_{1}^{(1)} )  \\
           &       & \displaystyle - \smallfrac{4}{3}  ( A_{29}^{(0)} + A_{29}^{(2)} ) + \smallfrac{1}{6} A_{1}^{(0)} + \smallfrac{13}{6}  A_{1}^{(1)} - \smallfrac{8\zeta(2)}{3} - \smallfrac{19}{6} \big].
\end{array}
\end{equation}
\end{subequations}

\begin{table}
\small
\begin{center}
\begin{tabular}{|c|c|c|}
\hline
$n$ & \vphantom{\Bigg|}$f_n(z)$                    &     $A_n(N)=\mathbf{M}[f_n(z)](N)$  \\
\hline\hline
1  & \vphantom{\Bigg|}$1$                           &     $\displaystyle\frac{1}{N}$     \\
\hline
2  & \vphantom{\Bigg|}$\ln(z)$                      &     $\displaystyle-\frac{1}{N^{2}}$     \\
\hline
3  & \vphantom{\Bigg|}$\ln^2(z)$                    &     $\displaystyle\frac{2}{N^{3}}$     \\
\hline
4  & \vphantom{\Bigg|}$\ln(1-z)$                    &     $\displaystyle-\frac{S_1(N)}{N}$     \\
\hline
5  & \vphantom{\Bigg|}$\ln^2(1-z)$                  &     $\displaystyle\frac{S_1^2(N)+S_2(N)}{N}$     \\
\hline
6  & \vphantom{\Bigg|}$\ln(z)\ln(1-z)$              &     $\displaystyle\frac{S_1(N)}{N^2}+\frac{S_2(N)-\zeta(2)}{N}$     \\
\hline
8 & \vphantom{\Bigg|}$\mbox{Li}_2(1-z)$            &     $\displaystyle-\frac{S_2(N)-\zeta(2)}{N}$     \\
\hline
10 & \vphantom{\Bigg|}$\mbox{Li}_2(-z)+\ln(z)\ln(1+z)$         &     $\displaystyle  -\frac{\zeta(2)}{2N}-\frac1{4N}\left[S_{2}\left(\frac{N-1}2\right)-S_{2}\left(\frac{N}2\right)\right] $     \\
\hline
11 & \vphantom{\Bigg|}$\ln^3(z)$                    &     $\displaystyle-\frac{6}{N^{4}}$     \\
\hline
12 & \vphantom{\Bigg|}$\ln^2(z)\ln(1-z)$            &     $\displaystyle\frac2{N}\left[\zeta(3)+\frac{\zeta(2)}{N}-\frac{S_1(N)}{N^2}-\frac{S_2(N)}{N}-S_3(N)\right]$     \\
\hline
13 & \vphantom{\Bigg|}$\ln(1-z)\mbox{Li}_2(1-z)
                  -\mbox{Li}_3(1-z)$               &    $\displaystyle\frac{S_1(N)S_2(N)-\zeta(2)S_1(N)+S_3(N)-\zeta(3)}{N}$ \\
\hline
14 & \vphantom{\Bigg|}$\ln(z)\ln^2(1-z)$            &     
$
\begin{array}{l}
\displaystyle \frac2{N}\bigg\{\zeta(3)+\zeta(2)S_1(N)-\frac{1}{2N}\left[S_1^2(N)+S_2(N)\right]\\
\displaystyle -S_1(N)S_2(N)-S_3(N)\bigg\}
\end{array}
$     \\
\hline
15 & \vphantom{\Bigg|}$S_{1,2}(1-z)$                 &     $\displaystyle-\frac1{N}\left[S_3(N)-\zeta(3)\right]$     \\
\hline
17 & \vphantom{\Bigg|}$\ln(1+z)$                    &     $\displaystyle\frac{\ln(2)}{N}+\frac{1}{2N}\left[S_1\left(\frac{N-1}2\right)-S_1\left(\frac{N}2\right)\right]$     \\
\hline
18 & \vphantom{\Bigg|}$\ln^3(1-z)$                  &     $\displaystyle-\frac{S_1^3(N)+3S_1(N)S_2(N)+2S_3(N)}{N}$     \\
\hline
19 & \vphantom{\Bigg|}$\mbox{Li}_3(-z)$            &      $\displaystyle-\frac{3\zeta(3)}{4N}+\frac{\zeta(2)}{2N^2}-\frac{\ln(2)}{N^3}-\frac{1}{2N^3}\left[S_1\left(\frac{N-1}2\right)-S_1\left(\frac{N}2\right)\right]$    
\\
\hline
\end{tabular}
\end{center}
\caption{\small Elementary Mellin transforms.\label{mellintranscc}}
\end{table}

\begin{table}
\small
\begin{center}
\begin{tabular}{|c|c|c|}
\hline
$n$ & \vphantom{\Bigg|}$f_n(z)$                     &     $A_n(N)=\mathbf{M}[f(z)](N)$  \\
\hline
\hline
20 & \vphantom{\Bigg|}$\ln(z)\mbox{Li}_2(-z)$       &     
$
\begin{array}{l}
\displaystyle -\frac{1}{2N^2}\left[\frac2{N}S_1\left(\frac{N-1}2\right)+\frac1{2}S_2\left(\frac{N-1}2\right)+\frac{4\ln(2)}{N}\right]\\
\displaystyle +\frac{1}{2N^2}\left[\frac2{N}S_1\left(\frac{N}2\right)+\frac1{2}S_2\left(\frac{N}2\right)+\zeta(2)\right]
\end{array}
$     \\
\hline
21 & \vphantom{\Bigg|}$\ln^2(z)\ln(1+z)$            &     
$
\begin{array}{l}
\displaystyle \frac1{2N}\bigg[\frac2{N^2}S_1\left(\frac{N-1}2\right)+\frac1{N}S_2\left(\frac{N-1}2\right)+\frac12S_3\left(\frac{N-1}2\right)+\frac{4\ln(2)}{N^2}\bigg]\\
\displaystyle -\frac1{2N}\bigg[\frac2{N^2}S_1\left(\frac{N}2\right)+\frac1{N}S_2\left(\frac{N}2\right)+\frac12S_3\left(\frac{N}2\right)\bigg]
\end{array}
$     \\
\hline
23 & \vphantom{\Bigg|}$\ln(z)\mbox{Li}_2(1-z)$      &     $\displaystyle\frac1{N^2}[S_2(N)-\zeta(2)]+\frac2{N}[S_3(N)-\zeta(3)]$     \\
\hline
27 & \vphantom{\Bigg|}$\displaystyle \left(\frac{1}{1-z}\right)_+$         &     $-S_1(N-1)$     \\
\hline
28 & \vphantom{\Bigg|}$\displaystyle \left(\frac{\ln(1-z)}{1-z}\right)_+$  &     $\displaystyle \frac12S_1^2(N-1)+\frac12S_2(N-1)$     \\
\hline
29 & \vphantom{\Bigg|}$\displaystyle \frac{\ln(z)}{1-z}$                   &     $\displaystyle S_{2}(N-1)-\zeta(2)$   \\
\hline
30 & \vphantom{\Bigg|}$\displaystyle \left(\frac{\ln^2(1-z)}{1-z}\right)_+$&     $\displaystyle -\frac13S_1^2(N-1)-S_1(N-1)S_2(N-1)-\frac23S_3(N-1)$     \\
\hline
31 & \vphantom{\Bigg|}$\displaystyle \frac{\ln^2(z)}{1-z}$                 &     $\displaystyle -2[S_{3}(N-1)-\zeta(3)]$   \\
\hline
32 & \vphantom{\Bigg|}$\displaystyle \frac{\ln(z)\ln(1-z)}{1-z}$           &     $\displaystyle \zeta(3)+\zeta(2)S_1(N-1)-S_1(N-1)S_2(N-1)-S_3(N+l-1)$   \\
\hline
\end{tabular}
\end{center}
\caption{Continuation of Table~\ref{mellintranscc}.\label{mellintranscc1}}
\end{table}

In addition to the elementary Mellin transforms listed in
Tables~\ref{mellintranscc}-\ref{mellintranscc1}, the coefficient
functions contain terms whose Mellin transform is known in closed
form, but is expressed in terms of generalized harmonic sums
$S_{i_1,\dots,i_n}(N)$. We have evaluated the Mellin transform of
these functions through suitable numerical approximations.  The
$x$--space expression whose Mellin transform will be evaluated in this
way are the following:
\begin{subequations}
\begin{equation}\label{expansion1}
f_{16}(z)=2\mbox{Li}_2(-z)\ln(1+z)+\ln(z)\ln^2(1+z)+2S_{1,2}(-z),
\end{equation}
\begin{equation}\label{expansion2}
f_{22}(z)=\mbox{Li}_3(1-z),
\end{equation}
\begin{equation}\label{expansion3}
f_{24}(z)=\mbox{Li}_3\left(\frac{1-z}{1+z}\right)-\mbox{Li}_3\left(-\frac{1-z}{1+z}\right),
\end{equation}
\begin{equation}\label{expansion4}
f_{25}(z)=\ln(z)\ln(1-z)\ln(1+z),
\end{equation}
\begin{equation}\label{expansion5}
f_{26}(z)=\ln(1-z)\mbox{Li}_2(-z),
\end{equation}
\begin{equation}\label{expansion6}
f_{33}(z)=\frac{\mbox{Li}_2(1-z)}{1-z}.
\end{equation}
\end{subequations}
We will now consider each of these functions in turn.

In order to determine the Mellin transform of Eq.~(\ref{expansion1}),
we  use Eq.~(32) of Ref.~\cite{Blumlein:1998if}, which  can be written as
\begin{equation}
\begin{array}{rl}
\displaystyle A_{16}(N)=\mathbf{M}[f_{16}(z)](N) & = \displaystyle \frac1{N}\mathbf{M}\left[\frac{z\widetilde{\Phi}(z)}{1+z}\right]-\frac1{2N}\mathbf{M}\left[\frac{z\ln^2(z)}{1+z}\right]\\
\\
&\displaystyle +\frac{\zeta(2)}{N}\mathbf{M}\left[\frac{z}{1+z}\right]-\frac{\zeta(2)\ln(2)}{N}+\frac{\zeta(3)}{4N}.
\end{array}
\end{equation}
The last two Mellin transforms are respectively\footnote{In general 
\begin{equation}\label{mastertrans}
\mathbf{M}\left[\frac{z\ln^n(z)}{1+z}\right] = \frac{(-1)^{n+1} n!}{2^{n+1}}\left[S_{n+1}\left(\frac{N-1}{2}\right)-S_{n+1}\left(\frac{N}{2}\right)\right].
\end{equation}
This Mellin transform is superficially different from the Mellin
transform number 15 in the appendix of Ref.~\cite{Blumlein:1998if},
but they turn out to be equivalent after suitable simplification. This
apparent difference is also responsible for the mismatch
between entries 13, 28, 43, and 62 in the Table in the appendix of
Ref.~\cite{Blumlein:1998if} and entries 17, 21, 20 and 19 in
Tables~\ref{mellintranscc} and \ref{mellintranscc1}, respectively. For
the same reason  entries number 4 and 57 of \cite{Blumlein:1998if}
look different from our Eqs.~(\ref{diff1})-(\ref{diff3}).  }:
\begin{equation}\label{diff1}
\displaystyle \mathbf{M}\left[\frac{z\ln^2(z)}{1+z}\right] =  -\frac{1}{4}\left[S_3\left(\frac{N-1}{2}\right)-S_3\left(\frac{N}{2}\right)\right],
\end{equation}
and 
\begin{equation}\label{diff2}
\mathbf{M}\left[\frac{z}{1+z}\right] =  -\frac{1}{2}\left[S_1\left(\frac{N-1}{2}\right)-S_1\left(\frac{N}{2}\right)\right].
\end{equation}
The Mellin transform of the term involving the special function
$\widetilde{\Phi}(z)$ is given in Ref.~\cite{Blumlein:1998if} in terms
of the generalized harmonic sum $S_{1,2}(N)$. To avoid evaluating this directly for complex $N$,
we instead rewrite
\begin{equation}\label{approx1}
\frac{z\widetilde{\Phi}(z)}{1+z}\cong\sum_{k=1}^{10}a_kz^k\widetilde{\Phi}(z),
\end{equation}
where the values of the coefficients $a_k$ are determined by  
fitting the polynomial $\sum_{k=1}^{10}a_kz^k$ to 
the function $z/(1+z)$ on the unit interval. One can then
use the Mellin transform of the function $z^k\widetilde{\Phi}(z)$, which reads
\begin{equation}\label{diff3}
\mathbf{M}[z^k\widetilde{\Phi}(z)]=\frac1{(N+k)^3}+\frac1{2(N+k)}\left[S_{2}\left(\frac{N+k-1}2\right)-S_{2}\left(\frac{N+k}2\right)\right].
\end{equation}
The desired Mellin transform $A_{16}(N)$ is immediately
found combining  Eqs. (\ref{diff1}), (\ref{diff2}), (\ref{approx1}) and~(\ref{diff3}).

In order to determine the Mellin transform of  Eq.~(\ref{expansion2}) 
we use the expansion
\begin{equation}
\mbox{Li}_3(z) = \sum_{k=1}^{\infty}\frac{z^k}{k^3}.
\end{equation}
It follows that
\begin{equation}
A_{22}(N)=\mathbf{M}[f_{22}(z)](N) =
\sum_{k=1}^{\infty}\frac1{k^3}\int_0^1z^{N-1}(1-z)^kdz, 
\end{equation}
but
\begin{equation}
\int_0^1z^{N-1}(1-z)^kdz =
\frac{\Gamma(N)\Gamma(k+1)}{\Gamma(N+k+1)} = \frac{k!}{(N+k)\dots
  (N+1)N}, 
\end{equation}
so that
\begin{equation}
A_{22}(N)=\sum_{k=1}^{\infty}\frac{k!}{k^3(N+k)\dots(N+1)N}.
\end{equation}
In our implementation we have truncated this series at $k=30$.

Next, 
we turn to the Mellin transform of Eq.~(\ref{expansion3}). 
In this case we fit  the function
\begin{equation}
g(z)=(1-z)^{b-1}\sum_{k=0}^{10}c_kz^k
\end{equation}
to the function $f_{24}(z)$.
However,  one can show that
\begin{equation}
f_{24}(0) = \mbox{Li}_3(1) - \mbox{Li}_3(-1) = \smallfrac{7}{4}\zeta(3) = c_0,
\end{equation}
so that $c_0$ is fixed, and we only have to fit
\begin{equation}
g(z)=(1-z)^{b-1}\big[\smallfrac{7}{4}\zeta(3)+\sum_{k=1}^{10}c_kz^k\big].
\end{equation}
The Mellin transform of Eq.~(\ref{expansion3}) follows immediately, because
\begin{equation}
\begin{array}{rl}
\displaystyle A_{24}(N)=\mathbf{M}[f_{24}(z)](N) & = \sum_{k=0}^{10}c_k\int_0^1z^{N+k-1}(1-z)^{b-1}dz\\
& \displaystyle =\frac{\Gamma(N)\Gamma(b)}{\Gamma(N+b)}\sum_{k=0}^{10}c_k\frac{(N+k)\dots(N+1)N}{(N+b+l+k)\dots(N+b+l)}.
\end{array}
\end{equation}

In order to determine  
the Mellin transform of Eq.~(\ref{expansion4}), we use the
representation given as  
Eq.~(21) of Ref.~\cite{Blumlein:1997vf}, in which 
the function $\ln(1+z)$ is approximated by the polynomial
\begin{equation}
\ln(1+z)\cong\sum_{k=1}^{8}d_kz^k.
\end{equation}
Using also Table~\ref{mellintranscc}, we then get
\begin{equation}
\begin{array}{c}
\displaystyle A_{25}(N)=\mathbf{M}[f_{25}(z)](N) = \sum_{k=1}^8 d_k\mathbf{M}[z^k\ln(z)\ln(1-z)] = \\
\\
\displaystyle \sum_{k=1}^8d_k\left[\frac{S_1(N+k)}{(N+k)^2}+\frac{S_2(N+k)-\zeta(2)}{N+k}\right].
\end{array}
\end{equation}

In order to determine the Mellin transform of 
Eq.~(\ref{expansion5}) we use the expansion
\begin{equation}
\mbox{Li}_2(-z) = \sum_{k=1}^{\infty}\frac{(-1)^kz^k}{k^2},
\end{equation}
so 
\begin{equation}
A_{26}(N)=\mathbf{M}[f_{26}(z)](N) = \sum_{k=1}^{\infty}\frac{(-1)^k}{k^2}\mathbf{M}[z^k\ln(1-z)]=\sum_{k=1}^{\infty}\frac{(-1)^{k+1}}{k^2}\frac{S_1(N+k)}{N+k}.
\end{equation}
In this case we have chosen to truncate the series at $k=100$.

Finally, we determine the Mellin transform of Eq.~(\ref{expansion6}). 
In this case we use the geometric series for $1/(1-z)$ so that
\begin{equation}
\begin{array}{c}
\displaystyle A_{33}(N)\mathbf{M}\left[f_{33}(z)\right](N) =
\sum_{k=0}^{\infty}\mathbf{M}[z^k\mbox{Li}_2(1-z)]= -
\sum_{k=0}^{\infty}\frac{S_2(N+k)-\zeta(2)}{N+k}.
\end{array}
\end{equation}
As in  the case above, the series is truncated at $k=100$.

Note that the Mellin transform of the $x$-space coefficient functions involve
terms of the form $z^lf_n(z)$. 
The Mellin transform $\mathbf{M}[z^lf_n(z)](N)$ can be obtained from
$\mathbf{M}[f_n(z)](N)$ using the identity $\mathbf{M}[z^l
  f_n(z)](N) = \mathbf{M}[f_n(z)](N+l)$. 
Thus the Mellin transform of any of the  terms in
Tables~\ref{mellintranscc}-\ref{mellintranscc1} and also any of
the terms Eqs.(\ref{expansion1})-(\ref{expansion6}) multiplied by a factor
$z^l$ can be obtained replacing $N$ with $N+l$. 

In conclusion, we have checked our calculation by comparing the inverse Mellin
transform of the $N$-space coefficients with the original
$x$-space results. We show in Table~\ref{tab:acc-ffns0} the accuracy of this
comparison for the various coefficient functions.
We find excellent accuracy for all coefficients and all values  of $x$.

\begin{table}[t]
\begin{center}
\vskip-0.1cm
\begin{tabular}{|c||c|c|c|c|}
\hline
 $x$ &      $\epsilon_{\rm rel}\lp C_{2,g}^{(n_l,0),2}\rp $    &
    $\epsilon_{\rm rel}\lp C_{2,q}^{(n_l,0),2}\rp $
 &      $\epsilon_{\rm rel}\lp C_{L,g}^{(n_l,0),2}\rp $
 &      $\epsilon_{\rm rel}\lp C_{L,q}^{(n_l,0),2}\rp $ \\
\hline
\hline
$10^{-7}$ & $7\times10^{-12}$& $4\times10^{-11}$ & $2\times10^{-10}$ & $7\times10^{-12}$\\
$10^{-6}$ & $3\times10^{-11}$& $1\times10^{-11}$ & $4\times10^{-11}$ & $3\times10^{-12}$\\
$10^{-5}$ & $2\times10^{-12}$& $3\times10^{-11}$ & $4\times10^{-13}$ & $2\times10^{-12}$\\
$10^{-4}$ & $6\times10^{-10}$& $1\times10^{-11}$ & $3\times10^{-11}$ & $4\times10^{-12}$\\
$10^{-3}$ & $6\times10^{-9}$& $2\times10^{-12}$ & $1\times10^{-10}$ & $2\times10^{-11}$\\
$10^{-2}$ & $7\times10^{-8}$& $7\times10^{-12}$ & $2\times10^{-10}$ & $1\times10^{-11}$\\
$10^{-1}$ & $1\times10^{-7}$& $4\times10^{-11}$ & $1\times10^{-11}$& $5\times10^{-13}$\\
$3\times10^{-1}$ & $9\times10^{-7}$& $1\times10^{-11}$ &$1\times10^{-11}$ & $5\times10^{-13}$\\
$5\times10^{-1}$ & $4\times10^{-6}$& $3\times10^{-11}$ & $8\times10^{-12}$ & $1\times10^{-13}$\\
$7\times10^{-1}$ & $1\times10^{-5}$& $9\times10^{-11}$ &$6\times10^{-12}$ & $5\times10^{-13}$\\
$9\times10^{-1}$ & $1\times10^{-5}$& $6\times10^{-8}$ & $7\times10^{-8}$ & $1\times10^{-8}$\\
\hline
\end{tabular}
\end{center}
\caption{\small Comparison of the inverse Mellin
 transforms of coefficient functions computed here
to the original $x$--space expressions of
Ref.~\cite{Buza:1995ie}: the percentage difference between the
original expression and the numerical Mellin inverse is shown in each case.
\label{tab:acc-ffns0}}
\vskip-0.1cm
\end{table}

%% file: sec-massive-nc.tex
\section{FastKernel implementation of FONLL-C}
\label{sec:massive-nc}

In this Appendix we discuss the implementation and benchmarking of
the FONLL-C neutral current structure functions in the 
FastKernel framework. At the same time we have implemented and benchmarked FONLL-B, 
since we expect to perform fits using this scheme in the future.

The implementation of FONLL-B and FONLL-C structure functions
in FastKernel requires the $\mathcal{O}\lp \alpha_s^2\rp$ 
massive heavy quark coefficient functions as well as their asymptotic $Q^2\to\infty$ limit 
in Mellin space. In Ref.~\cite{Ball:2011mu}
 we presented analytic results for the Mellin space
$\mathcal{O}\lp \alpha_s\rp$  heavy quark coefficient functions
for neutral current and charged current scattering (see
also Ref.~\cite{Blumlein:2011zu} for the latter).
For the  $\mathcal{O}\lp \alpha_s^2\rp$  heavy quark
coefficient functions in Mellin space we use the
parametrization of Ref.~\cite{Alekhin:2003ev}. 
The Mellin  transforms in the asymptotic limit as $Q^2\to\infty$ have
been determined as in 
Appendix~\ref{sec:massive-nc-comp}.

The $\mathcal{O}\lp \alpha_s^2\rp$ massless coefficient
functions were first computed in Refs.~\cite{CNNLOa,CNNLOb,CNNLOc,CNNLOd}. 
For the implementation in the FastKernel framework we have
used the fast Mellin space parametrizations of the exact
coefficient functions as given in Refs.~\cite{Moch:2004xu,Vermaseren:2005qc}.

\begin{figure}[t!]
\begin{center}
\includegraphics[width=0.40\textwidth]{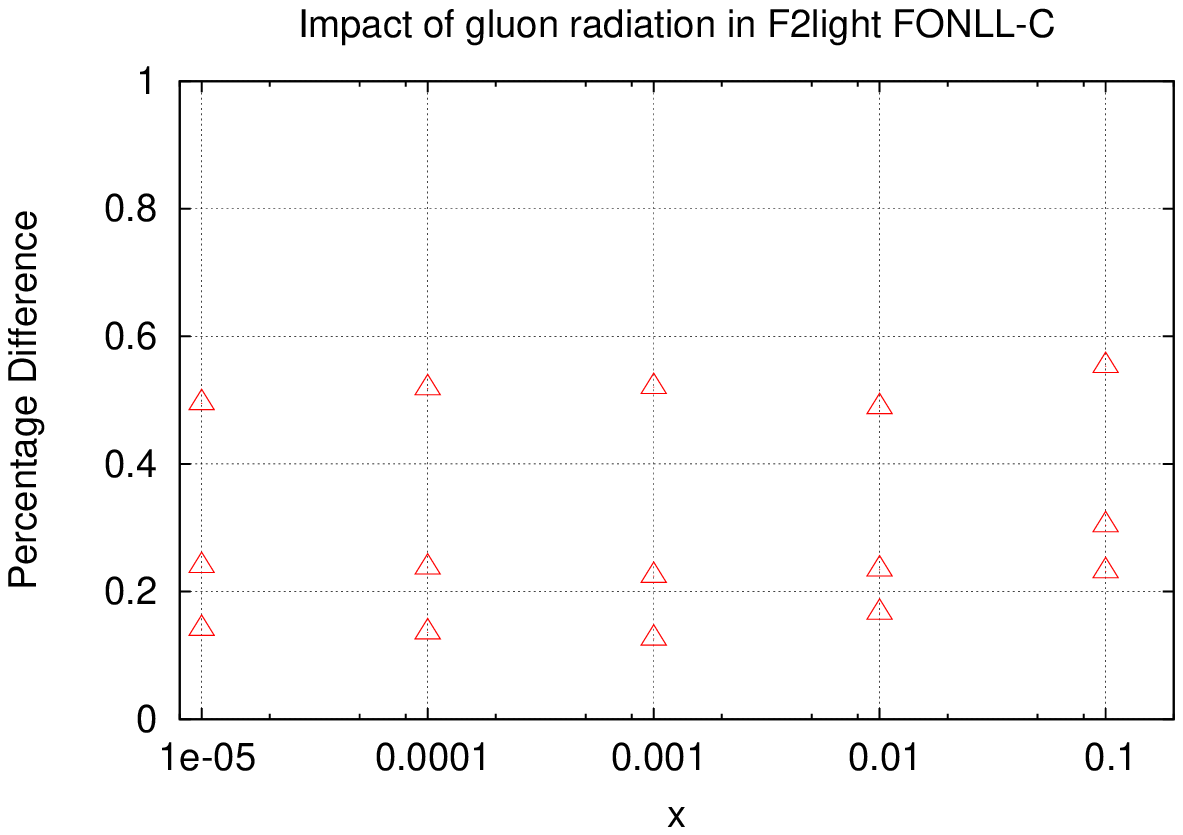}
\includegraphics[width=0.40\textwidth]{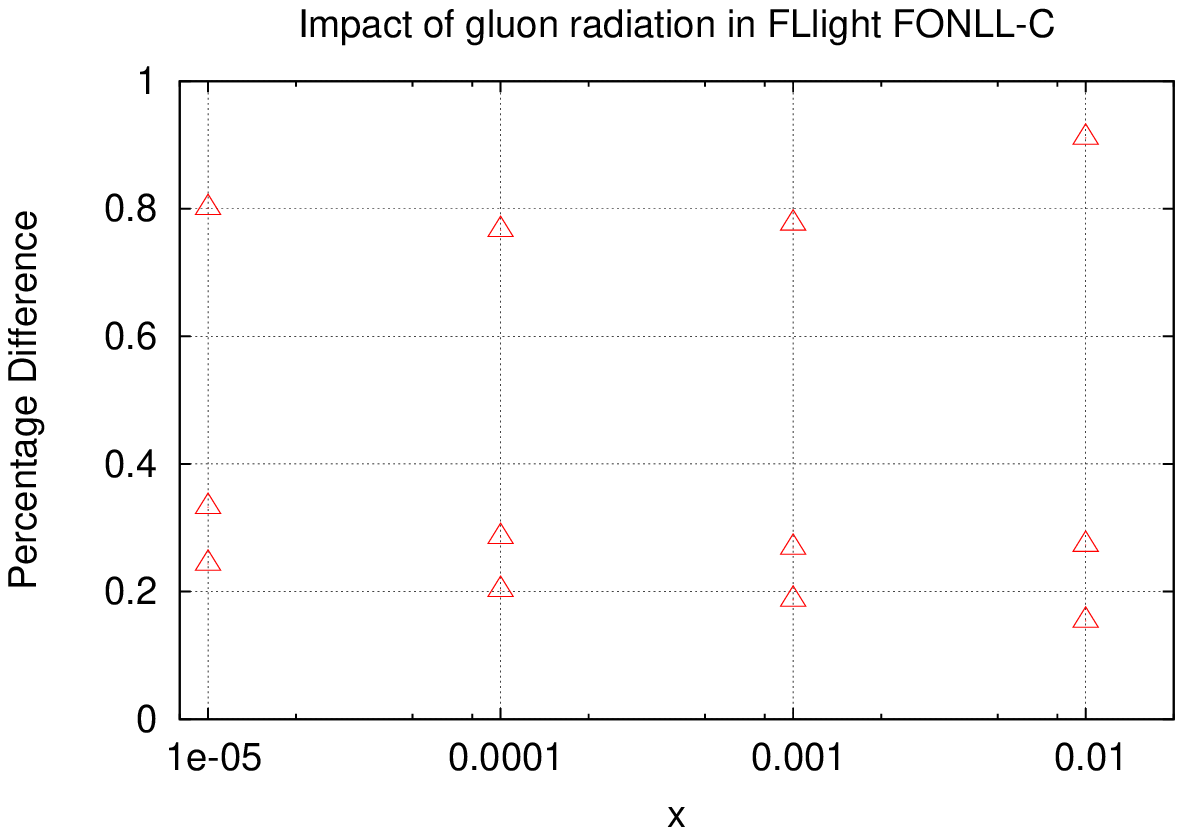}
\end{center}
\caption{\small Percentage difference between
 the NNLO light quark structure functions $F_2$ (left) and $F_L$ (right) with
and without gluon radiation contributions. From bottom to top the points
correspond to scales of $Q^2$=4, 10 and 100 GeV$^2$ respectively.}
\label{fig:glr-impact}
\end{figure}

The gluon radiation terms, namely  contributions with heavy quarks
in the final state but where the struck quark is light, have 
to be treated with care.
As discussed in Ref.~\cite{Forte:2010ta},  the gluon radiation
contribution (which first appears at $\mathcal{O}\lp \alpha_s^2\rp$)  
is part of the light quark structure functions.
We have checked (see Fig.~\ref{fig:glr-impact}) 
that the size of these terms is very
small both for  the $F_2$ and $F_L$ structure functions, typically
below 1\%.

Now we turn to the implementation and benchmarking of these
results in the FastKernel framework. 
Benchmarking has been performed  by comparing the FastKernel results
with the FONLLdis code~\cite{Forte:2010ta}, an 
$x$-space code that implements
all FONLL schemes. 
In Table~\ref{tab:tablebench-ffns}
we show the results of the benchmark comparison for the
$F_{2c}(x,Q^2)$  and $F_{Lc}(x,Q^2)$  structure functions in the
massive fixed-flavour number 
scheme at $\mathcal{O}\lp\alpha_s^2\rp$ for the FONLLdis code
and for the FastKernel code. Results are provided at the
reference points in the $(x,Q^2)$ plane and with the settings 
of the Les Houches heavy quark benchmarks~\cite{LHhq}. 
The accuracy is given as the percentage
difference between the FastKernel and FONLLdis calculations.
The accuracy is never worse than 1\%, which is amply
sufficient for our purposes. 
The accuracy of Table~\ref{tab:tablebench-ffns} is a little worse than
that of its  $\mathcal{O}\lp \alpha_s\rp$ counterpart, shown as
Tables~14-15 of Ref.~\cite{Ball:2011mu}. This may be
due to the  fact that the Mellin space parametrizations of the heavy quark
coefficient functions that we use~\cite{Alekhin:2003ev} are in turn  based
on a parametrization of the exact $x$-space coefficient functions,
while FONLLdis uses the original exact coefficient functions of
Ref~\cite{Laenen:1992zk}. This loss of accuracy is negligible for the
needs of current phenomenology, but more detailed studies of this
issue may be needed in the future when final combined HERA heavy quark
structure function data become available.

The same comparisons for the FONLL-B 
and FONLL-C are presented in Tables~\ref{tab:tablebench-fonllb}
and~\ref{tab:tablebench-fonllc} respectively. Comparable accuracy
is achieved for these two GM-VFN schemes, again sufficient for our purposes.

\begin{table}[ht] 
  \begin{center} 
    \begin{tabular}{|c|c|c|c|c|c|c|} 
      \hline 
      & \multicolumn{3}{|c|}{$F_{2c}$ FFNS} & \multicolumn{3}{|c|}{ $F_{Lc}$ FFNS} \\ 
      \hline 
      $     x$ & FONLLdis & FastKernel  
      & Accuracy (\%) & FONLLdis & FastKernel  
      & Accuracy (\%)\\
      \hline
      \hline
      \multicolumn{7}{|c|}{$Q^2=4$ GeV$^2$} \\
      \hline 
      $10^{-5}$ &    0.24591    &    0.24244&   1.43 &   0.02215    &    0.02184&   1.41 \\ 
      $10^{-4}$ &    0.13658    &    0.13481&   1.31 &   0.01306    &    0.01283&   1.75 \\ 
      $10^{-3}$ &    0.06384    &    0.06308&   1.20 &   0.00662    &    0.00653&   1.41 \\ 
      $10^{-2}$ &    0.02025    &    0.02007&   0.92 &   0.00238    &    0.00237&   0.38 \\ 
      \hline 
      \multicolumn{7}{|c|}{$Q^2=10$ GeV$^2$} \\
      \hline 
      $10^{-5}$ &    0.53701    &    0.53904&   0.38 &   0.08031    &    0.08105&   0.91 \\ 
      $10^{-4}$ &    0.29558    &    0.29550&   0.03 &   0.04611    &    0.04579&   0.71 \\ 
      $10^{-3}$ &    0.13909    &    0.13852&   0.24 &   0.02273    &    0.02254&   0.86 \\ 
      $10^{-2}$ &    0.04689    &    0.04664&   0.10 &   0.00832    &    0.00826&   0.70 \\ 
      \hline 
      \multicolumn{7}{|c|}{$Q^2=100$ GeV$^2$} \\
      \hline 
      $10^{-5}$ &    1.99594    &    2.00744&   0.57 &   0.44200    &    0.43976&   0.51 \\ 
      $10^{-4}$ &    1.00912    &    1.01479&   0.56 &   0.22148    &    0.21880&   1.22 \\ 
      $10^{-3}$ &    0.43527    &    0.43410&   0.27 &   0.09487    &    0.09380&   1.14 \\ 
      $10^{-2}$ &    0.13574    &    0.13492&   0.61 &   0.03019    &    0.03002&   0.57 \\ 
      \hline 
    \end{tabular} 
  \end{center}
  \caption{\small Benchmark comparisons for the
    $F_{2c}(x,Q^2)$  and $F_{Lc}(x,Q^2)$  structure functions in the
    FFN scheme at O($\alpha_s^2$) obtained using the FONLLdis code~\cite{Forte:2010ta}
    and the FastKernel code. Results are provided at the
    benchmark kinematical points in $x,Q^2$ and with the settings
    of the Les Houches heavy quark benchmarks~\cite{LHhq}. 
    The accuracy is given as the percentage 
    difference between the FastKernel and FONLLdis~\cite{Forte:2010ta} calculations.
    \label{tab:tablebench-ffns}}
\end{table} 

\begin{table}[ht] 
  \begin{center} 
  \begin{tabular}{|c|c|c|c|c|c|c|} 
    \hline 
    & \multicolumn{3}{|c|}{$F_{2c}$ FONLL-B} & \multicolumn{3}{|c|}{ $F_{Lc}$ FONLL-B} \\ 
    \hline 
    $     x$ & FONLLdis & FastKernel  
    & Accuracy (\%) & FONLLdis & FastKernel  
    & Accuracy (\%)\\
    \hline
    \hline
    \multicolumn{7}{|c|}{$Q^2=4$ GeV$^2$} \\
    \hline 
    $10^{-5}$ &    0.24787    &    0.24858&   0.29 &   0.02519    &    0.02524&   0.21 \\ 
    $10^{-4}$ &    0.13556    &    0.13598&   0.31 &   0.01435    &    0.01435&   0.01 \\ 
    $10^{-3}$ &    0.06360    &    0.06350&   0.15 &   0.00718    &    0.00715&   0.34 \\ 
    $10^{-2}$ &    0.02062    &    0.02051&   0.52 &   0.00258    &    0.00258&   0.08 \\ 
    \hline 
    \multicolumn{7}{|c|}{$Q^2=10$ GeV$^2$} \\
    \hline 
    $10^{-5}$ &    0.55100    &    0.55088&   0.02 &   0.09637    &    0.09679&   0.43 \\ 
    $10^{-4}$ &    0.30114    &    0.30150&   0.14 &   0.05229    &    0.05222&   0.14 \\ 
    $10^{-3}$ &    0.14371    &    0.14375&   0.02 &   0.02507    &    0.02499&   0.32 \\ 
    $10^{-2}$ &    0.05012    &    0.05015&   0.01 &   0.00908    &    0.00907&   0.06 \\ 
    \hline 
    \multicolumn{7}{|c|}{$Q^2=100$ GeV$^2$} \\
    \hline 
    $10^{-5}$ &    2.10034    &    2.08834&   0.57 &   0.48769    &    0.48716&   0.11 \\ 
    $10^{-4}$ &    1.04510    &    1.05096&   0.56 &   0.23569    &    0.23418&   0.65 \\ 
    $10^{-3}$ &    0.45879    &    0.45916&   0.08 &   0.09923    &    0.09888&   0.35 \\ 
    $10^{-2}$ &    0.15039    &    0.15030&   0.06 &   0.03170    &    0.03174&   0.15 \\ 
    \hline 
  \end{tabular} 
\end{center}
\caption{\small Same as Table~\ref{tab:tablebench-ffns} for the
  FONLL-B GM-VFN scheme. \label{tab:tablebench-fonllb}}
\end{table}

\begin{table}[ht] 
  \begin{center} 
    \begin{tabular}{|c|c|c|c|c|c|c|} 
      \hline 
      & \multicolumn{3}{|c|}{$F_{2c}$ FONLL-C} & \multicolumn{3}{|c|}{ $F_{Lc}$ FONLL-C} \\ 
      \hline 
      $     x$ & FONLLdis & FastKernel  
      & Accuracy (\%) & FONLLdis & FastKernel  
      & Accuracy  (\%)\\
      \hline
      \hline
      \multicolumn{7}{|c|}{$Q^2=4$ GeV$^2$} \\
      \hline 
      $10^{-5}$ &    0.27830    &    0.28163&   1.18 &   0.02468    &    0.02500&   1.30 \\ 
      $10^{-4}$ &    0.14709    &    0.14858&   1.00 &   0.01423    &    0.01441&   1.23 \\ 
      $10^{-3}$ &    0.06556    &    0.06591&   0.52 &   0.00733    &    0.00735&   0.24 \\ 
      $10^{-2}$ &    0.02034    &    0.02034&   0.00 &   0.00281    &    0.00283&   0.74 \\ 
      \hline 
      \multicolumn{7}{|c|}{$Q^2=10$ GeV$^2$} \\
      \hline 
      $10^{-5}$ &    0.69412    &    0.69873&   0.66 &   0.09909    &    0.10062&   1.52 \\ 
      $10^{-4}$ &    0.34662    &    0.34911&   0.88 &   0.05520    &    0.05550&   0.52 \\ 
      $10^{-3}$ &    0.15025    &    0.15114&   0.32 &   0.02682    &    0.02699&   0.63 \\ 
      $10^{-2}$ &    0.04986    &    0.05022&   0.13 &   0.01002    &    0.01008&   0.58 \\ 
      \hline 
      \multicolumn{7}{|c|}{$Q^2=100$ GeV$^2$} \\
      \hline 
      $10^{-5}$ &    2.36920    &    2.37887&   0.41 &   0.47822    &    0.47994&   0.36 \\ 
      $10^{-4}$ &    1.12695    &    1.13870&   1.03 &   0.23916    &    0.23914&   0.01 \\ 
      $10^{-3}$ &    0.47058    &    0.47317&   0.55 &   0.10262    &    0.10293&   0.30 \\ 
      $10^{-2}$ &    0.15175    &    0.15236&   0.40 &   0.03312    &    0.03327&   0.47 \\ 
      \hline 
    \end{tabular} 
  \end{center}
  \caption{\small  Same as Table~\ref{tab:tablebench-ffns} for the
    FONLL-C GM-VFN scheme. \label{tab:tablebench-fonllc}}
\end{table} 


%% file: sec-pdfevol-nnlo.tex
\section{Benchmarking the NNLO PDF evolution}

\label{sec:pdfevol-nnlo}

In this appendix we benchmark the accuracy of the NNLO
PDF evolution as implemented in the FastKernel
framework.
The NNLO anomalous dimensions for DGLAP evolution were
computed in Refs.~\cite{gNNLOa,gNNLOb}. 
In our code we use the accurate 
Mellin space  parametrizations also provided 
in Refs.~\cite{gNNLOa,gNNLOb}.

\begin{table}[t]
\begin{center}
\vskip-0.1cm
\begin{tabular}{|c||c|c|c|c|c|c|}
\hline
 $x$ &      $\epsilon_{\rm rel}\lp u_v\rp $    &
   $\epsilon_{\rm rel}\lp  xd_v  \rp $ &  $\epsilon_{\rm rel}\lp    L_- 
\rp $ &   $\epsilon_{\rm rel}\lp     L_+ \rp $ & 
 $\epsilon_{\rm rel}\lp      s_+  \rp $ &    $\epsilon_{\rm rel}\lp   g \rp $  \\
\hline
\hline
$10^{-7}$ & $2.5\,10^{-4}$ & $2.5\,10^{-4}$ & $3.4\,10^{-5}$ & $2.7\,10^{-5}$ & $4.7\,10^{-5}$ &  $5.7\,10^{-5}$ \\
$10^{-6}$ & $1.5\,10^{-4}$ &$1.0\,10^{-4}$ &$1.0\,10^{-4}$ & $6.1\,10^{-5}$ & $5.0\,10^{-5}$  & $5.8\,10^{-5}$   \\
$10^{-5}$ & $1.7\,10^{-4}$ &$1.6\,10^{-4}$ &$1.1\,10^{-4}$ & $5.2\,10^{-5}$ & $3.6\,10^{-5}$  & $7.8\,10^{-5}$    \\
$10^{-4}$ & $1.6\,10^{-4}$ &$1.7\,10^{-4}$ &$6.5\,10^{-5}$ & $4.3\,10^{-5}$ & $7.7\,10^{-5}$  & $8.6\,10^{-5}$     \\
$10^{-3}$ & $9.4\,10^{-5}$ &$6.9\,10^{-5}$ &$1.9\,10^{-5}$ & $7.6\,10^{-5}$  & $7.5\,10^{-5}$  &  $1.0\,10^{-4}$   \\
$10^{-2}$ & $2.9\,10^{-4}$ &$3.2\,10^{-4}$ &$4.5\,10^{-4}$ & $1.0\,10^{-4}$ & $1.3\,10^{-4}$  &  $1.2\,10^{-4}$     \\
$10^{-1}$ & $2.6\,10^{-4}$ &$3.8\,10^{-4}$ &$5.9\,10^{-4}$ & $5.2\,10^{-4}$ & $3.4\,10^{-4}$  & $7.5\,10^{-5}$    \\
$3\,10^{-1}$ & $1.3\,10^{-5}$ &$1.3\,10^{-5}$ &$4.7\,10^{-5}$ & $1.3\,10^{-5}$ & $9.9\,10^{-5}$  & $4.8\,10^{-5}$     \\
$5\,10^{-1}$ & $3.8\,10^{-5}$ &$5.8\,10^{-5}$ &$3.1\,10^{-5}$ & $4.2\,10^{-4}$ & $8.5\,10^{-4}$ & $2.3\,10^{-4}$     \\
$7\,10^{-1}$ & $1.8\,10^{-4}$ &$7.9\,10^{-5}$ &$2.4\,10^{-3}$ & $4.2\,10^{-3}$ & $9.4\,10^{-3}$ & $1.1\,10^{-3}$ \\
$9\,10^{-1}$ & $3.9\,10^{-3}$ &$1.3\,10^{-2}$ &$2.2\,10^{-1}$ & $2.3\,10^{-1}$ &$5.6\,10^{-1}$ 
&   $1.9\,10^{-1}$ \\
\hline
\end{tabular}
\end{center}
\caption{\small Relative accuracy of NNLO PDF evolution
in the ZM-VFNS
as implemented in the FastKernel framework in comparison to
the Les Houches benchmark tables~\cite{Dittmar:2005ed}.
\label{tab:lhacc}}
\vskip-0.1cm
\end{table}

The accuracy of our NNLO PDF evolution code,
based on the FastKernel framework, has been cross-checked
against the Les Houches PDF evolution
 benchmark tables~\cite{Dittmar:2005ed}. These were obtained
comparing the {\tt HOPPET}~\cite{Salam:2008qg}  $x$-space
and {\tt PEGASUS}~\cite{pegasus} $N$-space
evolution codes.
In order to perform a meaningful comparison, we use
the so-called iterated solution of the $N-$space evolution
equations, and use the same
initial PDFs and same
running coupling, following the procedure
described in detail in Ref.~\cite{Dittmar:2005ed}.

In Table~\ref{tab:lhacc} we show the relative difference
$\epsilon_{\rm rel}$
between our results 
and the benchmark tables of Refs.~\cite{Dittmar:2005ed} ,
for various combinations of PDFs 
at NLO in the zero-mass variable flavour number scheme. 
We see that the accuracy
is perfectly satisfactory for precision phenomenology.
Comparable accuracy is obtained in a fixed-flavour number
scheme. Finally, by benchmarking against  {\tt PEGASUS}, we
have also checked that comparable accuracy is obtained if the truncated
solution of the evolution equations is used, instead of the iterated one.